\documentclass[12pt]{article}
\usepackage{graphicx}
\usepackage{amssymb}

\input epsf
\newcommand{\setM}{{\mbox{\boldmath$M$}}}

\newcommand{\AAA}{\mbox{\boldmath$A$}}
\newcommand{\aaa}{\mbox{\boldmath$a$}}
\newcommand{\BB}{\mbox{\boldmath$B$}}
\newcommand{\bb}{\mbox{\boldmath$b$}}
\newcommand{\kk}{\mbox{\boldmath$k$}}
\newcommand{\CC}{\mbox{\boldmath$C$}}
\newcommand{\cc}{\mbox{\boldmath$c$}}
\newcommand{\DD}{\mbox{\boldmath$D$}}
\newcommand{\ee}{\mbox{\boldmath$e$}}
\newcommand{\FF}{\mbox{\boldmath$F$}}
\newcommand{\ff}{\mbox{\boldmath$f$}}
\newcommand{\HH}{\mbox{\boldmath$H$}}
\newcommand{\bg}{\mbox{\boldmath$g$}}
\newcommand{\jj}{\mbox{\boldmath$j$}}
\newcommand{\JJ}{\mbox{\boldmath$J$}}
\newcommand{\vl}{\mbox{\boldmath$l$}}
\newcommand{\LL}{\mbox{\boldmath$L$}}
\newcommand{\mm}{\mbox{\boldmath$m$}}
\newcommand{\MM}{\mbox{\boldmath$M$}}
\newcommand{\NN}{\mbox{\boldmath$N$}}
\newcommand{\nn}{\mbox{\boldmath$n$}}
\newcommand{\uu}{\mbox{\boldmath$u$}}

\newcommand{\qq}{\mbox{\boldmath$q$}}
\newcommand{\sss}{\mbox{\boldmath$\sigma$}}
\newcommand{\rr}{\mbox{\boldmath$r$}}

\newcommand{\QQ}{\mbox{\boldmath$Q$}}
\newcommand{\RR}{\mbox{\boldmath$R$}}

\newcommand{\vv}{\mbox{\boldmath$v$}}

\newcommand{\yy}{\mbox{\boldmath$y$}}
\newcommand{\xx}{\mbox{\boldmath$x$}}
\newcommand{\xxx}{\mbox{\boldmath$x$}}
\newcommand{\XX}{\mbox{\boldmath$X$}}

\newcommand{\UNIT}{\mbox{\boldmath$1$}}

\newcommand{\bOmega}{\mbox{\boldmath$\Omega$}}
\newcommand{\bPi}{\mbox{\boldmath$\Pi$}}

\newcommand{\bgamma}{\mbox{\boldmath$\gamma$}}
\newcommand{\bmu}{\mbox{\boldmath$\mu$}}
\newcommand{\bnabla}{\mbox{\boldmath$\nabla$}}

\newcommand{\balpha}{\mbox{\boldmath$\alpha$}}
\newcommand{\bbeta}{\mbox{\boldmath$\beta$}}

\newcommand{\nablah}{\overline{\nabla}}
\newcommand{\la}{\langle}
\newcommand{\ra}{\rangle}
\newcommand{\exm}{\mbox{\scriptsize ex}}
\newcommand{\dm}{\mbox{\scriptsize d}}
\newcommand{\hm}{\mbox{\scriptsize h}}
\newcommand{\rrm}{\mbox{\scriptsize r}}
\newcommand{\eqm}{\mbox{\scriptsize eq}}

\newcommand{\sm}{\mbox{\scriptsize s}}

\newcommand{\Hm}{\mbox{\scriptsize H}}
\newcommand{\Imm}{\mbox{\scriptsize\rm I}}

\newcommand{\Bm}{\mbox{\scriptsize B}}

\newcommand{\emm}{\mbox{\scriptsize e}}
\newcommand{\ppm}{\mbox{\scriptsize p}}
\newcommand{\ttau}{{\boldmath \mbox{$\tau$}}}
\newcommand{\QQh}{\widehat{\mbox{\bf Q}}}

\newcommand{\dd}{\mbox{\rm d}}
\newcommand{\FFh}{\widehat{\mbox{\bf F}}}
\newcommand{\Ab}{\mbox{\bf A}}
\newcommand{\kkh}{\widehat{\mbox{\bf k}}}
\newcommand{\kh}{\widehat{ k}}
\newcommand{\KK}{\mbox{\bf K}}

\newcommand{\II}{\mbox{\bf  1}}
\newcommand{\1}{\mbox{\bf 1}}
\newcommand{\0}{\mbox{\bf 0}}

\newcommand{\oMM}{\stackrel{\circ}{\mbox{\bf  M}}}

\newcommand{\otau}{\stackrel{\circ}{\boldmath \mbox{$\tau$}}}
\newcommand{\oQQ}{\stackrel{\circ}{\mbox{\bf QQ}}}
\newcommand{\oA}{\stackrel{\circ}{\mbox{\bf A}}}
\newcommand{\oB}{\stackrel{\circ}{\mbox{\bf B}}}
\newcommand{\oC}{\stackrel{\circ}{\mbox{\bf C}}}
\newcommand{\oxx}{\stackrel{\circ}{\mbox{\bf xx}}}

\newcommand{\LLambda}{{\bf \Lambda}}
\newcommand{\GGamma}{{\boldmath \mbox{$\Gamma$}}}

\newcommand{\dotgam}{{\boldmath \mbox{$ \dot{\gamma}$}}}
\newcommand{\dotgamh}{{\boldmath \hat{\mbox{$ \dot{\gamma}$}}}}
\newcommand{\dotga}{\dot{\gamma}}

\newcommand{\tr}{\mbox{tr}}
\newcommand{\dotgamhat}{\overline{\gamma}}
\newcommand{\dotepshat}{\overline{\varepsilon}}
\newcommand{\doteps}{\dot{\varepsilon}}
\newcommand{\nuhat}{\widehat{\nu}}
\newcommand{\tth}{\widehat{t}}
\newcommand{\nuelo}{{\vartheta}}

\newcommand{\ww}{\mbox{\boldmath$w$}}

\begin{document}
\title{\textbf{Constructive Methods \\ of Invariant Manifolds \\ for Kinetic Problems}}
\author{Alexander N. Gorban$^{1,2,3}$\thanks{agorban$@$mat.ethz.ch, $^{**}$ikarlin$@$mat.ethz.ch,
$^{***}$zinovyev$@$ihes.fr}, \and Iliya V. Karlin$^{1,2**}\!\!$, \and  and Andrei Yu.
Zinovyev$^{2,3***}$
\\ $^{1}$ ETH-Zentrum, Department of Materials, Institute of Polymers, \\ Sonneggstr. 3, ML J19,
CH-8092 Z{\"u}rich, Switzerland; \\ $^{2}$ Institute of Computational Modeling SB RAS, \\
Akademgorodok, Krasnoyarsk 660036, Russia; \\ $^{3}$ Institut des Hautes Etudes Scientifiques, \\ Le
Bois-Marie, 35, route de Chartres, F-91440, Bures-sur-Yvette, France}
\date{}

\maketitle

\begin{abstract}

The conception of the slow invariant manifold is recognized as the central idea
underpinning a transition from micro to macro and model reduction in kinetic theories.
We present the Constructive Methods of Invariant Manifolds for model reduction in physical and chemical
kinetics, developed during last two decades. The physical problem of reduced description is studied in
a most general form as a problem of constructing the slow invariant manifold. The invariance conditions
are formulated as the differential equation for a manifold immersed in the phase space ({\it the invariance equation}).
The equation of motion for immersed manifolds is obtained ({\it the film extension of the dynamics}).
Invariant manifolds are fixed points for this equation, and slow invariant manifolds are Lyapunov stable fixed points, thus
{\it slowness is presented as stability}.

A collection of methods to derive analytically
and to compute numerically the slow invariant manifolds is presented.
Among them, iteration methods based on incomplete linearization, relaxation
method and the method of invariant grids are developed.
The systematic use of thermodynamics structures and of the quasi--chemical representation allow
to construct approximations which are in concordance with physical restrictions.

The following examples of applications are presented: Nonperturbative deviation of physically consistent
hydrodynamics from the Boltzmann equation and from the reversible dynamics,
for Knudsen numbers $Kn\sim 1$; construction of the moment equations for
nonequilibrium media and their dynamical correction (instead of extension of list of variables) to
gain more accuracy in description of highly nonequilibrium flows; determination of
molecules dimension (as diameters of equivalent hard spheres) from experimental viscosity data;
model reduction in chemical kinetics; derivation and numerical implementation
of constitutive equations for polymeric fluids; the limits of
macroscopic description for polymer molecules, etc.



\end{abstract}

\clearpage

\tableofcontents

\section{Introduction}\label{intro}

In this review, we present a collection of constructive methods to study slow (stable) positively
invariant manifolds of dynamic systems. The main objects of our study are dissipative dynamic systems
(finite or infinite) which arise in various problems of kinetics. Some of the results and methods
presented herein may have a more general applicability, and can be useful not only for dissipative
systems but also, for example, for conservative systems.

Nonequilibrium statistical physics is a collection of ideas and methods to extract slow invariant
manifolds. Reduction of description for dissipative systems assumes (explicitly or implicitly) the
following picture: There exists a manifold of slow motions in the phase space of the system. From the
initial conditions the system goes quickly in a small neighborhood of the manifold, and after that
moves slowly along this manifold (see, for example, \cite{VanKampen1}). The manifold of slow motion
must be positively invariant: if the motion starts on the manifold at $t_0$, then it stays on the
manifold at $t>t_0$. Frequently used wording ``invariant manifold" is not really exact: For the
dissipative systems, the possibility of extending the solutions (in a meaningful way) backwards in
time is limited. So, in nonequilibrium statistical physics we study {\it positively invariant} slow
manifolds. The necessary invariance condition can be written explicitly as the differential equation
for the manifold immersed into the phase space.\footnote{This picture is directly applicable to {\it
dissipative} systems. Time separation for {\it conservative} systems and the way from the reversible
mechanics (for example, from the Liouville equation) to dissipative systems (for example, to the
Boltzmann equation) requires some additional ideas and steps. For any conservative system, a
restriction of its dynamics onto any invariant manifold is conservative again. We should represent a
dynamics of a large conservative system as a result of dynamics in its small subsystems, and it is
necessary to take into account that a {\it macroscopically} small interval of time can be considered
as an infinitely large interval for a small subsystem, i.e. {\it microscopically}. It allows us to
represent a relaxation of such a large systems as an ensemble of {\it indivisible events} (for
example, collision) which happen to its small subsystems. The Bogolyubov-Born-Green-Kirkwood-Yvon
(BBGKY) hierarchy and Bogolyubov method for derivation of the Boltzmann equation give us the
unexcelled realization of this approach \cite{Bogol}.}

A dissipative system may have many closed positively invariant sets. For example, for every set of
initial conditions $K$, unification of all the trajectories $\{x(t),t \geq 0\}$ with initial
conditions $x(0)\in K$ is positively invariant. Thus, selection of the slow (stable) positively
invariant manifolds becomes important problem\footnote{Nevertheless, there exists a different point
of view: ``Non--uniqueness, when it arises, is irrelevant for modeling" \cite{Roberts}, because the
differences between the possible manifolds are of the same order as the differences we set out to
ignore in establishing the low-dimensional model.}.

One of the difficulties in the problem of reducing the description is pertinent to the fact that
there exists no commonly accepted formal definition of slow (and stable) positively invariant
manifold. This difficulty is resolved in Section 4 of our review in the following way: First, we
consider manifolds immersed into a phase space and study their motion along trajectories. Second, we
subtract from this motion the motion of immersed manifolds along themselves, and obtain a new
equation for dynamics of manifolds in phase space: {\it the film extension of the dynamics}.
Invariant manifolds are fixed points for this extended dynamics, and {\bf slow} invariant manifolds
are {\bf Lyapunov stable fixed points}.

The main body of this review is about how to actually compute the slow invariant manifold. Here we
present three approaches to constructing slow (stable) positively invariant manifolds.

\begin{itemize}
\item {\it Iteration method} (Newton method subject to incomplete
linearization);
\item {\it Relaxation methods} based on a film
extension of the original dynamic system;
\item {\it The method of natural projector}
\end{itemize}

The Newton method (with incomplete linearization) is convenient for obtaining the explicit formulas -
even one iteration can give a good approximation.

Relaxation methods are oriented more at the numerical implementation. Nevertheless, several first
steps also can give appropriate analytical approximations, competitive with other methods.

Finally, the method of natural projector constructs not the manifold itself but a projection of slow
dynamics from the slow manifold onto some set of variables.

The Newton method subject to incomplete linearization was developed for the construction of slow
(stable) positively invariant manifolds for the following problems:

\begin{itemize}
\item Derivation of the post-Navier-Stokes hydrodynamics from
the Boltzmann equation \cite{GKAMSE92,GKTTSP94,KDNPRE97}.
\item Description of the dynamics of polymers solutions \cite{ZKD2000}.
\item Correction of the moment equations \cite{KGDNPRE98}.
\item Reduced description for the chemical kinetics \cite{InChLANL,GKZDPhA2000,Grids}.
\end{itemize}

Relaxation methods based on a film extension of the original dynamic system were applied for the
analysis of the Fokker-Planck equation \cite{KZFPLANL}. Applications of these methods in the theory
of the Boltzmann equation can benefit from the estimations, obtained in the papers
\cite{GKZNPhA96,GKZTTSP99}.

The method of natural projector was initially applied to derivation of the dissipative equations of
macroscopic dynamics from the conservative equations of microscopic dynamics
\cite{GKIOeNONNEWT2001,GKOeTPRE2001,GKMex2001,GKPRE02,GKGeoNeo,KaRiSu,KTGOePhA2003}. Using this
method, new equations were obtained for the post-Navier-Stokes hydrodynamics, equations of plasma
hydrodynamics and others \cite{GKOeTPRE2001,KTGOePhA2003}. This short-memory approximation is applied
to the Wigner formulation of quantum mechanics \cite{KaRiSu}. The dissipative dynamics of a single
quantum particle in a confining external potential is shown to take the form of a damped oscillator
whose effective frequency and damping coefficients depend on the shape of the quantum-mechanical
potential \cite{KaRiSu}. The method of natural projector can also be applied effectively  for the
dissipative systems: instead of Chapman-Enskog method in  theory of the Boltzmann equation, etc.

A natural initial approximation for the methods under consideration is a quasiequilibrium manifold.
It is the manifold of conditional maxima of the entropy. Most of the works on nonequilibrium
thermodynamics deal with corrections to quasi-equilibrium approximations, or with applications of
these approximations (with or without corrections). The construction of the quasi-equilibrium allows
for the following generalization: Almost every manifold can be represented as a set of minimizers of
the entropy under linear constrains. However, in contrast to the standard quasiequilibrium, these
linear constrains will depend on the point on the manifold. We describe the quasiequilibrium manifold
and a quasiequilibrium projector on the tangent space of this manifold. This projector is orthogonal
with respect to entropic scalar product (the bilinear form defined by the negative second
differential of the entropy). We construct the thermodynamical projector, which transforms the
arbitrary vector field equipped with the given Lyapunov function (the entropy) into a vector field
with the same Lyapunov function for an arbitrary anzatz manifold which is not tangent to the level of
the Lyapunov function. The uniqueness of this construction is demonstrated.

Here, a comment on the status of  most of the statements in this text is in order. Just like the
absolute majority of all claims concerning such things as general solutions of the Navier-Stokes and
Boltzmann equations, etc., they have the status of being plausible. They can become theorems only if
one restricts essentially the set of the objects under consideration. Among such restrictions we
should mention cases of exact reduction, i.e. exact derivation of the hydrodynamics from the kinetics
\cite{GKPRL96,KGAnPh2002}. In these (still infinite-dimensional) examples one can compare different
methods, for example, the Newton method with the methods of series summation in the perturbation
theory \cite{KGAnPh2002,GKJETP91}.

Also, it is necessary to stress here, that even if in the limit all the methods lead to the same
results, they can give rather different approximations ``on the way".

The rigorous grounds of the constructive methods of invariant manifolds should, in particular,
include the theorems about {\it persistence of invariant manifolds under perturbations}. The most
known result of this type is the Kolmogorov-Arnold-Moser theory about persistence of almost all
invariant tori of completely integrable system under small perturbation \cite{KAM,KAM1,KAM2}. Such
theorems exist for some classes of infinite dimensional dissipative systems too \cite{TitiPer}.
Unfortunately, it is not proven untill now that many important systems (the Boltzmann equation, 3D
Navier-Stokes equations, Grad equations, etc.) belong to these classes. So, it is necessary to act
with these systems without a rigorous basis.

Two approaches are widely known to the construction of the invariant manifolds:  the {\it Taylor
series expansion} \cite{Beyn1,Kaz1} and the method of {\it renormalization group}
\cite{Kun1,Kun2,Kun3,DEG}. The advantages and disadvantages of the Taylor expansion are well-known:
constructivity against the absence of physical meaning for the high-order terms (often) and
divergence in the most interesting cases (often).

In the paper \cite{Kun1} a geometrical formulation of the renormalization group method for global
analysis was given. It was shown that the renormalization group equation can be interpreted as an
envelope equation. Recently \cite{Kun2} the renormalization group method was formulated in terms of
the notion of invariant manifolds. This method was applied to derive kinetic and transport equations
from the respective microscopic equations \cite{Kun3}. The derived equations include Boltzmann
equation in classical mechanics (see also the paper \cite{RG}, where it was shown for the first time
that kinetic equations such as the Boltzmann equation can be understood naturally as renormalization
group equations), Fokker-Planck equation, a rate equation in a quantum field theoretical model etc.

The renormalization group approach was applied to the stochastic Navier-Stokes equation that
describes fully developed fluid turbulence \cite{FNS1,FNS2,Adzh}. For the evaluation of the relevant
degrees of freedom the renormalization group technique was revised  for discrete systems in the
recent paper Ref. \cite{DEG}.

The kinetic approach to subgrid modeling of fluid turbulence became more popular during last decade.
\cite{ChenTur,DegoTurb,SCI,AKSTurb}. A mean-field approach (filtering out subgrid scales) is applied
to the Boltzmann equation in order to derive a subgrid turbulence model based on kinetic theory. It
is demonstrated \cite{AKSTurb} that the only Smagorinsky type model which survives in the
hydrodynamic limit on the viscosity time scale is the so-called tensor-diffusivity model \cite{Smag}.

The new quantum field theory formulation of the problem of persistence of invariant tori in perturbed
completely integrable systems was obtained, and the new proof of the KAM theorem for analytic
Hamiltonians based on the renormalization group method was given \cite{BriGaKu}.

From the authors of the paper Ref. \cite{RG} point of view, the relation of renormalization group
theory and reductive perturbation theory has simultaneously been recognized: renormalization group
equations are actually the slow-motion equations which are usually obtained by reductive perturbation
methods.

The first systematic and (at least partially) successfull method of constructing invariant manifolds
for dissipative systems was the celebrated {\it Chapman-Enskog method} \cite{Chapman} for the
Boltzmann kinetic equation. The Chapman-Enskog method results in a series development of the
so-called normal solution (the notion introduced by Hilbert  \cite{Hilbert}) where the one-body
distribution function depends on time and space through its locally conserved moments. To the first
approximation, the Chapman-Enskog series leads to hydrodynamic equations with transport coefficients
expressed  in terms of molecular scattering cross-sections. However, next terms of the Chapman-Enskog
bring in the ``ultra-violet catastrophe" (noticed first by Bobylev \cite{Bob}) and negative
viscosity. These drawbacks pertinent  to the Taylor-series expansion disappear as soon as the Newton
method is used to construct the invariant manifold \cite{GKTTSP94}.

The Chapman-Enskog method was generalized many times \cite{Garsia2} and gave rise to a host of
subsequent works and methods, such as the famous method of the {\it quasi-steady state} in chemical
kinetics, pioneered by Bodenstein and Semenov and explored in considerable detail  by many authors
(see, for example, \cite{Bow,SeSlem,Fra,RouFra,Yab,InChLANL}), and the theory of {\it singularly
perturbed} differential equations \cite{Bow,Vas,Stry,Roos,Mishch,Novozh,Mil}.

There exist a group of methods to construct an ansatz for the invariant manifold based on the
spectral decomposition of the Jacobian. The idea to use the spectral decomposition of Jacobian fields
in the problem of separating the motions into fast and slow originates from methods of analysis of
stiff systems \cite{Gear}, and from methods of sensitivity analysis in control theory
\cite{RabiSens,Lam}. One of the currently  most popular methods based on the spectral decomposition
of Jacobian fields is the construction of the so-called {\it intrinsic low-dimensional manifold}
(ILDM) \cite{Maas}.

These methods were thoroughly analyzed in two papers \cite{Kaper1,Kaper2}. It was shown that the
successive applications of the Computational Singular Perturbation algorithm (developed in
\cite{Lam}) generate, order by order, the asymptotic expansion of a slow manifold, and the manifold
identified by the ILDM technique (developed in \cite{Maas}) agrees with the invariant manifold to
some order.

The theory of {\it inertial manifold} is based on the special linear dominance in higher dimensions.
Let an infinite-dimensional system have a form: $\dot{u}+Au=R(u)$, where $A$ is self-adjoint, and has
discrete spectrum $\lambda_i \rightarrow \infty$ with sufficiently big gaps between $\lambda_i$, and
$R(u)$ is continuous. One can build the slow manifold as the graph over a root space of $A$
\cite{Debush}. The textbook \cite{Chueshov} provides an exhaustive introduction to the scope of main
ideas and methods of this theory. Systems with linear dominance have limited utility in kinetics.
Often neither a big spectral gaps between $\lambda_i$ exists, no $\lambda_i \rightarrow \infty$ (for
example, for simplest model BGK equations, or for Grad equations). Nevertheless, the concept of
inertial attracting manifold has more wide field of applications than the theory, based on the linear
dominance assumption.

The Newton method with incomplete linearization as well as the relaxation method allow us to find an
approximate slow invariant manifolds without the preliminary stage of Jacobian field spectral
decomposition. Moreover, a necessary slow invariant subspace of Jacobian in equilibrium point appears
as a by-product of the Newton iterations (with incomplete linearization), or of the relaxation
method.

It is of importance to search for minimal (or subminimal) sets of natural parameters that uniquely
determine the long-time behaviour of a system. This problem was first discussed by Foias and Prodi
\cite{FoPro} and by Ladyzhenskaya \cite{Lady} for the 2D Navier-Stokes equations. They have proved
that the long-time behaviour of solutions is completely determined by the dynamics of sufficiently
large amount of the first Fourier modes. A general approach to the problem on the existence of a
finite number of determining parameters has been discussed \cite{Chue2,Chueshov}.

Past decade witnessed a rapid development of the so-called {\it set oriented} numerical methods
 \cite{ComIMDel1}. The purpose of these methods is to compute attractors,
invariant manifolds (often, computation of stable and unstable manifolds in hyperbolic systems
\cite{ComIMDel,ComIMBro,ComIMGar}). Also, one of the central tasks of these methods is to gain
statistical information, i. e. computations of physically observable invariant measures. The
distinguished feature of the modern set-oriented methods of numerical dynamics is the use of
ensembles of trajectories within a relatively short propagation time instead of  a long time single
trajectory.

In this paper we systematically consider a discrete analogue of the slow (stable) positively
invariant manifolds for dissipative systems, {\it invariant grids}. These invariant grids were
introduced in \cite{InChLANL}. Here we will describe the Newton method subject to incomplete
linearization and the relaxation methods for the invariant grids \cite{Grids}.

It is worth to mention, that the problem of the grid correction is fully decomposed into the tasks of
the grid's nodes correction. The edges between the nodes appear only in the calculation of the
tangent spaces at the nodes. This fact determines high computational efficiency of the invariant
grids method.

Let the (approximate) slow invariant manifold for a dissipative system be found. {\it What for have
we constructed it?} One important part of the answer to this question is: {\it We have constructed it
to create models of open system dynamics in the neighborhood of this manifold}. Different approaches
for this modeling are described. We apply these methods to the problem of reduced description in
polymer dynamics and derive the universal limit in dynamics of dilute polymeric solutions. It is
represented by the {\it revised Oldroyd 8 constants} constitutive equation for the polymeric stress
tensor. Coefficients of this constitutive equation are expressed in terms of the microscopic
parameters. This limit of dynamics of dilute polymeric solutions is universal in the same sense, as
Korteweg-De-Vries equation is universal in the description of the dispersive dissipative nonlinear
waves: any physically consistent equation should contain the obtained equation as a limit.

The phenomenon of {\it invariant manifold explosion} in driven open systems is demonstrated on the
example of dumbbell models of dilute polymeric solutions \cite{IK00}. This explosion gives us a
possible mechanism of drag reduction in dilute polymeric solutions \cite{drugs}.

Suppose that for the kinetic system the approximate invariant manifold has been constructed and the
slow motion equations have been derived. Suppose that we have solved the slow motion system and
obtained $x_{sl}(t)$. We consider the following two questions:

\begin{itemize}
\item{How well this solution approximates the true solution $x(t)$
given the same initial conditions?}
\item{How is it possible to use the solution $x_{sl}(t)$ for it's
refinement without solving the slow motion system (or it's modifications) again?}
\end{itemize}

These two questions are interconnected. The first question states the problem of the {\it accuracy
estimation}. The second one states the problem of {\it postprocessing}. We propose various algorithms
for postprocessing and accuracy estimation, and give an example of application.

Our collection of methods and algorithms can be incorporated into recently developed technologies of
computer-aided multiscale analysis which enable the ``level jumping" between microscopic and
macroscopic (system) levels. It is possible both for traditional technique based on transition from
microscopic equations to macroscopic equations and for the ``equation-free" approach \cite{TheoKev}.
This approach developed in recent series of work \cite{KevFree}, when successful, can bypass the
derivation of the macroscopic evolution equations when these equations conceptually exist but are not
available in closed form. The mathematics-assisted development of a computational superstructure may
enable alternative descriptions of the problem physics (e.g. Lattice Boltzmann (LB), kinetic Monte-
Carlo (KMC) or Molecular Dynamics (MD) microscopic simulators, executed over relatively short time
and space scales) to perform systems level tasks (integration over relatively large time and space
scales, coarse bifurcation analysis, optimization, and control) directly. In effect, the procedure
constitutes a system identification based, closure-on-demand computational toolkit. It is possible to
use macroscopic invariant manifolds in this environment without explicit equations.

The present paper comprises sections of the two kinds. Numbered sections contain basic notions,
methods and algorithmic realizations. Sections entitled ``Examples" contain various case studies
where the methods are applied to specific equations. Exposition in the ``Examples" sections is not as
consequent as in the numbered sections. Most of the examples can be read more or less independently.
Logical connections between sections is presented on Figs. \ref{flowchart},\ref{flowchart2}.

The list of cited literature is by no means complete although we spent effort in order to reflect at
least the main directions of studies related to computations of the invariant manifolds. We think
that this list is more or less exhaustive in the second-order approximation.

{\bf Acknowledgements.} First of all, we are grateful to our coauthors: M.S. S. Ansumali (Z\"urich),
Prof V. I. Bykov (Krasnoyarsk), Prof.  M. Deville (Lausanne), Dr. G. Dukek (Ulm),  Dr. P. Ilg
(Z\"urich-Berlin), Prof. T. F. Nonnenmacher (Ulm), Prof. H. C. \"{O}ttinger (Z\"urich), M.S. P. A.
Gorban (Krasnoyarsk--Omsk--Z\"urich),  M.S. A. Ricksen (Z\"urich), Prof. S. Succi (Roma), Dr. L. L.
Tatarinova (Krasnoyarsk--Z\"urich), Prof. G. S. Yablonskii (Novosibirsk--Saint--Louis), Dr. V. B.
Zmievskii (Krasnoyarsk--Lausanne--Montreal) for years of collaboration, stimulating discussion and
support. We thank Prof M. Grmela (Montreal) for detailed and encouraging discussion of the
geometrical foundations of nonequilibrium thermodynamics. Prof. M. Shubin (Moscow-Boston) explained
us some important chapters of the pseudodifferential operators theory. Finally, it is our pleasure to
thank Prof. Misha Gromov (IHES, Bures-sur-Yvette) for encouragement and the spirit of Geometry.

\clearpage

\section*{Mathematical notation and some terminology}

\begin{itemize}
\item{The {\it operator} $L$ from space $W$ to space $E$: $L:W\rightarrow E$.}

\item{The {\it kernel} of a linear operator $L:W\rightarrow E$ is a subspace $ \ker L \subset W $ that
transforms by $L$ into $0$: $ \ker L = \{ x \in W | Lx=0 \} $.}

\item{The {\it image} of a linear operator $L:W\rightarrow E$ is a subspace $ {\rm im} L = L(W) \subset E
$.}

\item{{\it Projector} is a linear operator $P: E \rightarrow E$ with the property $P^2=P$. Projector $P$ is
{\it orthogonal} one, if $ \ker P \bot {\rm im} P$ (the kernel of $P$ is orthogonal to the image of
$P$).}

\item{If $F: U \rightarrow V$ is a map of domains in normed spaces ($U \subset W, \: V \subset E$) then the
{\it differential} of $F$ at a point $x$ is a linear operator $D_x F: W \rightarrow E$ with the
property: $\|F(x+\delta x)-F(x) - (D_x F)(\delta x)\| =  o(\|\delta x\|)$. This operator (if it
exists) is the best linear approximation of the map $F(x+\delta x)-F(x)$.}

\item{The differential of the function $f(x)$ is the linear functional $D_x f$. The {\it gradient} of the
function $f(x)$ can be defined, if there is a given scalar product $\langle \: | \: \rangle$, and if
there exists  a Riesz representation for functional $D_x f$: $(D_x f)(a)=\langle {\rm grad}_x f | a
\rangle$. The gradient ${\rm grad}_x f$ is a vector.}

\item{The {\it second differential} of a map $F: U \rightarrow V$ is a bilinear operator $D^2_x F:W\times W
\rightarrow E$ which can be defined by Taylor formula: $F(x+\delta x)= F(x) + (D_x F)(\delta x) +
\frac{1}{2}(D^2_x F)(\delta x,\delta x)+ o(\|\delta x\|^2)$.}

\item{The differentiable map of domains in normed spaces $F: U \rightarrow V$ is an {\it immersion}, if for
any $x\in U$ the operator $D_x F$ is injective: $\ker D_x F = \{0\}$. In this case the image of $F$
($F(U)$) is called the {\it immersed manifold}, and the image of $D_x F$ is called the {\it tangent
space } to the immersed manifold $F(U)$. We shall use the notation $T_x$ for this tangent space: $
{\rm im}D_x F = T_x$.}

\item{The subset $U$ of the vector space $E$ is {\it convex}, if for every two points $x_{1},x_{2}\in U$ it
contains the segment between $x_{1}$ and $x_{2}$:  $\lambda x_{1}+(1-\lambda)x_{2}\in U$ for every
$\lambda\in[0,1]$.}

\item{The function $f$, defined on the convex set $U\subset E$, is {\it convex}, if its {\it epigraph}, i.e. the
set of pairs ${\rm Epi}f=\{(x,g)|x\in U, g\geq f(x)\}$, is the convex set in $E\times R$. The twice
differentiable function $f$ is convex  if and only if the quadratic form $(D^2_x f)(\delta x,\delta
x)$ is nonnegative.}

\item{The convex function $f$ is called {\it strictly convex} if in the domain of definition there is no line
segment on which it is constant and finite ($f(x)=const\neq\infty$). The sufficient condition for the
twice differentiable function $f$  to be {\it strictly convex} is that the quadratic form $(D^2_x
f)(\delta x,\delta x)$ is positive defined (i.e. it is positive for all $\delta x\neq 0$).}

\item{We use summation convention for vectors and tensors, $c_i g_i= \sum_i c_i g_i$, when it cannot cause
a confusion, in more complicated cases we use the sign $\sum$.}

\end{itemize}

\section{\textbf{The source of examples}}

In this section we follow, partially, the paper \cite{GKenc}, where nonlinear kinetic equations and
methods of reduced description are reviewed for a wide audience of specialists and postgraduate
students in physics, mathematical physics, material science, chemical engineering and
interdisciplinary research.

\subsection{\textbf{The Boltzmann equation}}

\subsubsection{The equation}

The {\bf Boltzmann equation} is the first and the most celebrated nonlinear kinetic equation
introduced by the great Austrian scientist Ludwig Boltzmann in 1872 \cite{Boltzmann}. This equation
describes the dynamics of a moderately rarefied gas, taking into account the two processes: the free
flight of the particles, and their collisions. In its original version, the Boltzmann equation has
been formulated for particles represented by hard spheres. The physical condition of rarefaction
means that only pair collisions are taken into account, a mathematical specification of which is
given by the {\bf Grad-Boltzmann limit}: If $N$ is the number of particles, and $\sigma$ is the
diameter of the hard sphere, then the Boltzmann equation is expected to hold when $N$ tends to
infinity, $\sigma$ tends to zero, $N\sigma^3$ (the volume occupied by the particles) tends to zero,
while $N\sigma^2$ (the total collision cross section) remains constant. The microscopic state of the
gas at time $t$ is described by the one-body distribution function $P(\xx,\vv,t)$, where $\xx$ is the
position of the center of the particle, and $\vv$ is the velocity of the particle. The distribution
function is the probability density of finding the particle at time $t$ within the infinitesimal
phase space volume centered at the phase point $(\xx,\vv)$. The collision mechanism of two hard
spheres is presented by a relation between the velocities of the particles before [$\vv$  and $\ww$ ]
and after [$\vv'$ and $\ww'$] their impact:
\begin{eqnarray}
\vv'=\vv-\nn (\nn, \vv -\ww), \nonumber\\ \ww'=\ww+\nn (\nn, \vv -\ww),\nonumber
\end{eqnarray}
where $\nn$ is the unit vector along $\vv-\vv'$. Transformation of the velocities conserves the total
momentum of the pair of colliding particles $(\vv'+\ww'=\vv+\ww)$, and the total kinetic energy
$(\vv'^2+\ww'^2=\vv^2+\ww^2)$ The Boltzmann equation reads:
\begin{eqnarray}
&&{\partial P\over \partial t}+\left(\vv,{\partial P\over
\partial \xx
}\right)= \label{S1} \\* &&
N\sigma^2\int_R\int_{B^-}(P(\xx,\vv',t)P(\xx,\ww',t)-P(\xx,\vv,t)P(\xx,\ww,t))\mid(\ww-\vv,\nn)\mid
d\ww d\nn, \nonumber
\end{eqnarray}

\noindent where integration in $\nn$ is carried over the unit sphere $R^3$, while integration in
$\ww$ goes over a hemisphere $B^-=\{\ww \mid (\ww-\vv,\nn)<0 \}$ . This hemisphere corresponds to the
particles entering the collison. The nonlinear integral operator in the right hand side of Eq.
(\ref{S1}) is nonlocal in the velocity variable, and local in space. The Boltzmann equation for
arbitrary hard-core interaction is a generalization of the Boltzmann equation for hard spheres under
the proviso that the true infinite-range interaction potential between the particles is cut-off at
some distance. This generalization amounts to a replacement,
\begin{eqnarray}
\sigma^2\mid(\ww-\vv,\nn)\mid d\nn\to B(\theta,\mid\ww-\vv\mid)d\theta d\varepsilon, \label{S2}
\end{eqnarray}
where function $B$ is determined by the interaction potential, and vector $\nn$ is identified with
two angles, $\theta$  and $\varepsilon$. In particular, for potentials proportional to the n-th
inverse power of the distance, the function $B$ reads,
\begin{eqnarray}
B(\theta,\mid\vv-\ww\mid)=\beta(\theta)\mid\vv-\ww\mid^{n-5\over n-1}.\label{S3}
\end{eqnarray}
In the special case $n=5$, function $B$ is independent of the magnitude of the relative velocity
(Maxwell molecules). Maxwell molecules occupy a distinct place in the theory of the Boltzmann
equation, they provide exact results. Three most important findings for the Maxwell molecules are
mentioned here: 1. The exact spectrum of the linearized Boltzmann collision integral, found by
Truesdell and Muncaster, 2. Exact transport coefficients found by Maxwell even before the Boltzmann
equation was formulated, 3. Exact solutions to the space-free model version of the nonlinear
Boltzmann equation. Pivotal results in this domain belong to Galkin who has found the general
solution to the system of moment equations in a form of a series expansion, to Bobylev, Krook and Wu
who have found an exact solution of a particular elegant closed form, and to Bobylev who has
demonstrated the complete integrability of this dynamic system.

The broad review of the Boltzmann equation and analysis of analytical solutions to kinetic models is
presented in the book of Cercignani \cite{Cercignani} A modern account of rigorous results on the
Boltzmann equation is given in the book \cite{CeIlPu}. Proof of the existence theorem for the
Boltzmann equation was done by DiPerna and Lions \cite{DPL}.

It is customary to write the Boltzmann equation using another normalization of the distribution
function, $f(\xx,\vv,t)d\xx d\vv$, taken in such a way that the function $f$ is compliant with the
definition of the hydrodynamic fields: the mass density
 $\rho$, the momentum density $\rho \uu$, and the energy density $\varepsilon$:
\begin{eqnarray}
\int f(\xx,\vv,t)md\vv &=&\rho(\xx,t),\nonumber \\ \int f(\xx,\vv,t)m \vv d\vv &=&\rho\uu(\xx,t),
\label{S4}
\\ \int f(\xx,\vv,t)m{v^2\over 2}d\vv &=&\varepsilon(\xx,t). \nonumber
\end{eqnarray}
Here $m$ is the particle's mass.

 The Boltzmann equation for the
distribution function $f$ reads,
\begin{eqnarray}
{\partial f\over \partial t}+\left(\vv,{\partial \over \partial \xx} f\right)=Q(f,f),\label{S5}
\end{eqnarray}
where the nonlinear integral operator in the right hand side is the Boltzmann collision integral,
\begin{eqnarray}
Q=\int_{R^3}\int_{B^-}(f(\vv')f(\ww')-f(\vv)f(\ww))B(\theta,\vv)d\ww d\theta d\varepsilon. \label{S6}
\end{eqnarray}

Finally, we mention the following form of the Boltzmann collision integral (sometimes referred to as
the scattering or the quasi-chemical representation),
\begin{eqnarray}
Q=\int W(\vv,\ww\mid\vv',\ww')[(f(\vv')f(\ww')-f(\vv)f(\ww))]d\ww d\ww' d\vv', \label{S7}
\end{eqnarray}
where $W$ is a generalized function which is called the probability density of the elementary event,
\begin{eqnarray}
W=w(\vv,\ww \mid\vv', \ww')\delta(\vv+\ww-\vv' -\ww')\delta(v^2+w^2-v'^2-w'^2).\label{S8}
\end{eqnarray}

\subsubsection{The basic properties of the Boltzmann equation}

Generalized function W has the following symmetries:
\begin{eqnarray}
W(\vv', \ww'\mid\vv,\ww)\equiv W(\ww', \vv'\mid\vv,\ww)\equiv W(\vv', \ww'\mid\ww,\vv)\equiv W(\vv,
\ww\mid\vv',\ww').\label{S9}
\end{eqnarray}

The first two identities reflect the symmetry of the collision process with respect to labeling the
particles, whereas the last identity is the celebrated {\bf detail balance} condition which is
underpinned by the time-reversal symmetry of the microscopic (Newton's) equations of motion. The
basic properties of the Boltzmann equation are:

1. {\bf Additive invariants of collision operator}:
\begin{eqnarray}
\int Q(f,f)\{1,\vv,v^2\}d\vv=0, \label{S10}
\end{eqnarray}
for any function $f$, assuming integrals exist. Equality (\ref{S10}) reflects the fact that the
number of particles, the three components of particle's momentum, and the particle's energy are
conserved by the collision. Conservation laws (\ref{S10}) imply that the local hydrodynamic fields
(\ref{S4}) can change in time only due to redistribution in the space.

2. Zero point of the integral ($Q=0$) satisfy the equation (which is also called {\bf the detail
balance}): For almost all velocities,
\begin{eqnarray}
f(\vv',\xx,t)f(\ww',\xx,t)=f(\vv,\xx,t)f(\ww,\xx,t). \nonumber
\end{eqnarray}

3. Boltzmann's {\bf local entropy production inequality}:
\begin{eqnarray}
\sigma (\xx,t)=- k_B \int  Q(f,f) \ln f d\vv \geq 0, \label{S11}
\end{eqnarray}
for any function $f$, assuming integrals exist. Dimensional {\bf Boltzmann's constant} ($k_B\approx
6\cdot 10^{-23}J/K$) in this expression serves for a recalculation of the energy units into the
absolute temperature units. Moreover, equality sign takes place if $\ln f$ is a linear combination of
the additive invariants of collision.

Distribution functions $f$ whose logarithm is a linear combination of additive collision invariants,
with coefficients dependent on $x$, are called {\bf local Maxwell distribution functions} $f_{LM}$,
\begin{eqnarray}
f_{LM}={\rho \over m} \left({2\pi k_BT\over m}\right)^{-3/2}\exp\left({-m(\vv-\uu)^2\over
2k_BT}\right).\label{S12}
\end{eqnarray}

Local Maxwellians are parametrized by values of five hydrodynamic variables, $\rho$ , $\uu$ and $T$.
This parametrization is consistent with the definitions of the hydrodynamic fields (\ref{S4}), $\int
f_{LM}(m,m\vv,mv^2/2)=(\rho, \rho\uu,\varepsilon)$ provided the relation between the energy and the
kinetic temperature $T$, holds, $\varepsilon ={3\rho \over 2 m k_BT}$.

4. Boltzmann's {\bf {\boldmath $H$} theorem}: The function
\begin{eqnarray}
S[f]=-k_B\int f \ln f d\vv, \label{S13}
\end{eqnarray}
is called the {\bf entropy density}. The {\bf local {\boldmath $H$} theorem} for distribution
functions independent of space states that the rate of the entropy density increase is equal to the
nonnegative entropy production,
\begin{eqnarray}
{dS\over dt}=\sigma\geq0. \label{S14}
\end{eqnarray}

Thus, if no space dependence is concerned, the Boltzmann equation describes relaxation to the unique
global Maxwellian (whose parameters are fixed by initial conditions), and the entropy density grows
monotonically along the solutions. Mathematical specifications of this property has been initialized
by Carleman, and many estimations of the entropy growth were obtained over the past two decades. In
the case of space-dependent distribution functions, the local entropy density obeys the {\bf entropy
balance equation}:
\begin{eqnarray}
{\partial S(\xx,t)\over \partial t}+\left({\partial \over
\partial \xx},{\bf J}_s(\xx,t)\right)=\sigma (\xx,t)\geq0,\label{S15}
\end{eqnarray}

\noindent where ${\bf J}_s$  is the entropy flux, ${\bf J}_s(\xx,t)=-k_B \int \ln f(\xx,t)\vv
f(\xx,t)d\vv.$ For suitable boundary conditions, such as, specularly reflecting or at the infinity,
the entropy flux gives no contribution to the equation for the {\bf total entropy}, $S_{tot}=\int
S(\xx,t)d\xx$
  and its rate of changes is then equal to the nonnegative total entropy
production $\sigma_{tot}=\int \sigma(\xx,t)d\xx$ (the {\bf global {\boldmath $H$} theorem)}. For more
general boundary conditions which maintain the entropy influx the global $H$ theorem needs to be
modified. A detailed discussion of this question is given by Cercignani. The local Maxwellian is also
specified as the maximizer of the Boltzmann entropy function (\ref{S13}), subject to fixed
hydrodynamic constraints (\ref{S4}). For this reason, the local Maxwellian is also termed as the
local equilibrium distribution function.

\subsubsection{Linearized collision integral}

Linearization of the Boltzmann integral around the local equilibrium results in the linear integral
operator,
\begin{eqnarray}
&&Lh(\vv)=\nonumber\\* &&\int W(\vv,\ww \mid\vv',\ww') f_{LM}(\vv)f_{LM}(\ww)\left[{h(\vv')\over
f_{LM}(\vv')}+{h(\ww')\over f_{LM}(\ww')}-{h(\vv)\over f_{LM}(\vv)}-{h(\ww)\over
f_{LM}(\ww)}\right]d\ww' d\vv' d\ww. \nonumber
\end{eqnarray}
Linearized collision integral is symmetric with respect to scalar product defined by the second
derivative of the entropy functional,
\begin{eqnarray}
\int f_{LM}^{-1}(\vv)g(\vv)L h(\vv)d \vv = \int f_{LM}^{-1}(\vv)h(\vv)Lg(\vv)d\vv,\nonumber
\end{eqnarray}
it is nonpositively definite,
\begin{eqnarray}
\int f_{LM}^{-1}(\vv)h(\vv)L h(\vv)d\vv\leq 0,\nonumber
\end{eqnarray}
where equality sign takes place if the function $hf_{LM}^{-1}$ is a linear combination of collision
invariants, which characterize the null-space of the operator $L$. Spectrum of the linearized
collision integral is well studied in the case of the small angle cut-off.

\subsection{\textbf{Phenomenology and Quasi-chemical representation of the Boltzmann equation}}

Boltzmann's original derivation of his collision integral was based on a phenomenological
``bookkeeping" of the gain and of the loss of probability density in the collision process. This
derivation postulates that the rate of gain $G$ equals
\begin{eqnarray}
G=\int W^+(\vv,\ww\mid\vv',\ww')f(\vv')f(\ww')d\vv' d\ww' d\ww, \nonumber
\end{eqnarray}
while the rate of loss is
\begin{eqnarray}
L=\int W^-(\vv,\ww\mid\vv',\ww')f(\vv)f(\ww)d\vv' d\ww' d\ww.\nonumber
\end{eqnarray}

The form of the gain and of the loss, containing products of one-body distribution functions in place
of the two-body distribution, constitutes the famous Stosszahlansatz. The Boltzmann collision
integral follows now as ($G-L$), subject to the detail balance for the rates of individual
collisions,
\begin{eqnarray}
W^+(\vv,\ww\mid\vv',\ww')=W^-(\vv,\ww\mid\vv',\ww').\nonumber
\end{eqnarray}

This representation for interactions different from hard spheres requires also the cut-off of
functions $\beta$  (\ref{S3}) at small angles. The gain-loss form of the collision integral makes it
evident that the detail balance for the rates of individual collisions is sufficient to prove the
local $H$ theorem. A weaker condition which is also sufficient to establish the $H$ theorem was first
derived by Stueckelberg (so-called {\bf  semi-detailed balance}), and later generalized {\bf to
inequalities of concordance}:
\begin{eqnarray}
\int d\vv' \int d\ww'(W^+(\vv,\ww \mid \vv',\ww')-W^-(\vv,\ww \mid \vv',\ww'))\geq 0,\nonumber\\ \int
d\vv \int d\ww (W^+(\vv,\ww \mid \vv',\ww')-W^-(\vv,\ww \mid \vv',\ww'))\leq 0.\nonumber
\end{eqnarray}

The semi-detailed balance follows from these expressions if the inequality signes are replaced by
equalities.

The pattern of Boltzmann's phenomenological approach is often used in order to construct nonlinear
kinetic models. In particular, {\bf nonlinear equations of chemical kinetics} are based on this idea:
If $n$ chemical species $A_i$  participate in a complex chemical reaction,
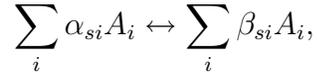
\begin{eqnarray}
\sum_i \alpha_{si}A_i\leftrightarrow \sum_i \beta_{si}A_i, \nonumber
\end{eqnarray}
where $\alpha_{si}$  and $\beta_{si}$  are nonnegative integers (stoichiometric coefficients) then
equations of chemical kinetics for the concentrations of species $c_j$  are written
\begin{eqnarray}
{dc_i\over dt}=\sum_{s=1}^n (\beta_{si}-\alpha_{si})\left[\varphi_s^+\exp \left(\sum_{j=1}^n{\partial
G\over \partial c_j}\alpha_{sj}\right) -\varphi_s^-\exp \left(\sum_{j=1}^n{\partial G\over \partial
c_j}\beta_{sj}\right)\right].\nonumber
\end{eqnarray}

Functions $\varphi_s^+$  and $\varphi_s^-$  are interpreted as constants of the direct and of the
inverse reactions, while the function $G$ is an analog of the Boltzmann's $H$-function. Modern
derivation of the Boltzmann equation, initialized by the seminal work of N. N. Bogoliubov, seek a
replacement condition, and which would be more closely related to many-particle dynamics. Such
conditions are applied to the $N$-particle Liouville equation should factorize in the remote enough
past, as well as in the remote infinity (the hypothesis of weakening of correlations). Different
conditions has been formulated by D. N. Zubarev, J. Lewis and others. The advantage of these
formulations is the possibility to systematically find corrections not included in the
Stosszahlansatz.

\subsection{\textbf{Kinetic models}}

Mathematical complications caused by the nonlinearly Boltzmann collision integral are traced back to
the Stosszahlansatz. Several approaches were developed in order to simplify the Boltzmann equation.
Such simplifications are termed kinetic models. Various kinetic models preserve certain features of
the Boltzmann equation, while sacrificing the rest of them. The most well known kinetic model which
preserve the $H$ theorem is the nonlinear Bhatnagar-Gross-Krook model (BGK) \cite{BGK}. The BGK
collision integral reads:
\begin{eqnarray}
Q_{BGK}=-{1\over \tau}(f-f_{LM}(f)).\nonumber
\end{eqnarray}
The time parameter $\tau > 0$ is interpreted as a characteristic relaxation time to the local
Maxwellian. The BGK is a nonlinear operator: Parameters of the local Maxwellian are identified with
the values of the corresponding moments of the distribution function $f$. This nonlinearly is of
``lower dimension" than in the Boltzmann collision integral because $f_{LM}(f)$ is a nonlinear
function of only the moments of $f$ whereas the Boltzmann collision integral is nonlinear in the
distribution function $f$ itself. This type of simplification introduced by the BGK approach is
closely related to the family of so-called mean-field approximations in statistical mechanics. By its
construction, the BGK collision integral preserves the following three properties of the Boltzmann
equation: additive invariants of collision, uniqueness of the equilibrium, and the $H$ theorem. A
class of kinetic models which generalized the BGK model to quasiequilibrium approximations of a
general form is described as follows: The quasiequilibrium  $f^*$ for the set of linear functionales
$M(f)$ is a distribution function $f^*(M)(\xx,\vv)$  which maximizes the entropy under fixed values
of functions $M$. The Quasiequilibrium (QE) models are characterized by the collision integral
\cite{GKMod},
\begin{eqnarray}
Q_{QE}(f)=-{1\over \tau}[f-f^*(M(f))]+Q_B(f^*(M(f)),f^*(M(f))).\nonumber
\end{eqnarray}
Same as in the case of the BGK collision integral, operator $Q_{QE}$ is nonlinear in the moments $M$
only. The QE models preserve the following properties of the Boltzmann collision operator: additive
invariants, uniqueness of the equilibrium, and the $H$ theorem, provided the relaxation time $\tau$
to the quasiequilibrium is sufficiently small. A different nonlinear model was proposed by Lebowitz,
Frisch and Helfand \cite{LFHMod}:
\begin{eqnarray}
Q_D=D\left({\partial \over \partial \vv}{\partial \over \partial \vv }f+{m\over k_BT}{\partial \over
\partial \vv}(\vv -\uu(f))f\right).\nonumber
\end{eqnarray}
The collision integral has the form of the self-consistent Fokker-Planck operator, describing
diffusion (in the velocity space) in the self-consistent potential. Diffusion coefficient $D>0$ may
depend on the distribution function $f$. Operator $Q_D$ preserves the same properties of the
Boltzmann collision operator as the BGK model. Kinetic BGK model has been used for obtaining exact
solutions of gasdynamic problems, especially its linearized form for stationary problems. Linearized
B GK collision model has been extended to model more precisely the linearized Boltzmann collision
integral.

\subsection{\textbf{Methods of reduced description}}

One of the major issues raised by the Boltzmann equation is the problem of the reduced description.
Equations of hydrodynamics constitute a closet set of equations for the hydrodynamic field (local
density, local momentum, and local temperature). From the standpoint of the Boltzmann equation, these
quantities are low-order moments of the one-body distribution function, or, in other words, the
macroscopic variables. The problem of the reduced description consists in giving an answer to the
following two questions:

1. What are the conditions under which the macroscopic description sets in?

2. How to derive equations for the macroscopic variables from kinetic equations?

The classical methods of reduced description for the Boltzmann equation are: the Hilbert method, the
Chapman-Enskog method, and the Grad moment method.

\subsubsection{The Hilbert method}

In 1911, David Hilbert introduced the notion of normal solutions,
$$f_H(\vv,\,n(\xx,t),\,\uu(\xx,t),\,T(\xx,t)),$$

\noindent that is, solution to the Boltzmann equation which depends on space and time only through
five hydrodynamic fields \cite{Hilbert}.

The normal solutions are found from a singularly perturbed Boltzmann equation,
\begin{eqnarray}
D_tf={1\over \varepsilon}Q(f,f),\label{S16}
\end{eqnarray}
where $\varepsilon$ is a small parameter, and $$D_t f\equiv {\partial \over \partial
t}f+(\vv,{\partial \over \partial \xx})f.$$ Physically, parameter $\varepsilon$ corresponds to the
Knudsen number, the ratio between the mean free path of the molecules between collisions, and the
characteristic scale of variation of the hydrodynamic fields. In the Hilbert method, one seeks
functions $n(\xx,t),\,\uu(\xx,t),\,T(\xx,t)$, such that the normal solution in the form of the
Hilbert expansion,
\begin{eqnarray}
f_H=\sum_{i=0}^{\infty}\varepsilon^i f_H^{(i)}\label{S17}
\end{eqnarray}
satisfies the Eq. (\ref{S16}) order by order. Hilbert was able to demonstrate that this is formally
possible. Substituting (\ref{S17}) into (\ref{S16}), and matching various order in $\varepsilon$, we
have the sequence of integral equations
\begin{eqnarray}
&&Q(f_H^{(0)},f_H^{(0)})=0,\label{S18}\\ &&Lf_H^{(1)}=D_tf_H^{(0)},\label{S19}\\
&&Lf_H^{(2)}=D_tf_H^{(1)}-Q(f_H^{(0)},f_H^{(1)}),\label{S20}
\end{eqnarray}
and so on for higher orders. Here $L$ is the linearized collision integral. From Eq.(\ref{S18}), it
follows that $f_H^{(0)}$  is the local Maxwellian with parameters not yet determined. The Fredholm
alternative, as applied to the second Eq. (\ref{S19}) results in

a) Solvability condition,
\begin{eqnarray}
\int D_tf_H^{(0)}\{1,\vv,v^2\}d\vv =0,\nonumber
\end{eqnarray}
which is the set of compressible Euler equations of the non-viscous hydrodynamics. Solution to the
Euler equation determine the parameters of the Maxwellian  $f_H^0$.

b) General solution $f_H^{(1)}=f_H^{(1)1}+f_H^{(1)2}$, where $f_H^{(1)1}$ is the special solution to
the linear integral equation (\ref{S19}), and $f_H^{(1)2}$ is yet undetermined linear combination of
the additive invariants of collision.

c) Solvability condition to the next equation (\ref{S19}) determines coefficients of the function
$f_H^{(1)2}$ in terms of solutions to the linear hyperbolic differential equations,
\begin{eqnarray}
\int D_t(f_H^{(1)1}+f_H^{(1)2})\{1,\vv,v^2\}d\vv=0.\nonumber
\end{eqnarray}
Hilbert was able to demonstrate that this procedure of constructing the normal solution can be
carried out to arbitrary order $n$, where the function $f_H^{(n)}$ is determined from the solvability
condition at the next, $(n+1)$-th order. In order to summarize, implementation of the Hilbert method
requires solutions for the function $n(\xx,t),\,\uu(\xx,t)$, and $T(\xx,t)$  obtained from a sequence
of partial differential equations.

\subsubsection{The Chapman-Enskog method}

A completely different approach to the reduced description was invented in 1917 by David Enskog
\cite{Ens}, and independently by Sidney Chapman \cite{Chapman}. The key innovation was to seek an
expansion of the time derivatives of the hydrodynamic variables rather than seeking the time-space
dependencies of these functions as in the Hilbert method.

The Chapman-Enskog method starts also with the singularly perturbed Boltzmann equation, and with the
expansion
\begin{eqnarray}
f_{CE}=\sum_{n=0}^{\infty} \varepsilon^n f_{CE}^{(n)}.\nonumber
\end{eqnarray}
However, the procedure of evaluation of the functions $f_{CE}^{(n)}$ differs from the Hilbert method:
\begin{eqnarray}
Q(f_{CE}^{(0)},f_{CE}^{(0)})&=&0,\label{S21}\\
Lf_{CE}^{(1)}&=&-Q(f_{CE}^{(0)},f_{CE}^{(0)})+{\partial^{(0)}\over
\partial t }f_{CE}^{(0)}+\left(\vv, {\partial \over \partial
\xx}\right)f_{CE}^{(0)}.\label{S22}
\end{eqnarray}
Operator $\partial^{(0)} /\partial t$  is defined from the expansion of the right hand side of
hydrodynamic equation,
\begin{eqnarray}
{\partial^{(0)}\over \partial t}\{\rho,\rho\uu,e\}\equiv -\int \left\{m,m\vv,{mv^2\over 2}\right\}
\left( \vv,{\partial \over \partial \xx} \right) f^{(0)}_{CE}d\vv.\label{S23}
\end{eqnarray}
From Eq. (\ref{S21}), function $f_{CE}^{(0)}$ is again the local Maxwellian, whereas (\ref{S23}) is
the Euler equations, and $\partial^{(0)} /\partial t$ acts on various functions $g(\rho,\rho\uu,e)$
according to the chain rule,
\begin{eqnarray}
{\partial^{(0)}\over
\partial t }g={\partial g\over \partial
\rho }{\partial^{(0)} \over \partial t}\rho +{\partial g\over
\partial (\rho \uu)}{\partial^{(0)} \over \partial t}\rho \uu +
{\partial g\over \partial e}{\partial^{(0)} e \over \partial t},\nonumber
\end{eqnarray}
while the time derivatives ${\partial^{(0)} \over \partial t}$ of the hydrodynamic fields are
expressed using the right hand side of Eq. (\ref{S23}).

 The result of the Chapman-Enskog definition of the
time derivative ${\partial^{(0)} \over \partial t}$, is that the Fredholm alternative is satisfied by
the right hand side of Eq. (\ref{S22}). Finally, the solution to the homogeneous equation is set to
be zero by the requirement that the hydrodynamic variables as defined by the function
$f^{(0)}+\varepsilon f^{(1)}$  coincide with the parameters of the local Maxwellian  $f^{(0)}$:
\begin{eqnarray}
\int \{1,\vv,v^2\}f_{CE}^{(1)}d\vv=0.\nonumber
\end{eqnarray}

The first correction $f_{CE}^{(1)}$ of the Chapman-Enskog method adds the terms
\begin{eqnarray}
{\partial^{(1)}\over \partial t }\{\rho,\rho\uu,e\}=-\int \left\{m,m\vv,{mv^2\over
2}\right\}\left(\vv,{\partial \over
\partial \xx}\right)f_{CE}^{(1)}d\vv \nonumber
\end{eqnarray}
to the time derivatives of the hydrodynamic fields. These terms correspond to the dissipative
hydrodynamics where viscous momentum transfer and heat transfer are in the Navier-Stokes and Fourier
form. The Chapman-Enskog method was the first true success of the Boltzmann equation since it had
made it possible to derive macroscopic equation without a priori guessing (the generalization of the
Boltzmann equation onto mixtures predicted existence of the thermodiffusion before it has been found
experimentally), and to express the kinetic coefficient in terms of microscopic particle's
interaction.

However, higher-order corrections of the Chapman-Enskog method, resulting in hydrodynamic equations
with derivatives (Burnett hydrodynamic equations) face serve difficulties both from the theoretical,
as well as from the practical sides. In particular, they result in unphysical instabilities of the
equilibrium.

\subsubsection{The Grad moment method}

In 1949, Harold Grad extended the basic assumption behind the Hilbert and the Chapman-Enskog methods
(the space and time dependence of the normal solutions is mediated by the five hydrodynamic moments)
\cite{Grad}. A physical rationale behind the Grad moment method is an assumption of the decomposition
of motions:

\noindent(i). During the time of order $\tau$, a set of distinguished moments $M'$ (which include the
hydrodynamic moments and a subset of higher-order moment) does not change significantly as compared
to the rest of the moments $M''$ (the fast evolution).

\noindent(ii). Towards the end of the fast evolution, the values of the moments $M''$ become
unambiguously determined by the values of the distinguished moments $M'$.

\noindent(iii). On the time of order $\theta \gg \tau$, dynamics of the distribution function is
determined by the dynamics of the distinguished moments while the rest of the moments remain to be
determined by the distinguished moments (the slow evolution period).

Implementation of this picture requires an ansatz for the distribution function in order to represent
the set of states visited in the course of the slow evolution. In Grad's method, these representative
sets are finite-order truncations of an expansion of the distribution functions in terms of Hermite
velocity tensors:
\begin{eqnarray}
f_G(M',\vv)=f_{LM}(\rho,\uu,E,\vv)[1+\sum_{(\alpha)}^N a_{(\alpha)}(M')H_{(\alpha)}(\vv-\uu)]
,\label{S24}
\end{eqnarray}
 where $H_{(\alpha)}(\vv-\uu)$ are various Hermite tensor polynomials,
orthogonal with the weight $f_{LM}$, while coefficient $a_{(\alpha)}(M')$ are known functions of the
distinguished moments $M'$, and $N$ is the highest order of $M'$. Other moments are functions of
$M'$: $M''=M''(f_G(M'))$.

Slow evolution of distinguished moments is found upon substitution of Eq. (\ref{S24}) into the
Boltzmann equation and finding the moments of the resulting expression (Grad's moment equations).
Following Grad, this extremely simple approximation can be improved by extending the list of
distinguished moments. The most well known is Grad's thirteen-moment approximation where the set of
distinguished moments consists of five hydrodynamic moments, five components of the traceless stress
tensor $\sigma_{ij}=\int m[(v_i-u_i)(v_j-u_j)-\delta_{ij}(\vv-\uu)^2/3]fd\vv,$ and of the three
components of the heat flux vector $q_i=\int (v_i-u_i)m(\vv-\uu)^2/2 fd\vv$.

 The time evolution hypothesis
cannot be evaluated for its validity within the framework of Grad's approach. It is not surprising
therefore that Grad's methods failed to work in situations where it was (unmotivatedly) supposed to,
primarily, in the phenomena with sharp time-space dependence such as the strong shock wave. On the
other hand, Grad's method was quite successful for describing transition between parabolic and
hyperbolic propagation, in particular, the second sound effect in massive solids at low temperatures,
and, in general, situations slightly deviating from the classical Navier-Stokes- Fourier domain.
Finally, the Grad method has been important background for development of phenomenological
nonequilibrium thermodynamics based on hyperbolic first-order equation, the so-called EIT (extended
irreversible thermodynamics).

\subsubsection{Special approximations}

Special approximation of the solutions to the Boltzmann equation has been found for several problems,
and which perform better than results of ``regular" procedures. The most well known is the ansatz
introduced independently by Mott-Smith and Tamm for the strong shock wave problem: The (stationary)
distribution function is thought as
\begin{eqnarray}
f_{TMS}(a(x))=(1-a(x))f_{+}+a(x)f_{-},\label{S25}
\end{eqnarray}
where $f_{\pm}$ are upstream and downstream Maxwell distribution functions, whereas $a(x)$ is an
undetermined scalar function of the coordinate along the shock tube.

Equation for function $a(x)$ has to be found upon substitution of Eq.(\ref{S25}) into the Bolltzmann
equation, and upon integration with some velocity-dependent function  $\varphi(\vv)$. Two general
problems arise with the special approximation thus constructed: Which function $\varphi(\vv)$ should
be taken, and how to find correction to the ansatz like Eq. (\ref{S25}).

\subsubsection{The method of invariant manifold}

The general approach to the problem of reduced description for dissipative system was recognized as
the problem of finding stable invariant manifolds in the space of distribution function
\cite{GKAMSE92,GK1,GKTTSP94}. The notion of invariant manifold generalizes the normal solution in the
Hilbert and in the Chapman-Enskog method, and the finite-moment sets of distribution function in the
Grad method: If $\Omega$ is a smooth manifold in the space of distribution function, and if
$f_{\Omega}$ is an element of $\Omega$, then $\Omega$ is invariant with respect to the dynamic
system,
\begin{eqnarray}
&&{  \partial f\over \partial t}=J(f), \label{S26}\\&& \mbox{if}  \: J(f_{\Omega})\in T\Omega,  \:
\mbox{for all} \: f_{\Omega}\in \Omega, \label{S27}
\end{eqnarray}
where $T\Omega$ is the tangent bundle of the manifold  $\Omega$. Application of the invariant
manifold idea to dissipative systems is based on iterations, progressively improving the initial
approximation, involves the following steps:

\paragraph{Thermodynamic projector}

 Given a manifold $\Omega$  (not obligatory invariant), the macroscopic
dynamics on this manifold is defined by the macroscopic vector field, which is the result of a
projection of vectors $J(f_{\Omega})$ onto the tangent bundle $T\Omega$. The thermodynamic projector
$ P^*_{f_{\Omega}}$ takes advantage of dissipativity:
\begin{eqnarray}
\mbox{ker} P^*_{f_{\Omega}}\subseteq \mbox{ker}D_fS\mid_{f_{\Omega}},\label{S28}
\end{eqnarray}

\noindent where $D_fS\mid_{f_{\Omega}}$ is the differential of the entropy evaluated in $f_{\Omega}$.

 This condition of thermodynamicity means that each state of the manifold $\Omega$ is
regarded as the result of decomposition of motions occurring near $\Omega$: The state $f_{\Omega}$ is
the maximum entropy state on the set of states $f_{\Omega}+\mbox{ker} P^*_{f_{\Omega}}$. Condition of
thermodynamicity does not define projector completely; rather, it is the condition that should be
satisfied by any projector used to define the macroscopic vector field,
$J'_{\Omega}=P^*_{f_{\Omega}}J(f_{\Omega})$. For, once the condition (\ref{S28}) is met, the
macroscopic vector field preserves dissipativity of the original microscopic vector field $J(f)$:
\begin{eqnarray}
D_fS\mid_{f_{\Omega}}\cdot P^*_{f_{\Omega}}(J(f_{\Omega}))\geq 0\mbox{ for all
}\,f_{\Omega}\in\Omega.\nonumber
\end{eqnarray}

The thermodynamic projector is the formalization of the assumption that $\Omega$ is the manifold of
slow motion: If a fast relaxation takes place at least in a neighborhood of $\Omega$, then the states
visited in this process before arriving at $f_{\Omega}$ belong to $\mbox{ker} P^*_{f_{\Omega}}$. In
general, $P^*_{f_{\Omega}}$ depends in a non-trivial way on $f_{\Omega}$.

\paragraph{Iterations for the invariance condition}

 The invariance
condition for the manifold $\Omega$ reads,
\begin{eqnarray}
P_{\Omega}(J(f_{\Omega}))-J(f_{\Omega})=0,\nonumber
\end{eqnarray}
here $P_{\Omega}$ is arbitrary (not obligatory thermodynamic) projector onto the tangent bundle of
$\Omega$. The invariance condition is considered as an equation which is solved iteratively, starting
with initial approximation $\Omega_0$. On the $(n+1)-$th iteration, the correction
$f^{(n+1)}=f^{(n)}+\delta f^{(n+1)}$ is found from linear equations,
\begin{eqnarray}
D_f J^*_n\delta f^{(n+1)}&=&P^*_{n}J(f^{(n)})-J(f^{(n)}),\nonumber
\\ P^*_{n}\delta f^{(n+1)}&=&0,\label{S29}
\end{eqnarray}
here $D_f J^*_n$ is the linear selfajoint operator with respect to the scalar product by the second
differential of the entropy $D^2_f S\mid_{f^{(n)}}$.

Together with the above-mentioned principle of thermodynamic projecting, the selfadjoint
linearization implements the assumption about the decomposition of motions around the $n$'th
approximation. The selfadjoint linearization of the Boltzmann collision integral $Q$ (\ref{S7})
around a distribution function $f$ is given by the formula,
\begin{eqnarray}
D_f Q^*\delta f&=& \int W(\vv,\ww, \mid \vv',\ww'){f(\vv)f(\ww)+f(\vv')f(\ww')\over 2}\times
\nonumber
\\ &&\left[{\delta f (\vv')\over f(\vv')}+{\delta f(\ww')\over f(\ww')}-{\delta f (\vv)\over
f(\vv)}-{\delta f(\ww)\over f(\ww)}\right]d\ww' d\vv' d\ww. \label{S30}
\end{eqnarray}

 If $f=f_{LM}$, the selfadjoint operator (\ref{S30}) becomes the
 linearized collision integral.

The method of invariant manifold is the iterative process: $$(f^{(n)},P^*_{n})\rightarrow
(f^{(n+1)},P^*_{n})\rightarrow (f^{(n+1)},P^*_{n+1})$$ On the each 1-st part of the iteration, the
linear equation (\ref{S29}) is solved with the projector known from the previous iteration. On the
each 2-nd part, the projector is updated, following the thermodynamic construction.

 The method of invariant
manifold can be further simplified if smallness parameters are known.

The proliferation of the procedure in comparison to the Chapman-Enskog method is essentially twofold:

First, the projector is made dependent on the manifold. This enlarges the set of admissible
approximations.

Second, the method is based on iteration rather than a series expansion in a smallness parameter.
Importance of iteration procedures is well understood in physics, in particular, in the
renormalization group approach to reducing the description in equilibrium statistical mechanics, and
in the Kolmogorov- Arnold-Moser theory of finite-dimensional Hamiltonian systems.

\subsubsection{Quasiequilibrium approximations}

Important generalization of the Grad moment method is the concept of the quasiequilibrium
approximations already mentioned above (we will discuss this approximation in detail in a separate
section). The quasiequilibrium distribution function for a set of distinguished moment $M$  maximizes
the entropy density $S$ for fixed $M$. The quasiequilibrium manifold $\Omega^*(M)$ is the collection
of the quasiequilibrium distribution functions for all admissible values of $M$. The quasiequilibrium
approximation is the simplest and extremely useful (not only in the kinetic theory itself)
implementation of the hypothesis about a decomposition of motions: If $M$ are considered as slow
variables, then states which could be visited in the course of rapid motion in the vicinity of
$\Omega^*(M)$ belong to the planes $\Gamma_{M}=\{f\mid m(f-f^*(M))=0\}$. In this respect, the
thermodynamic construction in the method of invariant manifold is a generalization of the
quasiequilibrium approximation where the given manifold is equipped with a quasiequilibrium structure
by choosing appropriately the macroscopic variables of the slow motion. In contrast to the
quasiequilibrium, the macroscopic variables thus constructed are not obligatory moments. A text book
example of the quasiequilibrium approximation is the generalized Gaussian function for $M=\{\rho,
\rho \uu,P\}$ where $P_{ij}=\int v_iv_j f d\vv$ is the pressure tensor.

The thermodynamic projector $P^*$ for a quasiequilibrium approximation was first introduced by B.
Robertson \cite{Robertson} (in a different context of conservative dynamics and for a special case of
the Gibbs-Shannon entropy). It acts on a function $\Psi$ as follows $$P^*_{M}\Psi =\sum_i{\partial
f^*\over
\partial M_i}\int m_i \Psi d\vv,$$ where $M=\int m_i f d\vv$. The quasiequilibrium approximation
does not exist if the highest order moment is an odd polynomial of velocity (therefore, there exists
no quasiequilibrium for thirteen Grad's moments). Otherwise, the Grad moment approximation is the
first-order expansion of the quasiequilibrium around the local Maxwellian.

\subsection{\textbf{Discrete velocity models}}

If the number of microscopic velocities is reduced drastically to only a finite set, the resulting
discrete velocity, continuous time and continuous space models can still mimic the gas-dynamic flows.
This idea was introduced in Broadwell's paper in 1963 to mimic the strong shock wave
\cite{Broadwell}.

Further important development of this idea was due to Cabannes and Gatignol in the seventies who
introduced a systematic class of discrete velocity models \cite{Discret2}. The structure of the
collision operators in the discrete velocity models mimics the polynomial character of the Boltzmann
collision integral. Discrete velocity models are implemented numerically by using the natural
operator splitting in which each update due to free flight is followed by the collision update, the
idea which dates back to Grad. One of the most important recent results is the proof of convergence
of the discrete velocity models with pair collisions to the Boltzmann collision integral.

\subsection{\textbf{Direct simulation}}

Besides the analytical approach, direct numerical simulation of Boltzmann-type nonlinear kinetic
equations have been developed since mid of 1960s \cite{GBird,Oran}. The basis of the approach is a
representation of the Boltzmann gas by a set of particles whose dynamics is modeled as a sequence of
free propagation and collisions. The modeling of collisions uses a random choice of pairs of
particles inside the cells of the space, and changing the velocities of these pairs in such a way as
to comply with the conservation laws, and in accordance with the kernel of the Boltzmann collision
integral. At present, there exists a variety of this scheme known under the common title of the
Direct Simulation Monte-Carlo method \cite{GBird,Oran}. The DSMC, in particular, provides data to
test various analytical theories.

\subsection{\textbf{Lattice Gas and Lattice Boltzmann models}}

Since mid of 1980s, the kinetic theory based approach to simulation of complex macroscopic phenomena
such as hydrodynamics has been developed. The main idea of the approach is construction of minimal
kinetic system in such a way that their long-time and large-scale limit matches the desired
macroscopic equations. For this purpose, the fully discrete (in time- space-velocity) nonlinear
kinetic equations are considered on sufficiently isotropic lattices, where the links represent the
discrete velocities of fictitious particles. In the earlier version of the lattice methods, the
particle-based picture has been exploited, subject to the exclusion rule (one or zero particle per
lattice link) [the Lattice gas model \cite{Lgas} ]. Most of the present versions use the distribution
function picture, where populations of the links are non-integer [the Lattice Boltzmann model
\cite{LB1,LB2,Mcnamara,Higuera,Benzi}]. Discrete-time dynamics consists of a propagation step where
populations are transmitted to adjacent links and collision step where populations of the links at
each node of the lattice are equilibrated by a certain rule. Most of the present versions use the
BGK-type equilibration, where the local equilibrium is constructed in such a way as to match desired
macroscopic equations. The Lattice Boltzmann method is a useful approach for computational fluid
dynamics, effectively compliant with parallel architectures. The proof of the $H$ theorem for the
Lattice gas models is based on the semi-detailed (or Stueckelberg's) balance principle. The proof of
the $H$ theorem in the framework of the Lattice Boltzmann method has been only very recently achieved
\cite{LB3,KGSBPRL,KFOeEPL,AKPRE00,AKJSP,AKOeEPL}.

\subsection{\textbf{Other kinetic equations}}

\subsubsection{The Enskog equation for hard spheres}

The Enskog equation for hard spheres is an extension of the Boltzmann equation to moderately dense
gases. The Enskog equation explicitly takes into account the nonlocality of collisions through a
two-fold modification of the Boltzmann collision integral: First, the one-particle distribution
functions are evaluated at the locations of the centers of spheres, separated by the non- zero
distance at the impact. This makes the collision integral nonlocal in space. Second, the equilibrium
pair distribution function at the contact of the spheres enhances the scattering probability. The
proof of the $H$ theorem for the Enskog equation has posed certain difficulties, and has led to a
modification of the collision integral.

Methods of solution of the Enskog equation are immediate generalizations of those developed for the
Boltzmann equation, but there is one additional difficulty. The Enskog collision integral is nonlocal
in space. The Chapman-Enskog method, when applied to the Enskog equation, is supplemented with a
gradient expansion around the homogeneous equilibrium state.

\subsubsection{The Vlasov equation}

The Vlasov equation (or kinetic equation for a self-consistent force) is the nonlinear equation for
the one-body distribution function, which takes into account a long-range interaction between
particles:
\begin{eqnarray}
{\partial \over \partial t}f+\left(\vv,{\partial \over \partial \xx}f\right)+\left({\bf F},{\partial
\over
\partial \vv}f\right)=0,\nonumber
\end{eqnarray}
where  ${\bf F}=\int \Phi(\mid \xx-\xx'\mid){\xx-\xx'\over \mid \xx-\xx'\mid}n(\xx')d\xx'$ is the
self-consistent force. In this expression $ \Phi(\mid \xx-\xx'\mid){\xx-\xx'\over \mid \xx-\xx'\mid}$
is the microscopic force between the two particles, and $n(\xx')$ is the density of particles,
defined self-consistently, $n(\xx')=\int f(\xx',\vv)d\vv.$

 The Vlasov equation is used for a description of collisionless plasmas in
which case it is completed by a set of Maxwell equation for the electromagnetic field \cite{LPi}. It
is also used for a description of the gravitating gas.

The Vlasov equation is an infinite-dimensional Hamiltonian system. Many special and approximate
(wave-like) solutions to the Vlasov equation are known and they describe important physical effects.
One of the most well known effects is the Landau damping: The energy of a volume element dissipates
with the rate $$Q\approx -\mid E\mid^2{\omega(k)\over k^2}\left.{df_0\over dv}\right|_{v={\omega
\over k}},$$ where $f_0$ is the Maxwell distribution function, $\mid E \mid$ is the amplitude of the
applied monochromatic electric field with the frequency $\omega(k)$ , $k$ is the wave vector. The
Landau damping is thermodynamically reversible effect, and it is not accompanied with an entropy
increase. Thermodynamically reversed to the Landau damping is the plasma echo effect.

\subsubsection{The Fokker-Planck equation}

The Fokker-Planck equation (FPE) is a familiar model in various problems of nonequilibrium
statistical physics \cite{VanKampen,Risk}. We consider the FPE of the form
\begin{equation}
\label{SFP} {\partial W(\xx,t) \over \partial t} ={\partial \over \partial \xx} \left\{D
\left[W{\partial \over
\partial \xx} U +{\partial \over \partial \xx} W\right]\right\}.
\end{equation}
Here $W(\xx,t)$ is the probability density over the configuration space $x$, at the time $t$, while
$U(\xx)$ and $D(\xx)$ are the potential and the positively semi-definite ($ (\yy,D\yy)\ge 0$)
diffusion matrix.

The FPE (\ref{SFP}) is particularly important in studies of polymer solutions \cite{Bird,Doi,HCO}.
Let us recall the two properties of the FPE (\ref{SFP}), important to what will follow: (i).
Conservation of the total probability: $ \int W(\xx,t)d x=1.$ (ii). Dissipation: The equilibrium
distribution, $W_{eq}\propto\exp(-U)$, is the unique stationary solution to the FPE (\ref{SFP}). The
entropy,
\begin{equation}
\label{entropy} S[W]=-\int W(\xx,t) \ln\left[\frac{W(\xx,t)}{W_{eq}(\xx)}\right]dx,
\end{equation}
is a monotonically growing function due to the FPE (\ref{SFP}), and it arrives at the global maximum
in the equilibrium. These properties are most apparent when the FPE (\ref{SFP}) is rewritten as
follows:
\begin{equation}
\label{GENERIC}
\partial_{t}W(\xx,t)=\hat{M}_W\frac{\delta S[W]}{\delta W(\xx,t)},
\end{equation}
where $$\hat{M}_W=-{\partial \over \partial \xx} \left[W(\xx,t)D(\xx)  {\partial \over \partial \xx}
\right]$$ is a positive semi--definite symmetric operator with kernel $1$. The form (\ref{GENERIC})
is the dissipative part of a structure termed GENERIC (the dissipative vector field is a metric
transform of the entropy gradient) \cite{GENERIC,GENERIC1}.

The entropy does not depend on kinetic constants. It is the same for different details of kinetics
and depends only on equilibrium data. Let us call this property ``universality". It is known that for
the Boltzmann equation there exists only one universal Lyapunov functional. It is the entropy (we do
not distinguish functionals which are connected by multiplication on a constant or adding a
constant). But for the FPE there exists a big family of universal Lyapunov functionals. Let $h(a)$ be
a convex function of one variable $a\geq 0$, $h''(a)>0$,

\begin{equation} \label{FPES}
S_h[W]=-\int W_{eq}(\xx) h\left[\frac{W(\xx,t)}{W_{eq}(\xx)}\right]dx.
\end{equation}

The production of the generalized entropy $S_h$,  $\sigma_h$ is nonnegative:

\begin{equation} \label{sigmah}
\sigma_h(\xx)=W_{eq}(\xx)h''\left[\frac{W(\xx,t)}{W_{eq}(\xx)}\right]\left({\partial \over \partial
\xx}\frac{W(\xx,t)}{W_{eq}(\xx)},D{\partial \over \partial
\xx}\frac{W(\xx,t)}{W_{eq}(\xx)}\right)\geq 0.
\end{equation}

The most important variants for choice of $h$:

\noindent $h(a)=a\ln a$, $S_h$ is the Boltzmann--Gibbs--Shannon entropy (in the Kullback form
\cite{Kull,Pla}),

\noindent $h(a)=a\ln a-\epsilon \ln a$, $S_h^\epsilon$ is the maximal family of {\it additive}
entropies \cite{ENTR1,ENTR2,ENTR3} (these entropies are additive for composition of independent
subsystems).

\noindent $h(a)=\frac{1-a^q}{1-q}$, $S_h^q$ is the family of Tsallis entropies \cite{Tsa,Abe}. These
entropies are not additive, but become additive after nonlinear monotonous transformation. This
property can serve as definition of the Tsallis entropies in the class of generalized entropies
(\ref{FPES}) \cite{ENTR3}.

\subsection{\textbf{Equations of chemical kinetics and their reduction}}

\subsubsection{Outline of the dissipative reaction kinetics} \label{KINETICS} We begin with an outline
of the reaction kinetics (for details see e.\ g.\ the book
 \cite{Yab}). Let us consider a closed system with $n$ chemical species ${\rm A}_1,\dots,{\rm
A}_n$, participating in a complex reaction. The complex reaction is represented by the following
stoichiometric mechanism:
\begin{equation}
\label{stoi} \alpha_{s1}{\rm A}_1+\ldots+\alpha_{sn}{\rm A}_n\rightleftharpoons \beta_{s1}{\rm
A}_1+\ldots+\beta_{sn}{\rm A}_n,
\end{equation}
where the index $s=1,\dots,r$ enumerates the reaction steps, and where integers, $\alpha_{si}$ and
$\beta_{si}$,  are stoichiometric coefficients. For each reaction step $s$, we introduce
$n$--component vectors $\balpha_s$ and $\bbeta_s$ with components $\alpha_{si}$ and $\beta_{si}$.
Notation $\mbox{\boldmath$\gamma$}_s$  stands for the vector with integer components
$\gamma_{si}=\beta_{si}-\alpha_{si}$ (the stoichiometric vector).

For every $A_i$ an {\it extensive variable} $N_i$, ``the number of particles of that species", is
defined. The concentration of $A_i$ is $c_i=N_i/V$, where $V$ is the volume.

Given the stoichiometric mechanism (\ref{stoi}), the reaction kinetic equations read:
\begin{equation}\label{reaction}
\dot{\NN}= V\JJ(\cc),\ \JJ(\cc)=\sum_{s=1}^{r}\bgamma_sW_s(\cc),
\end{equation}
where dot denotes the time derivative, and $W_s$  is the reaction rate function of the step $s$. In
particular, {\it the mass action law} suggests the polynomial form of the reaction rates:
\begin{equation}
\label{MAL} W_s(\cc)=W_s^+(\cc) -  W_s^-(\cc) = k^+_s(T) \prod_{i=1}^{n}c_i^{\alpha_i} - k^-_s(T)
\prod_{i=1}^{n} c_i^{\beta_i},
\end{equation}
where $k^+_s(T)$ and $k^-_s(T)$ are the constants of the direct and of the inverse reactions rates of
the $s$th reaction step, $T$ is the temperature. The (generalized) Arrhenius equation gives the most
popular form of dependence $k^+_s(T)$:

\begin{equation}\label{Arr}
k^{\pm}_s(T)=a^{\pm}_s T^{b^{\pm}_s} \exp(S^{\pm}_s/k_B) \exp(-H^{\pm}_s/k_BT),
\end{equation}
\noindent where $a^{\pm}_s, \: b^{\pm}_s$ are constants, $H^{\pm}_s$ are activation ethalpies,
$S^{\pm}_s$ are activation entropies.

The rate constants are not independent. The {\it principle of detail balance} gives the following
connection between these constants: There exists such a positive vector $\cc^{\rm eq}(T)$ that

\begin{equation}\label{dbchem}
W_s^+(\cc^{\rm eq})=W_s^+(\cc^{\rm eq}) \: \mbox{for all}  \: s=1,\dots,r .
\end{equation}

The necessary and sufficient conditions for existence of such $\cc^{\rm eq}$ can be formulate as the
system of polynomial equalities for $\{k^{\pm}_s\}$, if the the stoichiometric vectors
$\{\bgamma_{s}\}$ are linearly dependent (see, for example, \cite{Yab}).

The reaction kinetic equations (\ref{reaction}) do not give us a closed system of equations, because
dynamics of the volume $V$ is not defined still. Four classical conditions for closure of this system
are well studied: $U,\: V = const$ (isolated system, $U$ is the internal energy); $H,\: P = const$
(thermal isolated isobaric system, $P$ is the pressure, $H=U+PV$ is the enthalpy), $V, \: T = const$
(isochoric isothermal conditions); $P, \: T = const$ (isobaric isothermal conditions). For $V, \: T =
const$ we do not need additional equations and data. It is possible just to divide equation
(\ref{reaction}) on the constant volume and write

\begin{equation}\label{creaction}
\dot{\cc}= \sum_{s=1}^{r}\bgamma_sW_s(\cc).
\end{equation}

For non-isothermal and non-isochoric conditions we do need addition formulae to derive $T$ and $V$.
For all four classical conditions the thermodynamic Lyapunov functions $G$ for kinetic equations are
known:
\begin{eqnarray}\label{tdlya}
U,\: V = const, \: G_{U,V}=-S/k_B; && V, \: T = const, \: G_{V,T}=F/k_BT=U/k_BT-S/k_B; \nonumber \\
H,\: P = const, \: G_{H,P}=-S/k_B; && P, \: T = const, \: G_{P,T}=G/T=H/k_BT-S/k_B,
\end{eqnarray}
\noindent where $F=U-TS$ is the free energy (Helmholtz free energy), $G=H-TS$ is the free entalphy
(Gibbs free energy). All the thermodynamic Lyapunov functions are normalized to dimensionless  scale
(if one measures the number of particles in moles, then it is necessary to change $k_B$ to $R$). All
these function decrease in time. For classical conditions the correspondent thermodynamic Lyapunov
functions can be written in the form: $G_{\bullet}(const, \NN)$. The derivatives $\partial
G_{\bullet}(const, \NN) / \partial N_i$ are the same functions of $\cc$ and $T$ for all classical
conditions:
\begin{equation} \label{mu}
\mu_i(\cc,T)=\frac{\partial G_{\bullet}(const, \NN)}{\partial N_i}=\frac{\mu_{{\rm chem}
i}(\cc,T)}{k_BT},
\end{equation}
where $\mu_{{\rm chem} i}(\cc,T)$ is the chemical potential of $A_i$.

Usual $G_{\bullet}(const, \NN)$ are strictly convex functions of $\NN$, and the matrix  $\partial
\mu_i / \partial c_j$ is positively defined. The dissipation inequality (\ref{Htheorem}) holds
\begin{equation} \label{Htheorem}
\frac{dG_{\bullet}}{dt}=V(\bmu,\JJ) \leq 0.
\end{equation}
This inequality is the restriction on  possible kinetic low and on possible values of kinetic
constants.


The most important generalization of the mass action law (\ref{MAL}) is  the Marcelin-De Donder
kinetic function. This generalization \cite{Fein,ByGoYa} is based on ideas of the thermodynamic
theory of affinity \cite{DeDonder36}. We use the kinetic function suggested in its final form in
\cite{ByGoYa}. Within this approach, the functions $W_s$ are constructed as follows: For a given
$\bmu(\cc,T)$ (\ref{mu}), and for a given stoichiometric mechanism (\ref{stoi}), we define the gain
($+$) and the loss ($-$) rates of the $s$th step,
\begin{equation}
\label{MDD} W_s^{+}=\varphi_s^+ \exp( \bmu,\mbox{\boldmath$\alpha$}_s),\quad W_s^{-}=\varphi_s^-\exp(
\bmu,\mbox{\boldmath$\beta$}_s),
\end{equation}
where $\varphi_s^{\pm}>0$ are kinetic factors. The Marcelin-De Donder kinetic function reads:
$W_s=W_s^+-W_s^-$, and the right hand side of the kinetic equation (\ref{reaction}) becomes,
\begin{equation}
\label{KINETIC MDD} \JJ=\sum_{s=1}^{r}\bgamma_s \{\varphi_s^+ \exp(\bmu,\balpha_s)-
\varphi_s^-\exp(\bmu,\bbeta_s)\}.
\end{equation}
For the Marcelin-De Donder reaction rate (\ref{MDD}), the dissipation inequality (\ref{Htheorem})
reads:
\begin{equation}
\label{HMDD} \dot{G}=\sum_{s=1}^{r} [(\bmu,\bbeta_s) - (\bmu,\balpha_s)] \left\{\varphi_s^+
e^{(\mbox{\boldmath {\scriptsize $\mu$}},\mbox{\boldmath {\scriptsize $\alpha$}}_s)}-
\varphi_s^-e^{(\mbox{\boldmath {\scriptsize $\mu$}},\mbox{\boldmath {\scriptsize
$\beta$}}_s)}\right\}\le 0.
\end{equation}
The kinetic factors $\varphi_s^{\pm}$ should satisfy certain conditions in order to make valid the
dissipation inequality (\ref{HMDD}). A well known sufficient condition is the detail balance:
\begin{equation}
\label{DB} \varphi_s^+=\varphi_s^-,
\end{equation}
other sufficient conditions are discussed in detail elsewhere \cite{G1,Yab,DKN97}.

For ideal systems, function $G_{\bullet}$ is constructed from the thermodynamic data of individual
species. It is convenient to start from the isochoric isothermal conditions. The Helmholtz free
energy for ideal system is
\begin{equation}
\label{Freen} F=k_BT \sum_i N_i[\ln c_i - 1 + \mu_{0i}]+const_{T,V},
\end{equation}
where the internal energy is assumed to be a linear function: $$U=\sum_i N_iu_i(T)=\sum_i N_i (u_{0i}
+ C_{Vi}T)$$ in given interval of $\cc$, $T$,  $u_i(T)$ is the internal energy of $A_i$ per particle.
It is well known that $S=-(\partial F/ \partial T)_{V,N=const}$,  $U=F+TS=F-T(\partial F/ \partial
T)_{V,N=const}$, hence, $u_i(T)=-k_BT^2d\mu_{0i}/dT$ and
\begin{equation}
\label{mu0} \mu_{0i}= \delta_i + u_{0i}/k_BT - (C_{Vi}/k_B)\ln T,
\end{equation}
where $\delta_i=const$, $C_{Vi}$ is the $A_i$ heat capacity at constant volume (per particle).

In concordance with the form of ideal free energy (\ref{Freen}) the expression for $\bmu$ is:
\begin{equation}
\label{muid} \mu_{i}=  \ln c_i +  \delta_i + u_{0i}/k_BT - (C_{Vi}/k_B)\ln T.
\end{equation}

For the function $\bmu$ of the form (\ref{muid}), the Marcelin-De Donder equation casts into the more
familiar mass action law form (\ref{MAL}). Taking into account the principle of detail balance
(\ref{DB}) we get the ideal rate functions:
\begin{eqnarray}
\label{MALMD} W_s(\cc)&=&W_s^+(\cc)-W_s^-(\cc), \nonumber \\ W_s^+(\cc)&=&\varphi(\cc,T) T^{-\sum_i
\alpha_{si}C_{Vi}/k_B} e^{\sum_i \alpha_{si}(\delta_i + u_{0i}/k_BT)}\prod_{i=1}^{n}c_i^{\alpha_i},
\nonumber \\ W_s^-(\cc) &=&\varphi(\cc,T) T^{-\sum_i \beta_{si}C_{Vi}/k_B} e^{\sum_i
\beta_{si}(\delta_i + u_{0i}/k_BT)}\prod_{i=1}^{n} c_i^{\beta_i}.
\end{eqnarray}
where $\varphi(\cc,T)$ is an arbitrary positive function from thermodynamic point of view.

Let us discuss further the vector field $\JJ(\cc)$ in the concentration space (\ref{creaction}).
Conservation laws (balances) impose linear constrains on admissible vectors $d\cc/dt$:
\begin{equation}
\label{conser} (\bb_i, \cc)=B_i=const,\ i=1,\dots,l,
\end{equation}
where $\bb_i$ are fixed and linearly independent vectors. Let us denote as $\BB$ the set of vectors
which satisfy the conservation laws (\ref{conser}) with given $B_i$:
\[
\BB=\left\{\cc|(\bb_1,\cc)=B_1,\dots, (\bb_l,\cc)=B_l\right\}.
\]
The natural phase space $\XX$ of the system (\ref{creaction}) is the intersection of the cone of
$n$-dimensional vectors with nonnegative components, with the set $\BB$, and ${\rm dim}\XX=d=n-l$. In
the sequel, we term a vector $\cc\in\XX$ the state of the system. In addition, we assume that each of
the conservation laws is supported by each elementary reaction step, that is
\begin{equation}
\label{sep} (\bgamma_s,\bb_i)=0,
\end{equation}
for each pair of vectors $\bgamma_s$ and $\bb_i$.

Reaction kinetic equations describe variations of the states in time.  The phase space $\XX$ is
positive-invariant of the system (\ref{creaction}): If $\cc(0)\in\XX$, then $\cc(t)\in\XX$ for all
the times $t>0$.

In the sequel, we assume that the kinetic equation (\ref{creaction}) describes  evolution towards the
unique equilibrium state, $\cc^{\rm eq}$, in the interior of the phase space $\XX$. Furthermore, we
assume that there exists a strictly convex function $G(\cc)$ which decreases monotonically in time
due to Eq.\ (\ref{creaction}):

Here $\bnabla G$ is the vector of partial derivatives $\partial G/\partial c_i$, and the convexity
assumes that the $n\times n$ matrices
\begin{equation}
\label{MATRIX} \HH_{{\mbox{\boldmath {\scriptsize $\cc$}}}}=\|\partial^2G(\cc)/\partial c_i\partial
c_j\|,
\end{equation}
are positive definite for all $\cc\in\XX$. In addition, we assume that the matrices (\ref{MATRIX})
are invertible if $\cc$ is taken in the interior of the phase space.

The function $G$ is the Lyapunov function of the system (\ref{reaction}), and $\cc^{\rm eq}$ is the
point of global minimum of the function $G$ in the phase space $\XX$. Otherwise stated, the manifold
of equilibrium states $\cc^{\rm eq}(B_1,\dots,B_l)$ is the solution to the variational problem,
\begin{equation}
\label{EQUILIBRIUM} G\to{\rm min}\ {\rm for\ }(\bb_i,\cc)=B_i,\ i=1,\dots,l.
\end{equation}
For each fixed value of the conserved quantities $B_i$, the solution is unique. In many cases,
however, it is convenient to consider the whole equilibrium manifold, keeping the conserved
quantities as parameters.

For example, for perfect systems in a constant volume under a constant temperature, the Lyapunov
function $G$ reads:
\begin{equation}
\label{gfun} G=\sum_{i=1}^{n}c_i[\ln(c_i/c^{\rm eq}_i)-1].
\end{equation}

It is important to stress that $\cc^{\rm eq}$ in Eq.\ (\ref{gfun}) is an {\it arbitrary} equilibrium
of the system, under arbitrary values of the balances. In order to compute $G(\cc)$, it is
unnecessary to calculate the specific equilibrium $\cc^{\rm eq}$ which corresponds to the initial
state $\cc$. Let us compare the Lyapunov function $G$ (\ref{gfun}) with the classical formula for the
free energy (\ref{Freen}). This comparison gives a possible choice for $\cc^{\rm eq}$:
\begin{equation}
\ln c^{\rm eq}_i = -\delta_i - u_{0i}/k_BT + (C_{Vi}/k_B)\ln T.
\end{equation}

\subsubsection{The problem of reduced description in chemical kinetics}

\label{reduction_review} What does it mean, ``to reduce the description of a chemical system''? This
means the following:
\begin{enumerate}
\item To shorten  the list of species.
This, in turn, can be achieved in two ways:

(i) To eliminate inessential components from the list;

(ii) To lump some of the species into integrated components.

\item To shorten the list of reactions. This also can be done in several ways:

(i) To eliminate inessential reactions, those which do not significantly influence the reaction
process;

(ii) To assume that some of the reactions ``have been already completed'', and that the equilibrium
has been reached along their paths (this leads to dimensional reduction because the rate constants of
the ``completed'' reactions are not used thereafter, what one needs are equilibrium constants only).

\item To decompose the motions into fast and slow, into independent (almost-independent)
and slaved etc. As the result of such a decomposition, the system admits a study ``in parts''. After
that, results of this study are combined into a joint picture. There are several approaches which
fall into this category. The famous method of the {\it quasi-steady state} (QSS), pioneered by
Bodenstein and Semenov,  follows the Chapman-Enskog method. The {\it partial equilibrium
approximations} are predecessors of the Grad method and quasiequilibrium approximations in physical
kinetics. These two family of methods have different physical backgrounds and mathematical forms.

\end{enumerate}

\subsubsection{Partial equilibrium approximations}\label{partial_eq}

{\it Quasi-equilibrium with respect to reactions} is constructed as follows: From the list of
reactions (\ref{stoi}), one selects those which are assumed to equilibrate first. Let they be indexed
with the numbers $s_1,\dots,s_k$. The quasi-equilibrium manifold is defined by the system of
equations,
\begin{equation}
\label{st1} W^+_{s_i}=W^-_{s_i},\ i=1,\dots,k.
\end{equation}
This system of equations looks particularly elegant when written in terms of conjugated (dual)
variables,  $\bmu=\bnabla G$:
\begin{equation}
\label{st2} ( \bgamma_{s_i},\bmu)=0,\ i=1,\dots,k.
\end{equation}
In terms of conjugated variables, the quasi-equilibrium manifold forms a linear subspace. This
subspace, $L^{\perp}$, is the orthogonal completement to the linear envelope of vectors, $L={\rm
lin}\{\bgamma_{s_1},\dots,\bgamma_{s_k}\}$.

{\it Quasi-equilibrium with respect to species} is constructed practically  in the same way but
without selecting the subset of reactions. For a given set of species, $A_{i_1}, \dots, A_{i_k}$, one
assumes that they evolve fast to equilibrium, and remain there. Formally, this means that in the
$k$-dimensional subspace of the space of concentrations with the coordinates $c_{i_1},\dots,c_{i_k}$,
one constructs the subspace $L$ which is defined by the balance equations, $( \bb_i,\cc)=0$. In terms
of the conjugated variables, the quasi-equilibrium manifold, $L^{\perp}$, is defined by equations,
\begin{equation}
\label{qe1} \bmu\in L^{\perp},\ (\bmu=(\mu_1,\dots,\mu_n)).
\end{equation}
The same quasi-equilibrium manifold can be also defined with the help of fictitious reactions: Let
$\bg_1,\dots,\bg_q$ be a basis in $L$. Then Eq.\ (\ref{qe1}) may be rewritten as follows:
\begin{equation}
\label{qe2} ( \bg_i,\bmu)=0,\ i=1,\dots,q.
\end{equation}

{\it Illustration:} Quasi-equilibrium with respect to reactions in hydrogen oxidation: Let us assume
equilibrium with respect to dissociation reactions, ${\rm H}_2\rightleftharpoons 2{\rm H}$, and,
${\rm O}_2\rightleftharpoons 2{\rm O}$, in some subdomain of reaction conditions. This gives:
\[k_1^+c_{{\rm H}_2}=k_1^-c^2_{\rm H},\ k_2^+c_{{\rm O}_2}=k_2^-c_{\rm O}^2.\]
Quasi-equilibrium with respect to species: For the same reaction, let us assume equilibrium over
${\rm H}$, ${\rm O}$, ${\rm OH}$, and ${\rm H}_2{\rm O}_2$, in a subdomain of reaction conditions.
Subspace $L$ is defined by balance constraints:
\[ c_{\rm H}+c_{\rm OH}+2c_{{\rm H}_2{\rm O}_2}=0,\ c_{\rm O}+c_{\rm OH}
+2c_{{\rm H}_2{\rm O}_2}=0.\] Subspace $L$ is two-dimensional. Its basis, $\{\bg_1,\bg_2\}$ in the
coordinates $c_{\rm H}$, $c_{\rm O}$, $c_{\rm OH}$, and $c_{{\rm H}_2{\rm O}_2}$ reads:
\[
\bg_1=(1,1,-1,0),\quad \bg_2=(2,2,0,-1).
\]
Corresponding Eq.\ (\ref{qe2}) is:
\[ \mu_{\rm H}+\mu_{\rm O}=\mu_{\rm OH},\ 2\mu_{\rm H}+2\mu_{\rm O}=
\mu_{{\rm H}_2{\rm O}_2}.\]

{\it General construction of the quasi-equilibrium manifold}: In the space of concentration, one
defines a subspace $L$ which satisfies the balance constraints:
\[ ( \bb_i,L)\equiv0.\]
The orthogonal complement of $L$ in the space with coordinates $\bmu=\bnabla G$ defines then the
quasi-equilibrium manifold $\bOmega_{L}$. For the actual computations, one requires the inversion
from $\bmu$ to $\cc$. Duality structure $\bmu\leftrightarrow\cc$ is well studied by many authors
\cite{Orlov84,DKN97}.

{\it Quasi-equilibrium projector.} It is not sufficient to just derive the manifold, it is also
required to define a {\it projector} which would transform the vector field defined on the space of
concentrations to a vector field on the manifold. Quasi-equilibrium manifold consists of points which
minimize $G$ on the affine spaces of the form $\cc+L$. These affine  planes are hypothetic planes of
fast motions ($G$ is decreasing in the course of the fast motions). Therefore, the quasi-equilibrium
projector maps the whole space of concentrations on $\bOmega_L$ parallel to $L$. The vector field is
also projected onto the tangent space of $\bOmega_L$ parallel to $L$.

Thus, the quasi-equilibrium approximation implies the decomposition of motions into the fast -
parallel to $L$, and the slow - along the quasi-equilibrium manifold. In order to construct the
quasi-equilibrium approximation, knowledge of reaction rate constants of ``fast'' reactions is not
required (stoichiometric vectors of all these fast reaction are in $L$, $\bgamma_{{\rm fast}}\in L$,
thus, knowledge of $L$ suffices), one only needs some confidence in that they all are sufficiently
fast \cite{Volpert85}. The quasi-equilibrium manifold itself is constructed based on the knowledge of
$L$ and of $G$. Dynamics on the quasi-equilibrium manifold is defined as the quasi-equilibrium
projection of the ``slow component'' of kinetic equations (\ref{reaction}).

\subsubsection{Model equations}

The assumption behind  the quasi-equilibrium is the hypothesis of the decomposition of motions into
fast and slow. The quasi-equilibrium approximation itself describes slow motions. However, sometimes
it becomes necessary to restore to the whole system, and to take into account the fast motions as
well. With this, it is desirable to keep intact one of the important advantages of the
quasi-equilibrium approximation -
 its independence of the rate constants of fast reactions.
For this purpose, the detailed fast kinetics is replaced by a model equation ({\it single relaxation
time approximation}).

{\it Quasi-equilibrium models} (QEM) are constructed as follows: For each concentration vector $\cc$,
consider the affine manifold, $\cc+L$. Its intersection with the quasi-equilibrium manifold
$\bOmega_L$ consists of one point. This point delivers the  minimum to $G$ on $\cc+L$. Let us denote
this point as $\cc^*_L(\cc)$. The equation of the quasi-equilibrium model reads:
\begin{equation}
\label{QEmodel} \dot{\cc}=-\frac{1}{\tau}[\cc-\cc^*_L(\cc)]+\sum_{{\rm
slow}}\bgamma_{s}W_s(\cc^*_L(\cc)),
\end{equation}
where $\tau>0$ is the relaxation time of the fast subsystem. Rates of slow reactions are computed at
the points $\cc^*_L(\cc)$ (the second term in the right hand side of Eq.\ (\ref{QEmodel}), whereas
the rapid motion is taken into account by a simple relaxational term (the first term in the right
hand side of Eq.\ (\ref{QEmodel}). The most famous model kinetic equation is the BGK equation in the
theory of the Boltzmann equation \cite{BGK}. The general theory of the quasi-equilibrium models,
including proofs of their thermodynamic consistency, was constructed in the paper \cite{GKMod}.

{\it Single relaxation time gradient models} (SRTGM) were considered in the context of the lattice
Boltzmann method for hydrodynamics \cite{AKJSP,AKMod}. These models are aimed at improving the
obvious drawback of quasi-equilibrium models (\ref{QEmodel}): In order to construct the QEM, one
needs to compute the function,
\begin{equation}
\label{QEA}
 \cc^*_L(\cc)=\arg\min_{\mbox{\boldmath {\scriptsize $x$}}\in
 \mbox{\boldmath {\scriptsize $c$}}+L,\ \mbox{\boldmath {\scriptsize $x$}}>0}G(\xxx).
\end{equation}
This is a convex programming problem. It does not always has a closed-form solution.

Let $\bg_1,\dots,\bg_k$ is the orthonormal basis of $L$. We denote as $\DD(\cc)$ the $k\times k$
matrix with the elements $( \bg_i,\HH_{\mbox{\boldmath {\scriptsize $\cc$}}} \bg_j)$, where
$\HH_{\mbox{\boldmath {\scriptsize $\cc$}}}$ is the matrix of second derivatives of $G$
(\ref{MATRIX}). Let $\CC(\cc)$ be the inverse of $\DD(\cc)$. The single relaxation time gradient
model has the form:
\begin{equation}
\dot{\cc}=-\frac{1}{\tau}\sum_{i,j}\bg_i\CC(\cc)_{ij}( \bg_j,\bnabla G) +\sum_{{\rm
slow}}\bgamma_{s}W_s(\cc).\label{SRTGM}
\end{equation}
The first term drives the system to the minimum of $G$ on $\cc+L$, it does not require solving the
problem (\ref{QEA}), and its spectrum in the quasi-equilibrium is the same as in the
quasi-equilibrium model (\ref{QEmodel}). Note that the slow component is evaluated in the ``current''
state $\cc$.

The equation (\ref{SRTGM}) has a simple form
\begin{equation}
\dot{\cc}=-\frac{1}{\tau} {\rm grad} G,
\end{equation}
if one calculates ${\rm grad} G$ with the entropic scalar product\footnote{Let us remind that ${\rm
grad} G$ is the Riesz representation of the differential of $G$ in the phase space \XX : $G(\cc +
\Delta\cc)=G(\cc)+ \langle {\rm grad} G(\cc), \Delta\cc \rangle + o(\Delta\cc)$. It depends on the
scalar product, and from thermodynamic point of view there is only one distinguished scalar product
in concentration space. Usual definition of ${\rm grad} G$ as the vector of partial derivatives
corresponds to the standard scalar product $(\bullet,\bullet)$.} $\langle \xx , \yy
\rangle=(\xx,\HH_{\mbox{\boldmath {\scriptsize $\cc$}}}\yy)$.

The models (\ref{QEmodel}) and (\ref{SRTGM})
 lift the quasi-equilibrium approximation to a kinetic equation
by approximating the fast dynamics with a single ``reaction rate constant'' - relaxation time $\tau$.

\subsubsection{Quasi-steady state approximation}\label{QSS}
The quasi-steady state approximation (QSS) is a tool used in a major number of works. Let us split
the list of species in two groups: The basic and the intermediate (radicals etc). Concentration
vectors are denoted accordingly, $\cc^{\rm s}$ (slow, basic species), and $\cc^{\rm f}$ (fast,
intermediate species). The concentration vector $\cc$ is the direct sum, $\cc=\cc^{\rm
s}\oplus\cc^{\rm f}$. The fast subsystem is Eq.\ (\ref{reaction}) for the component $\cc^{\rm f}$ at
fixed values of $\cc^{{\rm s}}$. If it happens that this way defined fast subsystem relaxes to a
stationary state, $\cc^{\rm f}\to\cc^{\rm f}_{\rm qss}(\cc^{\rm s})$, then the assumption that
$\cc^{\rm f}=\cc^{\rm f}_{\rm qss}(\cc)$ is precisely  the QSS assumption. The slow subsystem is the
part of the system (\ref{reaction}) for $\cc^{\rm s}$, in the right hand side of which the component
$\cc^{\rm f}$ is replaced with $\cc^{\rm f}_{\rm qss}(\cc)$. Thus, $\JJ=\JJ_{\rm s}\oplus\JJ_{\rm
f}$, where
\begin{eqnarray}
\dot{\cc}^{\rm f}&=&\JJ_{\rm f}(\cc^{\rm s}\oplus\cc^{\rm f}), \ \cc^{\rm s}={\rm const}; \quad
\cc^{\rm f}\to\cc^{\rm f}_{\rm qss}(\cc^{\rm s});\label{fast}\\ \dot{\cc}^{\rm s}&=&\JJ_{\rm
s}(\cc^{\rm s}\oplus\cc_{\rm qss}^{\rm f}(\cc^{\rm s})). \label{slow}
\end{eqnarray}
Bifurcations in the system (\ref{fast}) under variation of $\cc^{\rm s}$ as a parameter are
confronted to kinetic critical phenomena. Studies of more complicated dynamic phenomena in the fast
subsystem (\ref{fast}) require various techniques of averaging, stability analysis of the averaged
quantities etc.

Various versions of the QSS method are well possible, and are actually used widely, for example, the
hierarchical QSS method. There, one defines not a single fast subsystem but a hierarchy of them,
$\cc^{{\rm f}_1},\dots,\cc^{{\rm f}_k}$. Each subsystem  $\cc^{{\rm f}_i}$ is regarded as a slow
system for all the foregoing subsystems, and it is regarded as a fast subsystem for the following
members of the hierarchy. Instead of one system of equations (\ref{fast}), a hierarchy of systems of
lower-dimensional equations is considered, each of these subsystem is easier to study analytically.

Theory of singularly perturbed systems of ordinary differential equations is used to provide a
mathematical background and further development of the QSS approximation. In spite of a broad
literature on this subject, it remains, in general, unclear, what is the smallness parameter that
separates the intermediate (fast) species from the basic (slow). Reaction rate constants cannot be
such a parameter (unlike in the case of the quasi-equilibrium). Indeed, intermediate species
participate in the {\it same} reactions, as the basic species (for example, ${\rm
H}_2\rightleftharpoons 2{\rm H}$, ${\rm H}+{\rm O}_2\rightleftharpoons {\rm OH}+{\rm O}$). It is
therefore incorrect to state that $\cc^{\rm f}$ evolve faster than $\cc^{\rm s}$. In the sense of
reaction rate constants, $\cc^{\rm f}$ is not faster.

For catalytic reactions, it is not difficult to figure out what is the smallness parameter that
separates the intermediate species from the basic, and which allows to upgrade the QSS assumption to
a singular perturbation theory rigorously \cite{Yab}. This smallness parameter is the ratio of
balances: Intermediate species include the catalyst, and their total amount is simply significantly
less than the amount of all the $\cc_i$'s. After renormalizing to the variables of one order of
magnitude, the small parameter appears explicitly. The simplest example gives the catalytic reaction
$A+Z\rightleftharpoons AZ \rightleftharpoons P+Z$ (here $Z$ is a catalyst, $A$ and $P$ are an initial
substrate and a product). The kinetic equations are (in obvious notations):
\begin{eqnarray}\label{MihMen}
\dot{c}_A&=&-k^+_1 c_A c_Z + k^-_1 c_{AZ}, \nonumber \\ \dot{c}_Z &=& -k^+_1 c_A c_Z + k^-_1 c_{AZ} +
k^+_2 c_{AZ} - k^-_2 c_{Z}c_P, \nonumber \\ \dot{c}_{AZ}&=&k^+_1 c_A c_Z - k^-_1 c_{AZ} - k^+_2
c_{AZ} + k^-_2 c_{Z}c_P, \nonumber \\ \dot{c}_P&=&k^+_2 c_{AZ} - k^-_2 c_{Z}c_P.
\end{eqnarray}
The constants and the reactions rates are the same for concentrations $c_A, c_P$, and for $c_Z,
c_{AZ}$, and can not give a reason for relative slowness of $c_A, c_P$ in comparison with $c_Z,
c_{AZ}$, but there may be another source of slowness. There are two balances for this kinetics:
$c_A+c_P+c_{AZ}=B_A,$ $c_Z+c_{AZ}=B_Z$. Let us go to dimensionless variables: $$\varsigma_A=c_A/B_A,
\: \varsigma_P=c_P/B_A, \: \varsigma_Z=c_Z/B_Z, \: \varsigma_{AZ}=c_{AZ}/B_Z;$$
\begin{eqnarray}\label{MihMenDL}
\dot{\varsigma_A}&=&B_Z\left[-k^+_1 \varsigma_A \varsigma_Z + \frac{k^-_1}{B_A}\varsigma_{AZ}\right],
\nonumber \\ \dot{\varsigma_Z}&=&B_A\left[ -k^+_1 \varsigma_A \varsigma_Z +
\frac{k^-_1}{B_A}\varsigma_{AZ} + \frac{k^+_2}{B_A}\varsigma_{AZ} - k^-_2 \varsigma_Z \varsigma_P
\right], \nonumber \\ && \varsigma_A+\varsigma_P + \frac{B_Z}{B_A}\varsigma_{AZ}=1, \:
\varsigma_Z+\varsigma_{AZ}=1; \: \varsigma_{\bullet}\geq 0.
\end{eqnarray}
For $B_Z \ll B_A$ the slowness of $\varsigma_A, \: \varsigma_P$ is evident from these equations
(\ref{MihMenDL}).

 For usual radicals, the origin of the smallness parameter is quite similar. There are much
less radicals than the basic species (otherwise, the QSS assumption is inapplicable). In the case of
radicals, however, the smallness parameter cannot be extracted directly from balances $B_i$
(\ref{conser}). Instead, one can come up with a thermodynamic estimate: Function $G$ decreases in the
course of reactions, whereupon we obtain the limiting estimate of concentrations of any specie:
\begin{equation}
\label{TDlim} c_i\le \max_{G(\mbox{\boldmath {\scriptsize $c$}})\le G(\mbox{\boldmath {\scriptsize
$c$}}(0))} c_i,
\end{equation}
where $\cc(0)$ is the initial composition. If the concentration $c_{\rm R}$ of the radical R is small
both initially and in the equilibrium, then it should remain small also along the path to the
equilibrium. For example, in the case of ideal $G$ (\ref{gfun}) under relevant conditions, for any
$t>0$, the following inequality is valid:
\begin{equation}
\label{INEQ_R} c_{\rm R}[\ln(c_{\rm R}(t)/c_{\rm R}^{\rm eq})-1]\le G(\cc(0)).
\end{equation}
Inequality (\ref{INEQ_R}) provides the simplest (but rather coarse) thermodynamic estimate of $c_{\rm
R}(t)$ in terms of $G(\cc(0))$ and $c_{\rm R}^{\rm eq}$ {\it uniformly for $t>0$}. Complete theory of
thermodynamic estimates of dynamics has been developed in the book \cite{G1}.

One can also do computations without a priori estimations, if one accepts the QSS assumption until
the values $\cc^{\rm f}$ stay sufficiently small. It is the simplest way to operate with QSS: Just
use it {\bf until} $\cc^{\rm f}$ are small.

Let us assume that an a priori estimate has been found, $c_i(t)\le c_{i\ {\rm max}}$, for each $c_i$.
These estimate may depend on the initial conditions, thermodynamic data etc. With these estimates, we
are able to renormalize the variables in the kinetic equations (\ref{reaction}) in such a way that
renormalized variables take their values from the unit segment $[0,1]$: $\tilde{c}_i=c_i/c_{i\ {\rm
max}}$. Then the system (\ref{reaction}) can be written as follows:
\begin{equation}
\label{reduced} \frac{d\tilde{c}_i}{dt}=\frac{1}{c_{i\ {\rm max}}}J_i(\cc).
\end{equation}
The system of dimensionless parameters, $\epsilon_i=c_{i\ {\rm max}}/\max_i c_{i\ {\rm max}}$ defines
a hierarchy of relaxation times, and with its help one can establish various realizations of the QSS
approximation. The simplest version is the standard QSS assumption: Parameters $\epsilon_i$ are
separated in two groups, the smaller ones, and of the order $1$. Accordingly, the concentration
vector is split into $\cc^{\rm s}\oplus\cc^{\rm f}$. Various hierarchical QSS are possible, with
this, the problem becomes more tractable analytically.

There exist a variety of ways to introduce the smallness parameter into kinetic equations, and one
can find applications to each of the realizations. However, the two particular realizations remain
basic for chemical kinetics: (i) Fast reactions (under a given thermodynamic data); (ii) Small
concentrations. In the first case, one is led to the quasi-equilibrium approximation, in the second
case - to the classical QSS assumption. Both of these approximations allow for hierarchical
realizations, those which include not just two but many relaxation time scales. Such a {\it
multi-scale approach}  essentially simplifies analytical studies of the problem.

\subsubsection{Thermodynamic criteria for selection of important reactions}
One of the problems addressed by the sensitivity analysis is the selection of the important and
discarding the unimportant reactions. In the paper \cite{BYA77} was suggested a simple principle to
compare importance of different reactions according to their contribution to the entropy production
(or, which is the same, according to their contribution to $\dot{G}$). Based on this principle,
Dimitrov \cite{Dimitrov82} described domains of parameters in which the reaction of hydrogen
oxidation, ${\rm H}_2+{\rm O}_2+{\rm M}$, proceeds due to different mechanisms. For each elementary
reaction, he has derived the domain inside which the contribution of this reaction is essential
(nonnegligible). Due to its simplicity, this entropy production principle is especially well suited
for analysis of complex problems. In particular, recently, a version of the entropy production
principle was used in the problem of selection of boundary conditions for Grad's moment equations
\cite{Struchtrup98,GKZ02}. For ideal systems (\ref{gfun}), as well, as for the Marcelin-De Donder
kinetics (\ref{HMDD}) the contribution of the $s$th reaction to $\dot{G}$ has a particularly simple
form:
\begin{equation}
\label{dotGs} \dot{G}_{s}=-W_s\ln\left(\frac{W_s^+}{W_s^-}\right),\ \dot{G}=\sum_{s=1}^{r}\dot{G}_s.
\end{equation}

\subsubsection{Opening}

One of the problems to be focused on when studying closed systems is to prepare extensions of the
result for open or driven by flows systems. External flows are usually taken into account by
additional terms in the kinetic equations (\ref{reaction}):
\begin{equation}
\label{external} \dot{\NN}=V\JJ(\cc)+\bPi(\cc,t).
\end{equation}
It is important to stress here that the vector field $\JJ(\cc)$ in equations (\ref{external})  is the
same, as for the closed system, with thermodynamic restrictions, Lyapunov functions,  ets. The
thermodynamic structures are important for analysis of open systems (\ref{external}), if the external
flow $\bPi$ is  small in some sense, is linear function of $\cc$,  has small time derivatives, etc.
There are some general results for such ``weakly open" systems, for example the Prigogine minimum
entropy production theorem \cite{Prig} and the estimations of possible of steady states and limit
sets for open systems, based on thermodynamic functions and stoihiometric equations \cite{G1}.

There are general results for another limit case: for very intensive flow the dynamics is very simple
again \cite{Yab}. Let the flow has a natural structure:
$\bPi(\cc,t)=v_{in}(t)\cc_{in}(t)-v_{out}(t)\cc(t)$, where $v_{in}$ and $v_{out}$ are the rates of
inflow and outflow, $\cc_{in}(t)$ is the concentration vector for inflow. If $v_{out}$ is
sufficiently big, $v_{out}(t)>v_0$ for some critical value  $v_0$ and all $t>0$, then for the open
system (\ref{external}) the Lyapunov norm exists: for any two solutions $\cc^1(t)$ and $\cc^2(t)$ the
function $\|\cc^1(t) - \cc^2(t)\|$ monotonically decreases in time. Such a critical value  $v_0$
exists for any norm, for example, for usual Euclidian norm $\|\bullet\|^2=(\bullet,\bullet)$.

For arbitrary form of $\bPi$ the system (\ref{external}) can lost all signs of being thermodynamic
one. Nevertheless, thermodynamic structures often can help in the study of open systems.

The seminal questions are: What happens with slow/fast motion separation after opening? Which slow
invariant manifold for the closed system can be deformed to the slow invariant manifold for the open
system? Which slow invariant manifold for the closed system can be used as approximate slow invariant
manifold for the open system? There is more or less useful technique to seek the answers for specific
systems under consideration.

The way to study an open system as the result of opening a closed system may be fruitful. In any
case, out of this way we have just a general dynamical system (\ref{external}) and no hints what to
do with.

\begin{centering}

***

\end{centering}

Basic introductory textbook on physical kinetics of the Landau L. D. and Lifshitz E. M. Course of
Theoretical Physics \cite{LPi} contains many further examples and their applications.

Modern development of kinetics follows the route of specific numerical methods, such as direct
simulations. An opposite tendency is also clearly observed, and the kinetic theory based schemes are
increasingly used for the development of numerical methods and models in mechanics of continuous
media.

\section{\textbf{Invariance equation in the differential form}}\label{diff}

The notions and notations of this section will apply elsewhere below.

Definition of the invariance in terms of motions and trajectories assumes, at least, existence and
uniqueness theorems for solutions of the original dynamic system. This prerequisite causes
difficulties when one studies equations relevant to physical and chemical kinetics, such as, for
example, equations of hydrodynamics. Nevertheless, there exists a necessary \textit{differential
condition of invariance}: The vector field of the original dynamic system touches the manifold in
every point. Let us write down this condition in order to set up notation.

Let   $E$ be a linear space, let $U$ (the phase space) be a domain in $E$, and let a vector field
$J:\ U\rightarrow E$ be defined in $U$. This vector field defines the original dynamic system,
\begin{equation}\label{sys}
 \frac{dx}{dt}=J(x),\ x\in U.
\end{equation}

In the sequel, we consider submanifolds in $U$ which are parameterized with a given set of
parameters. Let a linear space of parameters $L$ be defined, and let $W$ be a domain in $L$. We
consider differentiable maps, $F:\ W\rightarrow U$, such that, for every $y\in W$, the differential
of $F$, $D_yF:\ L\rightarrow E$, is an isomorphism of $L$ on a subspace of $E$. That is, $F$ are the
manifolds, immersed in the phase space of the dynamic system (\ref{sys}), and parametrized by
parameter set $W$.

\textit{Remark:} One never discusses the choice of norms and topologies is such a general setting. It
is assumed that the corresponding choice is made appropriately in each specific case.

We denote $T_y$ the tangent space at the point $y$, $T_y=(D_yF)(L)$. {\it The  differential condition
of invariance} has the following form: For every $y\in W$,
\begin{equation}\label{diffinv}
  J(F(y))\in T_y.
\end{equation}

Let us rewrite the differential condition of invariance (\ref{diffinv}) in a form of a differential
equation. In order to achieve this, one needs to define a projector $P_y:\ E\rightarrow T_y$ for
every $y\in W$. Once a projector $P_y$ is defined, then condition (\ref{diffinv}) takes the form:
\begin{equation}\label{defect}
  \Delta_y=(1-P_y)J(F(y))=0.
\end{equation}
Obviously, by  $P_y^2=P_y$ we have, $P_y\Delta_y=0$. We refer to the function $\Delta_y$ as {\it the
defect of invariance} at the point $y$. The defect of invariance will be encountered oft in what will
follow.

Equation (\ref{defect}) is the first-order differential equation for the function $F(y)$. Projectors
$P_y$ should be tailored to the specific physical features of the problem at hand. A separate section
below is devoted to the construction of projectors. There we shall demonstrate how to construct a
projector, $P(x,T):\ E\rightarrow T$, given a point $x\in U$ and a specified subspace $T$. We then
set $P_y=P(F(y),T_y)$ in equation (\ref{defect}) \footnote{One of the main routes to define the field
of projectors $P(x,T)$ would be to make use of a Riemannian structure. To this end, one defines a
scalar product in $E$ for every point $x\in U$, that is, a bilinear form $\langle p | q\rangle_x$
with a positive definite quadratic form, $\langle p | p\rangle_x>0$, if $p\ne 0$. A good candidate
for such a scalar product is the bilinear form defined by the negative second differential of the
entropy at the point $x$, $-D^2S(x)$. As we demonstrate it later in this review, this choice is
essentially the only correct one close to the equilibrium. However, far from the equilibrium, an
improvement is required in order to guarantee the thermodymamicity condition, $\ker
P_y\subset\ker(D_xS)_{x=F(y)}$, for the field of projectors, $P(x,T)$, defined for any $x$ and $T$,
if $T\not{\!\subset} \ker D_xS.$ The thermodymamicity condition provides the preservation of the type
of dynamics: if $dS/dt > 0$ for initial vector field (\ref{sys}) in point $x=F(y)$, then $dS/dt > 0$
in this point $x$ for projected vector field $P_y(J(F(y)))$ too.}.

There are two possible senses of the notion ``approximate solution of invariance equations"
(\ref{defect}):
\begin{enumerate}
\item{The approximation of the solution;}
\item{The map $F$ with small defect of invariance (the right hand side approximation).}
\end{enumerate}
If someone is looking for the approximation of the first kind, then he needs theorems about existence
of solutions, he should find the estimations of deviations from the exact solution, because the right
hand side  not always gives the good estimation, etc.  The second kind of approximations does not
requires hypothesis of existence of solutions. Moreover, the manifold with sufficiently small defect
of invariance can serve as a slow manifold by itself. So, we shall accept the concept of approximate
invariant manifold (the manifold with small defect of invariance) instead of the approximation of the
invariant manifold (see also \cite{Kev,GaAr} and other works about approximate inertial manifolds).
Sometime these approximate invariant manifolds will give approximations of the invariant manifolds,
sometimes not, but it is additional and often difficult problem to make a distinction betwen these
situations. In addition to defect of invariance, the key role in analysis of motion separation into
the fast and the slow components play Jacobians, the differentials of $J(x)$. Some estimations of
errors of this separation will be presented below in the subsection devoted to {\it post-processing}.

Our paper is focused on nonperturbative methods for computing invariant manifolds, but it should be
mentioned that in the huge amount of applications the Taylor expansion is in use, and sometimes it
works rather well. The main idea is the continuation of slow manifold with respect to a small
parameter: Let our system depends on the parameter $\varepsilon$, and let  a manifold of steady
states exist for $\varepsilon=0$, as well, as fibers of motions towards these steady states, for
example
\begin{equation}
\dot{x}=\varepsilon f(x,y); \: \: \dot{y}= g(x,y).
\end{equation}
For $\varepsilon=0$ a value of (vector) variable $x$ is a vector of conserved quantities. Let for
every $x$ the equation of fast motion, $\dot{y}= g(x,y)$, be globally stable: Its solution $y(t)$
tends to the unique (for given $x$) stable fixed point $y_x$. If the function $g(x,y)$ meet the
conditions of the implicit function theorem, then the graph of the map $x\mapsto y_x$ forms a
manifold $\Omega_0=\{(x, y_x)\}$ of steady states. For small $\varepsilon>0$ we can look for the slow
manifold in a form of a series in powers of $\varepsilon$: $\Omega_{\varepsilon} = \{(x,
y(x,\varepsilon)\}$, $y(x,\varepsilon)=y_x+\varepsilon y^1(x)+\varepsilon^2 y^2(x)+\ldots .$ The
fibers of fast motions can be constructed in a  form of a power series too (the zero term is the fast
motion $\dot{y}= g(x,y)$ in the affine planes $x= {\rm const}$). This analytic continuation with
respect to the parameter $\varepsilon$ for small $\varepsilon>0$ is studied in the ``Geometric
singular perturbation theory" \cite{Fenichel,JonesCKRT}. As it was mentioned above, the first
successful application of such an approach to construction of a slow invariant manifold in a form of
Taylor expansion in powers of small parameter of singular perturbation $\varepsilon$ was the
Chapman-Enskog expansion \cite{Chapman}.

It is well-known in various applications that there are many different ways to introduce a small
parameter into a system, there are many ways to present a given system as a member of a
one-parametric family with a manifold of fixed points for zero value of a parameter. And different
ways of specification of such a parameter result in different definitions of slowness of positively
invariant manifold. Therefore it is desirable to study the notion of separation of motions without
such an artificial specification. {\it The notion of slow positively invariant manifold should be
intrinsic}. At least we should try to invent such a notion.

\clearpage

\section{\textbf{Film extension of the dynamics: Slowness as stability}}\label{membrane}

\subsection{\textbf{Equation for the film motion}}

One of the difficulties in the problem of reducing the description is caused by the fact that there
exist no commonly accepted formal definition of slow (and stable) positively invariant manifolds.
Classical definitions of stability and of the asymptotic stability of the invariant sets sound as
follows: Let a dynamic system be defined in some metric space, (so that we can measure distances
between points), and let $x(t,x_0)$ be a motion of this system at time $t$ with the initial condition
$x(0)=x_0$ at time $t=0$. The subset $S$ of the phase space is called \textit{invariant} if it is
made of whole trajectories, that is, if $x_0\in S$ then $x(t,x_0)\in S$ for all $t\in
(-\infty,\infty)$.

Let us denote as $\rho(x,y)$ the distance between the points $x$ and $y$. The distance from $x$ to a
closed set $S$ is defined as usual: $\rho(x,S)=\inf\{\rho(x,y)|y\in S\}$. The closed invariant subset
$S$ is called \textit{stable}, if for every $\epsilon>0$ there exists $\delta>0$ such that if
$\rho(x_0,S)<\delta$, then for every $t>0$ it holds $\rho(x(t,x_0),S)<\epsilon$. A closed invariant
subset $S$ is called \textit{asymptotically stable} if it is stable and attractive, that is, there
exists $\epsilon>0$ such that if $\rho(x_0,S)<\epsilon$, then $\rho(x(t,x_0),S)\rightarrow 0$ as
$t\rightarrow \infty$.

Formally, one can reiterate the definitions of stability and of the asymptotic stability for
positively invariant subsets. Moreover, since in the definitions mentioned above it goes only about
$t \geq 0$ or $t\rightarrow \infty$, it might seem that positively invariant subsets can be a natural
object of study for stability issues. Such conclusion is misleading, however. The study of the
classical stability of the positively invariant subsets reduces essentially to the notion of
stability of invariant sets - maximal attractors.

Let $Y$ be a closed positively invariant subset of the phase space. {\it The maximal attractor} for
$Y$ is the set $M_Y$,

\begin{equation}\label{maxattr}
  M_Y=\bigcap_{t\geq 0}T_t(Y),
\end{equation}

\noindent where $T_t$ is the shift operator for the time $t$:

\[T_t(x_0)=x(t,x_0).\]

\noindent The maximal attractor $M_Y$ is invariant, and the stability of $Y$ defined classically is
equivalent to the stability of $M_Y$ under any sensible assumption about homogeneous continuity (for
example, it is so for a compact phase space).

For systems which relax to a stable equilibrium, the maximal attractor is simply one and the same for
any bounded positively invariant subset, and it consists of a single stable point.

It is important to note that in the definition (\ref{maxattr}) one considers motions of a positively
invariant subset to equilibrium \textit{along itself}: $T_tY\subset Y$ for $t\geq 0$. It is precisely
this motion which is uninteresting from the perspective of the comparison of stability of positively
invariant subsets. If one subtracts this \textit{motion along itself} out of the vector field $J(x)$
(\ref{sys}), one obtains a less trivial picture.

We again assume submanifolds in $U$ parameterized with a single parameter set $F:\ W\rightarrow U$.
Note that there exist a wide class of transformations which do not alter the geometric picture of
motion: For a smooth diffeomorphism $\varphi: W\rightarrow W$ (a smooth coordinate transform), maps
$F$ and $F\circ \varphi$ define the same geometric pattern in the phase space.

Let us consider motions of the manifold $F(W)$ along solutions of equation (\ref{sys}). Denote as
$F_t$ the time-dependent map, and write equation of motion for this map:

\begin{equation}\label{tmap}
  \frac{dF_t(y)}{dt}=J(F_t(y)).
\end{equation}

Let us now subtract the component of the vector field responsible for the motion of the map $F_t(y)$
along itself from the right hand side of equation (\ref{tmap}). In order to do this, we decompose the
vector field $J(x)$ in each point $x=F_t(y)$ as

\begin{equation}\label{decomp}
  J(x)=J_{\parallel}(x)+J_{\perp}(x),
\end{equation}
where $J_{\parallel}(x)\in T_{t,y}$ ($T_{t,y}=(D_yF_t(y)(L)$). If projectors are well defined,
$P_{t,y}=P(F_t(y),T_{t,y})$, then decomposition (\ref{decomp}) has the form:
\begin{equation}\label{decomp2}
J(x)=P_{t,y}J(x)+(1-P_{t,y})J(x).
\end{equation}
Subtracting the component $J_{\parallel}$ from the right hand side of equation (\ref{tmap}), we
obtain,

\begin{equation}\label{tmapsubtr}
\frac{dF_t(y)}{dt}=(1-P_{t,y})J(F_t(y)).
\end{equation}

Note that the geometric pictures of motion corresponding to equations (\ref{tmap}) and
(\ref{tmapsubtr}) are identical \textit{locally} in $y$ and $t$. Indeed, the infinitesimal shift of
the manifold $W$ along the vector field is easily computed:
\begin{equation}\label{shiftalong}
  (D_yF_t(y))^{-1}J_{\parallel}(F_t(y))=(D_yF_t(y))^{-1}(P_{t,y}J(F_t(y))).
\end{equation}

\noindent This defines a smooth change of the coordinate system (assuming all solutions exist). In
other words, the component $J_{\perp}$ defines the motion of the manifold in $U$, while we can
consider (locally) the component $J_{\parallel}$ as a component which locally defines motions in $W$
(a coordinate transform).

The positive semi-trajectory of motion (for $t>0$) of any submanifold in the phase space along the
solutions of initial differential equation (\ref{sys}) (without subtraction of $J_{\parallel}(x)$) is
the positively invariant manifold. The closure of such semi-trajectory is an invariant subset. The
construction of the invariant manifold as a trajectory of an appropriate initial edge may be useful
for producing invariant exponentially attracting set \cite{IneManCFTe88,Robi}.  Very recently, the
notion of exponential stability of invariants manifold for ODEs was revised by splitting motions into
tangent and transversal (orthogonal) components in the work \cite{Shnol}.

We further refer to equation (\ref{tmapsubtr}) as {\it the film extension} of the dynamic system
(\ref{sys}). The phase space of the dynamic system (\ref{tmapsubtr}) is the set of maps $F$ (films).
Fixed points of equation (\ref{tmapsubtr}) are solutions to the invariance equation in the
differential form (\ref{defect}). These include, in particular, all positively invariant manifolds.
Stable or asymptotically stable fixed points of equation (\ref{tmapsubtr}) are slow manifolds we are
interested in. It is the notion of stability associated with the film extension of the dynamics which
is relevant to our study. Below in section \ref{sec:relax}, we consider relaxation methods for
constructing slow positively invariant manifolds on the basis of the film extension
(\ref{tmapsubtr}).



\subsection{\textbf{Stability of analytical solutions}}

When studying the Cauchy problem for equation (\ref{tmapsubtr}), one should ask a question of how to
choose the boundary conditions: Which conditions the function $F$ must satisfy at the boundary of
$W$? Without fixing the boundary conditions, the general solution of the Cauchy problem for the film
extension equations (\ref{tmapsubtr}) in the class of smooth functions on $W$ is essentially
ambiguous.

The boundary of $W$, $\partial W$, splits in two pieces: $\partial W= \partial W_+ \bigcup \partial
W_-$. For a smooth boundary these parts can be defined as
\begin{eqnarray}\label{bound+-}
&&\partial W_+=\{y\in \partial W | (\nu(y),(DF(y))^{-1}(P_yJ(F(y))))< 0\}, \nonumber \\ &&\partial
W_-=\{y\in \partial W | (\nu(y),(DF(y))^{-1}(P_yJ(F(y)))) \geq 0\}.
\end{eqnarray}

\noindent where $\nu(y)$ denotes the unit outer normal vector in the boundary point $y$,
$(DF(y))^{-1}$ is the isomorphism of the tangent space $T_y$ on the linear space of parameters $L$.

One can understand the boundary splitting (\ref{bound+-}) in such a way: The projected vector field
$P_yJ(F(y))$ defines dynamics on the manifold $F(W)$, this dynamics is the image of some dynamics on
$W$. The corresponding vector field on $W$ is $v(y)=(DF(y))^{-1}(P_yJ(F(y)))$. The boundary part
$\partial W_+$ consists of points $y$, where the velocity  vector $v(y)$ is pointed inside $W$, and
for $y\in\partial W_-$ this vector $v(y)$ is directed outside of $W$ (or is tangent to $\partial W$).
The splitting $\partial W= \partial W_+ \bigcup \partial W_-$ depends on $t$ with the vector field
$v(y)$: $$v_t(y)=(DF_t(y))^{-1}(P_yJ(F_t(y))),$$ and dynamics of $F_t(y)$ is determined by Eq.
(\ref{tmapsubtr}).

If we would like to derive a solution of the film extension, (\ref{tmapsubtr}) $F(y,t)$ for $(y,t)
\in W \times [0,\tau]$, for some time $\tau>0$,  then it is necessary to fix some boundary conditions
on $\partial W_+$ (for the ``incoming from abroad"  part of the function $F(y)$).

Nevertheless, there is a way to study equation (\ref{tmapsubtr}) without introducing any boundary
conditions. It is in the spirit of the classical Cauchy-Kovalevskaya theorem \cite{CoKo1,CoKo2,CoKo3}
about analytical Cauchy problem solutions with analytical data, as well as in the spirit of the
classical Lyapunov auxilary theorem about analytical invariant manifolds in the neighborhood of a
fixed point \cite{Lya,Kaz1} and H. Poincar\'e \cite{Lya2} theorem about analytical linearization of
analytical non-resonant contractions (see \cite{Lya1}).

We note in passing that recently, the interest to the classical analytical Cauchy problem revived in
the mathematical physics literature \cite{LevAnalit,TitiAnalit}. In particular, analogs of the
Cauchy-Kovalevskaya theorem were obtained for generalized Euler equations \cite{LevAnalit}. A
technique to estimate the convergence radii of the series emerging therein was also developed.

Analytical solutions to equation (\ref{tmapsubtr}) do not require boundary conditions on the boundary
of $W$. The analycity condition itself allows finding unique analytical solutions of the equation
(\ref{tmapsubtr}) with the analytical right hand side $(1-P)J$ for analytical initial conditions
$F_0$ in $W$ (assuming that such solutions exist). Of course, the analytical continuation without
additional regularity conditions is an ill-posed problem. However, it may be useful to go from
functions to germs\footnote{The germ is the sequences of Taylor coefficients that represent an
analytical function near a given point.}: we can solve chains of ordinary differential equations for
Taylor coefficients instead of partial differential equations for functions (\ref{tmapsubtr}), and
after that it may be possible to prove the convergence of the Taylor series thus obtained. This is
the way to prove the Lyapunov auxilary theorem \cite{Lya}, and one of the known ways to prove the
Cauchy-Kovalevskaya theorem.

Let us consider the system (1) with stable equilibrium point $x^*$, real analytical right hand side
$J$, and real analytical projector field $P(x,T):\ E\rightarrow T$.  We shall study real analytical
sub-manifolds, which include the equilibrium point point $x^*$ ($0\in W, F(0)=x^*$). Let us expand
$F$ in a Taylor series in the neighborhood of zero:

\begin{equation}
F(y) = x^* + A_1(y)+A_2(y,y) + \ldots + A_k(y,y,\ldots,y)+...\:,
\end{equation}

\noindent where $A_k(y,y,\ldots,y)$ is a symmetric $k$-linear operator ($k=1,2,\ldots$).

Let us expand also the right hand side of the film equation (\ref{tmapsubtr}). Matching operators of
the same order, we obtain a chain of equations for $A_1, \ldots , A_k, \ldots$:

\begin{equation}\label{systexpanded}
\frac{dA_k}{dt} = \Psi_k(A_1,\ldots,A_k).
\end{equation}

It is crucially important, that the dynamics of $A_k$ does not depend on $A_{k+1}, \ldots$, and
equations (\ref{systexpanded}) can be studied in the following order: we first study the dynamics of
$A_1$, then the dynamics of $A_2$ with the $A_1$ motion already given, then $A_3$ and so on.

Let the projector $P_{y}$ in equation (\ref{tmapsubtr}) be analytical function of the derivative
$D_yF(y)$ and of the deviation $x-x^*$. Let the correspondent Taylor expansion at the point
$(A_1^0(\bullet),x^*)$ have the form:

\begin{eqnarray}\label{proexpanded}
&&D_yF(y)(\bullet)=A_1(\bullet)+\sum_{k=2}^{\infty}kA_k(y,\ldots,\bullet), \\
&&P_{y}=\sum_{k,m=0}^{\infty}P_{k,m}(\underbrace{D_yF(y)(\bullet)-A_1^0(\bullet), \ldots ,
D_yF(y)(\bullet)-A_1^0(\bullet)}_{k} ; \underbrace{F(y)-x^*, \ldots , F(y)-x^*}_{m}), \nonumber
\end{eqnarray}

\noindent where  $A_1^0(\bullet)$, $A_1(\bullet)$, $A_k(y,\ldots,\bullet)$ are linear operators.
$P_{k,m}$ is a $k+m$-linear operator ($k,m=0,1,2,\ldots$) with values in the space of linear
operators $E\rightarrow E$. The operators $P_{k,m}$ depend on the operator $A_1^0(\bullet)$ as on a
parameter. Let the point of expansion $A_1^0(\bullet)$ be the linear part of $F$:
$A_1^0(\bullet)=A_1(\bullet)$.

Let us represent the analytical vector field $J(x)$ as a power series:

\begin{equation}\label{fieldexpanded}
J(x)=\sum_{k=1}^{\infty}J_k(x-x^*,\ldots,x-x^*),
\end{equation}

\noindent where $J_k$ is a symmetric $k$-linear operator ($k=1,2,\ldots$).

Let us write, for example, the first two equations of the equation chain (\ref{systexpanded}):

\begin{eqnarray}\label{firstexpanded}
\frac{dA_1(y)}{dt}& =& (1-P_{0,0})J_1(A_1(y)), \nonumber \\
\frac{dA_2(y,y)}{dt}&=&(1-P_{0,0})[J_1(A_2(y,y))+J_2(A_1(y),A_1(y))]-  \nonumber \\
&&[2P_{1,0}(A_2(y,\bullet))+P_{0,1}(A_1(y))]J_1(A_1(y)).
\end{eqnarray}

\noindent Here operators $P_{0,0}, \: P_{1,0}(A_2(y,\bullet)), \: P_{0,1}(A_1(y))$ parametrically
depend on the operator $A_1(\bullet)$, hence, the first equation is nonlinear, and the second is
linear with respect to $A_2(y,y)$. The leading term in the right hand side has the same form for all
equations of the sequence (\ref{systexpanded}):

\begin{equation}\label{mainglied}
\frac{dA_n(y, \ldots , y)}{dt} = (1-P_{0,0})J_1(A_n(y, \ldots ,
y))-nP_{1,0}(A_n(\underbrace{y,\ldots,y}_{n-1}, \bullet))J_1(A_1(y)) + \ldots \: .
\end{equation}

There are two important conditions on $P_{y}$ and $D_yF(y)$: $P_{y}^2=P_{y}$, because $P_{y}$ is a
projector, and $\mbox{im}P_{y}=\mbox{im}D_yF(y),$ because $P_{y}$ projects on the image of $D_yF(y)$.
If we expand these conditions in the power series, then we get the conditions on the coefficients.
For example, from the first condition we get:
\begin{eqnarray}\label{addcon}
&&P_{0,0}^2=P_{0,0}, \nonumber \\
&&P_{0,0}[2P_{1,0}(A_2(y,\bullet))+P_{0,1}(A_1(y))]+[2P_{1,0}(A_2(y,\bullet))+P_{0,1}(A_1(y))]P_{0,0}=
\nonumber \\ && 2P_{1,0}(A_2(y,\bullet))+P_{0,1}(A_1(y)), \ldots \: .
\end{eqnarray}
After multiplication the second equation in (\ref{addcon}) with $P_{0,0}$  we get
\begin{equation}\label{P^2exp}
P_{0,0}[2P_{1,0}(A_2(y,\bullet))+P_{0,1}(A_1(y))]P_{0,0}=0.
\end{equation}
Similar identities can be obtained for any oder of the expansion.  These equalities allow us to
simplify the stationary equation for the sequence (\ref{systexpanded}). For example, for the first
two equations of this sequence (\ref{firstexpanded}) we obtain the following stationary equations:

\begin{eqnarray}\label{stati}
&&(1-P_{0,0})J_1(A_1(y))=0, \nonumber  \\ && (1-P_{0,0})[J_1(A_2(y,y))+J_2(A_1(y),A_1(y))]- \nonumber
\\ && [2P_{1,0}(A_2(y,\bullet))+P_{0,1}(A_1(y))]J_1(A_1(y))=0.
\end{eqnarray}

The operator $P_{0,0}$ is the projector on the space $\mbox{im}A_1$ (the image of $A_1$), hence, from
the first equation in (\ref{stati}) it follows: $J_1(\mbox{im}A_1)\subseteq \mbox{im}A_1$. So,
$\mbox{im}A_1$ is a $J_1$-invariant subspace in $E$ ($J_1=D_x J(x)|_{x^*}$) and
$P_{0,0}(J_1(A_1(y))\equiv J_1(A_1(y)$. It is equivalent to the first equation of (\ref{stati}). Let
us multiply the second equation of (\ref{stati}) with $P_{0,0}$ on the left. As a result we obtain
the condition: $$P_{0,0}[2P_{1,0}(A_2(y,\bullet))+P_{0,1}(A_1(y))]J_1(A_1(y))=0,$$ for solution of
equations (\ref{stati}), because $P_{0,0}(1-P_{0,0})\equiv 0$. If $A_1(y)$ is a solution of the first
equation of (\ref{stati}), then this condition becomes an identity, and we can write the second
equation of (\ref{stati}) in the form
\begin{eqnarray}\label{stati2}
&&(1-P_{0,0})\times   \\ && [J_1(A_2(y,y))+J_2(A_1(y),A_1(y))-
(2P_{1,0}(A_2(y,\bullet))+P_{0,1}(A_1(y)))J_1(A_1(y))]=0. \nonumber
\end{eqnarray}

It should be stressed, that the choice of projector field $P_{y}$ (\ref{proexpanded}) has impact only
on the $F(y)$ parametrization, whereas the invariant geometrical properties of solutions of
(\ref{tmapsubtr}) do not depend on projector field if some transversality and analycity conditions
hold. The conditions of thermodynamic structures preservation significantly reduce ambiguousness of
the projector choice. One of the most important condition is $\ker P_y \subset \ker D_xS$, where
$x=F(y)$ and $S$ is the entropy (see the section about the entropy below). The thermodynamic
projector is the unique operator which transforms the arbitrary vector field equipped with the given
Lyapunov function into a vector field with the same Lyapunov function on the arbitrary submanifold
which is not tangent to the level of the Lyapunov function. For the thermodynamic projectors $P_y$
the entropy $S(F(y))$ conserves on solutions $F(y,t)$ of the equation (\ref{tmapsubtr}) for any $y\in
W$.

If projectors $P_y$ in equations (\ref{proexpanded})-(\ref{stati2}) are thermodynamic, then $P_{0,0}$
is the orthogonal projector  with respect to the entropic scalar product\footnote{This scalar product
is the bilinear form defined by the negative second differential of the entropy at the point $x^*$,
$-D^2S(x)$.}. For orthogonal projectors the operator $P_{1,0}$ has a simple explicit form. Let $A:L
\rightarrow E$ be an isomorphic injection (an isomorphism on the image), and $P:E\rightarrow E$ be
the orthogonal projector on the image of $A$. The orthogonal projector on the image of perturbed
operator $A+ \delta A$ is $P+ \delta P$,
\begin{eqnarray} \label{P_1,0}
\delta P &  = & (1-P)\delta A A^{-1}P+(\delta A A^{-1}P)^+(1-P) + o(\delta A),  \nonumber \\
P_{1,0}(\delta A(\bullet)) & = & (1-P)\delta A(\bullet) A^{-1}P+(\delta A(\bullet) A^{-1}P)^+(1-P).
\end{eqnarray}
Here, in (\ref{P_1,0}), the operator $A^{-1}$ is defined on $\mbox{im}A$, $\mbox{im}A=\mbox{im}P$,
the operator $A^{-1}P$ acts on $E$.

Formula for $\delta P$ (\ref{P_1,0}) follows from the three conditions:
\begin{equation} \label{Pcon}
(P+\delta P)(A+ \delta A)=A+ \delta A, \: (P+\delta P)^2=P+\delta P, \:(P+\delta P)^+=P+\delta P.
\end{equation}

Every $A_k$ is driven by $A_1, \ldots ,A_{k-1}$. Stability of the germ of the positively invariant
analytical manifold $F(W)$ at the point $0$ ($F(0)=x^*$) is defined as stability of the solution of
corresponding equations sequence (\ref{systexpanded}). Moreover, the notion of the $k$-jet stability
can be useful: let's call $k$-jet stable such a germ of positively invariant manifold $F(M)$ at the
point $0$ ($F(0)=x^*$), if the corresponding solution of the equations sequence (\ref{systexpanded})
is stable for $k=1,\ldots,n$. The simple ``triangle" structure of the equation sequence
(\ref{systexpanded}) with the form (\ref{mainglied}) of principal linear part makes the problem of
jets stability very similar for all orders $n>1$.

Let us demonstrate the stability conditions for the 1-jets in a $n$-dimensional space $E$. Let the
Jacobian matrix $J_1=D_x J(x)|_{x^*}$ be selfadjoint with a simple spectrum $\lambda _1 , \ldots,
\lambda _n$, and the projector $P_{0,0}$ be orthogonal (this is a typical ``thermodynamic"
situation). Eigenvectors of $J_1$ form a basis in $E$: $\{e_i\}_{i=1}^n$. Let a linear space of
parameters $L$ be a $k$-dimensional real space, $k<n$. We shall study stability of a operator $A_1^0$
which is a fixed point for the first equation of the sequence (\ref{systexpanded}).   The operator
$A_1^0$ is a fixed point of this equation, if $\mbox{im}A_1^0$ is a $J_1$-invariant subspace in $E$.
We discuss full-rank operators, so, for some order of $\{e_i\}_{i=1}^n$ numbering, the matrix of
$A_1^0$ should have a form: $a_{1ij}^0=0$, if $i> k$. Let us choose the basis in $L$:
$l_j=(A_1^0)^{-1}e_j, \: (j=1, \ldots, k)$. For this basis $a_{1ij}^0=\delta _{ij}, \: (i=1, \ldots,
n, \: j=1, \ldots, k$, $\delta _{ij}$ is the Kronecker symbol). The corresponding projectors $P$ and
$1-P$ have the matrices:
\begin{equation}
P=\mbox{diag}(\underbrace{1,\ldots,1}_k,\underbrace{0,\ldots,0}_{n-k}), \:
1-P=\mbox{diag}(\underbrace{0,\ldots,0}_k,\underbrace{1,\ldots,1}_{n-k}),
\end{equation}
where $\mbox{diag}(\alpha _1 , \ldots, \alpha _n)$ is the $n \times n$ diagonal matrix with numbers
$\alpha _1 , \ldots, \alpha _n$ on the diagonal.

Equations of the linear approximation for the dynamics of the deviations $\delta A$ read:
\begin{equation}\label{dotA}
\frac{d \delta A}{dt}=\mbox{diag}(\underbrace{0,\ldots,0}_k,\underbrace{1,\ldots,1}_{n-k})
[\mbox{diag}(\lambda _1, \ldots, \lambda _n)\delta A - \delta A \mbox{diag}(\underbrace{\lambda
_1,\ldots,\lambda _k}_k)].
\end{equation}

The time derivative of $A$ is orthogonal to $A$: for any $y,z \in L$ the equality
$(\dot{A}(y),A(x))=0$ holds, hence, for the stability analysis it is necessary and sufficient to
study $\delta A$ with $\mbox{im}\delta A_1^0 \perp \mbox{im}A$. The matrix for  such a $\delta A$ has
a form: $$\delta a_{ij}=0, \: \mbox{if} \: i\leq k. $$ For $i=k+1, \ldots , n$, $j=1, \ldots , k$
equation (\ref{dotA}) gives:
\begin{equation}\label{lambdalambda}
\frac{d \delta a_{ij}}{dt}=(\lambda _i-\lambda _j)\delta a_{ij}.
\end{equation}

From equation (\ref{lambdalambda}) the stability condition follows:

\begin{equation}\label{stacon}
\lambda _i-\lambda _j < 0 \: \mbox{for all} \: i>k, \: j\leq k.
\end{equation}
This means that the relaxation {\it towards}  $\mbox{im}A$ (with the spectrum of relaxation times
$|\lambda _i|^{-1}$ $(i=k+1, \ldots, n)$) is  faster, then the relaxation {\it along} $\mbox{im}A$
(with the spectrum of relaxation times $|\lambda _j|^{-1}$ $(j=1, \ldots, k)$).

Let the condition (\ref{stacon}) holds. The relaxation time for the film (in the first approximation)
is: $$\tau=1/(min_{i>k}|\lambda _i|-max_{j\leq k}|\lambda _j|),$$ thus it depends on the {\bf
spectral gap} in the spectrum of the operator $J_1=D_x J(x)|_{x^*}$.

It is the gap between spectra of two restrictions of the operator $J_1$, $J_1^{\|}$ and $J_1^{\bot}
$, respectively. The operator $J_1^{\|}$ is the restriction of $J_1$ on the $J_1$-invariant subspace
$\mbox{im}A_1^0$ (it is the tangent space to the slow invariant manifold at the point $x^*$). The
operator $J_1^{\bot}$ is the restriction of $J_1$ on the orthogonal complement to $\mbox{im}A_1^0$.
This subspace is also $J_1$-invariant, because $J_1$ is selfadjoint. The spectral gap between spectra
of these two operators is the spectral gap between relaxation {\it towards} the slow manifold and
relaxation {\it along} this manifold.

The stability condition (\ref{stacon}) demonstrates that our formalization of the slowness of
manifolds as the stability of fixed points for the film extension (\ref{tmapsubtr}) of initial
dynamics met the intuitive expectations.

For the analysis of the system (\ref{systexpanded}) in the neighborhood of some manifold $F_0$
($F_0(0)=x^*$), the following parametrization can be convenient. Let's consider
$F_0(y) = A_1(y)+...,$  $T_0  = A_1(L)$ is a tangent space to $F_0(W)$ at the point $x^*$,  $E = T_0
\bigoplus H$ is the direct sum decomposition.


We shall consider analytical sub-manifolds in the form
\begin{equation}\label{submanif}
x = x^* + (y,\Phi(y)),
\end{equation}
\noindent where $y \in W_0 \subset T_0$, $W_0$ is neighborhood of zero in $T_0$, $\Phi(y)$ is an
analytical map of $W_0$ in $H$, $\Phi(0)=0$.

Any analytical manifold close to $F_0$ can be represented in this form.

Let us define the projector $P_y$ that corresponds to the decomposition (\ref{submanif}), as the
projector on $T_y$ parallel to $H$. Furthermore, let us introduce the corresponding decomposition of
the vector field $J = J_y\bigoplus J_z, J_y \in T_0, J_z \in H$. Then

\begin{equation}\label{MembrProjector}
P_y(J) = (J_y,(D_y\Phi(y))J_y).
\end{equation}

The corresponding equation of motion of the film (\ref{tmapsubtr})  has the following form:

\begin{equation}\label{MotEquat}
\frac{d\Phi(y)}{dt} = J_z(y,\Phi(y)) - (D_y\Phi(y))J_y(y,\Phi(y)).
\end{equation}

If $J_y$ and $J_z$ depend analytically on their arguments, then from (\ref{MotEquat}) one can easily
obtain a hierarchy of equations of the form (\ref{systexpanded}) (of course, $J_y(x^*)=0, \
J_z(x^*)=0$).

Using these notions, it is convenient to formulate the Lyapunov Auxiliary Theorem \cite{Lya}. Let
$T_0=R^m, H=R^p$, and in $U$ an analytical vector field is defined $J(y,z)=J_y(y,z)\bigoplus
J_z(y,z)$, ($y\in T_0, z\in H$), and the following conditions are satisfied:

1) $J(0,0) = 0$;

2) $D_zJ_y(y,z)\big|_{(0,0)} = 0$;

3) $0\notin conv\{k_1,..,k_m\}$, where $k_1,..,k_m$ are the eigenvalues of
$D_yJ_y(y,z)\big|_{(0.0)}$, and $conv\{k_1,..,k_m\}$ is the convex envelope of $\{k_1,..,k_m\}$;

4) the numbers $k_i$ and $\lambda_j$ are not related by any equation of the form

\begin{equation}\label{Res}
\sum_{i=1}^{m}m_ik_i = \lambda_j,
\end{equation}

\noindent where $\lambda_j \; (j=1,..,p)$ are eigenvalues of $D_zJ_z(y,z)\big|_{(0,0)}$, and $m_i\geq
0$ are integers, $\sum_{i=1}^{m} m_i >0$.

Let us consider analytical manifold $(y,\Phi(y))$ in $U$ in the neighborhood of zero ($\Phi(0)=0$)
and write for it the differential invariance equation with the projector (\ref{MembrProjector}):

\begin{equation}
(D_y\Phi(y))J_y(y,\Phi(y)) = J_z(y,\Phi(y)).
\end{equation}

{\it Lyapunov Auxiliary theorem.} Given conditions 1-4, equation (\ref{submanif}) has the unique
analytical in the neighborhood of zero solution, satisfying condition $\Phi(0)=0$.

Recently various new applications of this theorem  were developed \cite{Kaz1,Kaz2,Krener,Kaz3}.

Studying germs of invariant manifolds using Taylor expansion in a neighborhood of a fixed point is
definitely useful from the theoretical as well as from the practical perspective. But the well known
difficulties pertinent to this approach, of convergence, of small denominators (connected with
proximity to the resonances (\ref{Res})) and others call for development of different methods. A hint
can be found in the famous {\bf KAM} theory: one should use iterative methods instead of the Taylor
expansion \cite{KAM,KAM1,KAM2}. Below we present two such methods:

\begin{itemize}
\item The Newton method subject to incomplete linearization;
\item The relaxation method which is the Galerkin-type approximation
to Newton's method with projection on defect of invariance (\ref{defect}), i.e. on the right hand
side of equation (\ref{tmapsubtr}).
\end{itemize}

\section{\textbf{Entropy, quasiequilibrium and projectors field}}\label{qe}

Projector operators $P_y$ contribute both to the invariance equation (\ref{diffinv}), and to the film
extension of the dynamics (\ref{tmapsubtr}). Limiting results, exact solutions, etc. only weakly
depend on the particular choice of projectors, or do not depend at all on it. However, the validity
of approximations obtained in each iteration step towards the limit does strongly depend on the
choice of the projector. Moreover, if we want each approximate solution to be consistent with such
physically crucial conditions as the second law of thermodynamics (the entropy of the isolated
systems increases), then the choice of the projector becomes practically unique.

In this section we consider the main ingredients for constructing the projector, based on the two
additional structures: (a) The moment parameterization, and (b) The entropy and the entropic scalar
product.

\subsection{\textbf{Moment parameterization}}

Same as in the previous section, let a regular map (projection) is defined, $\Pi: U\rightarrow W$. We
consider only maps $F: W \rightarrow U$ which  satisfy $\Pi\circ F=1$. We seek slow invariant
manifolds among such maps. (A natural remark is in order here: sometimes one has to consider $F$
which are defined not on the whole $W$ but only on some subset of it.) In this case, the unique
projector consistent with the given structure is the superposition of the differentials:
\begin{equation}\label{momproj}
  P_yJ=(D_yF)_y\circ(D_x\Pi)_{F(y)}.
\end{equation}
In the language of differential equations, formula (\ref{momproj}) has the following significance:
First, equation (\ref{sys}) is projected,
\begin{equation}\label{fine2coarse}
  \frac{dy}{dt}=(D_x\Pi)_{F(y)}J(F(y)).
\end{equation}
Second, the latter equation is lifted back to $U$ with the help of $F$ and its differential,
\begin{equation}\label{coarse2fine}
 x(t)=F(y(t)),\ \left.\frac{dx}{dt}\right|_{x=F(y)}=(D_yF)_y \left(\frac{dy}{dt}\right)=(D_yF)_y((D_x\Pi)_{F(y)}J(F(y)))=P_yJ.
\end{equation}
The most standard example of the construction just described is as follows: $x$ is the distribution
density, $y=\Pi(x)$ is the set of selected moments of this density, $F: y\rightarrow x$ is a
``closure assumption", which constructs a distribution density parameterized by the values of the
moments $y$. Another standard example is relevant to problems of chemical kinetics: $x$ is a detailed
description of the reacting species (including all the intermediates and radicals), $y$ are
concentrations of stable reactants and products of the reaction.

The moment parameterization and moment projectors (\ref{momproj}) are often encountered in the
applications. However, they have some shortcomings. In particular, it is by far not always happen
that the moment projection  \textit{transforms a dissipative system into another dissipative system}.
Of course, for invariant $F(y)$ \textit{any} projector transforms the dissipative system into a
dissipative system. However, for various approximations to invariant manifolds (closure assumptions)
this is not readily the case\footnote{See, e.\ g. a discussion of this problem for the
Tamm--Mott-Smith approximation for the strong shock wave in \cite{GK1}.}. The property of projectors
to \textit{preserve the type of the dynamics} will be imposed below as one of the requirements.

\subsection{\textbf{Entropy and quasiequilibrium}}

The dissipation properties of the system (\ref{sys}) are described by specifying the \textit{entropy}
$S$, the distinguished \textit{Lyapunov function} which monotonically increases along solutions of
equation (\ref{sys}). In a certain sense, this Lyapunov function is more fundamental than the system
(\ref{sys}) itself. That is, usually, the entropy is known much better than the right hand side of
equation (\ref{sys}). For example, in chemical kinetics, the entropy is obtained from the
\textit{equilibrium} data. The same holds for other Lyapunov functions, which are defined by the
entropy and by specification of the reaction conditions (the free energy, $U-TS$, for the isothermal
isochoric processes, the free enthalpy, $U-TH$, for the isothermal isobaric processes etc.). On
physical grounds, all these entropic Lyapunov functions are proportional (up to additive constants)
to the entropy of the minimal isolated system which includes the system under study \cite{G1}. In
general, with some abuse of language, we term the Lyapunov functional $S$ the entropy elsewhere
below, although it is a different functional for non-isolated systems.

Thus, we assume that a concave functional $S$ is defined in $U$, such that it takes maximum in an
inner point $x^*\in U$. This point is termed the equilibrium.

For any dissipative system (\ref{sys}) under consideration in $U$, the derivative of $S$ due to
equation (\ref{sys}) must be nonnegative,

\begin{equation}\label{sign}
  \frac{dS}{dt}\bigg|_x=(D_xS)(J(x))\geq0,
\end{equation}
where $D_xS$ is the linear functional, the differential of the entropy, while the equality in
(\ref{sign}) is achieved only in the equilibrium $x=x^*$.

Most of the works on nonequilibrium thermodynamics deal with corrections to quasiequilibrium
approximations, or with applications of these approximations (with or without corrections). This
viewpoint is not the only possible but it proves very efficient for the construction of a variety of
useful models, approximations and equations, as well as methods to solve them\footnote{From time to
time it is  discussed in the literature, who was the first to introduce the quasiequilibrium
approximations, and how to interpret them. At least a part of the discussion is due to a different
r\^{o}le the quasiequilibrium plays in the entropy--conserving and the dissipative dynamics. The very
first use of the entropy maximization dates back to the classical work of G.\ W.\ Gibbs \cite{Gibb},
but it was first claimed for a principle of informational statistical thermodynamics by E.\ T.\
Jaynes  \cite{Janes1}. Probably the first explicit and systematic use of quasiequilibria to derive
dissipation from entropy--conserving systems is due to the works of D.\ N.\ Zubarev. Recent detailed
exposition is given in \cite{Zubarev}. The method of nonequilibrium ensemble was developed also by Eu
\cite{Eu}. For dissipative systems, the use of the quasiequilibrium to reduce description can be
traced to the works of H.\ Grad on the Boltzmann equation \cite{Grad}. A review of the ideas of the
underlying method behind informational statistical thermodynamics was presented in ref.
\cite{Garsia1}.  The connection between entropy maximization and (nonlinear) Onsager formalism was
also studied \cite{Nett,Orlov84}. The viewpoint of two of the present authors (ANG and IVK) was
influenced by the papers by L.\ I.\ Rozonoer and co-workers, in particular, \cite{KoRoz,Ko,Roz}. A
detailed exposition of the quasiequilibrium approximation for Markov chains is given in the book
\cite{G1} (Chapter 3, {\it Quasiequilibrium and entropy maximum}, pp.\ 92-122), and for the BBGKY
hierarchy in the paper ref. \cite{Kark}. We have applied maximum entropy principle to the description
the universal dependence the 3-particle distribution function $F_3$ on the 2-particle distribution
function $F_2$ in classical systems with binary interactions \cite{BGKTMF}. For a discussion the
quasiequilibrium moment closure hierarchies for the Boltzmann equation \cite{Ko} see the papers
\cite{MBCh,MBChLANL,Lever}. A very general discussion of the maximum entropy principle with
applications to dissipative kinetics is given in the review \cite{Bal}. Recently the quasiequilibrium
approximation with some further correction was applied to description of rheology of polymer
solutions \cite{IKOePhA02,IKOePhA03} and of ferrofluids \cite{IlKr,IKar2}.}. We shall now introduce
the quasiequilibrium approximation in the most general setting.

{Nett}

A \textit{linear moment parameterization} is a linear operator, $\Pi: E\rightarrow L$, where $L={\rm
im} \Pi = E/ \ker \Pi,$ $\ker \Pi$ is a closed linear subspace of space $E$, and $\Pi$ is the
projection of $E$ onto factor-space $L$. Let us denote $W=\Pi(U)$. \textit{Quasiequilibrium} (or
restricted equilibrium, or conditional equilibrium) is the embedding, $F^*: W\rightarrow U$, which
puts into correspondence to each $y\in W$ the solution to the entropy maximization problem:

\begin{equation}\label{emax}
  S(x)\rightarrow\max,\ \Pi(x)=y.
\end{equation}

We assume that, for each $y\in{\rm int} W$, there exists the unique solution $F^*(y)\in {\rm int} U$
to the problem (\ref{emax}). This solution, $F^*(y)$, is called the quasiequilibrium, corresponding
to the value $y$ of the macroscopic variables. The set of quasiequilibria $F^*(y)$, $y\in W$, forms a
manifold in ${\rm int} U$, parameterized by the values of the macroscopic variables $y\in W$.

Let us specify some notations:  $E^T$ is the adjoint to the $E$ space.  Adjoint spaces and operators
will be indicated by $^T$, whereas notation $^*$ is earmarked for equilibria and quasiequilibria.

Furthermore, $[l,x]$ is the result of application of the functional $l\in E^T$ to the vector $x\in
E$.  We recall that, for an operator $A:E_1\to E_2$, the adjoint operator, $A^T:E_1^T\to E_2^T$ is
defined by the following relation:  For any $l\in E_2^T$ and $x\in E_1$,

\[ [l,Ax]=[A^Tl,x]. \]

Next, $D_xS(x)\in E^T$ is the differential of the entropy functional $S(x)$, $D_x^2S(x)$ is the
second differential of the entropy functional $S(x)$. The corresponding quadratic functional
$D_x^2S(x)(z,z)$ on $E$ is defined by the Taylor formula,

\begin{equation}
\label{Taylor} S(x+z)=S(x)+[D_xS(x),z]+\frac{1}{2}D_x^2S(x)(z,z)+o(\|z\|^2).
\end{equation}

We keep the same notation for the corresponding symmetric bilinear form, $D_x^2S(x)(z,p)$, and also
 for the linear operator, $D_x^2S(x):E\to E^T$, defined by the formula,

\[ [D_x^2S(x)z,p]= D_x^2S(x)(z,p).\]

In the latter formula, on the left hand side, there is the operator, on the right hand side there is
the bilinear form. Operator $D_x^2S(x)$ is symmetric on $E$, $D_x^2S(x)^T=D_x^2S(x)$.

Concavity of the entropy $S$ means that, for any $z\in E$, the inequality holds, $$D_x^2S(x)(z,z)\le
0;$$ in the restriction onto the affine subspace parallel to $\ker \Pi$ we assume the strict
concavity, $$D_x^2S(x)(z,z)< 0, \: \mbox{if} \: z\in \ker \Pi,\: \mbox{and  if} \: z\ne0.$$

In the remainder of this subsection we are going to construct the important object, the projector
onto the tangent space of the quasiequilibrium manifold.

Let us compute the derivative $D_yF^*(y)$.  For this purpose, let us apply the method of Lagrange
multipliers:  There exists such a linear functional $\Lambda(y)\in (L)^T$, that

\begin{equation}
\label{Lagrange} D_xS(x)\big|_{F^*(y)}=\Lambda(y)\cdot \Pi,\ \Pi(F^*(y))=y,
\end{equation}

or

\begin{equation}
\label{Lagrange2} D_xS(x)\big|_{F^*(y)}=\Pi^T\cdot\Lambda(y),\ \Pi(F^*(y))=y.
\end{equation}
From equation (\ref{Lagrange2}) we get,

\begin{equation}
\Pi(D_{y}F^*(y))=1_{L},
\end{equation}
where we have indicated the space in which the unit operator acts. Next, using the latter expression,
we transform the differential of the equation (\ref{Lagrange}),

\begin{equation}
D_{y}\Lambda=(\Pi(D_x^2S)_{F^*(y)}^{-1}\Pi^T)^{-1},
\end{equation}
and, consequently,

\begin{equation}
\label{derivative} D_{y}F^*(y)=(D_x^2S)_{F^*(y)}^{-1}\Pi^T(\Pi(D_x^2S)_{F^*(y)}^{-1}\Pi^T)^{-1}.
\end{equation}
Notice that, elsewhere in equation (\ref{derivative}), operator $(D_x^2S)^{-1}$ acts on the linear
functionals from $L^T$. These functionals are precisely those which become zero on $\ker \Pi$ or,
which is the same, those which can be represented as linear functionals of macroscopic variables.

The tangent space to the quasiequilibrium manifold at the point $F^*(y)$ is the image of the operator
$D_{y}F^*(y)$:

\begin{equation}
\label{ann} {\rm im} \left(D_{y}F^*(y)\right)=(D_x^2S)_{F^*(y)}^{-1}L^T= (D_x^2S)_{F^*(y)}^{-1}{\rm
Ann} (\ker \Pi)
\end{equation}
where ${\rm Ann} (\ker \Pi)$ is the set of linear functionals which become zero on $\ker \Pi$.
Another way to write equation (\ref{ann}) is the following:
\begin{equation}
\label{ann2} x\in {\rm im} \left(D_{y}F^*(y)\right)\Leftrightarrow (D_x^2S)_{F^*(y)}(z,p)=0,\ p\in
\ker \Pi.
\end{equation}
This means that $ {\rm im}\left(D_{y}F^*(y)\right)$ is the orthogonal completement of $\ker \Pi$ in
$E$ with respect to the scalar product,

\begin{equation}
\label{eproduct} \langle z|p\rangle_{F^*(y)}=-(D_x^2S)_{F^*(y)}(z,p).
\end{equation}

The entropic scalar product (\ref{eproduct}) appears often in the constructions below. (Usually, it
becomes the scalar product indeed after the conservation laws are excluded). Let us denote as
$T_y={\rm im}(D_y F^*(y))$ the tangent space to the quasiequilibrium manifold at the point $F^*(y)$.
Important role in the construction of quasiequilibrium dynamics
 and its generalizations is played by the quasiequilibrium projector, an operator which projects
$E$ on $T_y$ parallel to $\ker \Pi$. This is the orthogonal projector with respect to the entropic
scalar product, $P_y^*:E\to T_y$:

\begin{equation}
\label{qeproj} P^*_y=D_yF^*(y)\cdot \Pi =\left(D_x^2S\big|_{F^*(y)}\right)^{-1} \Pi^T \left( \Pi
\left(D_x^2S\big|_{F^*(y)}\right)^{-1}\Pi^T\right)^{-1}\Pi.
\end{equation}
It is straightforward to check the equality $P_y^{*2}=P_y^*$, and the self-adjointness of $P_y^*$
with respect to the entropic scalar product (\ref{eproduct}). Thus, we have introduced the basic
constructions: the quasiequilibrium manifold, the entropic scalar product, and the quasiequilibrium
projector.

The construction of the quasiequilibrium allows for the following generalization: Almost every
manifold can be represented as a set of minimizers of the entropy under linear constrains. However,
in contrast to the standard quasiequilibrium, these linear constrains will depend, generally
speaking, on the point on the manifold.

So, let the manifold $\Omega=F(W)\subset  U$ be given. This is a parametric set of distribution
functions. However,  now macroscopic variables $y$ are not functionals on $R$ or $U$ but just
parameters defining points on the manifold. The problem is how to extend the definitions of $y$ onto
a neighborhood of $F(W)$ in such a way that $F(W)$ will appear as the solution to the variational
problem:

\begin{equation}
\label{smax2} S(x)\to\max,\ \Pi(x)=y.
\end{equation}

For each point $F(y)$, we identify $T_y\in E$, the tangent space to the manifold $\Omega$ in $F_y$,
and the subspace $Y_y\subset E$, which depends smoothly on $y$, and which has the property,
$Y_y\bigoplus T_y=E$. Let us define $\Pi(x)$ in the neighborhood of $F(W)$ in such a way, that

\begin{equation}
\label{plane1} \Pi(x)=y, \ {\rm if}\ x-F(y)\in Y_y.
\end{equation}

The point $F(y)$ is the solution of the quasiequilibrium problem (\ref{smax2}) if and only if
\begin{equation}
\label{plane2} D_xS(x)\big|_{F(y)}\in {\rm Ann}\ Y_y.
\end{equation}
That is, if and only if $Y_y\subset {\rm ker} D_xS(x)\big|_{F(y)}$. It is always possible to
construct subspaces $Y_y$ with the properties just specified, at least locally, if the functional
$D_xS\big|_{F(y)}$ is not identically equal to zero on $T_y$.

The construction just described allows to consider practically any manifold as a quasiequilibrium.
This construction is required when one seeks the induced dynamics on a given manifold. Then the
vector fields are projected on $T_y$ parallel to $Y_y$, and this preserves intact the basic
properties of the quasiequilibrium approximations.

Let us return to the usual linear moment parametrization. {\it Quasiequilibrium entropy} $S(y)$ is a
functional on $W$. It is defined as the value of the entropy on the corresponding quasiequilibrium
$x=F^*(y)$:

\begin{equation}
S(y)=S(F^*(y))
\end{equation}

{\it Quasiequilibrium dynamics} is a dynamics on $W$, defined by the equation (\ref{fine2coarse}) for
the quasiequilibrium $F^*(y)$:

\begin{equation}\label{qedynam}
\frac{dy}{dt} = \Pi J(F^*(y)).
\end{equation}

Here $\Pi$ is constant linear operator (in the general case (\ref{fine2coarse}), it may become
nonlinear). The corresponding quasiequilibrium dynamics on the quasiequilibrium manifold $F^*(W)$ is
defined using the projector (\ref{momproj}):

\begin{equation}
\label{eqrus34} \frac{dx}{dt} = P^*_y|_{x=F^*(y)}J(x) = (D_yF^*)_{x=F^*(y)}\Pi J(x), \: x \in F^*(W).
\end{equation}

The orthogonal projector $P^*_y$ in the right hand side of equation, (\ref{eqrus34}) can be
explicitly written using the second derivative of $S$ and the operator $\Pi$ (\ref{qeproj}). Let's
remind that the only distinguished scalar product in $E$ is the entropic scalar product
(\ref{eproduct}):

\begin{equation}
{\langle z,p \rangle}_x = -(D_x^2S)_x(z,p)
\end{equation}

It depends on the point $x \in U$. This dependence $\langle|\rangle_x$ endows $U$ with the structure
of a Riemann space.

The most important property of the quasiequilibrium system (\ref{qedynam}), (\ref{eqrus34}) is
highlighted by the {\it conservation of the dynamics type} theorem: if for the original dynamic
system (\ref{sys}) $\frac{dS}{dt} \geq 0$, then for the quasiequilibrium dynamics $\frac{dS}{dt} \geq
0$. If for the original dynamic system (\ref{sys}) $\frac{dS}{dt} = 0$ (conservative system), then
for the quasiequilibrium dynamics $\frac{dS}{dt} = 0$ as well.

\subsection{\textbf{Thermodynamic projector without a priori parameterization}}

Quasiequilibrium manifolds is a place where the entropy and the moment parameterization meet each
other. The projectors $P_y$ for a quasiequilibrium manifold is nothing but the orthogonal with
respect to the entropic scalar product $\langle | \rangle_x$ projector (\ref{qeproj}). The
quasiequilibrium projector preserves the type of dynamics. Note that in order to preserve the type of
dynamics we needed only one condition to be satisfied,

\begin{equation}\label{eprop}
  \ker P_y\subset\ker(D_xS)_{x=F(y)}.
\end{equation}

Let us require that the field of projectors, $P(x,T)$, is defined for any $x$ and $T$, if

\begin{equation}\label{transversality}
  T\not{\!\subset} \ker D_xS.
\end{equation}

It follows immediately from these conditions that in the equilibrium, $P(x^*,T)$ is the orthogonal
projector onto $T$ (ortogonality is with respect to the entropic scalar product $\langle |
\rangle_{x^*}$).

The field of projectors is constructed in the neighborhood of the equilibrium based on the
requirement of the maximal smoothness of $P$ as a function of $g_x=D_xS$ and $x$. It turns out that
to the first order in the deviations $x-x^*$ and $g_x-g_{x^*}$, the projector is defined uniquely.
Let us first describe the construction of the projector, and next discuss its uniqueness.

Let the subspace $T\subset E$, the point $x$, and the differential of the entropy at this point,
$g=D_xS$, be defined in such a way that the transversality condition (\ref{transversality}) is
satisfied. Let us define $T_0=T\bigcap\ker g_x$. By the condition (\ref{transversality}), $T_0\neq
T$. Let us denote, $e_g=e_g(T)\in T$ the vector in $T$, such that $e_g$ is orthogonal to $T_0$, and
is normalized by the condition $g(e_g)=1$. The vector $e_g$ is defined unambiguously. The projector
$P_{S,x}=P(x,T)$ is defined as follows: For any $z\in E$,

\begin{equation}\label{projgen}
  P_{S,x}(z)=P_0(z)+e_gg_x(z),
\end{equation}
where $P_0$ is the orthogonal projector on $T_0$ (orthogonality is with respect to the entropic
scalar product $\langle |\rangle_{x}$). The {\it entropic projector} (\ref{projgen}) depends on the
point $x$ through the $x$-dependence of the scalar product $\langle |\rangle_{x}$, and also through
the differential of $S$ in $x$, the functional $g_x$.

Obviously, $P(z)=0$ implies $g(z)=0$, that is, the thermodynamicity requirement (\ref{eprop}) is
satisfied. Uniqueness of the thermodynamic projector (\ref{projgen}) is supported by the requirement
of the \textit{maximal smoothness} (analyticity) \cite{InChLANL} of the projector as a function of
$g_x$ and $\langle |\rangle_{x}$, and is done in two steps which we sketch here (detailed proof is
given in ref. \cite{UNIMOLD}):

\begin{enumerate}
  \item Considering the expansion of the entropy in the
  equilibrium up to the quadratic terms, one demonstrates that in the
  equilibrium the thermodynamic projector is the orthogonal
  projector with respect to the scalar product $\langle|\rangle_{x^*}$.
  \item For a given $g$, one considers auxiliary dissipative
  dynamic systems (\ref{sys}), which satisfy the condition:
  For every $x'\in U$, it holds, $g_x(J(x'))=0$, that is,
  $g_x$ defines an additional linear conservation law for the
  auxiliary systems. For the auxiliary systems, the point $x$ is
  the equilibrium. Eliminating the linear conservation law $g_x$,
  and using the result of the previous point, we end up with the
  formula (\ref{projgen}).
\end{enumerate}

Thus, the entropic structure defines unambiguously the field of projectors (\ref{projgen}), for which
the dynamics of {\it any} dissipative system (\ref{sys}) projected on {\it any} closure assumption
remains dissipative.

\clearpage

\addcontentsline{toc}{subsection}{\textbf{Example 1: Quasiequilibrium projector and defect of
invariance for the Local Maxwellians manifold of the Boltzmann equation}}

\subsection*{\textbf{Example 1: Quasiequilibrium projector and defect of
invariance for the Local Maxwellians manifold of the Boltzmann equation}}

The Boltzmann equation is one of the everlasting equations. It remains the most inspiring source for
the model reduction problems. With this subsection we start a series of examples for the Boltzmann
equation.

\addcontentsline{toc}{subsubsection}{Difficulties of classical methods of the Boltzmann equation
theory}

\subsubsection*{\textbf{Difficulties of classical methods of the Boltzmann equation theory}}

As was mentioned above, the first systematic and (at least partially) successful method of
constructing invariant manifolds for dissipative systems was the celebrated {\it Chapman-Enskog
method} \cite{Chapman} for the Boltzmann kinetic equation. The main difficulty of the Chapman-Enskog
method  \cite{Chapman}  are "nonphysical" properties of high-order approximations. This was stated by
a number of authors and was discussed in detail in \cite{Cercignani}. In particular, as it was noted
in \cite{Bob}, the  Burnett approximation results in a short-wave instability of the acoustic
spectra. This fact contradicts the $H$-theorem (cf. in \cite{Bob}). The Hilbert expansion contains
secular terms \cite{Cercignani}. The latter contradicts the $H$-theorem.
\par
The other difficulties of both  of  these  methods  are:  the restriction upon the choice of initial
approximation  (the  local equilibrium approximation), the demand for a small parameter,  and the
usage  of  slowly   converging   Taylor   expansion.   These difficulties never allow a direct
transfer  of  these  methods  on essentially nonequilibrium situations.
\par
The  main  difficulty  of  the  Grad  method   \cite{Grad}   is   the uncontrollability of the chosen
approximation. An extension of the list of moments can result in a certain success, but it  can  also
give nothing. Difficulties of moment expansion in the problems of shock waves and sound propagation
can be seen in \cite{Cercignani}.
\par
Many attempts were made to refine these methods. For the Chapman-Enskog and  Hilbert methods  these
attempts  are based in general on some "good" rearrangement of expansions  (e.g. neglecting
high-order  derivatives  \cite{Cercignani},  reexpanding  \cite{Cercignani},  Pade approximations and
partial summing \cite{GKJETP91,MBCh,Slem2}, etc.). This type of work with formal series is wide
spread  in  physics. Sometimes  the results are surprisingly good - from the renormalization theory
in quantum fields to the Percus-Yevick equation and the ring-operator in statistical mechanics.
However,  one  should realize  that  a success is not at all guaranteed. Moreover, rearrangements
never remove the restriction  upon  the  choice  of  the  initial  local equilibrium approximation.
\par
Attempts  to  improve  the   Grad   method   are   based   on quasiequilibrium approximations
\cite{KoRoz,Ko}. It was found in \cite{Ko}  that Grad  distributions  are  linearized versions of
appropriate quasiequilibrium approximations (see also the late papers \cite{MBCh,MBChLANL,Lever}). A
method which treats fluxes (e.g. moments with respect to collision integrals) as  independent
variables in a quasiequilibrium description was introduced in \cite{Karlin1,MBCh,GKPRE96,MBChLANL}.
\par
An important feature of quasiequilibrium approximations is that they are always thermodynamic, i.e.
they are concordant  with the $H$-theorem  due  to  their  construction.  However, quasiequilibrium
approximations do not remove the uncontrollability  of  the  Grad method.
\par

\addcontentsline{toc}{subsubsection}{Boltzmann Equation (BE)}

\subsubsection*{\textbf{Boltzmann Equation (BE)}}

The phase space $E$  consists  of distribution functions $f(\vv,\xx)$ which depend on the spatial
variable $\xx$  and  on velocity variable \vv. The variable $\xx$ spans an  open domain $\Omega
^{3}_{\mbox{\boldmath{\scriptsize $x$}}} \subseteq {\bf R}_{\mbox{\boldmath{\scriptsize $x$}}}$, and
the variable $\vv$ spans the space ${\bf R}^{3}_{\mbox{\boldmath{\scriptsize $v$}}}$. We require that
$f(\vv,\xx)\in F$ are nonnegative functions, and also that the following integrals are finite for
every ${\xx\in \Omega }_{\mbox{\boldmath{\scriptsize $x$}}}$ (the existence of moments and  of the
entropy):
\par
\begin{eqnarray}
&&I^{(i_{1}i_{2}i_{3})}_{\mbox{\boldmath{\scriptsize $x$}}}(f)=\int
v^{i_{1}}_{1}v^{i_{2}}_{2}v^{i_{3}}_{3}f(\vv,\xx)d^{3}\vv, i_{1}\geq 0,i_{2}\geq 0,i_{3}\geq 0;
\label{3.1a} \\ &&H_{\mbox{\boldmath{\scriptsize $x$}}}(f)=\int f(\vv,\xx)(\hbox{ln{\it
f}}(\vv,\xx)-1)d^{3}\vv, H(f)=\int H_{\mbox{\boldmath{\scriptsize $x$}}}(f)d^{3}\xx \label{3.1b}
\end{eqnarray}
\noindent Here and below integration in $\vv$ is made over ${\bf R}^{3}_{\mbox{\boldmath{\scriptsize
$v$}}}$, and  it  is made over $\Omega _{\mbox{\boldmath{\scriptsize $x$}}}$ in $\xx$. For every
fixed ${ \\x \in \Omega }_{\mbox{\boldmath{\scriptsize $x$}}}, \:
I^{(\cdots)}_{\mbox{\boldmath{\scriptsize $x$}}}$ and $H_{\mbox{\boldmath{\scriptsize $x$}}}$ might
be treated as functionals defined in {\it F}.
\par
We write BE in the form of  (\ref{sys})  using  standard  notations \cite{Cercignani}:
\par
\begin{equation}
{\partial f\over \partial t}=J(f), \ J(f)=-v_{s}{\partial f\over \partial x_{s}} + Q(f,f)
\label{3.2}\end{equation} \noindent Here and further a summation in two repeated indices  is assumed,
and $Q(f,f)$ stands for the Boltzmann collision  integral  [1].  The latter represents the
dissipative part of the  vector  field $J(f)$ (\ref{3.2}).
\par
{\it In this paper we consider the case when  boundary  conditions for equation} (\ref{3.2}) {\it are
relevant to the local  with  respect  to} $\xx$ {\it form of the} $H${\it -theorem.}
\par
For every fixed $\xx$, we denote as $H^{0}_{\mbox{\boldmath{\scriptsize $x$}}}(f)$  the  space  of
linear functionals
\par
\noindent $\sum_{i=0}^4 a_{i}(\xx)\int \psi _{i}(\vv)f(\vv,\xx)d^{3}\vv$,
 where $\psi _{i}(\vv)$   represent   summational
invariants of a collision $[1,2] (\psi _{0}=1, \psi _{i}=v_{i}, i=1,2,3, \psi _{4}=v^{2})$.  We write
(mod{\it H}$^{0}_{\mbox{\boldmath{\scriptsize $x$}}}(f))$ if an expression is valid within the
accuracy of adding a functional from $H^{0}_{\mbox{\boldmath{\scriptsize $x$}}}(f)$. The local
$H$-theorem  states:  for any functional
\par
\begin{equation}
H_{\mbox{\boldmath{\scriptsize $x$}}}(f)=\int f(\vv,\xx)(\hbox{ln{\it f}}(\vv,\xx)-1)d^{3}\vv \
(\hbox{mod{\it H}}^{0}_{\mbox{\boldmath{\scriptsize $x$}}}(f)) \label{3.3}\end{equation}

\noindent the following inequality is valid:
\par
\begin{equation}
dH_{\mbox{\boldmath{\scriptsize $x$}}}(f)/dt\equiv \int Q(f,f)\bigr|_{f=f(\mbox{\boldmath{\scriptsize
$v$}},\mbox{\boldmath{\scriptsize $x$}})} \ln f(\vv,\xx)d^{3}\vv \leq 0 \label{3.4}\end{equation}
\noindent Expression (\ref{3.4}) is   equal to   zero   if and only if $\ln  f =\sum_{i=0}^4
a_{i}(\xx)\psi _{i}(\vv)$.
\par
Although all functionals (\ref{3.3}) are equivalent in the sense of the $H$-theorem, it  is
convenient  to  deal  with  the  functional
\par
$$ H_{\mbox{\boldmath{\scriptsize $x$}}}(f)=\int f(\vv,\xx)(\hbox{ln{\it f}}(\vv,\xx)-1)d^{3} \vv.$$
All what was said in previous sections can be  applied  to  BE (\ref{3.2}). Now we will discuss some
specific points.

\addcontentsline{toc}{subsubsection}{Local manifolds}

\subsubsection*{\textbf{Local manifolds}}

Although  the  general  description  of manifolds $\Omega \subset F$ (Section 2.1) holds as well for
BE, a specific class of manifolds might be defined due to the different  character  of spatial and of
velocity dependencies in  BE  vector  field  (\ref{3.2}). These manifolds will be  called  {\bf local
manifolds},  and  they  are constructed as follows. Denote as $F_{\hbox{\scriptsize loc}}$ the set of
functions $f(\vv)$ with finite integrals
\par
\begin{eqnarray}
&&a) I^{(i_{1}i_{2}i_{3})}(f)=\int v^{i_{1}}_{1}v^{i_{2}}_{2}v^{i_{3}}_{3}f(\vv)d^{3}\vv, i_{1}\geq
0,i_{2}\geq 0,i_{3}\geq 0; \nonumber \\ &&b) H(f)=\int f(\vv)\hbox{ln{\it f}}(\vv)d^{3}\vv
\end{eqnarray}

\noindent In order to construct a local manifold in $F$, we, firstly, consider a manifold in
$F_{\hbox{\scriptsize loc}}$. Namely, we define a domain $A\subset B$, where $B$  is  a linear space,
and consider a smooth  immersion $A \to  F_{\hbox{\scriptsize loc}}$: $a \to  f(a,\vv)$. The set of
functions $f(a,\vv)\in F_{\hbox{\scriptsize loc}}$, where $a$ spans the domain {\it A}, is a manifold
in $F_{\hbox{\scriptsize loc}}$. Secondly,  we  consider   {\it all}   bounded   and sufficiently
smooth functions $a(\xx)$: $\Omega_{\mbox{\boldmath{\scriptsize $x$}}} \to  A$, and we define the
local manifold in $F$ as the set of functions $f(a(\xx),\vv)$. Roughly speaking, the local manifold
is a set of functions which are parameterized with $\xx$-dependent functions $a(\xx)$. A local
manifold will be called a {\bf locally finite-dimensional} manifold if $B$ is a finite-  dimensional
linear space.
\par

Locally finite-dimensional manifolds are a natural source  of initial  approximations   for
constructing   dynamic   invariant manifolds in BE theory. For example, the Tamm-Mott-Smith  (TMS)
approximation  gives  us  locally two-dimensional manifold $\left\{f(a_{-},a_{+})\right\}$ which
consists of distributions

\begin{equation}
f(a_{-},a_{+}) = a_{-}f_{-} + a_{+}f_{+} \label{2.3}\end{equation}

\noindent Here $a_{-}$ and $a_{+}$ (the coordinates on the  manifold $\Omega _{\hbox{\scriptsize
TMS}}=\left\{f(a_{-},a_{+})\right\})$ are non-negative real functions of the position vector $\xx$,
and $f_{-}$ and $f_{+}$ are fixed Maxwellians.

Next example is locally five-dimensional manifold $\left\{f(n,\uu,T)\right\}$  which  consists of
local Maxwellians (LM). The $LM$  manifold  consists  of  distributions $f_{0}$  which  are labeled
with parameters $n, \uu$, and $T$:
\par
\begin{equation}\
f_{0}(n,\uu,T)=n\left({2\pi k_{B}T\over m}\right)^{-3/2} \exp \left(- {m(\vv-\uu)^{2}\over
2k_{B}T}\right) \label{4.1}\end{equation}

Parameters $n, \uu$, and $T$ in (\ref{4.1}) are functions depending on \xx. In this section we will
not indicate this dependency explicitly.
\par
Distribution $f_{0}(n,\uu,T)$  is  the  unique  solution   of   the variational problem:
\begin{eqnarray}
&&H(f)=\int   f \ln  f d^{3}\vv \to  \min \nonumber \\ \hbox{for:}\nonumber \\ && M_{0}(f)=\int 1
\cdot f d^3\vv; \nonumber \\ &&M_{i}(f)=\int v_{i}fd^{3}\vv=\hbox{{\it nu}}_{i}, i=1,2,3;\qquad
\nonumber \\ &&M_{4}(f)=\int v^{2}fd^{3}\vv= {3nk_B T \over m} + nu^2 \label{4.2}
\end{eqnarray}

\noindent Hence,  the $LM$  manifold   is   a   quasiequilibrium   manifold. Considering $n, \uu$,
and $T$ as five scalar parameters (see the  remark on  locality  in  Section  3),  we  see that $LM$
manifold   is parameterized with the  values  of $M_{s}(f), s=0,\ldots,4,$  which  are defined in the
neighborhood  of $LM$  manifold.  It  is  sometimes convenient to consider the variables
$M_{s}(f_{0}), s=0,\ldots ,4,$  as  new coordinates on $LM$ manifold.  The  relationship between the
sets $\left\{M_{s}(f_{0})\right\}$ and $\left\{n,\uu,T\right\}$ is:
\par
\begin{equation}
n=M_{0}; u_{i}=M^{-1}_{0}M_{i}, i=1,2,3; T={m\over 3k_{B}}M^{-1}_{0}(M_{4}-M^{-1}_{0}M_{i}M_{i})
\label{4.3}\end{equation}

This is a standard moment parametrization of a quasiequilibrium manifold.

\addcontentsline{toc}{subsubsection}{Thermodynamic quasiequilibrium projector}

\subsubsection*{\textbf{Thermodynamic quasiequilibrium projector}}

Thermodynamic quasiequilibrium projector $P_{f_{0}(n,\mbox{\boldmath{\scriptsize $u$}},T)}(J)$ onto
the tangent  space $T_{f_{0}(n,\mbox{\boldmath{\scriptsize $u$}},T)}$ is defined as:
\par
\begin{equation}
P_{f_{0}(n,\mbox{\boldmath{\scriptsize $u$}},T)}(J)=\sum_{s=0}^4 {\partial f_{0}(n,\uu,T)\over
\partial M_{s}} \int \psi _{s}Jd^{3}\vv \label{4.4}\end{equation} \noindent Here we have assumed that
$n, \uu$, and $T$ are functions  of $M_{0},\ldots ,M_{4}$ (see relationship (\ref{4.3})), and
\par
\begin{equation}
\psi _{0}=1, \psi _{i}=v_{i}, i=1,2,3, \psi _{4}=v^{2} \label{4.5}\end{equation} \noindent
Calculating derivatives in (\ref{4.4}), and next returning to  variables $n, \uu$, and $T$, we
obtain:

\begin{eqnarray}
P_{f_{0}(n,\mbox{\boldmath{\scriptsize $u$}},T)}(J)=f_{0}(n,\uu,T) \left\{ \left[ {1\over n}
-{mu_{i}\over nk_{B}T}(v_{i}-u_{i})+\left({mu^{2}\over 3nk_{B}} - {T\over n}\right) \left(
{m(\vv-\uu)^2 \over 2k_{B}T^{2}}- \right. \right. \right. \nonumber \\ \left. \left. -{3\over 2T}
\right) \right] \int 1\cdot Jd^{3}\vv+ \left[  {m\over nk_{B}T}(v_{i}-u_{i}) - {2mu_{i}\over 3nk_{B}}
\left({m(\vv-\uu)^2 \over 2k_{B}T^{2}} - {3\over 2T} \right) \right] \int v_{i}Jd^{3}\vv + \nonumber
\\ \left. +{m\over 3nk_{B}} \left({m(\vv-\uu)^2 \over 2k_{B}T^{2}}- {3\over 2T} \right) \int
\vv^{2}Jd^{3}\vv \right\} \label{4.6a}
\end{eqnarray}

\noindent It is sometimes convenient to rewrite (\ref{4.6a}) as:
\par
\begin{equation}
P_{f_{0}(n,\mbox{\boldmath{\scriptsize $u$}},T)}(J)=f_{0}(n,\uu,T) \sum_{s=0}^{4} \psi
^{(s)}_{f_{0}(n,\uu,T)}\int \psi ^{(s)}_{f_{0}(n,\mbox{\boldmath{\scriptsize $u$}},T)}Jd^{3}\vv
\label{4.6b}
\end{equation}
\noindent Here
\begin{eqnarray}
\psi ^{(0)}_{f_{0}(n,\mbox{\boldmath{\scriptsize $u$}},T)}=n^{-1/2}, \qquad
 \psi ^{(i)}_{f_{0}(n,\mbox{\boldmath{\scriptsize $u$}},T)}=({2 / n})^{1/2}c_{i}, i=1,2,3, \nonumber\\
\psi ^{(4)}_{f_{0}(n,\mbox{\boldmath{\scriptsize $u$}},T)}=({2/3n})^{1/2}(c^{2}-(3/2)); \qquad
c_{i}=(m/2k_{B}T)^{1/2}(v_{i}-u_{i}) \label{4.7}\end{eqnarray} \noindent It is easy to check that
\par
\begin{equation}
\int f_{0}(n,\uu,T) \psi ^{(k)}_{f_{0}(n,\mbox{\boldmath{\scriptsize $u$}},T)} \psi
^{(l)}_{f_{0}(n,\mbox{\boldmath{\scriptsize $u$}},T)}d^{3}\vv=\delta _{kl} \label{4.8}\end{equation}
\noindent Here $\delta _{kl}$ is the Kronecker delta.
\par

\addcontentsline{toc}{subsubsection}{Defect of invariance for the $LM$ manifold}

\subsubsection*{\textbf{Defect of invariance for the $LM$ manifold}}

The defect of invariance for the $LM$ manifold at the point $f_{0}(n,\uu,T)$ for the $BE$ is:
\par
\begin{eqnarray}
\Delta (f_{0}(n,\uu,T))=P_{f_{0}(n,\mbox{\boldmath{\scriptsize $u$}},T)}
 \left(-(v_{s}-u_{s}){\partial f_{0}(n,\uu,T)\over \partial x_{s}}+Q(f_{0}(n,\uu,T)) \right)-
\nonumber \\ -\left( -(v_{s}-u_{s}){\partial f_{0}(n,\uu,T)\over \partial x_{s}} +Q(f_{0}(n,\uu,T))
\right)= \nonumber \\ =P_{f_{0}(n,\mbox{\boldmath{\scriptsize $u$}},T)} \left(-(v_{s}-u_{s}){\partial
f_{0}(n,\uu,T) \over
\partial x_{s}} \right) +(v_{s}-u_{s}){\partial f_{0}(n,\uu,T) \over \partial x_{s}}
\label{4.9}\end{eqnarray} \noindent Substituting (\ref{4.6a}) into (\ref{4.9}), we obtain:
\par
\begin{eqnarray}
\Delta (f_{0}(n,\uu,T))=f_{0}(n,\uu,T) \left\{ \left( {m(\vv-\uu)^{2}\over 2k_{B}T} - {5\over 2}
\right)(v_{i}-u_{i}) {\partial \hbox{ln{\it T}}\over \partial x_{i}} +\nonumber\right.  \\ \left. +
{m\over k_{B}T}(((v_{i}-u_{i})(v_{s}-u_{s})-{1\over 3} \delta_{is}(\vv-\uu)^{2}){\partial u_{s}\over
\partial x_{i}} \right\} \label{4.10}\end{eqnarray}

{\it The $LM$ manifold is not a dynamic invariant  manifold  of  the Boltzmann equation and the
defect (\ref{4.10}) is not identical to zero}.

\clearpage

\addcontentsline{toc}{subsection}{\textbf{Example 2: Scattering rates versus moments: alternative
Grad equations}}

\subsection*{\textbf{Example 2: Scattering rates versus moments: alternative Grad equations}}

 In this subsection scattering rates (moments of collision integral) are treated as new independent variables, and as an
alternative to moments of the distribution function, to describe the rarefied gas near local
equilibrium. A  version of entropy maximum principle is used to derive the   Grad-like description in
terms of a finite number of scattering rates. New equations are compared to the Grad moment system in
the heat non-conductive case. Estimations for hard spheres demonstrate, in particular, some 10$\%$
excess of the viscosity coefficient resulting from the scattering rate description, as compared to
the Grad moment estimation.

In 1949, Harold Grad \cite{Grad} has extended the basic assumption behind the Hilbert and
Chapman-Enskog method (the space and time dependence of the normal solutions is mediated by the five
hydrodynamic moments \cite{Chapman}). A physical rationale behind the Grad moment method is an
assumption of the decomposition of motion (i). During the time of order $\tau$, a set of
distinguished moments $M'$ (which include the hydrodynamic moments and a subset of higher-order
moment) does not change significantly as compared to the rest of the moments $M''$ (the fast
evolution) (ii). Towards the end of the fast evolution, the values of the moments $M''$ become
unambiguously determined by the values of the distinguished moments $M'$, and (iii). On the time of
order $\theta \gg \tau$, dynamics of the distribution function is determined by the dynamics of the
distinguished moments while the rest of the moments remains to be determined by the distinguished
moments (the slow evolution period).

Implementation of this picture requires an ansatz for the distribution function in order to represent
the set of states visited in the course of the slow evolution. In Grad's method, these representative
sets are finite-order truncations of an expansion of the distribution functions in terms of Hermit
velocity tensors:
\begin{eqnarray}
f_C(M',\vv)=f_{LM}(\rho,\uu,E,\vv)[1+\sum_{(\alpha)}^N a_{\alpha}(M')H_{(\alpha)}(\vv-\uu)]
,\label{24}
\end{eqnarray}
 where $H_{(\alpha)}(\vv-\uu)$ are various Hermit tensor polynomials,
orthogonal with the weight $f_{LM}$, while coefficient $a_{(\alpha)}(M')$ are known functions of the
distinguished moments $M'$, and $N$ is the highest order of $M'$. Other moments are functions of
$M'$: $M''=M''(f_C(M'))$.

Slow evolution of distinguished moments is found upon substitution of Eq. (\ref{24}) into the
Boltzmann equation and finding the moments of the resulting expression (Grad's moment equations).
Following Grad, this extremely simple approximation can be improved by extending the list of
distinguished moments. The most well known is Grad's thirteen-moment approximation where the set of
distinguished moments consists of five hydrodynamic moments, five components of the traceless stress
tensor $\sigma_{ij}=\int m[(v_i-u_i)(v_j-u_j)-\delta_{ij}(\vv-\uu)^2/3]fd\vv,$ and of the three
components of the heat flux vector $q_i=\int (v_i-u_i)m(\vv-\uu)^2/2 fd\vv$.

 The time evolution hypothesis
cannot be evaluated for its validity within the framework of Grad's approach. It is not surprising
therefore that Grad's methods failed to work in situations where it was (unmotivatedly) supposed to,
primarily, in the phenomena with sharp time-space dependence such as the strong shock wave. On the
other hand, Grad's method was quite successful for describing transition between parabolic and
hyperbolic propagation, in particular the second sound effect in massive solids at low temperatures,
and, in general, situations slightly deviating from the classical Navier-Stokes- Fourier domain.
Finally, the Grad method has been important background for development of phenomenological
nonequilibrium thermodynamics based on hyperbolic first-order equation, the so-called EIT (extended
irreversible thermodynamics \cite{EIT}).

Important generalization of the Grad moment method is the concept of quasiequilibrium approximations
already mentioned above. The quasiequilibrium distribution function for a set of distinguished moment
$M'$  maximizes the entropy density $S$ for fixed $M'$. The quasiequilibrium manifold $\Omega^3(M)$
is the collection of the quasiequilibrium distribution functions for all admissible values of $M$.
The quasiequilibrium approximation is the simplest and extremely useful (not only in the kinetic
theory itself) implementation of the hypothesis about a decomposition: If $M'$ are considered as slow
variables, then states which could be visited in the course of rapid motion in the neighbored of
$\Omega^*(M')$ belong to the planes  $\Gamma_{M'}=\{f\mid m'(f-f^*(M'))=0\}$. In this respect, the
thermodynamic construction in the method of invariant manifold is a generalization of the
quasiequilibrium approximation where the given manifold is equipped with a quasiequilibrium structure
by choosing appropriately the macroscopic variables of the slow motion. In contrast to the
quasiequilibrium, the macroscopic variables thus constructed are not obligatory moments. A text book
example of the quasiequilibrium approximation is the generalized Gaussian function for $M'=\{\rho,
\rho \uu,P\}$ where
 $P_{ij}=\int v_iv_j f d\vv$ is the pressure tensor. The quasiequilibrium approximation does not exist if the highest
order moment is an odd polynomial of velocity (therefore, there exists no quasiequilibrium for
thirteen Grad's moments). Otherwise, the Grad moment approximation is the first-order expansion of
the quasiequilibrium around the local Maxwellian.

The classical Grad moment method \cite{Grad} provides an approximate solution to the Boltzmann
equation, and leads to  a closed system of equations where hydrodynamic variables $\rho$, \uu, and
$P$ (density, mean flux, and pressure) are coupled to a finite set of non-hydrodynamic variables. The
latter are usually the stress tensor $\sigma$ and the heat flux {\bf q} constituting 10 and 13 moment
Grad systems. The Grad method was originally introduced for diluted gases to describe regimes beyond
the normal solutions \cite{Chapman}, but later it was used, in particular, as a prototype of certain
phenomenological schemes in nonequilibrium thermodynamics \cite{EIT}.

However, the moments  do not constitute the unique system of non-hydrodynamic variables, and the
exact dynamics might be equally expressed in terms of other infinite sets of variables (possibly, of
a non-moment nature). Moreover, as long as one shortens the description to only a finite subset of
variables, the advantage of the moment description above other systems is not obvious.

\addcontentsline{toc}{subsubsection}{Nonlinear functionals instead of moments in the closure problem}

\subsubsection*{\textbf{Nonlinear functionals instead of moments in the closure problem}}

Here we consider a   new system of non-hydrodynamic variables, {\it scattering rates} $M^w(f)$:
\begin{eqnarray}
\label{MSc} M^w_{i_1 i_2 i_3}(f)&=&\int \mu_{i_1 i_2 i_3} Q^w(f)d\vv; \\\nonumber \mu_{i_1 i_2
i_3}&=&mv_1^{i_1}v_2^{i_2}v_3^{i_3},
\end{eqnarray}
which, by definition, are the moments of the Boltzmann collision integral $Q^w (f)$:
\begin{eqnarray*}
Q^w(f)=\int w(\vv^{\prime},\vv_1^{\prime},\vv,\vv_1) \left\{f(\vv^{\prime})f(\vv_1^{\prime}) -
f(\vv)f(\vv_1) \right\}d\vv^{\prime}d\vv_1^{\prime}d\vv_1.
\end{eqnarray*}

Here $w$ is the probability density of a change of the velocities, $(\vv,\vv_1)\rightarrow
(\vv^{\prime},\vv^{\prime}_1)$, of the two particles after their encounter, and $w$ is defined by a
model of pair interactions. The description in terms of the scattering rates $M^w$ (\ref{MSc}) is
alternative to the usually treated description in terms of the moments $M$: $M_{i_1 i_2 i_3}(f)=\int
\mu_{i_1 i_2 i_3} f d\vv$.

A reason to consider scattering rates instead of the moments is that $M^w$ (\ref{MSc}) reflect
features of the interactions because of the  $w$ incorporated in their definition, while the moments
do not. For this reason we can expect that, in general, a description with a {\it finite} number of
scattering rates will be more informative than a description provided by the same number of their
moment counterparts.

To come to  the Grad-like equations in terms of the scattering rates, we have to complete the
following two steps:

i). To derive a hierarchy of transport equations for $\rho$, $\uu$, $P$, and $M^w_{i_1 i_2 i_3}$ in a
neighborhood of the local Maxwell states $f_0 (\rho,\uu,P)$.

ii). To truncate this hierarchy, and to come to a closed set of equations with respect to $\rho$,
$\uu$, $P$, and a finite number of scattering rates.

In the step (i), we derive a description with infinite number of variables, which is formally
equivalent both to the Boltzmann equation near the local equilibrium, and to the description with an
infinite number of moments. The approximation comes into play in the step (ii) where we reduce the
description to a finite number of variables. The difference between the moment and the alternative
description occurs at this point.

The  program (i) and (ii) is similar to what is done in the Grad method \cite{Grad}, with the only
exception (and this is important) that we should always use scattering rates as independent variables
and not to expand them into series in moments. Consequently, we will use a method of a closure in the
step (ii) that does not refer to the moment expansions. Major steps of the computation will be
presented below.

\addcontentsline{toc}{subsubsection}{Linearization}

\subsubsection*{\textbf{Linearization}}

To complete the step (i), we represent $f$ as $f_0 (1+\varphi)$, where $f_0$ is the local Maxwellian,
and we linearize the scattering rates (\ref{MSc}) with respect to $\varphi$:

\begin{eqnarray}
\label{DM} \Delta M^w_{i_1 i_2 i_3}(\varphi)&=&\int \Delta \mu^w_{i_1 i_2 i_3} f_0 \varphi d\vv;
\\\nonumber \Delta \mu^w_{i_1 i_2 i_3}&=& L^w ( \mu_{i_1 i_2 i_3} ).
\end{eqnarray}

Here $L^w$ is the usual linearized collision integral, divided by $f_0$. Though $\Delta M^w$ are
linear in $\varphi$, they are not moments because  their microscopic densities, $\Delta \mu^w$, are
not velocity polynomials  for a general case of $w$.

It is not difficult to derive the corresponding hierarchy of transport equations for variables
$\Delta M^w_{i_1 i_2 i_3}$, $\rho$, $\uu$, and $P$  (we will further refer to this hierarchy as to
the alternative chain): one has to calculate the time derivative of the scattering rates (\ref{MSc})
due to the Boltzmann equation, in the linear approximation (\ref{DM}), and to complete the system
with the five known balance equations for the hydrodynamic moments (scattering rates of the
hydrodynamic moments are equal to zero due to conservation laws). The structure of the alternative
chain is quite similar to that of the usual moment transport chain, and for this reason we do not
reproduce it here (details of calculations can be found in \cite{book}). One should only keep in mind
that the stress tensor and the heat flux vector in the balance equations for $\uu$ and $P$ are no
more independent variables, and they are expressed in terms of $\Delta M^w_{i_1 i_2 i_3}$, $\rho$,
$\uu$, and $P$.

\addcontentsline{toc}{subsubsection}{Truncating the chain}

\subsubsection*{\textbf{Truncating the chain}}

To truncate the alternative chain (step (ii)), we have, first, to choose a finite set of "essential"
scattering rates (\ref{DM}), and, second, to obtain the distribution functions which depend
parametrically only on $\rho$, $\uu$, $P$, and on the chosen set of scattering rates. We will
restrict our consideration to a single non-hydrodynamic variable, $\sigma_{ij}^w$, which is the
counterpart of the stress tensor $\sigma_{ij}$. This choice corresponds to the polynomial $mv_i v_j $
in the expressions (\ref{MSc}) and (\ref{DM}), and the resulting equations will be alternative to the
10 moment Grad system \footnote{To get the alternative to the 13 moment Grad equations, one should
take into account the scattering counterpart of the  heat flux, $q^w_i = m \int v_i \frac {v^2}{2}
Q^w (f) d\vv$.}. For a spherically symmetric interaction, the expression for $\sigma_{ij}^w$ may be
written:

\begin{eqnarray}
\label{es} \sigma ^w_{ij}(\varphi)&=&\int \Delta \mu^w_{ij} f_0 \varphi d\vv;\\\nonumber \Delta
\mu^w_{ij}=
 L^w (mv_i v_j )&=&\frac {P}{\eta_0^w (T)}S^w (c^2)\left\{c_i c_j - \frac
 {1}{3}\delta_{ij}c^2\right\}.
\end{eqnarray}

Here $\eta_0^w (T)$ is the first Sonine polynomial approximation of the Chapman-Enskog viscosity
coefficient (VC) \cite{Chapman}, and, as usual, ${\bf c}=\sqrt {\frac {m}{2kT}} (\vv-\uu)$. The
scalar dimensionless function $S^w$ depends only on $c^2$, and its form depends on the  choice of
interaction $w$.

\addcontentsline{toc}{subsubsection}{Entropy maximization}

\subsubsection*{\textbf{Entropy maximization}}

Next, we find the functions $f^* (\rho,\uu,P,\sigma_{ij}^w)=f_0 (\rho,\uu,P)(1+ \varphi^*
(\rho,\uu,P,\sigma_{ij}^w))$ which maximize the Boltzmann entropy $S(f)$ in a neighborhood of $f_0$
(the quadratic approximation to the entropy is valid within the accuracy of our consideration), for
fixed values of $\sigma_{ij}^w$. That is, $\varphi^*$ is a solution to the following conditional
variational problem:

\begin{eqnarray}
\label{var} \Delta S(\varphi )  =  -\frac {k_B}{2}\int f_0 \varphi^2 d\vv\rightarrow
\mbox{max},\\\nonumber {\rm i)} \int \Delta \mu^w_{ij} f_0 \varphi d\vv=\sigma_{ij}^w; \quad {\rm
ii)} \int \left\{1,\vv,v^2 \right\} f_0 \varphi d\vv=0.
\end{eqnarray}
The second (homogeneous) condition in (\ref{var}) reflects that a deviation $\varphi$ from the state
$f_0$ is due only to non-hydrodynamic degrees of freedom, and it is straightforwardly satisfied for
$\Delta \mu_{ij}^w$ (\ref{es}).

Notice, that if we turn to the usual moment description, then condition (i) in (\ref{var}) would fix
the stress tensor $\sigma_{ij}$ instead of its scattering counterpart $\sigma_{ij}^w$. Then the
resulting function $f^* (\rho,\uu,P, \sigma_{ij})$ will be exactly the 10 moment Grad approximation.
It can be shown that a choice of any finite set of higher moments as the constraint (i) in
(\ref{var}) results in the corresponding Grad approximation. In that sense our method of constructing
$f^*$ is a direct generalization of the Grad method onto the alternative description.

The Lagrange multipliers method gives straightforwardly the solution to the problem (\ref{var}).
After   the alternative chain is closed with the functions $f^* (\rho,\uu,P,\sigma_{ij}^w)$, the step
(ii) is completed, and we arrive at  a set of equations with respect to the variables $\rho$, $\uu$,
$P$, and $\sigma_{ij}^w$. Switching to the variable $\zeta_{ij} = n^{-1} \sigma^w_{ij}$, we have:

\begin{eqnarray}
\partial_t n + \partial_i (nu_i)=0;\label{a} \\
\rho (\partial_t u_k + u_i \partial_i u_k )+ \partial_k P +
\partial_i \left\{ \frac {\eta_0^w (T) n}{2r^w P} \zeta_{ik} \right\}=0;
\label{b}\\ \frac {3}{2} (\partial_t P + u_i \partial_i P)+\frac {5}{2} P \partial_i u_i + \left\{
\frac {\eta_0^w (T) n}{2r^w P} \zeta_{ik} \right\}\partial_i u_k =0;\label{c}\\
\partial_t \zeta_{ik} + \partial_s (u_s \zeta_{ik})
+\{\zeta_{ks} \partial_s u_i + \zeta_{is} \partial_s u_k  - \frac {2}{3} \delta_{ik} \zeta_{rs}
\partial_s u_r \}\\\nonumber+ \left\{ \gamma^w - \frac {2\beta^w}{r^w}\right\}\zeta_{ik}\partial_s
u_s
-
\frac {P^2}{\eta_0^w (T)n}(\partial_i u_k + \partial_k u_i \\\nonumber - \frac {2}{3} \delta_{ik}
\partial_s u_s ) - \frac{\alpha^w P}{r^w \eta_0^w (T)}\zeta_{ik} =0\label{d}.
\end{eqnarray}
Here $\partial_t = \partial /\partial t, \partial_i = \partial / \partial x_i$, summation in two
repeated indices is assumed, and the coefficients $r^w$, $\beta^w$, and $\alpha^w$ are defined with
the help of the  function $S^w$ (\ref{es}) as follows:
\begin{eqnarray}
\label{constants} r^w = \frac {8}{15\sqrt{\pi}}\int_0^{\infty} \mbox{e}^{-c^2}c^6 \left( S^w (c^2)
\right)^2 dc; \\\nonumber \beta^w = \frac {8}{15\sqrt{\pi}}\int_0^{\infty} \mbox{e}^{-c^2}c^6 S^w
(c^2)\frac{dS^w (c^2)}{d(c^2)} dc;\\\nonumber \alpha^w = \frac {8}{15\sqrt{\pi} }\int_0^{\infty}
\mbox{e}^{-c^2}c^6 S^w (c^2)R^w (c^2) dc.
\end{eqnarray}
The function $R^w (c^2)$ in the last expression is defined due to the action of the operator $L^w$ on
the function $S^w (c^2)(c_i c_j - \frac{1}{3}\delta_{ij}c^2 )$:
\begin{equation}
\label{action} \frac{P}{\eta_0^w}R^w (c^2)(c_i c_j - \frac {1}{3}\delta_{ij}c^2)= L^w (S^w (c^2)(c_i
c_j - \frac{1}{3}\delta_{ij}c^2 )).
\end{equation}
Finally, the parameter $\gamma^w$ in (\ref{a}-\ref{d}) reflects the temperature dependence of the VC:
\[
\gamma^w = \frac {2}{3}\left( 1- \frac {T}{\eta_0^w (T)} \left( \frac {d\eta_0^w (T)}{dT} \right)
\right).
\]
The set of ten equations (\ref{a}-\ref{d}) is alternative to the 10 moment Grad equations.

\addcontentsline{toc}{subsubsection}{A new determination of molecular dimensions (revisit)}

\subsubsection*{\textbf{A new determination of molecular dimensions (revisited)}}

The first observation to be made is that for Maxwellian molecules we have: $S^{MM} \equiv 1$, and
$\eta_0^{MM} \propto T$; thus $\gamma^{MM}= \beta^{MM}=0$, $r^{MM}=\alpha^{MM}=\frac{1}{2}$, and
(\ref{a}-\ref{d}) becomes the 10 moment Grad system under a simple change of variables $
\lambda\zeta_{ij}=\sigma_{ij}$, where $\lambda$ is the proportionality coefficient in the temperature
dependence of $\eta_0^{MM}$.

These properties (the function $S^w$ is a constant, and the VC is proportional to $T$) are true only
for Maxwellian molecules. For all other interactions, the function $S^w$ is not identical to one, and
the VC $\eta_0^w (T)$ is not proportional to $T$. Thus, the shortened alternative description is not
equivalent indeed to the Grad moment description. In particular, for hard spheres, the exact
expression for the function $S^{HS}$ (\ref{es}) reads:

\begin{eqnarray}
\label{RS} S^{HS} = \frac {5\sqrt 2}{16} \int_0^1 \exp (-c^2 t^2)(1-t^4 ) \left( c^2 (1-t^2 ) + 2
\right) dt, \\\nonumber \eta_0^{HS} \propto \sqrt T.
\end{eqnarray}

Thus, $\gamma^{HS}=\frac {1}{3}$, and $\frac{\beta^{HS}}{r^{HS}}\approx 0.07$, and the equation for
the function $\zeta_{ik}$  (\ref{d}) contains a nonlinear term,
\begin{equation}
\label{nonlin} \theta^{HS}\zeta_{ik}\partial_s u_s ,
\end{equation}
where $\theta^{HS}\approx 0.19$. This term is missing in the  Grad 10 moment equation.

Finally, let us  evaluate the  VC which results from the alternative description (\ref{a}-\ref{d}).
Following Grad's arguments \cite{Grad}, we see that, if the relaxation of $\zeta_{ik}$ is fast
compared to the hydrodynamic variables, then the two last terms in the equation for $\zeta_{ik}$
(\ref{a}-\ref{d}) become dominant, and the equation for $\uu$ casts into the standard Navier-Stokes
form with an effective VC $\eta^w_{{\rm eff}}$:
\begin{equation}
\label{effective} \eta^w_{{\rm eff}} = \frac {1}{2\alpha^w} \eta_0^w .
\end{equation}

For Maxwellian molecules, we easily derive that the coefficient $\alpha^w$ in eq. (\ref{effective})
is equal to $\frac{1}{2}$. Thus, as one expects, the effective VC (\ref{effective}) is equal to the
Grad value, which, in turn, is equal to the exact value in the frames of the Chapman-Enskog method
for this model.

For all interactions different from the Maxwellian molecules, the VC $\eta^w_{{\rm eff}}$
(\ref{effective}) is not equal to $\eta_0^w $. For hard spheres, in particular, a computation of the
VC (\ref{effective}) requires  information about the function $R^{HS}$ (\ref{action}). This is
achieved upon a substitution of the function $S^{HS}$ (\ref{RS}) into the eq. (\ref{action}).
Further, we have to  compute  the action of the operator $L^{HS}$ on the function $S^{HS}(c_i c_j -
\frac{1}{3}\delta_{ij}c^2 )$, which is rather complicated. However, the VC $\eta^{HS}_{{\rm eff}}$
can be relatively easily estimated  by using a function $S^{HS}_a = \frac {1}{\sqrt 2} (1 + \frac
{1}{7} c^2 )$, instead of the function $S^{HS}$, in eq. (\ref{action}). Indeed, the function
$S^{HS}_a$  is tangent to the function $S^{HS}$ at $c^2 = 0$, and  is its majorant (see Fig.\
\ref{Hard}). Substituting  $S^{HS}_a$ into eq. (\ref{action}), and computing the action of the
collision integral, we find the approximation $R_a^{HS}$; thereafter we evaluate the integral
$\alpha^{HS} $ (\ref{constants}), and finally come to the following expression:

\begin{equation}
\label{visc} \eta^{HS}_{{\rm eff}} \ge \frac{75264}{67237}\eta_0^{HS} \approx 1.12\eta_0^{HS}.
\end{equation}

Thus, for hard spheres, the description in terms of scattering rates results in  the VC of  more than
10$\%$ higher than in the Grad moment description.

A discussion of the results concerns the following two items.

1. Having two not equivalent descriptions which were obtained within one method, we may ask: which is
more relevant? A simple test is to compare characteristic times of an approach to hydrodynamic
regime. We have $\tau_G \sim \eta_0^{HS}/P$ for 10-moment description, and $\tau_a \sim
\eta^{HS}_{{\rm eff}}/P$ for alternative description. As $\tau_a > \tau_G $, we see that scattering
rate decay slower than corresponding moment, hence, at least for rigid spheres, alternative
description is more relevant. For Maxwellian molecules both the descriptions are, of course,
equivalent.

\begin{table}
\caption{Three virial coefficients: experimental $B_{{\rm exp}}$, classical $B_0 $
\cite{Hirschfelder}, and reduced $B_{{\rm eff}}$ for three gases at $T=500 K$} \label{TabVir}
\begin{center}
\begin{tabular}{|c|c|c|c|}\hline
&$B_{{\rm exp}}$&$B_0 $&$B_{{\rm eff}}$\\\hline Argon&8.4&60.9&50.5\\ Helium&10.8&21.9&18.2\\
Nitrogen &168&66.5&55.2\\ \hline
\end{tabular}
\end{center}
\end{table}

2. The VC $\eta^{HS}_{{\rm eff}} $ (\ref{visc}) has the same temperature dependence as $\eta_0^{HS}
$, and also the same dependence on a scaling parameter (a diameter of the sphere). In th classical
book \cite{Chapman} (pp.\ 228-229), "sizes" of molecules are presented, assuming that a molecule is
represented with an equivalent sphere and VC is estimated as $\eta_0^{HS} $. Since our estimation of
VC differs only by a dimensionless factor from $\eta_0^{HS} $, it is straightforward to conclude that
effective sizes of molecules will be reduced by the factor $b$, where $$b= \sqrt{\eta_0^{HS}/
\eta^{HS}_{{\rm eff}} } \approx 0.94 .$$ Further, it is well known that sizes of molecules estimated
via viscosity in \cite{Chapman} disagree with the estimation via the virial expansion of the equation
of state. In particular, in book \cite{Hirschfelder}, p.\ 5 the measured second virial coefficient
$B_{{\rm exp}}$ was compared with the calculated $B_0 $, in which the diameter of the sphere was
taken from the viscosity data.  The reduction of the diameter  by factor $b$ gives $B_{{\rm eff}}=b^3
B_0 $. The values $B_{{\rm exp}}$ and $B_0 $ \cite{Hirschfelder} are compared with $B_{{\rm eff}}$ in
the Table \ref{TabVir} for three gases at $T=500 K$. The results for argon and helium are better for
$B_{{\rm eff}}$, while for nitrogen $B_{{\rm eff}}$ is worth than $B_0 $. However, both  $B_0$ and
$B_{{\rm eff}}$ are far from the experimental values.

Hard spheres is, of course, an oversimplified model of interaction, and the comparison presented does
not allow for a decision between $\eta_0^{HS}$ and  $ \eta^{HS}_{{\rm eff}}$. However, this simple
example illustrates to what extend the correction to the VC can affect a comparison with experiment.
Indeed, as is well known, the first-order Sonine polynomial computation for the Lennard-Jones (LJ)
potential gives a very good fit of the temperature dependence of the VC for all noble gases
\cite{Dorf}, subject to a proper choice of the two unknown scaling parameters of the LJ potential
\footnote{A comparison of molecular parameters of the LJ potential, as derived from the viscosity
data, to those obtained from independent sources, can be found elsewhere, e.g. in  Ref.
\cite{Chapman}, p.\ 237.}. We may expect that a dimensionless correction of the VC for the LJ
potential might be of the same order as above for rigid spheres. However, the functional character of
the temperature dependence will not be affected, and a fit will be obtained subject to a different
choice of the molecular parameters of the LJ potential.

There remains, however, a general question how the estimation of the VC (\ref{effective}) responds to
the exact value \cite{Chapman}, \cite{Resibois}. Since the analysis performed above does not
immediately appeal to the exact Chapman-Enskog expressions just mentioned, this question remains open
for a further work.

\section{\textbf{Newton method with incomplete linearization}}\label{newton}

 Let us come back to the invariance equation (\ref{defect}),

\[\Delta_y=(1-P_y)J(F(y))=0.\]

One of the most efficient methods to solve this equation is the Newton method with incomplete
linearization. Let us linearize the vector field $J$ around $F(y)$:

\begin{equation}\label{linfield}
 J(F(y)+\delta F(y))=J(F(y))+(DJ)_{F(y)}\delta F(y)+o(\delta
 F(y)).
\end{equation}

Equation of the Newton method with incomplete linearization makes it possible to determine $\delta
F(y)$:

\begin{equation}\label{Nm1}
 \left\{\begin{array}{l}
   P_y\delta F(y)=0 \\
   (1-P_y)(DJ)_{F(y)}\delta F(y)=(1-P_y)J(F(y)).\
 \end{array}\right.
\end{equation}

The crucial point here is that the same projector $P_y$ is used as in the equation (\ref{defect}),
that is, without computing the variation of the projector $\delta P$ (hence, the linearization of
equation (\ref{defect}) is incomplete). We recall that projector $P_y$ depends on the tangent space
$T_y=\textrm{Im}(DF)_y$. If the thermodynamic projector (\ref{projgen}) is used here, then $P_y$
depends also on $\langle|\rangle_{F(y)}$ and on $g=(DS)_{F(y)}$.

Equations of the Newton method with incomplete linearization (\ref{Nm1}) are not differential
equations in $y$ anymore, they do not contain derivatives of the unknown $\delta F(y)$ with respect
to $y$ (which would be the case if the variation of the projector $\delta P$ has been taken into
account). The absence of the derivatives in equation (\ref{Nm1}) significantly simplifies its
solving. However, even this is not the main advantage of the incomplete linearization. More essential
is the fact that iterations of the Newton method with incomplete linearization is expected to
converge to slow invariant manifolds, unlike the usual Newton method. This has been demonstrated in
\cite{GKTTSP94} in the linear approximation.

In order to illustrate the nature of the Eq.\ (\ref{Nm1}), let us consider the case of linear
manifolds for linear systems. Let a linear evolution equation be given in the finite-dimensional real
space: $\dot{\xx}={\bf A} \xx$, where ${\bf A}$ is negatively definite symmetric matrix with a simple
spectrum. Let us further assume quadratic Lyapunov function, $S(\xx)=\langle \xx,\xx\rangle$. The
manifolds we consider are lines, $\vl(y)=y\ee$, where $\ee$ is the unit vector, and $y$ is a scalar.
The invariance equation for such manifolds reads: $\ee\langle\ee,{\bf A} \ee\rangle-{\bf A}\ee=0$,
and is simply the eigenvalue problem for the operator ${\bf A}$. Solutions to the latter equation are
eigenvectors $\ee_{i}$, corresponding to eigenvalues $\lambda_{i}$.

Assume that we have chosen a line, $\vl_0=y\ee_0$, defined by the unit vector $\ee_0$, and that
$\ee_0$ is not an eigenvector of ${\bf A}$. We seek another line, $\vl_1=a\ee_1$, where $\ee_1$ is
another unit vector, $\ee_1=\xx_1/\|\xx_1\|$, $\xx_1=\ee_0+\delta\xx$. The additional condition in
(\ref{Nm1}) reads: $P_y\delta F(y)=0$, i.e. $\langle \ee_0, \delta\xx \rangle = 0$. Then the Eq.\
(\ref{Nm1}) becomes $[1-\ee_0\langle\ee_0,\cdot\rangle]{\bf A }[\ee_0+\delta\xx]=0$. Subject to the
additional condition, the unique solution is as follows: $\ee_0+\delta\xx= \langle\ee_0,{\bf
A}^{-1}\ee_0\rangle^{-1}{\bf A}^{-1}\ee_0$. Rewriting the latter expression in the eigen--basis of
${\bf A}$, we have: $\ee_0+\delta\yy\propto \sum_{i}\lambda_i^{-1}\ee_i\langle\ee_i,\ee_0\rangle$.
The leading term in this sum corresponds to the eigenvalue with the minimal absolute value. The
example indicates that the  method (\ref{Nm1}) seeks the direction of the {\it slowest relaxation}.
For this reason, the Newton method with incomplete linearization (\ref{Nm1}) can be recognized as the
basis of an iterative  construction of the manifolds of slow motions.

In an attempt to simplify computations, the question which always can be asked is as follows: To what
extend is the choice of the projector essential in the equation (\ref{Nm1})? This question is a valid
one, because, if we accept that iterations converge to a relevant slow manifold, and also that the
projection on the true invariant manifold is insensible to the choice of the projector, then should
one care of the projector on each iteration? In particular, for the moment parameterizations, can one
use in equation (\ref{Nm1}) the projector (\ref{momproj})? Experience gained from some of the
problems studied by this method indicates that this is possible. However, in order to derive
physically meaningful equations of motion along the approximate slow manifolds, one has to use the
thermodynamic projector (\ref{projgen}). Otherwise we are not guaranteed from violating the
dissipation properties of these equations of motion.

\clearpage

\addcontentsline{toc}{subsection}{\textbf{Example 3: Non-perturbative correction of Local Maxvellian
manifold and derivation of nonlinear hydrodynamics from Boltzmann equation (1D)}}

\subsection*{\textbf{Example 3: Non-perturbative correction of Local Maxvellian
manifold and derivation of nonlinear hydrodynamics from Boltzmann equation (1D)}}

This section is a continuation of {\bf Example 1}. Here we apply the method of invariant manifold to
a particular situation when the initial manifold consists of local Maxwellians (\ref{4.1})  (the $LM$
manifold). This manifold and  its corrections play the central role in the problem of derivation of
hydrodynamics from  BE. Hence, any   method   of   approximate investigation of  BE  should  be
tested  with  the $LM$ manifold. Classical  methods  (Chapman-Enskog  and  Hilbert methods)   use
Taylor-type expansions into powers of a small  parameter  (Knudsen number expansion). However,  as we
have mentioned above,  the method of invariant manifold, generally speaking, assumes no small
parameters, at  least in  its formal  part  where   convergency properties are not discussed.  We
will  develop an appropriate technique  to  consider  the  invariance  equation  of the  first
iteration.  This involves ideas of parametrix expansions  of  the  theory  of pseudodifferential and
Fourier integral operators \cite{25,26}. This approach will make it possible to reject the
restriction of using small parameters.

We search for a correction to the $LM$ manifold as:

\begin{equation}
f_{1}(n,\uu,T)=f_{0}(n,\uu,T) + \delta f_{1}(n,\uu,T). \label{4.11}\end{equation}

\noindent We will use the  Newton method  with incomplete linearization for obtaining the correction
$\delta f_{1}(n,\uu,T)$, because  we  search  for  a manifold  of  slow (hydrodynamic)  motions. We
introduce   the representation:

\begin{equation}
\delta f_{1}(n,\uu,T)=f_{0}(n,\uu,T)\varphi (n,\uu,T). \label{4.12}\end{equation}

\addcontentsline{toc}{subsubsection}{Positivity  and  normalization}

\subsubsection*{\textbf{Positivity  and  normalization}}

When  searching  for   a correction, we should be ready to face two  problems  that  are typical for
any method of successive approximations in BE  theory. Namely, the first of this problems is that the
correction
\par
$$ f_{\Omega _{k+1}}=f_{\Omega _{k}}+\delta f_{\Omega _{k+1}} $$ \noindent obtained from the
linearized invariance  equation  of  the $k+1$-th iteration may be not a non-negatively defined
function and thus it can not be used directly to define the thermodynamic projector for the $k+1$-th
approximation. In order to overcome this difficulty, we can treat the procedure  as  a  process  of
correcting  the  dual variable $\mu _{f}=D_{f} H(f)$ rather than the process of immediate correcting
the distribution functions.

The dual variable $\mu _{f}$ is:

\begin{equation}
\mu _{f} \bigr|_{f=f(\mbox{\boldmath {\scriptsize $x,v$}})} =D_{f}H(f)\bigr|_{f=f(\mbox{\boldmath
{\scriptsize $x,v$}})} =D_{f}H_{\mbox{\boldmath {\scriptsize $x$}}}(f)\bigr|_{f=f(\mbox{\boldmath
{\scriptsize $x,v$}})} = \ln  f (\vv,\xx) .\label{3.7}
\end{equation}

Then,  at  the $k+1$-th  iteration,  we search for new dual variables $\mu _{f}\bigr|_{\Omega
_{k+1}}$:
\par
\begin{equation}
\mu _{f}\bigr|_{\Omega _{k+1}}=\mu _{f}\bigr|_{\Omega _{k}} + \delta \mu _{f}\bigr|_{\Omega _{k+1}}.
\label{3.16}\end{equation}

\noindent Due to the relationship $\mu _{f} \longleftrightarrow f$, we have:
\begin{equation}
\delta \mu_{f}\bigr|_{\Omega _{k+1}}=\varphi _{\Omega _{k+1}}+O(\delta f^{2}_{\Omega _{k+1}}),
 \varphi _{\Omega _{k+1}}=f^{-1}_{\Omega _{k}}\delta f_{\Omega _{k+1}}.
\label{3.17}\end{equation} \noindent Thus, solving the linear invariance equation of the $k$-th
iteration with respect to the unknown function $\delta f_{\Omega _{k+1}}$, we find a  correction to
the dual variable $\varphi _{\Omega _{k+1}} (\ref{3.17})$, and  we  derive  the  corrected
distributions $f_{\Omega _{k+1}}$ as

\begin{equation}
f_{\Omega _{k+1}}=\exp (\mu _{f}\bigr|_{\Omega _{k}}+\varphi _{\Omega _{k+1}})=f_{\Omega _{k}}\exp
(\varphi _{\Omega _{k+1}}) . \label{3.18}\end{equation} \noindent

Functions (\ref{3.18}) are positive, and  they  satisfy  the  invariance equation and the additional
conditions  within the accuracy of $\varphi _{\Omega _{k+1}}$.

However, the second difficulty  which  might  occur  is  that functions  (\ref{3.18})  might  have no
finite  integrals  (\ref{3.1b}).   In particular, this difficulty can be a result of some
approximations used in solving equations.  Hence,  we  have  to "regularize" the functions
(\ref{3.18}). A sketch of an approach to make this regularization might be as follows: instead of
$f_{\Omega _{k+1}} (\ref{3.18})$, we consider functions:

\begin{equation}
f^{(\beta )}_{\Omega _{k+1}}=f_{\Omega _{k}}\exp (\varphi _{\Omega _{k+1}}+\varphi
^{\hbox{\scriptsize reg}}(\beta )). \label{3.19}\end{equation}

\noindent Here $\varphi ^{\hbox{\scriptsize reg}}(\beta )$ is a function labeled with $\beta \in B$,
and $B$  is a  linear space.  Then we derive $\beta _{*}$ from the  condition  of coincidence of
macroscopic parameters. Further consideration of this procedure \cite{GKTTSP94} remains out of frames
of this paper.

The   two difficulties mentioned here are not specific for  the  approximate method developed. For
example, corrections to the $LM$ distribution in the Chapman-Enskog method  \cite{Chapman}  and  the
thirteen-moment  Grad approximation \cite{Grad} are not non-negatively defined functions,  while the
thirteen-moment quasiequilibrium approximation \cite{Ko}  has  no finite integrals $(\ref{3.1a})$ and
$(\ref{3.1b})$.

\addcontentsline{toc}{subsubsection}{Galilean invariance of invariance equation}

\subsubsection*{\textbf{Galilean invariance of invariance equation}}

In some cases, it is convenient to consider BE vector field  in  a reference system which moves  with
the  flow  velocity.  In  this reference system, we define the BE vector field as:
\par
\begin{equation}
{df\over dt}=J_{u}(f), {df\over dt}={\partial f\over \partial t}+u_{\mbox{\boldmath {\scriptsize
$x$}},s}(f){\partial f\over
\partial x_{s}}; J_{u}(f)=
-(v_{s}-u_{\mbox{\boldmath {\scriptsize $x$}},s}(f)){\partial f\over \partial x_{s}} +Q(f,f).
\label{3.13}\end{equation} \noindent Here $u_{\mbox{\boldmath {\scriptsize $x$}},s}(f)$ stands for
the $s$-th component of the flow velocity:
\par
\begin{equation}
u_{\mbox{\boldmath {\scriptsize $x$}},s}(f)=n^{-1}_{\mbox{\boldmath {\scriptsize $x$}}}(f)\int
v_{s}f(\vv,\xx)d^{3}\vv; \; n_{\mbox{\boldmath {\scriptsize $x$}}}(f)=\int f(\vv,\xx)d^{3}\vv .
\label{3.14}\end{equation}

\noindent In particular, this form of BE vector field is convenient when the initial  manifold
$\Omega _{0}$ consists  of  functions $f_{\Omega _{0}}$ which depend explicitly on
$(\vv-u_{\mbox{\boldmath {\scriptsize $x$}}}(f))$ (i.e., if functions $f_{\Omega _{0}}\in \Omega
_{0}$ do not change under velocity shifts: $\vv\to \vv+c$, where $c$ is a constant vector).
\par
Substituting $J_{u}(f)$ (\ref{3.13}) instead  of $J(f)$ (\ref{3.2})
 into  all expressions which
depend on the BE vector field, we  transfer  all procedures developed above into the moving reference
system.  In particular, we obtain  the  following  analog  of  the  invariance equation of the first
iteration:
\begin{eqnarray}
 (P^{0*}_{a(\mbox{\boldmath {\scriptsize $x$}})}(\cdot )-1)J^{0}_{u,\mbox{\scriptsize lin},a(\mbox{\boldmath
 {\scriptsize $x$}})} (\delta f_{1}(a({\xx}),\vv))+\Delta (f_{0} (a({\xx}),{\vv}))=0; \qquad \nonumber
\\
 J^{0}_{u,\mbox{\scriptsize lin},a(\mbox{\boldmath {\scriptsize $x$}})}(g)=
 \left\{ n^{-1}_{\mbox{\boldmath {\scriptsize $x$}}}(f_{0}(a(\xx)))
\int v_{s} g d^{3}\vv+\right. \qquad \nonumber \\
 \left.+u_{\mbox{\boldmath {\scriptsize $x$}},s}(f_{0}(a(\xx)))n^{-1}_{\mbox{\boldmath {\scriptsize $x$}}}(f_{0}(a(\xx)))
\int g d^{3}\vv \right\} {\partial f_{0}(a(\xx),\vv)\over \partial x_{s}} - \nonumber \qquad\\
 - \left( v_{s}-u_{\mbox{\boldmath {\scriptsize $x$}},s}(f_{0}(a(\xx)))\right)
 {\partial g \over \partial x_{s}} +
 L_{f_{0}(a({\mbox{\boldmath {\scriptsize $x$}}}),{\mbox{\boldmath {\scriptsize $v$}}})}(g);
\nonumber \qquad \\
 \Delta (f_{0}(a(\xx),\vv))=(P^{*}_{a(\mbox{\boldmath {\scriptsize $x$}})}(\cdot )-1)
 J_{u}(f_{0}(a(\xx),\vv)).
\label{3.15}
\end{eqnarray}
\noindent Additional conditions do not depend on the  vector  field, and thus they remain valid for
equation (\ref{3.15}).

\addcontentsline{toc}{subsubsection}{The equation of the first iteration}

\subsubsection*{\textbf{The equation of the first iteration}}

The equation of the first iteration in the form of (\ref{3.17}) for the correction $\varphi
(n,\uu,T)$ is:
\par
\begin{eqnarray}
\left\{ P_{f_{0}(n,{\mbox{\boldmath {\scriptsize $u$}}},T)} (\cdot )-1\right\} \left\{ -(v_{s}-u_{s})
{\partial f_{0}(n,\uu,T)\over
\partial x_{s}} + f_{0}(n,\uu,T)L_{f_{0}(n,{\uu},T)}(\varphi ) - \nonumber \right.
\\ - (v_{s}-u_{s}){\partial (f_{0}(n,\uu,T)\varphi )\over \partial x_{s}} -
n^{-1}\left(f_{0}(n,\uu,T))  \left(\int v_{s}f_{0}(n,\uu,T) \varphi d^{3}\vv +  \right. \right.
\nonumber \\ \left. \left. \left. + u_{s}(f_{0}(n,\uu,T) \right) \int f_{0}(n,\uu,T)\varphi d^{3}\vv
\right) {\partial f_{0}(n,\uu,T)\over \partial x_{s}} \right\}=0 . \label{4.13a}
\end{eqnarray}

\noindent Here $f_{0}(n,\uu,T)L_{f_{0}(n,\mbox{\boldmath {\scriptsize $u$}},T)} (\varphi )$ is the
linearized Boltzmann collision integral:

\begin{eqnarray}
f_{0}(n,\uu,T) L_{f_{0}(n,\mbox{\boldmath {\scriptsize $u$}},T)} (\varphi )= \int w({\bf v'},\vv |
\vv,\vv_{1})f_{0}(n,\uu,T)\times \nonumber \\ \times \left\{\varphi '+\varphi'_1 -\varphi_{1}-\varphi
\right\}d^{3}{\bf v'}d^{3}{\bf v'_1} d^{3}\vv_{1} . \label{4.14}\end{eqnarray}

\noindent and $w({\bf v'},\vv'_1 |\vv,\vv_{1})$  is  the  kernel  of  the  Boltzmann collision
integral, standard notations label the velocities before and after a collision.
\par
Additional condition for  equation $(\ref{4.13a})$  has  the form:
\par
\begin{equation}
P_{f_{0}(n,\mbox{\boldmath {\scriptsize $u$}},T)} (f_{0}(n,\uu,T)\varphi )=0 .
\label{4.15}\end{equation} \noindent In detail notation:
\begin{eqnarray}
\int 1\cdot f_{0}(n,\uu,T)\varphi d^{3}\vv=0, \qquad
 \int v_{i}f_{0}(n,\uu,T)\varphi d^{3}\vv=0,\  i=1,2,3, \nonumber \\
\int \vv^{2}f_{0}(n,\uu,T)\varphi d^{3}\vv=0 . \label{4.16}\end{eqnarray}

\noindent Eliminating in (\ref{4.13a}) the terms containing $\int v_{s}f_{0}(n,\uu,T)\varphi
d^{3}\vv$  and $\int f_{0}(n,\uu,T)\varphi d^{3}\vv$ with the aid of (\ref{4.16}),  we obtain the
following form of equation (\ref{4.13a}):

\begin{eqnarray}
&&\{P_{f_{0}(n,\mbox{\boldmath {\scriptsize $u$}},T)}(\cdot )-1 \} \times  \label{4.13b} \\  &&
\left( -(v_{s}-u_{s}){\partial f_{0}(n,\uu,T)\over
\partial x_{s}} + f_{0}(n,\uu,T)L_{f_{0}(n,\mbox{\boldmath {\scriptsize $u$}},T)}(\varphi ) -
(v_{s}-u_{s}){\partial (f_{0}(n,\uu,T)\varphi )\over \partial x_{s}} \right)=0 . \nonumber
\end{eqnarray}

In order to consider the properties of equation $(\ref{4.13b})$,  it is useful to introduce real
Hilbert spaces $G_{f_{0}(n,\mbox{\boldmath {\scriptsize $u$}},T)}$ with  scalar products:
\par
\begin{equation}
(\varphi ,\psi )_{f_{0}(n,\mbox{\boldmath {\scriptsize $u$}},T)} = \int f_{0}(n,\uu,T)\varphi \psi
d^{3}\vv . \label{4.17}\end{equation} \noindent Each  Hilbert  space  is  associated  with  the
corresponding $LM$ distribution $f_{0}(n,\uu,T)$.

The projector $P_{f_{0}(n,\mbox{\boldmath {\scriptsize $u$}},T)}$ (\ref{4.6b}) is associated with a
projector $\Pi _{f_{0}(n,\mbox{\boldmath {\scriptsize $u$}},T)}$ which acts in the space
$G_{f_{0}(n,\mbox{\boldmath {\scriptsize $u$}},T)}$:
\begin{equation}
\Pi _{f_{0}(n,\mbox{\boldmath {\scriptsize $u$}},T)}(\varphi
)=f^{-1}_{0}(n,\uu,T)P_{f_{0}(n,\mbox{\boldmath {\scriptsize $u$}},T)}(f_{0}(n,\uu,T)\varphi ).
\label{4.18}\end{equation}

\noindent It is an orthogonal projector, because
\begin{equation}
\Pi _{f_{0}(n,\mbox{\boldmath {\scriptsize $u$}},T)}(\varphi )=\sum_{s=0}^4 \psi
^{(s)}_{f_{0}(n,\mbox{\boldmath {\scriptsize $u$}},T)} (\psi ^{(s)}_{f_{0}(n,\mbox{\boldmath
{\scriptsize $u$}},T)},\varphi )_{f_{0}(n,\mbox{\boldmath {\scriptsize $u$}},T)}.
\label{4.19}\end{equation}

\noindent Here $\psi ^{(s)}_{f_{0}(n,\mbox{\boldmath {\scriptsize $u$}},T)}$ are given by the
expression (\ref{4.7}).

We can rewrite the equation of the first iteration $(\ref{4.13b})$ in the form:
\begin{equation}
L_{f_{0}(n,\mbox{\boldmath {\scriptsize $u$}},T)}(\varphi ) + K_{f_{0}(n,\mbox{\boldmath {\scriptsize
$u$}},T)}(\varphi ) = D_{f_{0}(n,\mbox{\boldmath {\scriptsize $u$}},T)} . \label{4.20}\end{equation}

\noindent Notations used here are:
\begin{eqnarray}
&&D_{f_{0}(n,\mbox{\boldmath {\scriptsize $u$}},T)}=f^{-1}_{0}(n,\uu,T)\Delta (f_{0}(n,\uu,T)); \\
&&K_{f_{0}(n,\mbox{\boldmath {\scriptsize $u$}},T)}(\varphi )= \left\{\Pi _{f_{0}(n,\mbox{\boldmath
{\scriptsize $u$}},T)}(\cdot )-1 \right\} f^{-1}_{0}(n,\uu,T)(v_{s}-u_{s}){\partial
(f_{0}(n,\uu,T)\varphi) \over \partial x_{s}}. \nonumber\label{4.21}
\end{eqnarray}
\noindent The additional condition for equation (\ref{4.20}) is:
\begin{eqnarray}
(\psi ^{(s)}_{f_{0}(n,\mbox{\boldmath {\scriptsize $u$}},T)},\varphi )_{f_{0}(n,\mbox{\boldmath
{\scriptsize $u$}},T)}=0, s=0,\ldots ,4 .\label{4.22}\end{eqnarray}

Now we will list the properties of the  equation  (\ref{4.20})  for usual models of a collision
\cite{Chapman}:

\noindent a) The linear integral operator $L_{f_{0}(n,\mbox{\boldmath {\scriptsize $u$}},T)}$ is
selfadjoint  with respect to the scalar product $(\cdot ,\cdot )_{f_{0}(n,\mbox{\boldmath
{\scriptsize $u$}},T)}$, and the quadratic form $(\varphi,L_{f_{0}(n,\mbox{\boldmath {\scriptsize
$u$}},T)}(\varphi ))$ is negatively defined in $\mbox{Im} _{f_{0}(n,\mbox{\boldmath {\scriptsize
$u$}},T)}$.

\noindent b) The kernel of $L_{f_{0}(n,\mbox{\boldmath {\scriptsize $u$}},T)}$ does not depend on
$f_{0}(n,\uu,T)$, and it is the linear envelope of the polynomials $\psi _{0}=1, \psi _{i}=v_{i},
i=1,2,3,$  and $\psi _{4}=v^{2}$.
\par
\noindent c) The RHS $D_{f_{0}(n,\mbox{\boldmath {\scriptsize $u$}},T)}$ is orthogonal to ker{\it
L}$_{f_{0}(n,\mbox{\boldmath {\scriptsize $u$}},T)}$ in the sense of the scalar product $(\cdot
,\cdot )_{f_{0}(n,\mbox{\boldmath {\scriptsize $u$}},T)}$.

 \noindent d) The projecting operator
$\Pi _{f_{0}(n,\mbox{\boldmath {\scriptsize $u$}},T)}$ is the selfadjoint projector onto ker{\it
L}$_{f_{0}(n,\mbox{\boldmath {\scriptsize $u$}},T)}$:
\begin{equation}
\Pi _{f_{0}(n,\mbox{\boldmath {\scriptsize $u$}},T)}(\varphi ) \in \hbox{ ker{\it
L}}_{f_{0}(n,\mbox{\boldmath {\scriptsize $u$}},T)} \label{4.23}\end{equation} \noindent Projector
$\Pi _{f_{0}(n,\mbox{\boldmath {\scriptsize $u$}},T)}$ projects orthogonally.

\noindent e)  The  image  of  the  operator $K_{f_{0}(n,\mbox{\boldmath {\scriptsize $u$}},T)}$  is
orthogonal   to ker{\it L}$_{f_{0}(n,\mbox{\boldmath {\scriptsize $u$}},T)}$.

\noindent f) Additional condition (\ref{4.22}) require the  solution  of  equation (\ref{4.20}) to be
orthogonal to ker{\it L}$_{f_{0}(n,\mbox{\boldmath {\scriptsize $u$}},T)}$.

These  properties  result  in  the  {\it necessity  condition}  for solving the equation (\ref{4.20})
with the additional constraint (\ref{4.22}). This  means  the  following:  equation   (\ref{4.20}),
provided   with constraint (\ref{4.22}), satisfies the necessary condition for  to  have an unique
solution in Im{\it L}$_{f_{0}(n,\mbox{\boldmath {\scriptsize $u$}},T)}$.

\noindent {\bf Remark.} Because  of  the  {\it differential}  part  of   the   operator
$K_{f_{0}(n,\mbox{\boldmath {\scriptsize $u$}},T)}$, we are not able to apply the Fredholm
alternative to obtain the {\it necessary and sufficient} conditions for solvability  of equation
(\ref{4.22}). Thus, the condition mentioned here is, rigorously speaking, only the  necessity
condition. Nevertheless,  we will still develop a formal procedure for solving the equation
(\ref{4.20}).

To this end, we paid no attention to the  dependency  of  all functions, spaces, operators, etc, on
\xx. It is useful  to  rewrite once again the equation (\ref{4.20}) in order to separate the local in
$\xx$ operators from those differential. Furthermore,  we  shall  replace the subscript
$f_{0}(n,\uu,T)$ with the subscript $\xx$ in  all  expressions. We represent (\ref{4.20}) as:
\begin{eqnarray}
A_{\mbox{\scriptsize loc}}(\xx,\vv)\varphi -A_{\mbox{\scriptsize diff}}\left(\xx,{\partial \over
\partial \xx},\vv\right)\varphi =-D(\xx,\vv); \nonumber \\ A_{\mbox{\scriptsize loc}}(\xx,\vv)\varphi
=-\left\{L_{\mbox{\boldmath {\scriptsize $x$}}}(\vv)\varphi +(\Pi _{\mbox{\boldmath {\scriptsize
$x$}}}(\vv)-1)r_{\mbox{\boldmath {\scriptsize $x$}}}\varphi \right\}; \nonumber
\\ A_{\mbox{\scriptsize diff}}\left(\xx,{\partial \over \partial \xx},\vv\right)\varphi =
(\Pi _{\mbox{\boldmath {\scriptsize $x$}}}(\cdot )-1)\left((v_{s}-u_{s}){\partial \over \partial
x_{s}}\varphi \right); \nonumber
\\ \Pi _{\mbox{\boldmath {\scriptsize $x$}}}(\vv) g=\sum_{s=0}^4 \psi ^{(s)}_{\mbox{\boldmath {\scriptsize
$x$}}} (\psi ^{(s)}_{\mbox{\boldmath {\scriptsize $x$}}},g); \nonumber
\\ \psi ^{(0)}_{\mbox{\boldmath {\scriptsize $x$}}}=n^{-1/2},\qquad  \psi ^{(s)}_{\mbox{\boldmath {\scriptsize $x$}}}=(2/n)^{1/2}c_{s}
(\xx,\vv),\ s=1,2,3, \nonumber \\ \psi ^{(4)}_{\mbox{\boldmath {\scriptsize
$x$}}}=(2/3n)^{1/2}(c^2(\xx,\vv)-3/2); \qquad
c_{i}(\xx,\vv)=(m/2k_{B}T(\xx))^{1/2}(v_{i}-u_{i}(\xx)), \nonumber
\\ r_{\mbox{\boldmath {\scriptsize $x$}}}=(v_{s}-u_{s}) \left({\partial \hbox{ln{\it n}}\over \partial x_{s}} + {m\over
k_{B}T}(v_{i}-u_{i}){\partial u_{i}\over \partial x_{s}} + \left({m(\vv-\uu)^2 \over 2k_{B}T}-
{3\over 2} \right) {\partial \hbox{ln{\it T}}\over
\partial x_{s}} \right); \nonumber
\\ D(\xx,\vv)=\left\{ \left({m(\vv-\uu)^{2}\over 2k_{B}T} - {5\over 2} \right)
(v_{i}-u_{i}){\partial \hbox{ln{\it T}}\over \partial x_{i}} + \nonumber \right. \\ \left. + {m\over
k_{B}T}\left(((v_{i}-u_{i})(v_{s}-u_{s})-{1\over 3}\delta _{is} (\vv-\uu)^{2}\right){\partial
u_{s}\over
\partial x_{i}}\right\} . \label{4.24}\end{eqnarray} \noindent Here we have omitted the dependence on
$\xx$ in  the  functions $n(\xx)$, $u_{i}(\xx)$, and $T(\xx)$. Further, if no discrepancy might
occur,  we  will always assume  this  dependence,  and  we  will  not  indicate  it explicitly.

The additional condition for this equation is:
\begin{equation}
\Pi _{\mbox{\boldmath {\scriptsize $x$}}}(\varphi )=0.
\end{equation}
Equation (\ref{4.24}) is linear in $\varphi $. However, the main  difficulty in solving this equation
is caused  with  the  differential  in $\xx$ operator $A_{\mbox{\scriptsize diff}}$ which does  not
commutate with  the  local  in $\xx$ operator $A_{\mbox{\scriptsize loc}}$.

\addcontentsline{toc}{subsubsection}{Parametrix Expansion}

\subsubsection*{\textbf{Parametrix Expansion}}

In  this  subsection  we  introduce  a  procedure  to  construct approximate solutions of equation
(\ref{4.23}). This procedure  involves an expansion similar to the parametrix expansion in the theory
of pseudo-differential (PDO) and Fourier integral operators (FIO).

Considering $\varphi \in \mbox{Im} L_{\mbox{\boldmath {\scriptsize $x$}}}$, we write a formal
solution of equation (\ref{4.24}) as:
\par
\begin{equation}
\varphi (\xx,\vv)=\left(A_{\mbox{\scriptsize loc}}(\xx,\vv)-A_{\mbox{\scriptsize
diff}}\left(\xx,{\partial \over
\partial \xx},\vv\right)\right)^{-1}(-D(\xx,\vv)) \label{4.25}\end{equation} \noindent It is
useful to extract the differential operator ${\partial \over \partial \mbox{\boldmath {\scriptsize
$x$}}}$ from the operator $A_{\mbox{\scriptsize diff}}(\xx,{\partial \over \partial \mbox{\boldmath
{\scriptsize $x$}}},\vv)$:

\begin{equation}
\varphi (\xx,\vv)=\left(1-B_{s}(\xx,\vv){\partial \over \partial x_{s}}\right)^{-1}\varphi
_{\mbox{\scriptsize loc}}(\xx,\vv). \label{4.26}\end{equation} \noindent Notations used here are:
\begin{eqnarray}
\varphi _{\mbox{\scriptsize loc}}(\xx,\vv)=A^{-1}_{\mbox{\scriptsize loc}}(\xx,\vv)(-D(\xx,\vv))=
\nonumber
\\ =[-L_{\mbox{\boldmath {\scriptsize $x$}}}(\vv)-(\Pi _{\mbox{\boldmath {\scriptsize $x$}}}
(\vv)-1)r_{\mbox{\boldmath {\scriptsize $x$}}}]^{-1}(-D(\xx,\vv)); \nonumber \\
B_{s}(\xx,\vv)=A^{-1}_{\mbox{\scriptsize loc}}(\xx,\vv)(\Pi _{\mbox{\boldmath {\scriptsize
$x$}}}(\vv)-1)(v_{s}-u_{s})= \label{4.27}
\\ =[-L_{\mbox{\boldmath {\scriptsize $x$}}}(\vv)-(\Pi _{\mbox{\boldmath {\scriptsize $x$}}}
(\vv)-1)r_{\mbox{\boldmath {\scriptsize $x$}}}]^{-1}(\Pi_{\mbox{\boldmath {\scriptsize
$x$}}}(\vv)-1)(v_{s}-u_{s}) . \nonumber
\end{eqnarray}

We  will  now  discuss  in  more  detail  the  character   of expressions in (\ref{4.27}).

For every $\xx$, the function $\varphi _{\mbox{\scriptsize loc}}(\xx,\vv)$, considered as a function
of $\vv$, is an element of the Hilbert space $G_{\mbox{\boldmath {\scriptsize $x$}}}$. It gives  a
solution to the integral equation:
\par
\begin{equation}
-L_{\mbox{\boldmath {\scriptsize $x$}}}(\vv)\varphi _{\mbox{\scriptsize loc}}-(\Pi _{\mbox{\boldmath
{\scriptsize $x$}}}(\vv)-1)(r_{\mbox{\boldmath {\scriptsize $x$}}}\varphi _{\mbox{\scriptsize
loc}})=(-D(\xx,\vv)) \label{4.28}\end{equation} \noindent This latter linear integral equation has an
unique solution in $\hbox{Im} L_{\mbox{\boldmath {\scriptsize $x$}}}(\vv)$. Indeed,
\begin{eqnarray}
\hbox{ker{\it A}}^{+}_{\mbox{\scriptsize loc}}(\xx,\vv)=\ker (L_{\mbox{\boldmath {\scriptsize
$x$}}}(\vv)+(\Pi_{\mbox{\boldmath {\scriptsize $x$}}}(\vv)-1)r_{\mbox{\boldmath {\scriptsize
$x$}}})^{+}= \nonumber \\ =\ker (L_{\mbox{\boldmath {\scriptsize $x$}}}(\vv))^{+}\bigcap_{}\ker
((\Pi_{\mbox{\boldmath {\scriptsize $x$}}}(\vv)-1)r_{\mbox{\boldmath {\scriptsize $x$}}})^{+}=
\nonumber
\\  \ker (L_{\mbox{\boldmath {\scriptsize $x$}}} (\vv))^{+}\bigcap_{}\ker (r_{\mbox{\boldmath {\scriptsize $x$}}}
(\Pi_{\mbox{\boldmath {\scriptsize $x$}}}(\vv)-1)),\hbox{ and }G_{\mbox{\boldmath {\scriptsize
$x$}}}\bigcap\Pi_{\mbox{\boldmath {\scriptsize $x$}}}(\vv)G_{\mbox{\boldmath {\scriptsize
$x$}}}=\left\{0\right\} .
\end{eqnarray}
\noindent Thus, the existence of the  unique  solution  of  equation  (\ref{4.28}) follows from the
Fredholm alternative.
\par
Let us consider the operator $R(\xx,{\partial \over \partial \xx},\vv)$:
\begin{equation}
R\left(\xx,{\partial \over \partial \xx},\vv\right)=\left(1-B_{s}(\xx,\vv){\partial \over
\partial x_{s}}\right)^{-1}. \label{4.29}\end{equation} \noindent One can represent it as a formal series:
\par
\begin{equation}
R\left(\xx,{\partial \over \partial \xx},\vv\right)=\sum_{m=0}^{\infty} \left[B_{s}(\xx,\vv){\partial
\over \partial x_{s}}\right]^{m}. \label{4.30}\end{equation} Here
\begin{equation}
\left[B_{s}(\xx,\vv){\partial \over \partial x_{s}}\right]^{m}=B_{s_{1}}(\xx,\vv){\partial \over
\partial x_{s_{1}}} \ldots B_{s_{m}}(\xx,\vv){\partial \over \partial x_{s_{m}}}.
\label{4.31}\end{equation} \noindent Every term of the type (\ref{4.31}) can be represented as a
finite  sum of  operators  which  are  superpositions  of  the  following  two operations: of the
integral in $\vv$ operations with kernels depending on $\xx$, and of differential in $\xx$
operations.

Our goal is to  obtain  an  explicit  representation  of  the operator $R(\xx,{\partial \over
\partial \xx},\vv) (\ref{4.29})$ as an integral operator. If the operator $B_{s}(\xx,\vv)$ would not depend on $\xx ($i.e., if no dependence  on  spatial variables  would  occur  in
kernels  of  integral  operators,  in $B_{s}(\xx,\vv))$,  then  we  could  reach  our  goal via usual
Fourier transformation.  However,  operators $B_{s}(\xx,\vv)$  and ${\partial \over
\partial x_{k}}$   do   not commutate, and thus this elementary approach  does  not  work.  We will
develop  a  method   to   obtain   the   required   explicit representation using the ideas of PDO
and IOF technique.

We start with the representation (\ref{4.30}). Our strategy  is  to transform every summand
(\ref{4.31}) in order to  place  integral  in $\vv$ operators $B_{s}(\xx,\vv)$  left  to differential
operators ${\partial \over \partial x_{k}}$.   The transposition  of  every  pair ${\partial \over
\partial x_{k}}B_{s}(\xx,\vv)$  yields  an  elementary transform:

\begin{equation}
{\partial \over \partial x_{k}}B_{s}(\xx,\vv) \to B_{s}(\xx,\vv){\partial \over
\partial x_{k}} - \left[B_{s}(\xx,\vv),{\partial \over \partial x_{k}}\right].
\label{4.32}\end{equation} \noindent Here $[M,N]=MN-NM$ denotes the commutator of operators $M$ and
{\it N}.  We can represent (\ref{4.31}) as:
\par
\begin{eqnarray}
\left[B_{s}(\xx,\vv){\partial \over \partial x_{s}}\right]^{m} = B_{s_{1}}(\xx,\vv)\ldots
B_{s_{m}}(\xx,\vv){\partial \over \partial x_{s_{1}}}\ldots {\partial \over
\partial x_{s_{m}}} + \nonumber \\ + O\left(\left[B_{s_{i}}(\xx,\vv),{\partial \over \partial
x_{s_{k}}}\right]\right) . \label{4.33}\end{eqnarray}

\noindent Here $O([B_{s_{i}}(\xx,\vv),{\partial \over \partial x_{s_{k}}}])$ denotes the terms which
contain  one  or more pairs of brackets $[\cdot ,\cdot ]$. The first term in (\ref{4.33}) contains no
these brackets. We can continue  this  process  of selection  and extract the first-order in the
number of pairs of brackets  terms, the second-order terms, etc. Thus, we arrive at the {\bf
expansion into powers of commutator} of the expressions (\ref{4.31}).

In this paper we will consider  explicitly  the  zeroth-order term of this  commutator  expansion.
Neglecting  all  terms  with brackets in (\ref{4.33}), we write:

\begin{equation}
\left[B_{s}(\xx,\vv){\partial \over \partial x_{s}}\right]^{m}_{0} = B_{s_{1}}(\xx,\vv)\ldots
B_{s_{m}}(\xx,\vv){\partial \over \partial x_{s_{1}}}\ldots {\partial \over
\partial x_{s_{m}}} . \label{4.33a}
\end{equation}
\noindent Here the subscript zero indicates the zeroth order with respect to the number of brackets.
\par
We now substitute expressions $[B_{s}(\xx,\vv){\partial \over \partial x_{s}}]^{m}_{0} (\ref{4.33a})$
instead of expressions $[B_{s}(\xx,\vv){\partial \over \partial x_{s}}]^{m}$ (\ref{4.31}) into the
series (\ref{4.30}):
\begin{equation}
R_{0}\left(\xx,{\partial \over \partial x},\vv\right) =\sum_{m=0}^{\infty}
\left[B_{s}(\xx,\vv){\partial \over \partial x_{s}}\right]^{m}_{0} . \label{4.30a}
\end{equation}
\noindent The action of every summand $(\ref{4.33a})$  might  be  defined  via  the Fourier transform
with respect to spatial variables.

Denote as $F$  the  direct  Fourier  transform  of  a  function $g(\xx,\vv)$:
\begin{equation}
Fg(\xx,\vv)\equiv \hat{g}(\kk,\vv)= \int g(\xx,\vv)\exp (-ik_{s}x_{s})d^{p}\xx . \label{4.34a}
\end{equation}
\noindent Here $p$  is  the  spatial  dimension.  Then  the  inverse  Fourier transform is:

\begin{equation}
g(x,\vv)\equiv F^{-1} \hat{g}( \kk,\vv)=(2\pi )^{-p}\int \hat{g}(\kk,\vv)\exp (ik_{s}x_{s})d^{p}\kk .
\label{4.34b}
\end{equation}
\noindent The action of the operator $(\ref{4.33a})$ on a function $g(\xx,\vv)$ is defined as:

\begin{eqnarray}
\left[B_{s}(\xx,\vv){\partial \over \partial x_{s}}\right]^{m}_{0}g(\xx,\vv)= \nonumber
\\ =\left(B_{s_{1}}(\xx,\vv)\ldots B_{s_{m}}(\xx,\vv){\partial \over \partial
x_{s_{1}}}\ldots {\partial \over \partial x_{s_{m}}}\right)(2\pi )^{-p}\int
\hat{g}(\kk,\vv)e^{ik_{s}x_{s}}d^{p}\kk= \nonumber \\ =(2\pi )^{-p}\int \exp
(ik_{s}x_{s})[ik_{l}B_{l}(\xx,\vv)]^{m} \hat{g}(\kk,\vv)d^{p}\kk . \label{4.35}\end{eqnarray}
\noindent The account of (\ref{4.35}) in the formula $(\ref{4.30a})$ yields the  following definition
of the operator $R_{0}$:
\begin{equation}
R_{0}g(\xx,\vv)=(2\pi )^{-p}\int e^{ik_{s}x_{s}}(1-ik_{l}B_{l}(\xx,\vv))^{-1}\hat{g}(\kk,\vv)d^{p}\kk
. \label{4.36}\end{equation} \noindent This is the {\bf Fourier integral operator} (note that  the
kernel  of this integral operator depends on $\kk$ and  on $\xx)$.  The  commutator expansion
introduced  above  is  a  version  of  the  {\bf parametrix expansion} \cite{25,26}, while expression
(\ref{4.36}) is the leading term  of this expansion. The kernel $(1-ik_{l}B_{l}(\xx,\vv))^{-1}$  is
called  the  {\bf main symbol of the parametrix}.

The account of  (\ref{4.36})  in  the  formula  (\ref{4.26})  yields  the zeroth-order term of
parametrix expansion $\varphi _{0}(\xx,\vv)$:
\begin{equation}
\varphi _{0}(\xx,\vv)=F^{-1}(1-ik_{l}B_{l}(\xx,\vv))^{-1}F\varphi _{\mbox{\scriptsize loc}}.
\label{4.37}\end{equation} \noindent In detail notation:
\par
\begin{eqnarray}
\varphi _{0}(\xx,\vv)=(2\pi )^{-p}\int \int \exp (ik_{s}(x_{s}-y_{s}))\times \nonumber \\ \times
(1-ik_{s}[-L_{\mbox{\boldmath {\scriptsize $x$}}}(\vv)-(\Pi_{\mbox{\boldmath {\scriptsize
$x$}}}(\vv)-1)r_{\mbox{\boldmath {\scriptsize $x$}}}]^{-1} (\Pi_{\mbox{\boldmath {\scriptsize
$x$}}}(\vv)-1)(v_{s}-u_{s}(\xx)))^{-1}\times \nonumber \\ \times [-L_{\mbox{\boldmath {\scriptsize
$y$}}}(\vv)-(\Pi_{\mbox{\boldmath {\scriptsize $y$}}}(\vv)-1)r_{\mbox{\boldmath {\scriptsize
$y$}}}]^{-1}(-D(\yy,\vv))d^{p} \yy d^{p} \kk . \label{4.38} \end{eqnarray}

We now will list the steps to calculate the function $\varphi _{0}(\xx,\vv)$ (\ref{4.38}).

\noindent {\bf Step 1}. Solve the linear integral equation
\begin{equation}
[-L_{\mbox{\boldmath {\scriptsize $x$}}}(\vv)-(\Pi _{\mbox{\boldmath {\scriptsize
$x$}}}(\vv)-1)r_{\mbox{\boldmath {\scriptsize $x$}}}]\varphi _{\mbox{\scriptsize
loc}}(\xx,\vv)=-D(\xx,\vv) . \label{4.39a}
\end{equation}
\noindent and obtain the function $\varphi _{\mbox{\scriptsize loc}}(\xx,\vv)$.

\noindent {\bf Step 2.} Calculate the Fourier transform $\hat{\varphi}_{\mbox{\scriptsize
loc}}(\kk,\vv)$:
\begin{equation}
\hat{\varphi}_{\mbox{\scriptsize loc}} (\kk,\vv)=\int \varphi _{\mbox{\scriptsize loc}}(\yy,\vv)\exp
(-ik_{s}y_{s})d^{p}\yy . \label{4.39b}
\end{equation}
\noindent {\bf Step 3}. Solve the linear integral equation
\begin{eqnarray}
[-L_{\mbox{\boldmath {\scriptsize $x$}}}(\vv)-(\Pi_{\mbox{\boldmath {\scriptsize
$x$}}}(\vv)-1)(r_{\mbox{\boldmath {\scriptsize
$x$}}}+ik_{s}(v_{s}-u_{s}(\xx))]\hat{\varphi}_{0}(\xx,\kk,\vv)=-\hat{D}(\xx,\kk,\vv); \nonumber \\
-\hat{D}(\xx,\kk,\vv)=[-L_{\mbox{\boldmath {\scriptsize $x$}}}(\vv)-(\Pi_{\mbox{\boldmath
{\scriptsize $x$}}}(\vv)-1)r_{\mbox{\boldmath {\scriptsize $x$}}}]\hat{\varphi}_{\mbox{\scriptsize
loc}}(\kk,\vv) . \label{4.39c}
\end{eqnarray}
\noindent and obtain the function $\hat{\varphi}_{0}(\xx,\kk,\vv)$.

\noindent {\bf Step 4.} Calculate the inverse Fourier transform $\varphi _{0}(\xx,\vv)$:
\begin{equation}
\varphi _{0}(\xx,\vv)=(2\pi )^{-p}\int \hat{\varphi}_{0}(\xx,\kk,\vv)\exp (ik_{s}x_{s})d^{p}\kk .
\label{4.39d}
\end{equation}
Completing these four steps, we obtain an explicit expression for the zeroth-order term of parametrix
expansion $\varphi _{0}(\xx,\vv) (\ref{4.37})$.

As we have already mentioned it above,  equation $(\ref{4.39a})$  of Step 1 has an unique solution in
$\hbox{Im} L_{\mbox{\boldmath {\scriptsize $x$}}} (\vv)$.
 Equation $(\ref{4.39c})$ of Step
3 has the same property. Indeed, for every $\kk$, the right hand side $-\hat{D}(\xx,\kk,\vv)$ is
orthogonal to $\hbox{Im} \Pi_{\mbox{\boldmath {\scriptsize $x$}}}(\vv)$, and thus the existence and
the uniqueness of formal solution $\hat{\varphi}_{0}(\xx,\kk,\vv)$  follows  again  from  the
Fredholm alternative.

Thus, in Step 3, we obtain the unique solution $\hat{\varphi}_{0}(\xx,\kk,\vv)$. For every $\kk$,
this is a function which belongs to  $\hbox{Im}  L_{\mbox{\boldmath {\scriptsize $x$}}}(\vv)$.
Accounting that $f_{0}(\xx,\vv)=f_{0}(n(\xx),\uu(\xx),T(\xx),\vv)$ expose no explicit dependency on
$\xx$, we see that the inverse  Fourier  transform  of  Step  4  gives $\varphi _{0}(\xx,\vv)\in
\hbox{Im} L_{\mbox{\boldmath {\scriptsize $x$}}}(\vv)$.
\par
Equations (\ref{4.39a})-(\ref{4.39d})  provide  us  with  the  scheme  of constructing  the
zeroth-order  term  of  parametrix  expansion. Finishing this  section,  we  will  outline  briefly
the  way  to calculate the first-order term of this expansion.

Consider a formal operator $R=(1-AB)^{-1}$. Operator $R$ is  defined by a formal series:

\begin{equation}
R=\sum_{m=0}^{\infty}(AB)^{m} . \label{4.40}\end{equation} \noindent In every term of this series, we
want to place operators {\it A} left to operators {\it B}. In order to do this, we have to commutate
$B$  with  {\it A} from left to right. The commutation of every pair $BA$  yields  the elementary
transform $BA \to  AB - [A,B]$   where $[A,B]=AB- BA$. Extracting the terms with no commutators
$[A,B]$ and with  a  single commutator $[A,B]$, we arrive at the following representation:
\begin{equation}
R=R_{0}+R_{1}+(\mbox{terms with more than two brackets})  . \label{4.41a}\end{equation} \noindent
Here
\begin{eqnarray}
R_{0}=\sum_{m=0}^{\infty} A^{m}B^{m}; \label{4.41b} \\ R_{1}= -\sum_{m=2}^{\infty}
\sum_{i=2}^{\infty} iA^{m-i}[A,B]A^{i-1}B^{i-1}B^{m-i} . \label{4.41c}
\end{eqnarray}
\noindent Operator $R_{0}$ (\ref{4.41b})  is  the  zeroth-order  term  of  parametrix expansion
derived above. Operator $R_{1}$ (the  {\bf first-order  term  of parametrix expansion}) can be
represented as follows:
\begin{equation}
R_{1}=-\sum_{m=1}^{\infty}mA^{m}[A,B](\sum_{i=0}^{\infty} A^{i}B^{i})B^{m}= -\sum_{m=1}^{\infty}
mA^{m}CB^{m},\ \  C=[A,B]R_{0} . \label{4.41d}
\end{equation}
\noindent This expression can be considered as  an  {\it ansatz}  for  the  formal series
(\ref{4.40}), and it gives the most convenient way  to  calculate $R_{1}$. Its structure is similar
to that of $R_{0}$.  Continuing  in  this manner, we can derive the second-order term $R_{2}$, etc.
We  will  not discuss these questions in this paper.

In the next subsection we will consider in  more  detail  the first-order term of parametrix
expansion.

\addcontentsline{toc}{subsubsection}{Finite-Dimensional Approximations to Integral Equations}

\subsubsection*{\textbf{Finite-Dimensional Approximations to Integral Equations}}

Dealing  further  only  with the zeroth-order   term   of parametrix expansion (\ref{4.38}), we have
to solve two linear integral equations, $(\ref{4.39a})$  and $(\ref{4.39c})$.  These  equations
satisfy  the Fredholm alternative, and thus they  have  unique  solutions.  The problem we face here
is exactly of the same level of complexity as that of the Chapman-Enskog method \cite{Chapman}. The
usual approach is  to replace integral operators     with     some     appropriate finite-dimensional
operators.
\par
First  we  will  recall  standard   objectives   of   finite- dimensional  approximations,
considering  equation (\ref{4.39a}).  Let $p_{i}(\xx,\vv)$, where $i=1,2,\ldots $, be a basis in
$\mbox{Im} L_{\mbox{\boldmath {\scriptsize $x$}}}(\vv)$.  Every  function $\varphi (\xx,\vv)\in
\mbox{Im}L_{\mbox{\boldmath {\scriptsize $x$}}}(\vv)$ might be represented in this basis as:

\begin{equation}
\varphi (\xx,\vv)=\sum_{i=1}^{\infty} a_{i}(\xx)p_{i}(\xx,\vv); a_{i}(\xx)=(\varphi
(\xx,\vv),p_{i}(\xx,\vv))_{\mbox{\boldmath {\scriptsize $x$}}} . \label{4.42}\end{equation} \noindent
Equation (\ref{4.39a}) is equivalent  to  an  infinite  set  of  linear algebraic equations with
respect to unknowns $a_{i}(\xx)$:
\par
\begin{equation}
\sum_{i=1}^{\infty} m_{ki}(\xx)a_{i}(\xx)=d_{k}(\xx),\ \  k=1,2,\ldots . \label{4.43}
\end{equation} \noindent Here
\begin{eqnarray}
m_{ki}(\xx)=(p_{k}(\xx,\vv),A_{\mbox{\scriptsize loc}}(\xx,\vv)p_{i}(\xx,\vv))_{\mbox{\boldmath
{\scriptsize $x$}}}; \nonumber
\\ \qquad d_{k}(\xx)=-(p_{k}(\xx,\vv),D(\xx,\vv))_{\mbox{\boldmath {\scriptsize
$x$}}} . \label{4.44}
\end{eqnarray}
For a finite-dimensional approximation of equation (\ref{4.43})  we use a projection onto a finite
number of basis  elements $p_{i}(\xx,\vv), i=i_{1},\ldots ,i_{n}$. Then, instead of (\ref{4.42}), we
search for  the  function $\varphi_{\mbox{\scriptsize fin}}$:
\begin{equation}
\varphi _{\mbox{\scriptsize fin}}(\xx,\vv)=\sum_{s=1}^{n} a_{i_{s}} (\xx)p_{i_{s}}(\xx,\vv) .
\label{4.45a}
\end{equation}
\noindent Infinite set of equations (\ref{4.43}) is replaced with a finite set  of linear  algebraic
equations  with  respect   to $a_{i_{s}}(\xx)$,   where $s=1,\ldots ,n$:
\begin{equation}
\sum_{l=1}^{n} m_{i_{s}i_{l}}(\xx)a_{i_{l}}(\xx)= d_{i_{s}}(\xx),\ \  s=1,\ldots ,n . \label{4.45b}
\end{equation}
There are no a priori restrictions upon  the  choice  of  the basis, as well  as  upon  the  choice
of  its  finite-dimensional approximations. In  this  paper  we  use  the  standard  basis  of
unreducible Hermite tensors (see, for example, \cite{Cercignani,Grad}. The  simplest appropriate
version of a finite-dimensional  approximation  occurs if the finite set of Hermite tensors is chosen
as:
\begin{eqnarray}
p_{k}(\xx,\vv)=c_{k}(\xx,\vv)(c^{2}(\xx,\vv)-(5/2)), k=1,2,3; \nonumber
\\ p_{ij}(\xx,\vv)=c_{i}(\xx,\vv)c_{j}(\xx,\vv)-{1\over 3}\delta
_{ij}c^{2}(\xx,\vv),\ \  i,j=1,2,3; \nonumber \\
c_{i}(\xx,\vv)=\vv^{-1}_{T}(\xx)(v_{i}-u_{i}(\xx)),\quad v_{T}(\xx)=(2k_{B}T(\xx)/m)^{1/2} .
\label{4.46}\end{eqnarray} It is important to stress  here  that  "good"  properties  of
orthogonality of Hermite tensors,  as well  as  of  other similar polynomial systems in BE theory,
have the local  in $\xx$ character, i.e. when these functions are treated  as  polynomials  in
$c(\xx,\vv)$ rather than polynomials in \vv. For example, functions $p_{k}(\xx,\vv)$  and
$p_{ij}(\xx,\vv) (\ref{4.46})$ are orthogonal in the sense of the scalar  product $(\cdot ,\cdot
)_{\mbox{\boldmath {\scriptsize $x$}}}$:
\begin{equation}
(p_{k}(\xx,\vv),p_{ij}(\xx,\vv))_{\mbox{\boldmath {\scriptsize $x$}}} \propto \int
e^{-c^{2}(\mbox{\boldmath {\scriptsize $x,v$}})}p_{k}(\xx,\vv)p_{ij}(\xx,\vv)d^{3}c(\xx,\vv)=0 .
\end{equation}
\noindent On contrary, functions $p_{k}(\yy,\vv)$ and $p_{ij}(\xx,\vv)$  are  not orthogonal neither
in the sense of the scalar  product $(\cdot ,\cdot )_{\mbox{\boldmath {\scriptsize $y$}}}$,  nor  in
the sense of the scalar product $(\cdot ,\cdot )_{\mbox{\boldmath {\scriptsize $x$}}}$, if $\yy \neq
\xx$.  This  distinction  is important for constructing the parametrix expansion. Further,  we will
omit the  dependencies  on $\xx$ and $\vv$ in  the dimensionless velocity $c_{i}(\xx,\vv)
(\ref{4.46})$ if no misunderstanding might occur.
\par
In this section we will consider the case of one-dimensional in $\xx$ equations. We assume that:
\begin{equation}
u_{1}(\xx)=u(x_{1}),\quad u_{2}=u_{3}=0,\quad T(\xx)=T(x_{1}),\quad n(\xx)=n(x_{1}) .
\label{4.47}\end{equation} \noindent We write $x$ instead of $x_{1}$ below.  Finite-dimensional
approximation (\ref{4.46}) requires only two functions:
\begin{eqnarray}
p_{3}(x,\vv)=c^{2}_{1}(x,\vv)-{1\over 3}c^{2}(x,\vv), \quad
p_{4}(x,\vv)=c_{1}(x,\vv)(c^{2}(x,\vv)-(5/2)),\nonumber \\ c_{1}(x,\vv)=v^{-1}_{T}(x)(v_{1}-u(x)),
c_{2,3}(x,\vv)=v^{-1}_{T}(x)v_{2,3} . \label{4.48}\end{eqnarray} We now  will  make  a step-by-step
calculation of   the zeroth-order term of parametrix expansion, in the one-dimensional case, for the
finite-dimensional approximation (\ref{4.48}).

\textbf{Step 1.} {\it Calculation of} $\varphi _{\mbox{\scriptsize loc}}(x,\vv)$ {\it from equation}
(\ref{4.39a}).

We search for the function $\varphi _{\mbox{\scriptsize loc}}(x,\vv)$  in  the  approximation
(\ref{4.48}) as:
\begin{equation}
\varphi _{\mbox{\scriptsize loc}}(x,\vv)=a_{\mbox{\scriptsize
loc}}(x)(c^{2}_{1}-(1/3)c^{2})+b_{\mbox{\scriptsize loc}}(x)c_{1}(c^{2}-(5/2)) .
\label{4.49}\end{equation} \noindent Finite-dimensional  approximation $(\ref{4.45b})$  of  integral
equation (\ref{4.39a}) in the basis (\ref{4.48}) yields:
\begin{eqnarray}
m_{33}(x)a_{\mbox{\scriptsize loc}}(x)+m_{34}(x)b_{\mbox{\scriptsize loc}}(x)=\alpha
_{\mbox{\scriptsize loc}}(x); \nonumber
\\ m_{43}(x)a_{\mbox{\scriptsize loc}}(x)+m_{44}(x)b_{\mbox{\scriptsize loc}}(x)=\beta _{\mbox{\scriptsize
loc}}(x). \label{4.50}\end{eqnarray} \noindent Notations used are:
\begin{eqnarray}
m_{33}(x)=n(x)\lambda _{3}(x)+{11\over 9} {\partial u\over \partial x};\qquad m_{44}(x)=n(x)\lambda
_{4}(x)+{27\over 4} {\partial u\over \partial x}; \nonumber \\ m_{34}(x)=m_{43}(x)={v_{T}(x)\over 3}
\left({\partial \hbox{ln{\it n}}\over \partial x} + {11\over 2} {\partial \hbox{ln{\it T}}\over
\partial x}  \right); \nonumber \\ \lambda _{3,4}(x)=-{1\over \pi ^{3/2}}
\int e^{-c^{2}(\mbox{\boldmath {\scriptsize $x,v$}})}p_{3,4}(x,\vv)L_{\mbox{\boldmath {\scriptsize
$x$}}}(\vv)p_{3,4}(\xx,\vv)d^{3}c(\xx,\vv)>0 ; \nonumber
\\ \alpha _{\mbox{\scriptsize loc}}(x)=-{2\over 3} {\partial u\over \partial x};\qquad \beta
_{\mbox{\scriptsize loc}}(x)=-{5\over 4} v_{T}(x){\partial \hbox{ln{\it T}}\over \partial x} .
\label{4.51}\end{eqnarray} \noindent Parameters $\lambda _{3}(x)$  and $\lambda _{4}(x)$  are  easily
expressed  via  Enskog integral brackets, and they are calculated in \cite{Chapman} for a wide class
of molecular models.

Solving equation (\ref{4.50}), we obtain coefficients $a_{\mbox{\scriptsize loc}}(x)$  and
$b_{\mbox{\scriptsize loc}}(x)$ in the expression (\ref{4.49}):
\begin{eqnarray}
a_{\mbox{\scriptsize loc}}={A_{\mbox{\scriptsize loc}}(x)\over Z(x,0)};\qquad b_{\mbox{\scriptsize
loc}}={B_{\mbox{\scriptsize loc}}(x)\over Z(x,0)};\qquad Z(x,0)=m_{33}(x)m_{44}(x)-m^{2}_{34}(x);
\nonumber \\ A_{\mbox{\scriptsize loc}}(x)=\alpha _{\mbox{\scriptsize loc}}(x)m_{44}(x)-\beta
_{\mbox{\scriptsize loc}}(x)m_{34}(x); \nonumber \\ B_{\mbox{\scriptsize loc}}(x)=\beta
_{\mbox{\scriptsize loc}}(x)m_{33}(x)-\alpha _{\mbox{\scriptsize loc}}(x)m_{34}(x); \nonumber \\
a_{\mbox{\scriptsize loc}}=\displaystyle \frac{ \displaystyle -{2\over 3} {\partial u\over \partial
x} \left( n\lambda _{4}+{27\over 4} {\partial u \over \partial x} \right) +{5\over
12}v^{2}_{T}{\partial \ln T \over \partial x} \left( {\partial \ln n \over \partial x}+{11\over 2}
{\partial \ln T\over
\partial x} \right)} {\displaystyle \left( n\lambda_3 +{11 \over 9}{\partial u \over \partial x}
\right) \left( n\lambda_4 + {27 \over 4}{\partial u \over \partial x} \right) - {v_T^2 \over 9}
\left( {\partial \ln n \over
\partial x} + {11 \over 2} {\partial \ln T \over \partial x} \right)^2}; \nonumber \\
 b_{\mbox{\scriptsize loc}}=
\displaystyle \frac{ \displaystyle -{5\over 4}v_{T}{\partial \ln T \over \partial x} \left( n\lambda
_{3}+{11\over 9} {\partial u \over \partial x} \right) + {2\over 9}v_{T}{\partial u\over \partial x}
\left( {\partial \ln n \over \partial x} +{11\over 2} {\partial \ln  T \over \partial x} \right)}
{\displaystyle \left( n\lambda_3 + {11 \over 9}{\partial u \over x} \right) \left( n\lambda_4 + {27
\over 4} {\partial u \over \partial x} \right) - {v_T^2 \over 9} \left( {\partial \ln n \over x} +
{11 \over 2} {\partial \ln T \over x} \right)^2 } . \label{4.52}
\end{eqnarray}
\noindent These expressions complete Step 1.
\par
\textbf{ Step 2.} {\it Calculation of Fourier  transform  of {$\varphi _{\mbox{\scriptsize
loc}}(x,\vv)$} and its expression in the local basis}.

In this step we make two operations:

 \noindent i) The Fourier transformation of the function $\varphi
_{\mbox{\scriptsize loc}}(x,\vv)$:
\begin{equation}
\hat{\varphi}_{\mbox{\scriptsize loc}}(k,\vv)=\int_{-\infty}^{+\infty}
 \exp (-\hbox{{\it iky}})\varphi _{\mbox{\scriptsize loc}}(y,\vv)dy .
\label{4.53}\end{equation} \noindent ii)  The  representation  of $\hat{\varphi}_{\mbox{\scriptsize
loc}}(k,\vv)$ in   the   local   basis $\left\{p_{0}(x,\vv),\ldots ,p_{4}(x,\vv)\right\}$:
\begin{eqnarray}
p_{0}(x,\vv)=1, p_{1}(x,\vv)=c_{1}(x,\vv), p_{2}(x,\vv)=c^{2}(x,\vv)-(3/2), \nonumber
\\ p_{3}(x,\vv)=c^{2}_{1}(x,\vv)-(1/3)c^{2}(x,\vv), p_{4}(x,\vv)=c_{1}(x,\vv)(c^{2}(x,\vv)-(5/2)) .
\label{4.54}\end{eqnarray} \noindent Operation (ii) is necessary for completing Step 3 because there
we deal with $x$-dependent operators. Obviously, the function $\hat{\varphi} _{\mbox{\scriptsize
loc}}(k,\vv)$ (\ref{4.53}) is a finite-order polynomial in $\vv$, and thus the operation (ii) is
exact.

We obtain in (ii):
\begin{equation}
\hat{\varphi}_{\mbox{\scriptsize loc}}(x,k,\vv)\equiv \hat{\varphi}_{\mbox{\scriptsize
loc}}(x,k,c(x,\vv))=\sum_{i=0}^4 \hat{h}_{i}(x,k)p_{i}(x,\vv) . \label{4.55}\end{equation} \noindent
Here
\begin{equation}
\hat{h}_{i}(x,k)=(p_{i}(x,\vv),p_{i}(x,\vv))^{-2}_{x}(\hat{\varphi}_{\mbox{\scriptsize
loc}}(k,\vv),p_{i}(x,\vv))_{x} \label{4.56}\end{equation} . \noindent Let us introduce notations:
\begin{equation}
\vartheta \equiv \vartheta (x,y)=(T(x)/T(y))^{1/2}, \gamma \equiv \gamma (x,y)={u(x)-u(y)\over
v_{T}(y)} .  \label{4.57}\end{equation} \noindent Coefficients $\hat{h}_{i}(x,k) (\ref{4.56})$ have
the following explicit form:
\begin{eqnarray}
\hat{h}_{i}(x,k)= \int^{+\infty }_{-\infty } \exp (-\hbox{{\it iky}})h_{i}(x,y)dy;
h_{i}(x,y)=Z^{-1}(y,0)g_{i}(x,y) \nonumber \\ g_{0}(x,y)= B_{\mbox{\scriptsize loc}}(y)(\gamma
^{3}+{5\over 2}\gamma (\vartheta ^{2}-1)) + {2\over 3}A_{\mbox{\scriptsize loc}}(y)\gamma ^{2};
\nonumber \\ g_{1}(x,y)=B_{\mbox{\scriptsize loc}}(y)(3\vartheta \gamma ^{2}+ {5\over 2}\vartheta
(\vartheta ^{2}-1)) + {4\over 3}A_{\mbox{\scriptsize loc}}(y)\vartheta \gamma ; \nonumber \\
g_{2}(x,y)={5\over 3}B_{\mbox{\scriptsize loc}}(y)\vartheta ^{2}\gamma ; \nonumber \\
g_{3}(x,y)=B_{\mbox{\scriptsize loc}}(y)2\vartheta \gamma + A_{\mbox{\scriptsize loc}}(y)\vartheta
^{2}; \nonumber
\\ g_{4}(x,y)=B_{\mbox{\scriptsize loc}}(y)\vartheta ^{3} . \label{4.58}\end{eqnarray} \noindent Here $Z(y,0),
B_{\mbox{\scriptsize loc}}(y)$ and $A_{\mbox{\scriptsize loc}}(y)$ are functions defined in
(\ref{4.52})

\textbf{Step 3.} {\it Calculation  of  the  function  $\hat{\varphi}_{0}(x,k,\vv)$  from equation}
(\ref{4.39c}).

Linear integral equation (\ref{4.39c})  has  character  similar  to that of equation (\ref{4.39a}).
We search for the function $\hat{\varphi}_{0}(x,k,\vv)$  in the basis (\ref{4.48}) as:
\begin{equation}
\hat{\varphi}_{0}(x,k,\vv)=\hat{a}_{0}(x,k)p_{3}(x,\vv)+\hat{b}_{0}(x,k)p_{4}(x,\vv).
\label{4.59}\end{equation} \noindent Finite-dimensional approximation of the integral equation
$(\ref{4.39c})$ in the basis (\ref{4.48}) yields the following  equations  for  unknowns
$\hat{a}_{0}(x,k)$ and $\hat{b}_0 (x,k)$:
\begin{eqnarray}
m_{33}(x)\hat{a}_{0}(x,k)+\left[ m_{34}(x)+{1\over 3}\hbox{{\it ikv}}_{T}(x)\right]
\hat{b}_{0}(x,k)=\hat{\alpha}_{0}(x,k);\nonumber \\ \left[ m_{43}(x)+{1\over 3}\hbox{{\it
ikv}}_{T}(x) \right] \hat{a}_{0}(x,k)+m_{44}(x) \hat{b}_{0}(x,k)=\hat{\beta}_{0}(x,k) . \label{4.60}
\end{eqnarray}
\noindent Notations used here are:
\begin{eqnarray}
& \hat{\alpha}_{0}(x,k)=m_{33}(x)\hat{h}_{3}(x,k)+ m_{34}(x)\hat{h}_{4}(x,k)+\hat{s}_{\alpha }(x,k);
\nonumber & \\ & \hat{\beta}_{0}(x,k)=m_{43}(x)\hat{h}_{3}(x,k)+m_{44}(x)\hat{h}_{4}(x,k)+
\hat{s}_{\beta }(x,k); \nonumber & \\ & \hat{s}_{\alpha ,\beta }(x,k)= \int_{-\infty }^{+\infty }
\exp (-\hbox{{\it iky}})s_{\alpha ,\beta }(x,y)dy; \nonumber & \\ & s_{\alpha }(x,y)= {1\over
3}v_{T}(x)  \left({\partial \hbox{ln{\it n}}\over \partial x}+ 2{\partial \hbox{ln{\it T}}\over
\partial x} \right) h_{1}(x,y)+ {2\over 3} {\partial u\over \partial x}(h_{0}(x,y)+2h_{2}(x,y));
\label{4.61} & \\ & s_{\beta }(x,y)={5\over 4}v_{T}(x) \left( {\partial \hbox{ln{\it n}}\over
\partial x}h_{2}(x,y)+ {\partial \hbox{ln{\it T}}\over \partial x}(3h_{2}(x,y)+h_{0}(x,y)) \right)+
{2\partial u\over 3\partial x}h_{1}(x,y) . &
\end{eqnarray}
\noindent Solving equations (\ref{4.60}), we obtain functions $\hat{a}_{0}(x,k)$ and
$\hat{b}_{0}(x,k)$ in (\ref{4.59}):
\begin{eqnarray}
\hat{a}_{0}(x,k)=\frac{\hat{\alpha}_{0}(x,k)m_{44}(x)-\hat{\beta}_{0}(x,k)(m_{34}(x)+{1\over
3}\hbox{{\it ikv}}_{T}(x))}{  Z(x,{1\over 3}\hbox{{\it ikv}}_{T}(x))}; \nonumber \\
\hat{b}_{0}(x,k)=\frac{\hat{\beta}_{0}(x,k)m_{33}(x)-\hat{\alpha}_{0}(x,k)(m_{34}(x)+{1\over
3}\hbox{{\it ikv}}_{T}(x))}{ Z(x,{1\over 3}\hbox{{\it ikv}}_{T}(x))} . \label{4.62}\end{eqnarray}
\noindent Here
\begin{eqnarray}
Z(x,{1\over 3}\hbox{{\it ikv}}_{T}(x)) = Z(x,0) + {k^{2}v^{2}_{T}(x)\over 9} + {2\over 3}\hbox{{\it
ikv}}_{T}(x)m_{34}(x) = \nonumber \\ =\left(n\lambda _{3}+{11\partial u\over 9\partial x} \right)
 \left(n\lambda _{4}+{27\partial u\over 4\partial x} \right)
- {v^{2}_{T}(x)\over 9} \left({\partial \ln  n \over \partial x}+{11\partial \hbox{ln{\it T}}\over 2
\partial x} \right)^{2} + {k^{2}v^{2}_{T}(x)\over 9} + \nonumber \\ +{2\over 9}\hbox{{\it
ikv}}^{2}_{T}(x) \left({\partial \hbox{ln{\it n}}\over \partial x} + {11\partial \hbox{ln{\it
T}}\over 2 \partial x} \right) . \label{4.63}\end{eqnarray}

\noindent {\bf Step 4}. {\it Calculation  of  the inverse  Fourier  transform  of  the function
$\hat{\varphi}_{0}(x,k,\vv)$}.

The inverse  Fourier  transform  of  the  function  $\hat{\varphi}_{0}(x,k,\vv)$ (\ref{4.59}) yields:
\begin{equation}
\varphi _{0}(x,\vv)=a_{0}(x)p_{3}(x,\vv)+b_{0}(x)p_{4}(x,\vv) . \label{4.64}\end{equation} \noindent
Here
\begin{eqnarray}
a_{0}(x)={1\over 2\pi }\int^{+\infty }_{-\infty } \exp (ikx)\hat{a}_{0}(x,k)dk, \nonumber \\
b_{0}(x)={1\over 2\pi }\int^{+\infty }_{-\infty } \exp (ikx)\hat{b}_{0}(x,k)dk .
\label{4.65}\end{eqnarray} \noindent Taking into account expressions (\ref{4.52}),
(\ref{4.61})-(\ref{4.63}), and (\ref{4.58}), we obtain  the  explicit  expression  for  the  {\bf
finite-dimensional approximation of the zeroth-order term  of  parametrix  expansion} (\ref{4.64}):
\begin{eqnarray}
a_{0}(x)={1\over 2\pi }\int^{+\infty }_{-\infty }dy \int^{+\infty }_{-\infty }dk \exp
(ik(x-y))Z^{-1}(x,{1\over 3}\hbox{{\it ikv}}_{T}(x))\times \nonumber \\ \times
\left\{Z(x,0)h_{3}(x,y)+[s_{\alpha }(x,y)m_{44}(x)-s_{\beta }(x,y)m_{34}(x)]- \right. \nonumber \\
\left. -{1\over 3}\hbox{{\it ikv}}_{T}(x)[m_{34}(x)h_{3}(x,y)+m_{44}(x)h_{4}(x,y)+s_{\beta }(x,y)]
\right\}; \nonumber \\ b_{0}(x)={1\over 2\pi }\int^{+\infty }_{-\infty }dy\int^{+\infty }_{-\infty }
dk \exp (ik(x-y))Z^{-1}(x,{1\over 3}\hbox{{\it ikv}}_{T}(x))\times \nonumber \\ \times \left\{
Z(x,0)h_{4}(x,y)+[s_{\beta }(x,y)m_{33}(x)-s_{\alpha }(x,y)m_{34}(x)]- \right. \nonumber \\ \left.
-{1\over 3}{\it ikv}_{T}(x)[m_{34}(x)h_{4}(x,y)+m_{33}(x)h_{3}(x,y)+s_{\alpha }(x,y)] \right\} .
 \label{4.66} \end{eqnarray}

\addcontentsline{toc}{subsubsection}{Hydrodynamic Equations}

\subsubsection*{\textbf{Hydrodynamic Equations}}

Now we will discuss briefly the utility of obtained results for hydrodynamics.
\par
The correction to $LM$ functions $f_{0}(n,\uu,T) (\ref{4.1})$  obtained  has the form:
\par
\begin{equation}
f_{1}(n,\uu,T)=f_{0}(n,\uu,T)(1+\varphi _{0}(n,\uu,T)) \label{4.67}\end{equation} \noindent Here the
function $\varphi _{0}(n,\uu,T)$ is given explicitly  with  expressions (\ref{4.64})-(\ref{4.66}).
\par
The usual form of closed hydrodynamic equations for $n, \uu$, and $T$, where the traceless stress
tensor $\sigma _{ik}$ and the heat flux vector $q_{i}$ are expressed via hydrodynamic variables, will
be  obtained  if we substitute the function (\ref{4.67}) into balance  equations  of the density, of
the momentum, and of the energy. For $LM$ approximation, these balance equations result in Euler
equation of the  nonviscid liquid (i.e. $\sigma _{ik}(f_{0})\equiv 0,$  and $q_{i}(f_{0})\equiv 0)$.
For  the  correction $f_{1}$ (\ref{4.67}), we obtain  the  following expressions  of $\sigma =\sigma
_{xx}(f_{1})$  and $q=q_{x}(f_{1})$ (all other components are  equal to  zero  in  the  one-
dimensional situation under consideration):
\begin{equation}
\sigma ={1\over 3}na_{0}, \; q={5\over 4}nb_{0} . \label{4.68}\end{equation} \noindent Here $a_{0}$
and $b_{0}$ are given by expression (\ref{4.66}).

From the geometrical viewpoint,  hydrodynamic equations with the stress tensor and the heat flux
vector  (\ref{4.68}) have the following interpretation: we take the corrected  manifold $\Omega _{1}$
which consists of functions $f_{1}$ (\ref{4.67}), and we  project  the  BE vectors
$J_{\mbox{\boldmath {\scriptsize $u$}}}(f_{1})$ onto the tangent spaces $T_{f_{1}}$ using the $LM$
projector $P_{f_{0}}$ (\ref{4.6a}).
\par
Although  a  detailed  investigation  of  these  hydrodynamic equations is a subject of a special
study and it is not  the  goal of this paper, some points should be mentioned.

\addcontentsline{toc}{subsubsection}{Nonlocality}

\subsubsection*{\textbf{Nonlocality}}

Expressions (\ref{4.66})  expose  a  nonlocal  spatial dependency, and, hence, the corresponding
hydrodynamic  equations are nonlocal. This nonlocality appears through two contributions. The   first
of   these   contributions   might   be   called   a {\it frequency-response} contribution, and  it
comes  through  explicit non- polynomial $k$-dependency of integrands in (\ref{4.66}). This  latter
dependency has the form:
\par
\begin{equation}
\int^{+\infty}_{-\infty} {A(x,y)+\hbox{{\it ikB}}(x,y)\over C(x,y)+\hbox{{\it ikD}}(x,y)+k^{2}E(x,y)}
\exp (ik(x-y))dk . \label{4.69}\end{equation} \noindent Integration over $k$  in  (\ref{4.69})  can
be completed  via  auxiliary functions.
\par
The second nonlocal contribution might be called {\it correlative}, and it is due to relationships
via $(u(x)-u(y))$ (the difference  of flow velocities in points $x$ and $y$) and via $T(x)/T(y)$ (the
ratio of temperatures in points $x$ and $y$).

\addcontentsline{toc}{subsubsection}{Acoustic spectra}

\subsubsection*{\textbf{Acoustic spectra}}

The purely frequency-response  contribution to hydrodynamic equations is relevant to  small
perturbations  of equilibria. The stress tensor $\sigma $ and the heat flux $q (\ref{4.68})$ are:
\par
\begin{eqnarray}
\sigma  = -(2/3)n_{0}T_{0}R \left(2\varepsilon  {\partial u\over \partial \xi }^{'}- 3\varepsilon
^{2} {\partial ^{2}T\over \partial \xi ^{2}} \right); \nonumber \\ q = -(5/4)T^{3/2}_{0}n_{0}R \left(
3\varepsilon  {\partial T\over \partial \xi }^{'}-(8/5)\varepsilon ^{2} {\partial ^{2}u\over
\partial \xi ^{2}} \right) . \label{4.70}\end{eqnarray} \noindent Here

\begin{equation}
R = \left( 1 - (2/5)\varepsilon ^{2} {\partial ^{2} \over \partial \xi ^{2}} \right)-1 .
\label{4.71}\end{equation}

\noindent In (\ref{4.70}),  we  have  expressed  parameters $\lambda _{3}$ and $\lambda _{4}$  via
the viscosity coefficient $\mu $ of the Chapman-Enskog method  \cite{Chapman}  (it  is easy to see
from (\ref{4.51}) that $\lambda _{3}=\lambda _{4}\propto  \mu ^{-1}$ for spherically symmetric models
of a collision), and we have used the following  notations: $T_{0}$  and $n_{0}$ are  the equilibrium
temperature   and   density, $\xi =(\eta T^{1/2}_{0})^{-1}n_{0}x$  is  the dimensionless coordinate,
$\eta =\mu (T_{0})/T_{0}, u'=T^{-1/2}_{0}\delta u, T'=\delta T/T_{0}, n'=\delta n/n_{0}$, and $\delta
u, \delta T, \delta n$ are the  deviations of the flux velocity, of the temperature and of the
density  from their equilibrium values $u=0, T=T_{0}$  and $n=n_{0}$.  We also  use  the system of
units with $k_{B}=m=1.$
\par
In the linear case, the  parametrix  expansion  degenerates, and its zeroth-order term
$(\ref{4.39d})$ gives the solution  of  equation (\ref{4.24}).
\par
The dispersion relationship for the approximation (\ref{4.70}) is:
\par
\begin{eqnarray}
\omega ^{3}+(23k^{2}/6D)\omega ^{2}+\left\{k^{2}+(2k^{4}/D^{2})+(8k^{6}/5D^{2})\right\}\omega
+(5k^{4}/2D)=0; \nonumber \\ D=1+(4/5)k^{2} . \label{4.72}\end{eqnarray}

\noindent Here $k$ is the wave vector.

Acoustic spectra  given  by  dispersion  relationship  (\ref{4.72}) contains no nonphysical
short-wave instability  characteristic  to the Burnett  approximation (Fig. \ref{Disp}).  The
regularization of the Burnett approximation  \cite{GKJETP91,GKTTSP92}  has  the  same  feature. Both
of   these approximations predict a limit of  the  decrement $\mbox{Re}\omega $  for  short waves.

\addcontentsline{toc}{subsubsection}{Nonlinearity}

\subsubsection*{\textbf{Nonlinearity}}

Nonlinear dependency on ${\partial u\over \partial x}$,  on ${\partial \hbox{ln{\it T}}\over \partial
x}$,  and  on ${\partial \hbox{ln{\it n}}\over \partial x}$ appears already in the local
approximation $\varphi _{\mbox{\scriptsize loc}} (\ref{4.52})$.  In order to outline  some
peculiarities  of this nonlinearity,  we represent the zeroth-order term of the expansion  of
$a_{\mbox{\scriptsize loc}} (\ref{4.52})$ into powers of ${\partial \hbox{ln{\it T}}\over \partial
x}$ and ${\partial \hbox{ln{\it n}}\over
\partial x}$:
\par
\begin{equation}
a_{\mbox{\scriptsize loc}}= -{2\over 3} {\partial u\over \partial x} \left(n\lambda _{3}+{11\over 9}
{\partial u\over \partial x}  \right)^{-1}+ O \left({\partial \hbox{ln{\it T}}\over \partial x},
{\partial \hbox{ln{\it n}}\over \partial x} \right) . \label{4.73a}
\end{equation}
\noindent This expression describes the asymptotic of the "purely nonlinear" contribution  to  the
stress  tensor $\sigma  (\ref{4.68})$  for  a   strong divergency of a flow. The account of
nonlocality yields instead of $(\ref{4.70})$:
\par
\begin{eqnarray}
a_{0}(x)=-{1\over 2 \pi } \int^{+\infty}_{-\infty }dy\int^{+\infty }_{-\infty} dk \exp
(ik(x-y)){2\over 3} {\partial u\over \partial y}  \left(n\lambda _{3}+{11\over 9} {\partial u\over
\partial y} \right)^{-1}\times \nonumber \\ \times  \left[ \left( n\lambda _{3}+{11\over 9} {\partial
u\over \partial x} \right) \left(n\lambda _{4}+{27\over 4} {\partial u\over \partial x} \right) +
{k^2 v^2_{T} \over 9} \right]^{-1} \left[ \left( n\lambda _{3}+{11\over 9} {\partial u\over \partial
x} \right) \left(n\lambda _{4}+{27\over 4} {\partial u\over \partial x} \right)+ \right. \nonumber \\
\left. + {4\over 9} \left(n\lambda _{4}+{27\over 4} {\partial u\over dy} \right) {\partial u\over
\partial x} v^{-2}_{T}(u(x)-u(y))^{2}-{2\over 3}ik{\partial u\over \partial x} (u(x)-u(y)) \right] +
\nonumber \\ + O \left( {\partial \ln  T \over \partial x}, {\partial \ln  n \over \partial x}
\right) . \label{4.74b}
\end{eqnarray}
\noindent Both expressions, (\ref{4.73a}) and (\ref{4.74b}) become singular when
\par
\begin{equation}
{\partial u\over \partial y} \rightarrow \left({\partial u \over \partial y} \right)^{*}= -{9n
\lambda _{3} \over 11} . \label{4.75}\end{equation} \noindent Hence, the stress tensor (\ref{4.69})
becomes infinite if ${\partial u\over \partial y}$  tends  to ${\partial u\over \partial y}^{*}$ in
any point {\it y}. In other words, the  flow  becomes  infinitely viscid when ${\partial u\over
\partial y}$ approaches the negative value $-{9n\lambda _{3}\over 11}$. This  infinite viscosity
threshold prevents  a transfer of  the   flow   into nonphysical region of negative viscosity if
${\partial u\over
\partial y} > {\partial u\over \partial y}^{*}$  because  of the  infinitely  strong  dumping  at
${\partial u\over \partial y}^{*}$.  This  peculiarity  was detected  in  \cite{GKJETP91,GKTTSP92} as
a  result  of partial  summing  of   the Chapman-Enskog expansion. In particular, partial summing for
the simplest nonlinear situation \cite{KTTSP92,MBCh} yields the  following  expression for the stress
tensor $\sigma $:
\par
\begin{eqnarray}
\sigma =\sigma _{\mbox{\small I}R}+\sigma _{\mbox{\small II}R}; \quad \sigma _{\mbox{\small
I}R}=-{4\over 3} \left(1- {5\over 3}\varepsilon ^{2}{\partial ^{2}\over \partial \xi ^{2}}
\right)^{-1} \left( \varepsilon {\partial u\over \partial \xi }^{'}+ \varepsilon ^{2}{\partial
^{2}\theta' \over
\partial \xi ^{2}} \right); \quad \theta'=T'+n'; \nonumber \\ \sigma _{\mbox{\small II}R}={28\over
9} \left(1+{7\over 3}\varepsilon {\partial u' \over \partial \xi } \right)^{-1} {\partial ^{2}
u'\over
\partial \xi ^{2}} . \label{4.76}\end{eqnarray}

\noindent Notations here follow (\ref{4.70}) and (\ref{4.71}). Expression  (\ref{4.76})  might be
considered as a "rough draft"  of  the  "full" stress  tensor defined by $a_{0} (\ref{4.66})$. It
accounts both the frequency-response  and the nonlinear contributions $(\sigma _{\mbox{\small I}R}$
and $\sigma _{\mbox{\small II}R}$, respectively)  in  a simple  form  of  a  sum.  However,  the
superposition  of  these contributions in  (\ref{4.66})  is  more  complicated.  Moreover,   the
explicit correlative nonlocality of expression  (\ref{4.66})  was  never detected neither in
\cite{KTTSP92},  nor  in  numerous examples of partial summing \cite{MBCh}.
\par
Nevertheless, approximation (\ref{4.76}) contains  the  peculiarity of  viscosity  similar  to  that
in $(\ref{4.73a})$  and $(\ref{4.74b})$.   In dimensionless variables and $\varepsilon =1,$
expression (\ref{4.76})  predicts  the infinite threshold at velocity divergency equal to -(3/7),
rather than $-(9/11)$ in (\ref{4.73a}) and (\ref{4.74b}). Viscosity tends  to  zero  as the
divergency tends to positive infinity in both  approximations. Physical interpretation of these
phenomena was given in \cite{KTTSP92}: large {\it positive} values of ${\partial u\over \partial x}$
means that the gas diverges rapidly, and the flow becomes nonviscid because the particles  retard  to
exchange their momentum. On contrary, its {\it negative} values (such  as  -(3/7) for (\ref{4.76})
and -(9/11)) for $(\ref{4.73a})$ and $(\ref{4.74b}))$ describe a  strong compression of the flow.
Strong  deceleration results  in  "solid fluid" limit with an infinite viscosity (Fig.
\ref{Visc.eps}).
\par
\medskip

\par
Thus, hydrodynamic equations  for  approximation  (\ref{4.67})  are both nonlinear  and  nonlocal.
This  result  is  not  surprising, accounting the integro-differential character of equation
(\ref{4.24}).
\par
It is important that no small parameters  were  used  neither when we were deriving equation
(\ref{4.24}) nor when we  were  obtaining the correction (\ref{4.67}).

\clearpage

\addcontentsline{toc}{subsection}{\textbf{Example 4: Non-perturbative derivation of linear
hydrodynamics from Boltzmann equation (3D)}}

\subsection*{\textbf{Example 4: Non-perturbative derivation of linear hydrodynamics from Boltzmann
equation (3D)}}

Using Newton method instead of power series, a model of linear hydrodynamics is derived from the
Boltzmann equation for regimes where Knudsen number is of order unity. The model demonstrates no
violation of stability of acoustic spectra in contrast to Burnett hydrodynamics.

Knudsen number $\varepsilon$ (a ratio  between the mean free path, $l_c$, and a scale of hydrodynamic
flows, $l_h$) is a recognized order parameter when hydrodynamics is derived from the Boltzmann
equation \cite{Dorf}. The Chapman--Enskog method \cite{Chapman} establishes Navier-Stokes
hydrodynamic equations as the first-order correction to Euler hydrodynamics at
$\varepsilon\rightarrow 0$, and it also derives formal corrections of order $\varepsilon^2$,
$\varepsilon^3$, ... (known as Burnett and super-Burnett corrections). These corrections are
important outside the  strictly hydrodynamic domain $\varepsilon\ll 1$, and has to be considered for
an exension of hydrodynamic description into a highly nonequilidrium domain $\varepsilon\le 1$. Not
much is known about high-order in $\varepsilon$ hydrodynamics, especially in a nonlinear case.
Nonetheless, in a linear case, some definite information can be obtained. On the one hand,
experiments on sound propagation in noble gases are considerably better explained with Burnett and
super-Burnett hydrodynamics rather than with Navier-Stokes approximation alone \cite{Ford}. On the
other hand, a direct calculation shows a non-physical behavior of Burnett hydrodynamics for
ultra-short waves: acoustic waves increase instead of decay \cite{Bob}. The latter failure of Burnett
approximation cannot be rejected on a basis that for such regimes they might be not applicable
because for Navier-Stokes approximation, which is formally still less valid, no such violation is
observed.

These two results indicate that, at least in a linear regime, it makes sense to consider
hydrodynamics at $\varepsilon\le 1$, but Enskog way of deriving such hydrodynamics is problematic.
The problem of constructing solutions to the Boltzmann equation valid when $\varepsilon$ is of order
unity is one of the main open problems of classical kinetic theory \cite{Dorf}.

In this Example we suggest a new approach to derive a hydrodynamics at $\varepsilon\le 1$. The main
idea is to pose a problem of a finding a correction to Euler hydrodynamics in such a fashion that
expansions in $\varepsilon$ do not appear as a necessary element of analysis. This will be possible
by using Newton method instead of Taylor expansions to get such correction. We restrict our
consideration to a linear case. Resulting hydrodynamic equations do not exhibit the mentioned
violation.

The starting point is the set of local Maxwell distribution functions (LM)
$f_0(n,\mbox{\boldmath$u$},T;\vv)$, where $\vv$ is the particle's velocity, and $n$, $\uu$, and $T$
are local number density, average velocity, and temperature. We write the Boltzmann equation as:
\begin{equation}
\label{BE} \frac{df}{dt}=J(f), \quad J(f)=-(v-u)_i \cdot \partial_i f + Q(f),
\end{equation}
where $d/dt=\partial/\partial t + u_i \cdot \partial_i$ is the material derivative,
$\partial_i=\partial/\partial x_i$, while $Q$ is the Boltzmann collision integral \cite{Dorf}.

On the one hand, calculating r.h.s. of eq. (\ref{BE}) in LM-states, we obtain $J(f_0)$, a time
derivative  of LM-states due to Boltzmann equation. On the other hand, calculating a time derivative
of LM-states due to Euler dynamics, we obtain $P_0J(f_0)$, where $P_0$ is a  projector operator onto
the LM manifold (see \cite{GKTTSP94}):
\begin{equation}
\label{projLM} P_0 J=\frac {f_0}{n} \left\{ \int J d\cc+ 2c_i \cdot \int c_i J d\cc +
\frac{2}{3}\left(c^2 -\frac{3}{2}\right)\int \left(c^2 - \frac{3}{2}\right) J d\cc \right\},
\end{equation}
Since the LM functions are not solutions to the Boltzmann equation (\ref{BE}) (except for constant
$n$, $\uu$, and $T$), a difference of $J(f_0 )$ and $P_0 J(f_0 )$ is not equal to zero:
\begin{equation}
\label{delta}\Delta (f_0 ) = J(f_0 ) -P_0 J(f_0 )=-f_0 \left\{ 2(\partial_i u_k)  \left(c_i c_k
-\frac{1}{3}\delta_{ik} c^2 \right)+v_T \frac{\partial_i T}{T} c_i \left(c^2 -
\frac{5}{2}\right)\right\}.
\end{equation}
here $\cc=v_T^{-1}(\vv-\uu)$, and $v_T =\sqrt{2k_B T/m}$ is the thermal velocity. Note that the
latter expression gives a complete non-exactness of the linearized local Maxwell approximation, and
it is neither big nor small in itself. An unknown hydrodynamic solution of eq.(\ref{BE}),
$f_{\infty}(n,\uu,T;\vv)$, satisfies the following equation:
\begin{equation}
\label{inv} \Delta(f_{\infty})=J(f_{\infty})-P_{\infty}J(f_{\infty})=0,
\end{equation}
where $P_{\infty}$ is an unknown projecting operator. Both $P_{\infty}$ and $f_{\infty}$ are unknown
in eq. (\ref{inv}), but, nontheless, one is able to consider a sequence of corrections $\{f_1, f_2,
\dots\}$, $\{P_1 , P_2 , \dots \}$ to the initial approximation $f_0 $ and $P_0 $.    A method to
deal with equations of a form (\ref{inv}) was developed in \cite{GKTTSP94} for a general case of
dissipative systems. In particular, it was shown, how to ensure the $H$-theorem on every step of
approximations by choosing appropriate projecting operators $P_n $. In the present illustrative
example we will not consider projectors other than $P_0 $, rather, we will use an iterative procedure
to find $f_1 $.

Let us apply the Newton method with incomplete linearization to eq. (\ref{inv}) with $f_0 $ as
initial approximation for $f_{\infty}$ and with $P_0 $ as an initial approximation for $P_{\infty}$.
Writing $f_1 = f_0 + \delta f $, we get the first Newton iterate:
\begin{equation}
\label{it} L(\delta f/f_0 )+ (P_0 -1)(v-u)_i \partial_i \delta f + \Delta (f_0 ) = 0,
\end{equation}
where $L$ is a linearized collision integral.
\begin{equation}
\label{L} L(g)=f_0 (\vv)\int w(\vv_1^{\prime},\vv^{\prime};\vv_1 ,\vv)f_0 (\vv_1 ) \{
g(\vv_1^{\prime})+g({\vv}^{\prime})-g(\vv_1 )-g(\vv)\} d\vv_1^{\prime}d\vv^{\prime}d\vv_1.
\end{equation}
Here $w$ is a probability density of a change of velocities, $(\vv,\vv_1)\leftrightarrow
(\vv^{\prime},\vv^{\prime}_1)$, of a pair of molecules after their encounter. When deriving
(\ref{it}), we have accounted $P_0 L =0$, and an additional condition which fixes the same values of
$n$, $\uu$, and $T$ in states $f_1 $ as in LM states $f_0 $:

\begin{equation}
\label{cond} P_0 \delta f = 0.
\end{equation}

Equation (\ref{it}) is basic in what follows. Note that it contains no Knudsen number explicitly. Our
strategy will be to treat equation (\ref{it}) in such a way that the Knudsen number will appear
explicitly only at the latest stage of computations.

The two further approximations will be adopted. The first concerns a linearization of eq. (\ref{it})
about a global equilibria $F_0 $. The second concerns a finite-dimensional approximation of integral
operator in (\ref{it}) in velocity space. It is worthwhile noting here that none of these
approximations concerns an assumption about the Knudsen number.

Following the first of the approximations mentioned, denote as $\delta n $, $\delta \uu$, and $\delta
T$ deviations of hydrodynamic variables from their equilibrium values $n_0 $, $\uu_0 =0 $, and $T_0
$. Introduce also not-dimensional variables $\Delta n = \delta n/n_0 $, $\Delta \uu=\delta \uu/v_T^0
$, and $\Delta T = \delta T/T_0 $, where $v_T^0 $ is a heat velocity in equilibria, and a
not-dimensional relative velocity \boldmath $ \xi $ \unboldmath $=\vv/v_T^0 $. Correction $f_1 $ in a
linear in deviations from $F_0 $ approximation reads: $$f_1 = F_0 (1+\varphi_0 + \varphi_1 ),$$ where
$$\varphi_0 = \Delta n + 2\Delta u_i \xi_i  + \Delta T (\xi^2 - 3/2 )$$ is a linearized deviation of
LM from $F_0 $, and $\varphi_1 $ is an unknown function.    The latter is to be obtained from a
linearized version of eq.(\ref{it}).

Following the second approximation, we search for $\varphi_1 $ in a form:
\begin{equation}
\label{ansatz} \varphi_1 = A_i (\xx) \xi_i \left(\xi^2 - \frac{5}{2}\right) +B_{ik}(\xx) \left(\xi_i
\xi_k - \frac{1}{3} \delta_{ik} \xi^2 \right)+\dots
\end{equation}
where dots denote terms of an expansion of $\varphi_1 $ in velocity polynomials, orthogonal to $\xi_i
(\xi^2 - \nolinebreak 5/2)$ and $\xi_i \xi_k - 1/3 \delta_{ik} \xi^2 $, as well as to $1$, to
$\mbox{\boldmath$\xi$}$, and to $\xi^2 $. These terms do not contribute to shear stress tensor and
heat flux vector in hydrodynamic equations. Independency of functions $A$ and $B$ from $\xi^2 $
amounts to a first Sonine polynomial approximation of viscosity and heat transfer coefficients. Put
another way, we consider a projection onto a finite-dimensional subspace spanned by $\xi_i (\xi^2 -
\nolinebreak 5/2)$ and $\xi_i \xi_k - 1/3 \delta_{ik} \xi^2$. Our goal is to derive functions $A$ and
$B$ from a linearized version of eq.(\ref{it}). Knowing $A$ and $B$, we get the following expressions
for shear stress tensor $\sss$ and heat flux vector $\qq$:
\begin{equation}
\label{sigma} \sigma = p_0 B,\ \qq=\frac{5}{4}p_0 v_T^0 A,
\end{equation}
where $p_0 $ is equilibrium pressure of ideal gas.

Linearizing eq. (\ref{it}) near $F_0 $,  using an ansatz for $\varphi_1 $ cited above, and turning to
Fourier transform in space, we derive:
\begin{eqnarray}
\label{3d} \frac{5p_0}{3\eta_0}a_i(\kk)+iv_T^0 b_{ij}(\kk) k_j & =&-\frac{5}{2}iv_T^0 k_i
\tau(\kk);\\\nonumber \frac{p_0}{\eta_0}b_{ij}(\kk)+iv_T^0\overline{k_i a_j(\kk)}
&=&-2iv_T^0\overline{k_i\gamma_j(\kk)} ,
\end{eqnarray}
where $i=\sqrt{-1}$, $\kk$ is the wave vector, $\eta_0$ is the first Sonine polynomial approximation
of shear viscosity coefficient, $\aaa(\kk)$, $\bb(\kk)$, $\tau(\kk)$ and
\boldmath$\gamma$\unboldmath$(\kk)$ are Fourier transforms of $\AAA(\xx)$, $\BB(\xx)$, $\Delta
T(\xx)$, and $\Delta\uu (\xx)$, respectively, and the over-bar denotes a symmetric traceless dyad:
$$\overline{a_i b_j}=2a_i b_j - \frac{2}{3}\delta_{ij}a_s b_s.$$ Introducing a dimensionless wave
vector $\ff=[(v_T^0\eta_0)/(p_0)] \kk$, solution to Eq.\ (\ref{3d}) may be written:
\begin{eqnarray}
\label{SOLUTION3D}  b_{lj}(\kk)&=&-\frac{10}{3}i\overline{\gamma_l (\kk)
f_j}[(5/3)+(1/2)f^2]^{-1}\\\nonumber && +\frac{5}{3}i(\gamma_s (\kk) f_s)\overline{f_l
f_j}[(5/3)+(1/2)f^2]^{-1}[5+2f^2]^{-1}- \frac{15}{2}\tau(\kk)\overline{f_l f_j
}[5+2f^2]^{-1};\\\nonumber  a_l (\kk)&=&-\frac{15}{2}i f_l \tau (\kk) [5+2f^2]^{-1} \\\nonumber && -
[5+2f^2]^{-1}[(5/3)+(1/2)f^2]^{-1}[(5/3)f_l (\gamma_s (\kk) f_s)+\gamma_l (\kk) f^2(5+2f^2)].
\end{eqnarray}


Considering $z$-axis as a direction of propagation and denoting $k_z $ as $k$, $\gamma $ as $\gamma_z
$, we obtain from (\ref{3d}) the $k$-dependence of $a = a_z $ and $b = b_{zz} $:
\begin{eqnarray}
\label{1d-dim} a(k) & = & - \frac {\frac{3}{2} p_0^{-1}\eta_0 v_T^0 ik \tau(k) +\frac{4}{5}
p_0^{-2}\eta_0^2 (v_T^0 )^2 k^2  \gamma(k) }{1+\frac{2}{5} p_0^{-2}\eta_0^2 (v_T^0 )^2 k^2
},\\\nonumber b(k) & = & -\frac {\frac{4}{3} p_0^{-1}\eta_0 v_T^0 ik \gamma(k) + p_0^{-2}\eta_0^2
(v_T^0 )^2 k^2 \tau(k) } {1+\frac{2}{5} p_0^{-2}\eta_0^2 (v_T^0 )^2 k^2 }.
\end{eqnarray}

Using expressions for $\sigma$ and ${\bf q}$ cited above, and also using (\ref{1d-dim}), it is an
easy matter to close the linearized balance equations (given in Fourier terms):
\begin{eqnarray}
\label{hydro} \frac {1}{v_T^0 }\partial_t \nu(k) +  ik \gamma_k & = & 0, \\\nonumber \frac {2}{v_T^0
}
\partial_t \gamma(k) + ik(\tau(k) +\nu(k) ) + ik b(k) & = & 0, \\\nonumber \frac {3}{2v_T^0 }\partial
\tau +  ik\gamma(k) + \frac{5}{4} ika(k) & = & 0.
\end{eqnarray}

Eqs. (\ref{hydro}), together with expressions (\ref{1d-dim}), complete our derivation of hydrodynamic
equations.

To this end, the Knudsen number was not penetrating our derivations. Now it is worthwhile to
introduce it. The Knudsen number will appear most naturally if we turn to dimensionless form of eq.
(\ref{1d-dim}). Taking  $l_c = v_T^0 \eta_0 /p_0 $ ($l_c $ is of order of a mean free path), and
introducing a hydrodynamic scale $l_h $, so that $k = \kappa /l_h $, where $\kappa$ is a
not-dimensional wave vector, we obtain in (\ref{1d-dim}):

\begin{eqnarray}
\label{1d} a(\kappa)  &=& - \frac {\frac{3}{2} i\varepsilon\kappa \tau(\kappa) +
\frac{4}{5}\varepsilon^{2} \kappa^ {2} \gamma_{\kappa} } {1 + \frac{2}{5}\varepsilon^{2} \kappa^{2}
},\\\nonumber b(\kappa)&=&- \frac {\frac{4}{3}i\varepsilon\kappa \gamma(\kappa) + \varepsilon^2
\kappa^2 \tau(\kappa) } {1 + \frac{2}{5}\varepsilon^2 \kappa^2   },
\end{eqnarray}
where $\varepsilon=l_c /l_h $. Considering a limit $\varepsilon\rightarrow 0 $ in (\ref{1d}), we come
back to familiar Navier-Stokes expressions: $\sigma_{zz}^{NS}=-\frac{4}{3} \eta_0
\partial_z \delta u_z $, $q_z^{NS} = -\lambda_0 \partial_z \delta
T$, where $\lambda_0 = 15k_B \eta_0 /4m $ is the first Sonine polynomial approximation of heat
conductivity coefficient.

Since we were not assuming smallness of the Knudsen number $\varepsilon$ while deriving (\ref{1d}),
we are completely legal to put $\varepsilon=1 $. With all the approximations mentioned above, eqs.
(\ref{hydro}) and (\ref{1d-dim}) (or, equivalently, (\ref{hydro}) and (\ref{1d})) may be considered
as a model of a linear hydrodynamics at $\varepsilon$ of order unity. The most interesting feature of
this model is a non-polynomial dependence on $\kappa $. This amounts to that share stress tensor and
heat flux vector  depend on spatial derivatives of $\delta\uu$ and of $\delta T$ of an arbitrary high
order.

To find out a result of non-polynomial behavior (\ref{1d}), it is most informative to calculate a
dispersion relation for planar waves. It is worthwhile introducing dimensionless frequency $\lambda =
\omega l_h / v_T^0 $, where $\omega$ is a complex variable of a wave $\sim \exp(\omega t + ikz)$
($\mbox{Re}\omega $ is a damping rate, and $\mbox{Im}\omega $
 is a circular frequency). Making use of eqs. (\ref{hydro}) and
(\ref{1d}), writing $\varepsilon=1$, we obtain the following dispersion relation $\lambda(\kappa)$:

\begin{equation}
\label{disp} 12(1+\frac{2}{5}\kappa^2 )^2 \lambda^3 + 23\kappa^2 (1+\frac{2}{5}\kappa^2 )\lambda^2
+2\kappa^2 (5+5\kappa^2 +\frac{6}{5}\kappa^4 )\lambda + \frac{15}{2}\kappa^4 (1+\frac{2}{5}\kappa^2
)=0.
\end{equation}

 Fig. \ref{lindisp} presents a dependence ${\rm Re}\lambda (\kappa^2 )$ for acoustic waves obtained
from (\ref{disp}) and for the Burnett approximation \cite{Bob}.  The violation in the latter occurs
when the curve overcomes the horizontal axis.  In contrast to the Burnett approximation \cite{Bob},
the acoustic spectrum (\ref{disp}) is stable for all $\kappa $.  Moreover, ${\rm Re}\lambda(\kappa^2
)$ demonstrates a finite limit, as $\kappa^2 \rightarrow \infty $.

A discussion of results concerns the following two items:

1. The approach used avoids expansion into powers of the Knudsen number,
 and thus we obtain a hydrodynamics valid (at least formally) for moderate Knudsen numbers as an immediate
correction to Euler hydrodynamics. This is in a contrast to a usual treatment of a high-order
hydrodynamics as "(a well established) Navier-Stokes approximation + high-order terms". Navier-Stokes
hydrodynamics is recovered a posteriori, as a limiting case, but not as a necessary intermediate step
of computations.

2. Linear hydrodynamics derived is stable for all $k$, same as the Navier-Stokes hydrodynamics alone.
The $(1+\alpha k^2 )^{-1} $ "cut-off", as in (\ref{1d-dim}) and (\ref{1d}), was earlier found in a
"partial summing" of Enskog series \cite{GKJETP91,KGAnPh2002}.

Thus, we come to the following two conclusions:

1. A preliminary positive answer is given to the question of whether is it possible to construct
solutions of the Boltzmann equation valid for the Knudsen number of order unity.

2. Linear hydrodynamics derived can be used as a model for $\varepsilon=1$ with no danger to get a
violation of acoustic spectra at large $k$.

\clearpage

\addcontentsline{toc}{subsection}{\textbf{Example 5: Dynamic correction to moment approximations}}

\subsection*{\textbf{Example 5: Dynamic correction to moment
approximations}}

\addcontentsline{toc}{subsubsection}{Dynamic correction or extension of the list of variables?}

\subsubsection*{\textbf{Dynamic correction or extension of the list of variables?}}

Considering the Grad moment ansatz as a suitable first approximation to a closed finite-moment
dynamics, the correction is derived from the Boltzmann equation. The correction consists of two
parts, local and nonlocal. Locally corrected thirteen-moment equations are demonstrated to contain
exact transport coefficients. Equations resulting from the nonlocal correction give a microscopic
justification to some phenomenological theories of extended hydrodynamics.

A considerable part of the modern development of nonequilibrium thermodynamics is based on the idea
of extension of the list of relevant variables. Various phenomenological and semi-phenomenological
theories in this domain are known under the common title of the extended irreversible thermodynamics
(EIT) \cite{EIT}. With this, the question of a microscopic justification of the EIT becomes
important. Recall that a justification for some of the versions of the EIT was found witin the well
known Grad moment method \cite{Grad}.

Originally, the Grad moment approximation was introduced for the purpose of solving the
Boltzmann-like equations of the classical kinetic theory. The Grad method is used in various kinetic
problems, e.g., in plasma and in phonon transport. We mention also  that Grad equations assist in
understanding asymptotic features of gradient expansions, both in linear and nonlinear domains
\cite{MBCh,K3,K2,GKPRL96,KGAnPh2002}.

The essence of the Grad method is to introduce an approximation to the one-particle distribution
function $f$ which would depend only on a finite number $N$ of moments, and, subsequently, to use
this approximation to derive a closed system of $N$ moment equations from the kinetic equation. The
number $N$ (the level at which the moment transport hierarchy is truncated) is not specified in the
Grad method. One particular way to choose $N$ is to obtain an estimation of the  transport
coefficients (viscosity and heat conductivity) sufficiently close to their exact values provided by
the Chapman--Enskog method (CE) \cite{Chapman}. In particular, for the thirteen-moment (13M) Grad
approximation it is well known that transport coefficients are equal to the first Sonine polynomial
approximation to the exact CE values. Accounting for higher moments with $N>13$ can improve this
approximation (good for neutral gases but poor for plasmas
 \cite{Bal}).
However, what should be done, starting with the 13M approximation, to come to the exact CE transport
coefficients is an open question. It is also well known \cite{WidTit} that the Grad method provides a
poorly converging approximation when applied to strongly nonequilibrium problems (such as shock and
kinetic layers).

Another question comes from  the approximate character of the Grad equations, and  is discussed in
frames of the EIT: while the Grad equations are strictly hyperbolic at any level $N$ (i.e.,
predicting a finite speed of propagation), whether this feature will be preserved in the further
corrections.

These two questions are special cases of a more general one, namely, how to derive a closed
description with a given number of moments? Such a description is sometimes called mesoscopic
\cite{Ded} since it occupies an intermediate level between the hydrodynamic (macroscopic) and the
kinetic (microscopic) levels of description.

Here we aim at deriving the mesoscopic dynamics of thirteen moments \cite{KGDNPRE98} in the simplest
case when the kinetic description satisfies the linearized Boltzmann equation. Our approach will be
based on the two assumptions: (i). The mesoscopic dynamics of thirteen moments exists, and is
invariant with respect to the microscopic dynamics, and (ii). The 13M Grad approximation is a
suitable first approximation to this mesoscopic dynamics. The assumption (i) is realized as the
invariance equation for the (unknown) mesoscopic distribution function. Following the assumption
(ii), we solve the invariance equation iteratively, taking the 13M Grad approximation for the input
approximation, and consider the first iteration (further we refer to this as to the dynamic
correction, to distinguish from constructing another ansatz). We demonstrate that the correction
results in the exact CE transport coefficients. We also demonstrate how the dynamic correction
modifies the hyperbolicity of the Grad equations. A similar viewpoint on derivation of hydrodynamics
was earlier developed in \cite{GKTTSP94} (see previous Examples). We will return to a comparison
below.

\addcontentsline{toc}{subsubsection}{Invariance equation for 13M parameterization}

\subsubsection*{\textbf{Invariance equation for 13M parameterization}}

We denote as $n_0$, $\uu_0=0$, and $p_0$ the equilibrium values of the hydrodynamic parameters ($n$
is the number density, $\uu$ is the average velocity, and $p=nk_BT$ is the pressure). The global
Maxwell distribution function $F$ is $$F=n_0(v_T)^{-3}\pi^{-3/2}\exp(-c^2),$$ where
$v_T=\sqrt{2k_{\rm B}T_0m^{-1}}$ is the equilibrium thermal velocity, and $\cc=\vv/v_T$ is the
peculiar velocity of a particle. The near-equilibrium dynamics of the distribution function,
$f=F(1+\varphi)$, is due to the linearized Boltzmann equation:

\begin{eqnarray*}
\partial_t\varphi=\hat{J}\varphi\equiv
-v_Tc_i\partial_i\varphi +\hat{L}\varphi,\\\nonumber \hat{L}\varphi\!=\! \int\!w\!F(\vv_1 )
[\varphi(\vv_1^{\prime})\!+\!\varphi(\vv^{\prime})\!-\!\varphi(\vv_1)\!-\!\varphi(\vv)]
d\vv_1^{\prime}d\vv^{\prime}d\vv_1,
\end{eqnarray*}
where $\hat{L}$ is the linearized collision operator, and  $w$ is the probability density of pair
encounters. Furthermore, $\partial_i=\partial/\partial x_i$, and summation convention in two repeated
indices is assumed.

Let $n=\delta n/n_0$, $\uu=\delta\uu/v_T$, $p=\delta p/p_0$ ($p=n+T$, $T=\delta T/T_0$), be
dimensionless deviations of the hydrodynamic variables, while
$\mbox{\boldmath$\sigma$}=\delta\mbox{\boldmath$\sigma$}/p_0$ and $\qq=\delta\qq/(p_0v_T)$ are
dimensionless deviations of the stress tensor $\mbox{\boldmath$\sigma$}$, and of the heat flux $\qq$.
The linearized 13M Grad distribution function is $f_0=F(\cc)\left[1+\varphi_0 \right]$, where
\begin{eqnarray}
\label{13M} \varphi_0=\varphi_1+\varphi_2,
\\\nonumber
\varphi_1=n+2u_ic_i+T\left[c^2-(3/2)\right],\\\nonumber \varphi_2
=
\sigma_{ik}\overline{c_ic_k}+ (4/5)q_ic_i\left[c^2-(5/2)\right].
\end{eqnarray}
The overline denotes a symmetric traceless dyad.  We use the following convention:
\begin{eqnarray*}
\overline{a_ib_k}&=&a_ib_k+a_kb_i-\frac{2}{3}\delta_{ik}a_lb_l,\\
\overline{\partial_if_k}&=&\partial_if_k+\partial_kf_i-\frac{2}{3}\delta_{ik}\partial_lf_l.
\end{eqnarray*}

The 13M Grad's equations are derived in two steps: first, the 13M Grad's distribution function
(\ref{13M}) is inserted into the linearized Boltzmann equation to give a formal expression,
$\partial_t
 \varphi_0
=
\hat{J}\varphi_0$, second, projector $P_0$ is applied to this expression, where $P_0=P_1+P_2$, and
operators $P_1$ and $P_2$ act as follows:
\begin{eqnarray}
\label{PG} P_{1}J\!=\!\frac{F}{n_0} \!\left\{X_0\!\int\!X_0\!J\!d\vv\!+\!X_i\!
\!\int\!X_i\!J\!d\vv\!+\!X_4\!\int\!X_4\!J\!d\vv \right\},\\\nonumber
P_{2}J\!=\!\frac{F}{n_0}\!\left\{Y_{ik}\int Y_{ik} Jd\vv +Z_i\int Z_i J d\vv\right\}.
\end{eqnarray}
Here $X_0 =1$, $X_i=\sqrt{2}c_i$, where $i=1,2,3$, $X_4 =\sqrt{2/3}\left(c^2 -\frac{3}{2}\right)$,
$Y_{ik}=\sqrt{2}\overline{c_ic_k}$, and $Z_i=\frac{2}{\sqrt{5}}c_i\left(c^2 -\frac{5}{2}\right)$. The
resulting equation, $$P_0[F\partial_t\varphi_0]=P_0 [F\hat{J}\varphi_0],$$ is a compressed
representation for the 13M Grad equations for the macroscopic variables
$M_{13}=\{n,\uu,T,\mbox{\boldmath$\sigma$},\qq\}$.

Now we turn to the main purpose of this paper, and derive the dynamic correction to the 13M
distribution function (\ref{13M}). The assumption (i) [existence of closed dynamics of thirteen
moments] implies the invariance equation for the true mesoscopic distribution function,
$\tilde{f}(M_{13},\cc)=F[1+\tilde{\varphi}(M_{13},\cc)]$, where we have stressed that this function
depends parametrically on the same thirteen macroscopic parameters, as the original Grad
approximation. The invariance condition for $\tilde{f}(M_{13},\cc)$ reads \cite{GKTTSP94}:
\begin{equation}
\label{invariance1} (1-\tilde{P})[F\hat{J}\tilde{\varphi}]=0,
\end{equation}
where $\tilde{P}$ is the projector associated with $\tilde{f}$. Generally speaking, the projector
$\tilde{P}$ depends on the distribution function $\tilde{f}$ \cite{GKTTSP94,Bal}. In the following,
we use the projector $P_0$ (\ref{PG}) which will be consistent with our approximate treatment of the
Eq.\ (\ref{invariance1}).

Following the assumption (ii) [13M Grad's distribution function (\ref{13M}) is a good initial
approximation], the Grad's function $f_0$, and the projector $P_0$, are chosen as the input data for
solving  the equation (\ref{invariance1}) iteratively. The dynamic correction amounts to the first
iterate. Let us consider these steps in a more detail.

Substituting $\varphi_0$ (\ref{13M}) and $P_0$ (\ref{PG}) instead of $\varphi$ and $P$ in the
equation (\ref{invariance1}), we get: $(1-P_0)[F\hat{J}\varphi_0]\equiv \Delta_0\ne 0$, which
demonstrates that (\ref{13M}) is not a solution to the equation (\ref{invariance1}). Moreover,
$\Delta_0$ splits in two natural pieces: $\Delta_0=\Delta^{\rm loc}_0+ \Delta_0^{\rm nloc}$, where
\begin{eqnarray}
\label{D13M} \Delta_0^{\rm loc}&=&(1-P_{2})[F\hat{L}
 \varphi_{2}],\\\nonumber
\Delta^{\rm nloc}_0&=&(1-P_0) [-v_TFc_i\partial_i\varphi_0].
\end{eqnarray}
Here we have accounted for $P_{1}[F\hat{L}\varphi]=0$, and $\hat{L}\varphi_1=0$. The first piece of
Eq.\ (\ref{D13M}), $\Delta^{\rm loc}_0$, can be termed {\it local} because it does not account for
spatial gradients.  Its origin is twofold. In the first place, recall that  we are performing our
analysis in a non-local-equilibrium state (the 13M approximation is not a zero point of the Boltzmann
collision integral, hence $\hat{L}\varphi_0\ne 0$). In the second place, specializing to the
linearized case under consideration, functions $\overline{\cc \cc}$ and $\cc[c^2 -(5/2)]$, in
general, are not the eigenfunctions of the linearized collision integral, and hence
$P_2[F\hat{L}\varphi_0]\ne F\hat{L}\varphi_0$, resulting in $\Delta_0^{\rm loc}\ne0$ \footnote{Except
for Maxwellian molecules (interaction potential $U\sim r^{-4}$) for which $\hat{L}\varphi_0\ne 0$ but
$P_{2}[F\hat{L}\varphi_{\mbox{{\rm G}}}]= F\hat{L}\varphi_0$. Same goes for the relaxation time
approximation of the collision integral ($\hat{L}=-\tau^{-1}$).}.

The nonlocal part may be written as:

\begin{equation}
\label{nonloc} \Delta_0^{\rm nloc}=-v_TF(\Pi_{1|krs}\partial_k \sigma_{rs}
+\Pi_{2|ik}\overline{\partial_k q_i} +\Pi_{3}\partial_k q_k),
\end{equation}
where  $\Pi$ are velocity polynomials:
\begin{eqnarray*}
\Pi_{1|krs}&=&c_k\left[c_rc_s-(1/3)\delta_{rs}c^2\right]-(2/5) \delta_{ks}c_rc^2,
\\\nonumber
\Pi_{2|ik}&=&(4/5)\left[c^2-(7/2)\right]\left[c_ic_k-(1/3) \delta_{ik}c^2\right],
\\\nonumber
\Pi_3&=&(4/5)\left[c^2-(5/2)\right]\left[c^2-(3/2)\right]-c^2.
\end{eqnarray*}

We seek the  dynamic correction of the form: $$f= F[1+\varphi_0+\phi].$$ Substituting
$\varphi=\varphi_0+\phi$, and $P=P_0$, into Eq.\ (\ref{invariance1}), we derive an equation for the
correction $\phi$:
\begin{equation}
\label{I13M} (1\!-\!P_{2})[F\hat{L}(\varphi_{2}\!+\!\phi)]\!=\! (1\!-\!P_0)[v_TF\!c_i\partial_i
(\varphi_0\!+\!\phi\!)].
\end{equation}
Eq.\ (\ref{I13M}) should be supplied with the additional condition, $P_0[F\phi]=0$.

\addcontentsline{toc}{subsubsection}{Solution of the invariance equation}

\subsubsection*{\textbf{Solution of the invariance equation}}

Let us apply the usual ordering to solve the Eq.\ (\ref{I13M}), introducing a small parameter
$\epsilon$, multiplying the collision integral $\hat{L}$ with $\epsilon^{-1}$, and expanding
$\phi=\sum_{n}\epsilon^{n}\phi^{(n)}$. Subject to the additional condition, the resulting sequence of
linear integral equations is uniquely soluble. Let us consider the first two orders in $\epsilon$.

Because $\Delta_0^{\rm loc}\ne0$, the leading correction is of the order $\epsilon^0$, i.e. of the
same order as the initial approximation $\varphi_0$. The function $\phi^{(0)}$ is due the following
equation:
\begin{equation}
\label{I13Mloc} (1-P_{2})[F\hat{L}(\varphi_{2}+\phi^{(0)})]=0,
\end{equation}
subject to the  condition, $P_0[F\phi^{(0)}]=0$. Eq.\ (\ref{I13Mloc}) has the unique solution:
$\varphi_{2}+\phi^{(0)}=\sigma_{ik}Y^{(0)}_{ik}+q_iZ^{(0)}_i$, where functions, $Y^{(0)}_{ik}$ and
$Z^{(0)}_i$, are solutions to the integral equations:
\begin{equation}
\label{CE} \hat{L}Y_{ik}^{(0)}=bY_{ik}, \quad \hat{L}Z_i^{(0)}=aZ_i,
\end{equation}
subject to the  conditions, $P_{1}[F\mbox{{\sf Y}}^{(0)}]=0$ and $P_{1}[F\mbox{{\bf Z}}^{(0)}]=0$.
Factors $a$ and $b$ are:
\begin{eqnarray*}
a=\pi^{-3/2}\int e^{-c^2}{Z}_i^{(0)}
 \hat{L}{ Z}^{(0)}_id\cc,
\\\nonumber
b=\pi^{-3/2}\int e^{-c^2}{Y}^{(0)}_{ik} \hat{L}{Y}^{(0)}_{ik}d\cc.
\end{eqnarray*}

Now we are able to notice that the equation (\ref{CE}) coincides with the CE equations \cite{Chapman}
for the {\it exact transport coefficients} (viscosity and temperature conductivity). Emergency of
these well known equations in the present context is  important and rather unexpected: {\it when the
moment transport equations are closed with the locally corrected function $f^{\rm loc}
=F(1+\varphi_0+\phi^{(0)})$, we come to a closed set of thirteen equations containing the exact CE
transport coefficients.}

Let us analyze  the next order ($\epsilon^1$), where $\Delta_0^{\rm nloc}$ comes into play. To
simplify matters, we neglect the difference between the exact and the approximate CE transport
coefficients. The correction $\phi^{(1)}$ is due to the equation,
\begin{equation}
\label{I13Mnonloc2} (1-P_{2})[F\hat{L}\phi^{(1)}]+ \Delta_0^{\rm nloc}=0,
\end{equation}
the additional condition is: $P_0[F\phi^{(1)}]=0$. The problem (\ref{I13Mnonloc2}) reduces to three
integral equations of a familiar form:
\begin{equation}
\label{functions}
                 \hat{L}\Psi_{1|krs}=\Pi_{1|krs}, \quad
                \hat{L}\Psi_{2|ik}=\Pi_{2|ik}, \quad
                  \hat{L}\Psi_{3}=\Pi_{3}, \quad
\end{equation}
subject to conditions: $P_{1}[F\Psi_{1|krs}]=0$, $P_{1}[F\Psi_{2|ik}]=0$, and $P_{1}[F\Psi_{3}]= 0$.
Integral equations (\ref{functions}) are of the same structure as are the  integral equations
appearing in the CE method, and the methods to handle them are well developed \cite{Chapman}. In
particular, a reasonable and simple approximation is to take
$\Psi_{\alpha|\dots}=-A_{\alpha}\Pi_{\alpha|\dots}$. Then
\begin{equation}
\label{nonlocal} \phi^{(1)}=-v_T(A_1\Pi_{1|krs}\partial_k \sigma_{rs}
+A_2\Pi_{2|ik}\overline{\partial_k q_i} +A_3\Pi_{3}\partial_kq_k),
\end{equation}
where $A_{\alpha}$ are the approximate values of the kinetic coefficients, and which are expressed
via matrix elements of the linearized collision integral:

\begin{equation}
\label{coefficients} A_{\alpha}^{-1}\propto-\int
\exp(-c^2)\Pi_{\alpha|\dots}\hat{L}\Pi_{\alpha|\dots}d\cc> 0.
\end{equation}

The estimation can be  extended to a computational scheme for any given molecular model (e.\ g., for
the Lennard-Jones potential), in the manner of the transport coefficients computations in the CE
method.

\addcontentsline{toc}{subsubsection}{Corrected 13M equations}

\subsubsection*{\textbf{Corrected 13M equations}}

To summarize the results of the dynamic correction, we quote first the  unclosed  equations for the
variables $M_{13}=M_{13}=\{n,\uu,T, \mbox{\boldmath$\sigma$},\qq\}$:
\begin{eqnarray}
(1/v_T^{0})\partial_t n+\partial_iu_i=0,\label{chain1}\\ (2/v_T^{0})\partial_t u_i
+\partial_i(T+n)+\partial_k \sigma_{ik}=0,\label{chain2}\\ (1/v_T^{0})\partial_t T+(2/3)\partial_iu_i
+(2/3)\partial_iq_i=0,\label{chain3}\\ (1/v_T^{0})
\partial_t \sigma_{ik}+
2\overline{\partial_i u_k}-(2/3)\overline{\partial_iq_k}+\partial_l h_{ikl} =R_{ik}, \label{chain4}\\
(2/v_T)\partial_t q_i-(5/2)\partial_i p -(5/2)\partial_k\sigma_{ik} +\partial_k
g_{ik}=R_i\label{chain5}.
\end{eqnarray}
Terms spoiling the closure are: the higher moments of the distribution function,
\begin{eqnarray*}
h_{ikl}=2\pi^{-3/2}\!\int\!e^{-c^2}\!\varphi c_ic_kc_ld\cc,\\ g_{ik}=2\pi^{-3/2}\!\int\!
e^{-c^2}\!\varphi c_ic_kc^2d\cc,
\end{eqnarray*}
and the "moments" of the collision integral,
\begin{eqnarray*}
R_{ik}=\frac{2}{v_T} \pi^{-3/2}\int e^{-c^2}c_ic_k\hat{L}\varphi d\cc, \\
R_i=\frac{2}{v_T}\pi^{-3/2}\int e^{-c^2}c_ic^2 \hat{L}\varphi d\cc.
\end{eqnarray*}

The 13M Grad's distribution function (\ref{13M}) provides the zeroth-order closure approximation to
both the higher-order  moments and the "moments" of the collision integral:

\begin{eqnarray}
\label{13Mclosure} R^{(0)}_{ik}=-\mu_0^{-1}\sigma_{ik}, \ R^{(0)}_i=-\lambda_0^{-1}q_i, \\\nonumber
\partial_{l}h^{(0)}_{ikl}=(2/3)\delta_{ik}\partial_lq_l
+(4/5)\overline{\partial_iq_k}, \\\nonumber
\partial_l g_{lk}^{(0)}=(5/2)\partial_k(p+T)
+(7/2)\partial_l\sigma_{lk},
\end{eqnarray}
where $\mu_0$ and $\lambda_0$ are the first Sonine polynomial approximations to the viscosity and the
temperature conductivity coefficients \cite{Chapman}, respectively.

The local correction improves the closure of the "moments" of collision integral:
\begin{equation}
\label{13Mclosure1} R_{ik}=-\mu_{\rm CE}^{-1}\sigma_{ik}, \quad R_i=-\lambda_{\rm CE}^{-1}q_i,
\end{equation}
where index CE corresponds to exact Chapman--Enskog values of the transport coefficients.

The nonlocal correction adds the following terms to the higher moments:
\begin{eqnarray}
\label{13Mclosure2}
\partial_lg_{lk}&=&
\partial_l g_{lk}^{(0)}
-A_3\partial_k\partial_l q_{l}-A_2\partial_l\overline{\partial_lq_k},\\\nonumber
\partial_l h_{ikl}&=&
\partial_l h^{(0)}_{ikl}-A_1\partial_l\partial_l\sigma_{ik},,
\end{eqnarray}
where $A_i$ are the kinetic coefficients derived above.

In order to illustrate what changes in Grad equations with the nonlocal correction, let us consider a
model with two scalar variables, $T(x,t)$ and $q(x,t)$ (a simplified case of the one-dimensional
corrected 13M system where one retains only the variables responsible for heat conduction):
\begin{equation}
\label{simple}
\partial_t T+\partial_x q=0, \quad \partial_t q +\partial_x T
-a\partial_x^2 q+q=0.
\end{equation}
Parameter $a\ge 0$ controls "turning on" the nonlocal correction. Using  $\{q(k,\omega),
T(k,\omega)\}\exp(\omega t+ikx)$, we come  to a dispersion relation for the two roots
$\omega_{1,2}(k)$. Without the correction ($a=0$), there are two domains of $k$: for
$0\le\!k\!<\!k_{-}$, dispersion is diffusion-like ($\mbox{{\rm Re}}\omega_{1,2}(k)\le 0$, $\mbox{{\rm
Im}}\omega_{1,2}(k)=0$), while as $k\!\ge\!k_{-}$, dispersion is wave-like ($\omega_{1}(k)=
\omega_{2}^{\ast}(k)$,  $\mbox{{\rm Im}}\omega_{1}(k)\ne 0$). For $a$ between $0$ and $1$, the
dispersion modifies in the following way: The wave-like domain becomes bounded, and exists for
$k\!\in]k_-(a),k_+(a)[$, while the diffusion-like domain consists of two peaces, $k<k_-(a)$ and
$k>k_+(a)$.

The dispersion relation for $a=1/2$ is shown in the Fig.\ \ref{Figmom}. As $a$ increases to 1, the
boundaries of the wave-like domain, $k_{-}(a)$ and $k_{+}(a)$, move towards each other, and collapse
at $a=1$. For $a>1$, the dispersion relation becomes purely diffusive ($\mbox{{\rm
Im}}\omega_{1,2}=0$) for all $k$.

\addcontentsline{toc}{subsubsection}{Discussion: transport coefficients, destroying of the
hyperbolicity, etc.}

\subsubsection*{\textbf{Discussion: transport
coefficients, destroying of the hyperbolicity, etc.}}

(i). Considering the 13M Grad ansatz as a suitable approximation to the closed dynamics of thirteen
moments, we have found  that the first correction leads to exact Chapman-Enskog transport
coefficients. Further, the nonlocal part of this correction extends the Grad equations with terms
containing spatial gradients of the heat flux and of the stress tensor, destroying the hyperbolic
nature of the former. Corresponding kinetic coefficients are explicitly derived for the Boltzmann
equation.

(ii). Extension of Grad equations with terms like in (\ref{13Mclosure2}) was mentioned in many
versions of the EIT \cite{Jou}. These derivations were based on phenomenological  and
semi-phenomenological argument. In particular, the extension of the heat flux with appealing to
nonlocality effects in {\it dense} fluids. Here we have derived the  similar contribution from the
{\it simplest} (i.\ e.\ dilute gas) kinetics, in fact, from the assumption about existence of the
mesoscopic dynamics. The advantage of using the simplest kinetics is that corresponding kinetic
coefficients (\ref{coefficients}) become a matter of a {\it computation} for any molecular model.
This computational aspect will be discussed elsewhere, since it affects the dilute gas contribution
to dense fluids fits. Here we would like to stress a formal support of relevancy of the above
analysis: the nonlocal peace of dynamic correction is intermediated by the local correction, {\it
improving} the 13M Grad estimation to the ordinary transport coefficients.

(iii). When the invariance principle is applied to derive hydrodynamics (closed equations for the
variables $n$, $\uu$ and $T$) then  \cite{GKTTSP94} the local Maxwellian $f_{lm}$ is chosen as the
input distribution function for the invariance equation. In the linear domain, $f_{lm}
=F[1+\varphi_1]$, and the projector is $P_{lm}=P_1$, see eqs.\ (\ref{13M}) and (\ref{PG}). When the
latter expressions are substituted  into the invariance equation (\ref{invariance1}), we obtain
$\Delta_{lm}=\Delta_{lm}^{\rm nloc}=-v_TF\{2\partial_iu_k\overline{c_ic_k}+\partial_i
Tc_i[c^2-(5/2)]\}$, while $\Delta^{\rm loc}_{lm}\equiv0$
 because the local
Maxwellians are zero points of the Boltzmann collision integral. Consequently, the dynamic correction
begins with the order $\epsilon$, and the analog of the equation (\ref{I13Mnonloc2}) reads:
\[\hat{L}\phi_{lm}^{(1)}= v_T\{2\partial_iu_k\overline{c_ic_k}+
\partial_i Tc_i[c^2-(5/2)]\},\]
subject to a condition, $P_1[F\phi_{lm}^{(1)}]=0$. The latter is the familiar Chapman-Enskog
equation, resulting in the Navier-Stokes correction to the Euler equations \cite{Chapman}. Thus, {\it
the nonlocal dynamic correction is related to the 13M Grad equations entirely in the same way as the
Navier-Stokes are related to the Euler equations.} As the final comment to this point, it was
recently demonstrated with simple examples \cite{KGAnPh2002} that the invariance principle, as
applied to derivation of hydrodynamics, is equivalent to the summation of the Chapman-Enskog
expansion.

(iv). Let us discuss briefly the further corrections. The first local correction (the functions
$\mbox{{\sf Y}}_{1}$ and $\mbox{{\bf Z}}_{1}$ in the Eq.\ (\ref{CE})) is not the  limiting point of
our iterational procedure. When the latter is continued, the subsequent local corrections are found
from integral equations, $\hat{L}\mbox{{\sf Y}}_{n+1}=b_{n+1} \mbox{{\sf Y}}_{n}$, and
$\hat{L}\mbox{{\bf Z}}_{n+1}=a_{n+1} \mbox{{\bf Z }}_{n}$. Thus, we are led  to the following two
eigenvalue problems: $\hat{L}\mbox{{\sf Y}}_{\infty}=b_{\infty}\mbox{{\sf Y}}_{\infty}$, and
$\hat{L}\mbox{{\bf Z}}_{\infty}=a_{\infty}\mbox{{\bf Z}}_{\infty}$, where, in accord with general
argument \cite{GKTTSP94}, $a_{\infty}$ and $b_{\infty}$ are the closest to zero eigenvalues among all
the eigenvalue problems with the given tensorial structure \cite{GKPRE96}.

(v). Approach of this Example \cite{KGDNPRE98} can be extended to derive dynamic corrections to other
(non-moment) approximations of interest in the kinetic theory. The above analysis has demonstrated,
in particular, the importance of the local correction, generically relevant to an approximation which
is not a zero point of the collision integral. Very recently, this approach was successfully applied
to improve the nonlinear Grad's 13 moment equations \cite{StruTor}.

\section{\textbf{Decomposition of motions, non-uniqueness of
selection of fast motions, self-adjoint linearization, Onsager filter and quasi-chemical
representation}}\label{Onsager}

In section \ref{qe} we have used second law of thermodynamics - existence of the entropy - in order
to equip the problem of constructing slow invariant manifolds with a geometric structure. The
requirement of the entropy growth (universally, for all the reduced models) restricts significantly
the form of the projectors (\ref{projgen}).

In this section we introduce a different but equally important argument - the micro-reversibility
($T$-invariance), and its macroscopic consequences, the reciprocity relations. As first discussed by
Onsager in 1931, the implication of the micro-reversibility is the self-adjointness of the linear
approximation of the system (\ref{sys}) in the equilibrium $x^*$:
\begin{equation}\label{trueOns}
  \langle(D_xJ)_{x^*}z|p\rangle_{x^*}\equiv\langle
  z|(D_xJ)_{x^*}p\rangle_{x^*}.
\end{equation}

The main idea in the present section is to use the \textit{reciprocity relations} (\ref{trueOns})
\textit{for the fast motions}. In order to appreciate this idea, we should mention that the
decomposition of motions into fast and slow is not unique. Requirement (\ref{trueOns}) for any
equilibrium point of fast motions means the selection (filtration) of the fast motions. We term this
\textit{Onsager filter}. Equilibrium points of fast motions are all the points on manifolds of slow
motions.

There exist a trivial way to symmetrization, linear operator $A$ is decomposed into symmetric and
skew-symmetric parts, $A=\frac{1}{2}(A+A^{\dag})+\frac{1}{2}(A-A^{\dag})$. Here $A^{\dag}$ is adjoint
to $A$ with respect to a fixed scalar product (entropic scalar product in present context). However,
replacement of an operator with its symmetric part can lead to catastrophic (from the physical
standpoint) consequences such as, for example, loss of stability. In order to construct a sensible
Onsager filter, we shall use the \textit{quasi-chemical representation}.

The formalism of the quasi-chemical representation is one of the most developed means of modelling,
it makes it possible to "assemble" complex processes out of elementary processes. There exist various
presentations of the quasi-chemical formalism. Our presentation here is a generalization of an
approach suggested first by Feinberg \cite{Fein} (see also \cite{VanRys,ByGoYa,Yab}).

Symbol $A_i$ ("quasi-substance") is put into correspondence to each variable $x_i$. The
\textit{elementary reaction} is defined according to the \textit{stoichiometric equation},
\begin{equation}\label{stoichiomety}
  \sum_{i}\alpha_iA_i\rightleftharpoons\sum_{i}\beta_iA_i,
\end{equation}
where $\alpha_i$ (\textit{loss stoichiometric coefficients}) and $\beta_i$ (\textit{gain
stoichiometric coefficients}) are real numbers. Apart from the entropy, one specifies a monotonic
function of one variable, $\Psi(a)$, $\Psi'(a)>0$. In particular, function $\Psi(a)=\exp(\lambda a)$,
$\lambda=\textrm{const}$, is frequently encountered in applications.

Given the elementary reaction (\ref{stoichiomety}), one defines the rates of the direct and of the
inverse reactions:
\begin{eqnarray}
  W^+&=&w^*\Psi\left(\sum_{i}\alpha_i\mu_i\right),\nonumber\\
  W^-&=&w^*\Psi\left(\sum_{i}\beta_i\mu_i\right),\label{rates}
\end{eqnarray}
where $\mu_i=\frac{\partial S}{\partial x_i}$, $x^*=\textrm{const}$, $x^*>0$. The rate of the
elementary reaction is then defined as, $W=W^+-W^-$.

The equilibrium of the elementary reaction (\ref{stoichiomety}) is given by the following equation:
\begin{equation}\label{equilibrium}
  W^+=W^-.
\end{equation}
Thanks to the strict monotonicity of the function $\Psi$, equilibrium of the elementary reaction is
reached when the arguments of the functions coincide in equation (\ref{rates}), that is, whenever
\begin{equation}\label{equilibrium2}
  \sum_{i}(\beta_i-\alpha_i)\mu_i=0.
\end{equation}
Vector with the components $\gamma_i=\beta_i-\alpha_i$ is termed the \textit{stoichiometric vector of
the reaction}.

Let $x^0$ be a point of equilibrium of the reaction (\ref{stoichiomety}). The linear approximation of
the reaction rate has a particularly simple form:

\begin{equation}\label{linearW}
  W(x^0+\delta)=-w^*\Psi'(a(x^0))\langle\gamma|\delta\rangle_{x^0}+o(\delta),
\end{equation}
where $a(x^0)=\sum_i\alpha_i\mu_i(x^0)=\sum_i\beta_i\mu_i(x^0)$, and $\langle|\rangle_{x^0}$ is the
entropic scalar product in the equilibrium. In other words,

\begin{equation}\label{linearW1}
  (D_xW)_{x^0}=-w^*\Psi'(a(x^0))\langle\gamma|.
\end{equation}

Let us write down the kinetic equation for one elementary reaction:

\begin{equation}\label{chemkin1}
  \frac{dx}{dt}=\gamma W(x).
\end{equation}
Linearization of this equation in the equilibrium $x^0$ has the following form:

\begin{equation}\label{linchemkin}
 \frac{d\delta}{dt}=-w^*\Psi'(a(x^0))\gamma\langle\gamma|\delta\rangle_{x^0}.
\end{equation}

That is, the matrix of the linear approximation has the form,
\begin{equation}\label{K}
  K=-k^*|\gamma\rangle\langle\gamma|,
\end{equation}
where
\[k^*=w^*\Psi'(a(x^0))>0,\]
while the entropic scalar product of bra- and ket vectors is taken in the equilibrium point $x^0$.

If there are several elementary reactions, then the stoichiometric vectors $\gamma^r$ and the
reaction rates $W_r(x)$ are specified for each reaction, while the kinetic equation is obtained by
summing the right hand sides of equation (\ref{chemkin1}) for individual reactions,

\begin{equation}\label{chemkin2}
  \frac{dx}{dt}=\sum_r\gamma^r W_r(x).
\end{equation}

Let us assume that under the inversion of motions, the direct reaction transforms into the inverse
reaction. Thus, the $T$-invariance of the equilibrium means that it is reached in the point of the
\textit{detailed balance}, where all the elementary reaction come to equilibrium simultaneously:
\begin{equation}\label{db}
  W_r^+(x^*)=W_r^-(x^*).
\end{equation}
This assumption is nontrivial if vectors $\gamma^r$ are linearly dependent (for example, if the
number of reactions is greater than the number of species minus the number of conservation laws).

In the detailed balance case, the linearization of equation (\ref{chemkin2}) about $x^*$ has the
following form ($x=x^*+\delta$):
\begin{equation}\label{linchemkin2}
  \frac{d\delta}{dt}=-\sum_{r}k^*_r\gamma^r\langle\gamma^r|\delta\rangle_{x^*},
\end{equation}
where
\begin{eqnarray*}
k^*_r&=&w_r^*\Psi'_r(a^*_r)>0,\\ a^*_r&=&\sum_i\alpha^r_i\mu_i(x^*)=\sum_i\beta^r_i\mu_i(x^*).
\end{eqnarray*}
The following matrix of the linear approximation is obviously self-adjoint and stable:
\begin{equation}\label{K2}
  K=-\sum_rk^*_r|\gamma^r\rangle\langle\gamma^r|.
\end{equation}
Note that matrix $K$ is the sum of matrices of rank one.

Let us now extract the self-adjoint part of the form (\ref{K2}) in the \textit{arbitrary} point $x$.
After linearizing the reaction rate about $x$, we obtain:
\begin{equation}\label{lin1}
  W(x+\delta)=w^*\left(\Psi'(a(x))\langle\alpha|\delta\rangle_x-\Psi'(b(x))\langle\beta|\delta\rangle_x\right)
  +o(\delta),
\end{equation}
where
\begin{eqnarray*}
a(x)&=&\sum_i\alpha_i\mu_i(x),\\ b(x)&=&\sum_i\beta_i\mu_i(x).
\end{eqnarray*}
Let us introduce notation,
\begin{eqnarray*}
k^{\rm SYM}(x)&=&\frac{1}{2}w^*\left(\Psi'(a(x))+\Psi'(b(x))\right)>0,\\ k^{\rm
A}(x)&=&\frac{1}{2}w^*\left(\Psi'(a(x))-\Psi'(b(x))\right).
\end{eqnarray*}
In terms of this notation, equation (\ref{lin1}) may be rewritten,
\begin{equation}\label{lin2}
  W(x+\delta)=-k^{\rm SYM}(x)\langle\gamma|\delta\rangle_x+k^{\rm A}(x)\langle\alpha+\beta|\delta\rangle_x
  +o(\delta).
\end{equation}
The second term vanishes in the equilibrium ($k^{\rm A}(x^*)=0$, due to detailed balance).

\textit{Symmetric linearization} (Onsager filter) consists in using only the first term in the
linearized vector field (\ref{lin2}) when analyzing the fast motion towards the (approximate) slow
manifolds, instead of the full expression (\ref{lin1}). Matrix $K(x)$ of the linear approximation
becomes then the form similar to equation (\ref{K2}):
\begin{equation}\label{Kx}
  K(x)=-\sum_r k^{\rm SYM}_r(x)|\gamma^r\rangle\langle\gamma^r|,
\end{equation}
where
\begin{eqnarray*}
 k^{\rm SYM}_r(x)&=&\frac{1}{2}w^*_r\left(\Psi'_r(a(x))+\Psi'_r(b(x))\right)>0,\\
 a_r(x)&=&\sum_i\alpha^r_i\mu_i(x),\\
 b_r(x)&=&\sum_i\beta^r_i\mu_i(x),
\end{eqnarray*}
while the entropic scalar product $\langle|\rangle_x$ is taken at the point $x$. For each index of
the elementary reaction $r$, function $k^{\rm SYM}_r(x)$ is positive. Thus, stability of the
symmetric matrix (\ref{Kx}) is evident.

Symmetric linearization (\ref{Kx}) is distinguished also by the fact that it preserves the rank of
the elementary processes contributing to the complex mechanism: Same as in the equilibrium, matrix
$K(x)$ is the sum of rank one operators corresponding to each individual process. This is not the
case of the standard symmetrization.

Using the symmetric operator (\ref{Kx}) in the above Newton method with incomplete linearization can
be consider as a version of a heuristic strategy of "we act in such a way as if the manifold $F(W)$
were already slow invariant manifold". If this were the case, then, in particular, the fast motions
were described by the self-adjoint linear approximation.

We describe the quasi-chemical formalism for finite-dimensional systems. Infinite-dimensional
generalizations are almost obvious in many important cases, and are achieved by a mere replacement of
summation by integration. The best example give us collisions in the Boltzmann equation: each
velocity $v$ corresponds to a quasi-substance $A_v$, and a collision has a stoichiometric equation:
$$A_v+A_w \rightleftharpoons A_{v'}+A_{w'}.$$  In the Example to this section we consider the
Boltzmann collision integral from this standpoint in more details.

\clearpage

\addcontentsline{toc}{subsection}{\textbf{Example 6: Quasi-chemical representation and Self-adjoint
linearization of the Boltzmann collision operator}}

\subsection*{\textbf{Example 6: Quasi-chemical representation and Self-adjoint
linearization of the Boltzmann collision operator}}

 A decomposition of motions near thermodynamically nonequilibrium
states results in a linear relaxation towards this state. A linear operator of this relaxation is
explicitly constructed in the case of the Boltzmann equation.

An entropy-related specification of an equilibrium state is due to the two points of view. From the
first, thermodynamic viewpoint, equilibria  is a state in which the entropy is maximal. From the
second,
 kinetic
viewpoint, a quadratic form of entropy increases in a course of a linear regression towards this
state. If an underlying microscopic dynamics is time-reversible, the kinetic viewpoint is realized
due to known symmetric properties of a linearized kinetic operator.

In most of near-equilibrium studies, a principle of a decomposition of motions into rapid and slow
occupies a distinct place. In some special cases, decomposition of motions is taken into account
explicitly, by introducing a small parameter into dynamic equations. More frequently, however, it
comes into play implicitly, for example, through an assumption of a rapid decay of memory in
projection operator formalism \cite{Grabert}. Even in presence of long-living dynamic effects (mode
coupling), a decomposition of motions appears as a final instance to get a closed set of equations
for slow variables.

However, for closed systems, there remains a question: whether and to what extend the two
aforementioned entropy-related points of view are applicable to non-equilibrium states? Further, if
an answer is positive, then how to make explicitly a corresponding specification?

This Example is aimed at answering the questions just mentioned, and it is a straightforward
continuation of results \cite{GKAMSE92,GKTTSP94}. Namely, in \cite{GKAMSE92,GKTTSP94}, it was
demonstrated that a principle of a decomposition of motions alone constitutes a necessary and
sufficient condition for the thermodynamic specification of a non-equilibrium state (this will be
briefly reviewed in the next section). However, in a general situation, one deals with states $f$
other than $f_0$. A question is, whether these two ideas can be applied to $f\not= f_0$ (at least
approximately), and if so, then how to make the presentation explicit.

A positive answer to this question was given partially in frames of the method of invariant manifolds
\cite{GKAMSE92,GK1,GKTTSP94}. Objects studied in \cite{GKAMSE92,GK1,GKTTSP94} were manifolds in a
space of distribution functions, and the goal was to construct iteratively a manifold that is tangent
in all its points to a vector field of a dissipative system (an invariant manifold), beginning with
some initial manifold with no such property. It was natural to employ methods of KAM-theory
(Newton-type linear iterations to improve the initial manifold). However, an extra idea of a
decomposition of motions into rapid and slow near the manifold was strongly necessary to adapt
KAM-theory to dissipative systems. A geometrical formulation of this idea
\cite{GKAMSE92,GK1,GKTTSP94} results in a definition of a hyperplane of rapid motion, $\Gamma_f$ ,
associated with the state $f$, and orthogonal to the gradient of the entropy in $f$. In a physical
interpretation, $\Gamma_f$ contains all those states from a neighborhood of $f$, which come into $f$
in the course of rapid relaxation (as if $f$ were the final state of rapid processes occuring in its
neighborhood). Usually, $\Gamma_f$ contains more states than can come into $f$ in a rapid relaxation
because of conservation of some macroscopic quantities (e.g. density, momentum, and energy, as well
as, possibly, higher moments of $f$ which practically do not vary in rapid processes). Extra states
are eliminated by imposing additional restrictions, cutting out "thinner" linear manifolds, planes of
rapid motions $P_f$, inside $\Gamma_f$. Extremal property of $f$ on $\Gamma_f$ is preserved on $P_f$
as well (cf.\cite{GKAMSE92,GK1,GKTTSP94}).

Thus, decomposition of motions near a manifold results in the thermodynamical viewpoint: states $f$
belonging to the manifold are described as unique points of maximum of entropy on corresponding
hyperplanes of rapid motions $\Gamma_f$. This formulation defines a slow dynamics on manifolds in
agreement with the $H$-theorem for the Boltzmann equation, or with its analogs for other systems (see
\cite{GKAMSE92,GK1,GKTTSP94} for details). As it was shown in \cite{GKAMSE92,GK1,GKTTSP94},
decomposition of motions in a neighborhood of $f$ is a criteria (a necessary and sufficient
condition) of an existence of the thermodynamic description of $f$.

Newton iteration gives a correction, $f+\delta f$, to states of a non-invariant manifold, while
$\delta f$ is thought on $\Gamma_f$. Equation for $\delta f$ involves a linearization of the
collision integral in state $f$. Here, if $f\not= f_0$, we come to a problem of how to perform a
linearization of collision integral in concordance with the $H$-theorem (corrections to the manifold
of local equilibrium states were studied in a detail in \cite{GKTTSP94}).

Here we show that the aforementioned decomposition of motions results in the kinetic description of
states on manifolds of slow motions, and that Onsager's principle can be applied in a natural way to
linearize the Boltzmann collision integral.

Due to definition of $\Gamma_f$, the state $f$ is the unique point of minimum of the $H$-function on
$\Gamma_f$. In the first non-vanishing approximation, we have the following expression for $H$ in the
states on $\Gamma_f$:
\[
H(f+\delta f) \approx H(f)+\frac{1}{2} \langle \delta f|\delta f \rangle_f
\]
Here $\langle \cdot | \cdot \rangle_f$ denotes a scalar product generated by the second derivative of
$H$ in the state $f$: $\langle g_1| g_2 \rangle_f=\int f^{-1} g_1 g_2 d\vv$.

Decomposition of motions means that quadratic form $\langle\delta f |\delta f\rangle_f$ decays
monotonically in the course of the linear relaxation towards the state $f$. It is natural, therefore,
to impose a request that this linear relaxation should obey Onsager's principle. Namely, the
corresponding linear operator should be symmetric (formally self-adjoint) and non-positively definite
in scalar product $\langle\cdot|\cdot\rangle_f$, and its kernel should consist of linear combinations
of conserved quantities ($1$, $\vv$, and $v^2$). In other words, decomposition of motions should give
a picture of linear relaxation in a small neighborhood of $f$ similar to that in a small neighborhood
of $f_0$. Following this idea, we will now decompose the linearized collision integral $L_f$ in two
parts: $L_f^{\rm SYM}$ (satisfying  Onsager's principle), and $L_f^{\rm  A}$ (non-thermodynamic
part).

In the state $f$, each direct encounter, $(\vv,\vv_1)\rightarrow (\vv^{\prime},\vv_1^{\prime})$,
together with the reverse encounter, $(\vv^{\prime},\vv_1^{\prime}) \rightarrow (\vv,\vv_1)$,
contribute a rate, $G(f)-L(f)$, to the collision integral, where (see section 2):

\begin{displaymath}
W(f)= W(\vv^{\prime}, \vv_1^{\prime};\vv,\vv_1) \exp\left\{D_f H|_{f=f(\vv)}+D_f
H|_{f=f(\vv_1)}\right\};
\end{displaymath}
\begin{displaymath}
W^{\prime}(f)= W(\vv^{\prime}, \vv_1^{\prime};\vv,\vv_1) \exp\left\{D_f H|_{f=f(\mbox{\boldmath
{\scriptsize $v$}}^{\prime})}+ D_f H|_{f=f(\mbox{\boldmath {\scriptsize $v$}}_1^{\prime})}\right\};
\end{displaymath}

A deviation $\delta f$ from the state $f$ will change the rates of both the direct and the reverse
processes. Resulting deviations of rates are:

\begin{displaymath}
\delta W=W(f) \left\{D_f^2 H|_{f=f(\vv)}\cdot \delta f(\vv) +D_f^2 H|_{f=f(\vv_1)}\cdot \delta
f(\vv_1)\right\};
\end{displaymath}
\begin{displaymath}
\delta W^{\prime}= W^{\prime}(f) \left\{D_f^2 H|_{f=f(\vv^{\prime})} \cdot \delta f(\vv^{\prime})+
D_f^2 H|_{f=f(\vv_1^{\prime})}\cdot \delta f(\vv_1^{\prime})\right\};
\end{displaymath}

Symmetrization with respect to direct and reverse encounters will give a term proportional to a
balanced rate, $W^{\rm SYM}(f)= \frac {1}{2}(W(f)+W^{\prime}(f))$, in both of the expressions $\delta
W$ and $\delta W^{\prime}$. Thus, we come to the decomposition of the linearized collision integral
$L_f=L_f^{\rm SYM}+L_f^{\rm A} $, where

\begin{equation}
L_f^{\rm SYM}\delta f = \int w \frac {f^{\prime}f_1^{\prime}+ff_1}{2} \left\{\frac{\delta
f^\prime}{f^{\prime}}+ \frac{\delta f_1^{\prime}}{f_1^{\prime}} -\frac{\delta f_1}{f_1}-\frac{\delta
f}{f}\right\} d\vv_1^{\prime}d\vv^{\prime}d\vv_1 \label{SYM};
\end{equation}
\begin{equation}
L_f^{\rm  A}\delta f  = \int w \frac {f^{\prime}f_1^{\prime}-ff_1}{2}
 \left\{ \frac{\delta f^\prime}{f^{\prime}}+
\frac{\delta f_1^{\prime}}{f_1^{\prime}} +\frac{\delta f_1}{f_1}+\frac{\delta f}{f} \right\}
d\vv_1^{\prime}d\vv^{\prime}d\vv_1 \label{NT};
\end{equation}
$f=f(\vv), f_1=f(\vv_1), f^{\prime}=f(\vv^{\prime}), f_1^{\prime}=f(\vv_1^{\prime}), \delta f= \delta
f(\vv), \delta f_1= \delta f(\vv_1), \delta f^{\prime}= \delta f(\vv^{\prime}), \delta f_1^{\prime}=
\delta f(\vv_1^{\prime}).$

Operator $L_f^{\rm SYM}$ \ (\ref{SYM}) has the complete set of the aforementioned properties
corresponding to the Onsager's principle, namely:
\begin{enumerate}
\item[i) ]
$\langle g_1|L_f^{\rm SYM}|g_2\rangle_f= \langle g_2|L_f^{SYM}|g_1\rangle_f$ (symmetry);
\item[ii) ]
$\langle g|L_f^{\rm SYM}|g\rangle_f\leq 0 $ (local entropy production inequality);
\item[iii)]
$f,\vv f,v^2 f \in \ker L_f^{\rm SYM}$ (conservation laws).
\end{enumerate}
For an unspecified $f$, non-thermodynamic operator $L_f^{\rm A}$ \ (\ref{NT}) satisfies none of these
properties. If $f=f_0$, then the part  (\ref{NT}) vanishes, while operator $L_{f_0}^{\rm SYM}$
becomes the usual linearized collision integral due to the balance $W(f_0)=W^{\prime}(f_0)$.

Non-negative definite form $\langle\delta f|\delta f\rangle_f$ decays monotonically due to an
equation of linear relaxation, $\partial_t \delta f  = L_f^{\rm SYM}\delta f $, and the unique point
of minimum, $\delta f=0$, of $\langle\delta f|\delta f\rangle_f$ corresponds to the equilibrium point
of vector field $L_f^{\rm SYM} \delta f$.

Operator $L_f^{\rm SYM}$ describes the state $f$ as the equilibrium state of a linear relaxation.
Note that the method of extracting the symmetric part (\ref{SYM}) is strongly based on the
representation of direct and reverse processes, and it is not a simple procedure like, e.g., $\frac
{1}{2}(L_f + L_f^+ )$. The latter expression cannot be used as a basis for Onsager's principle since
it would violate conditions (ii) and (iii).

Thus, if motions do decompose into a rapid motion towards the manifold and a slow motion along the
manifold, then states on this manifold can be described from both the thermodynamical and kinetic
points of view. Our consideration results in an explicit construction of operator $L_f^{\rm SYM}$
(\ref{SYM}) responsible for the rapid relaxation towards the state $f$. It can be used, in
particular, for obtaining corrections to such approximations as Grad moment approximations and
Tamm--Mott-Smith approximation, in frames of the method \cite{GKAMSE92,GK1,KGDNPRE98}. The
non-thermodynamic part (\ref{NT}) is always present in $L_f$, when $f \not= f_0$, but if trajectories
of an equation $\partial_t \delta f = L_f \delta f $ are close to trajectories of an equation
$\partial_t \delta f  = L_f^{\rm SYM}\delta f $, then $L_f^{\rm SYM}$ gives a good approximation to
$L_f$. A conclusion on a closeness of trajectories depends on particular features of $f$, and
normally it can be made on a base of a small parameter. On the other hand, the explicit thermodynamic
and kinetic presentation of states on a manifold of slow motions (the extraction of $L_f^{\rm SYM}$
performed above and construction of hyper-planes $\Gamma_f$ \cite{GKAMSE92,GK1,GKTTSP94}) is based
only the very idea of a decomposition of motions, and can be obtained with no consideration of a
small parameter. Finally, though we have considered only the Boltzmann equation, the method of
symmetrization can be applied to other dissipative systems with the same level of generality as the
method \cite{GKAMSE92,GK1,GKTTSP94}.

\section{\textbf{Relaxation methods}}\label{sec:relax}

\textit{Relaxation method} is an alternative to the Newton iteration method described in section
\ref{newton}: The initial approximation to the invariant manifold $F_0$ is moved with the film
extension, equation (\ref{tmapsubtr}),
\[\frac{dF_t(y)}{dt}=(1-P_{t,y})J(F_t(y))=\Delta_{F(y)},\]
till a fixed point is reached. Advantage of this method is a relative freedom in its implementation,
because equation (\ref{tmapsubtr}) need not be solved exactly, one is interested only in finding
fixed points. Therefore, ``large stepping" in the direction of the defect, $\Delta_{F(y)}$ is
possible, the termination point is defined by the condition that the vector field becomes orthogonal
to $\Delta_{F(y)}$. For simplicity, let us consider the procedure of termination in the linear
approximation of the vector field. Let  $F_0(y)$ be the initial approximation to the invariant
manifold, and we seek the first correction,
\[F_1(y)=F_0(y)+\tau_1(y)\Delta_{F_0(y)},\]
where function $\tau(y)$ has dimension of time, and is found from the condition that the linearized
vector field attached to the points of the new manifold is orthogonal to the initial defect,
\begin{equation}\label{terminator1}
  \langle\Delta_{F_0(y)}|(1-P_{y})[J(F_0(y))+
  \tau_1(y)(D_xJ)_{F_0(y)}\Delta_{F_0(y)}]\rangle_{F_0(y)}=0.
\end{equation}
Explicitly,
\begin{equation}\label{terminator2}
 \tau_1(y)=-\frac{\langle\Delta_{F_0(y)}|\Delta_{F_0(y)}\rangle_{F_0(y)}}
 {\langle\Delta_{F_0(y)}|(D_xJ)_{F_0(y)}|\Delta_{F_0(y)}\rangle_{F_0(y)}}.
\end{equation}
Further steps $\tau_k(y)$ are found in the same way. It is clear from the latter equations that the
step of the relaxation method for the film extension is equivalent to the Galerkin approximation for
solving the step of the Newton method with incomplete linearization. Actually, the relaxation method
was first introduced in these terms in \cite{KZFPLANL}. A partially similar idea of using the
explicit Euler method to approximate the finite-dimensional invariant manifold on the basis of
spectral decomposition was proposed earlier in the work ref. \cite{Kev}.

An advantage of the equation (\ref{terminator2}) is the explicit form of the size of the steps
$\tau_k(y)$. This method was successfully applied to the Fokker-Plank equation \cite{KZFPLANL}.

\clearpage

\addcontentsline{toc}{subsection}{\textbf{Example 7: Relaxation method for the Fokker-Planck
equation}}

\subsection*{\textbf{Example 7: Relaxation method for the Fokker-Planck equation}}

Here we address the problem of closure for the FPE (\ref{SFP}) in a general setting. First, we review
the maximum entropy principle as a source of suitable quasiequilibrium initial approximations for the
closures. We also discuss a version of the maximum entropy principle, valid for a near-equilibrium
dynamics, and which results in explicit formulae for arbitrary $U$ and $D$.

In this Example we consider the FPE of the form (\ref{SFP}):
\begin{equation}
\label{FP}
\partial_t W(\xx,t)
=\partial_x \cdot\!\left\{D\cdot\left[W\partial_x U +\partial_x W\right]\right\}.
\end{equation}
Here $W(\xx,t)$ is the probability density over the configuration space $\xx$, at the time $t$, while
$U(\xx)$ and $D(\xx)$ are the potential and the positively semi-definite ($ y\cdot D\cdot y\ge 0$)
diffusion matrix.

\addcontentsline{toc}{subsubsection}{Quasi-equilibrium approximations for the Fokker-Planck equation}

\subsubsection*{\textbf{Quasi-equilibrium approximations for the Fokker-Planck equation}}

The quasi-equilibrium closures are almost never invariants of the true moment dynamics. For
corrections to the quasi-equilibrium closures, we apply the method of invariant manifold
\cite{GKTTSP94}, which is carried out (subject to certain approximations explained below) to explicit
recurrence formulae for one--moment near--equilibrium closures for arbitrary $U$ and $D$. These
formulae give a method for computing the lowest eigenvalue of the problem, and which dominates the
near-equilibrium FPE dynamics. Results are tested with model potential, including the FENE-like
potentials \cite{Bird,Doi,HCO}.

Let us denote as $M$ the set of linearly independent moments $\{M_0,M_1,\dots,M_k\}$, where
$M_i[W]=\int m_i(x)W(x)dx$, and where $m_0=1$. We assume that there exists a function $W^*(M,x)$
which extremizes the entropy $S$ (\ref{entropy}) under the constrains of fixed $M$. This
quasi-equilibrium distribution function may be written
\begin{equation}
\label{MEP} W^*=W_{eq}\exp\left[\sum_{i=0}^{k}\Lambda_im_i(x)-1\right],
\end{equation}
where $\Lambda=\{\Lambda_0,\Lambda_1,\dots,\Lambda_k\}$ are Lagrange multipliers. Closed equations
for moments $M$ are derived in two steps. First, the quasi-equilibrium distribution (\ref{MEP}) is
substituted into the FPE (\ref{FP}) or (\ref{GENERIC}) to give a formal expression: $\partial_t
W^*=\hat{M}_{W^*}(\delta S/\delta W)\big|_{W=W^*}$. Second, introducing a projector $\Pi^*$,
\[
\Pi^*\bullet=\sum_{i=0}^{k}(\partial W^{*}/\partial M_i)\int m(x)\bullet dx,
\]
and applying $\Pi^*$ on both sides of the formal expression, we derive closed equations for $M$ in
the quasi-equilibrium approximation. Further processing requires an explicit solution to the
constrains, $\int W^*(\Lambda,x)m_i(x)dx=M_i$, to get the dependence of Lagrange multipliers
$\Lambda$ on the moments $M$. Though typically the functions $\Lambda(M)$ are not known explicitly,
one general remark about the moment equations is readily available. Specifically, the moment
equations in the quasi-equilibrium approximation have the form:
\begin{equation}
\label{GENERIC2} \dot{M}_i=\sum_{j=0}^k M^*_{ij}(M)(\partial S^*(M)/
\partial M_j),
\end{equation}
where $S^*(M)=S[W^*(M)]$ is the macroscopic entropy, and where $M^*_{ij}$ is an $M$-dependent
$(k+1)\times(k+1)$ matrix:
\[
M^*_{ij}=\int W^*(M,x)[\partial_x m_i(x)]\cdot D(x)\cdot [\partial_x m_j(x)] dx.
\]
The matrix $M^*_{ij}$ is symmetric, positive semi--definite, and its kernel is the vector
$\delta_{0i}$. Thus, {\it the quasi-equilibrium closure reproduces the GENERIC structure on the
macroscopic level}, the vector field of macroscopic equations (\ref{GENERIC2}) is a metric transform
of the gradient of the macroscopic entropy.

The following version of the quasi-equilibrium closures makes it possible to derive more explicit
results in the general case \cite{Karlin1,Karlin11,MBCh,GKPRE96}: In many cases, one can split the
set of moments $M$ in two parts, $M_{I}=\{M_0,M_1,\dots,M_l\}$ and $M_{II}=\{M_{l+1},\dots,M_k\}$, in
such a way that the quasi-equilibrium distribution can be constructed explicitly for $M_{I}$ as
$W_{I}^*(M_{I},x)$. The full quasi-equilibrium problem for $M=\{M_{I},M_{II}\}$ in the "shifted"
formulation reads: extremize the functional $S[W_I^*+\Delta W]$ with respect to $\Delta W$, under the
constrains $M_{I}[W_I^*+\Delta W]=M_{I}$ and $M_{II}[W_I^*+\Delta W]=M_{II}$. Let us denote as
$\Delta M_{II}=M_{II}-M_{II}(M_{I})$ deviations of the moments $M_{II}$ from their values in the MEP
state $W_I^*$. For small deviations, the entropy is well approximated with its quadratic part

\[\Delta S=-\int\Delta W \left[1+\ln\frac{W_I^*}{W_{eq}}\right]dx
-\frac{1}{2}\int\frac{\Delta W^2}{W_I^*}dx.
\]
Taking into account the fact that $M_I[W^*_I]=M_I$, we come to the following maximizaton problem:
\begin{equation}
\label{triangle} \Delta S[\Delta W]\to max,\ M_I[\Delta W]=0,\ M_{II}[\Delta W]=\Delta M_{II}.
\end{equation}
The solution to the problem (\ref{triangle}) is always explicitly found from a $(k+1)\times(k+1)$
system of linear algebraic equations for Lagrange multipliers. This method was applied to systems of
Boltzmann equations for chemical reacting gases \cite{Karlin1,Karlin11}, and for an approximate
solution to the Boltzmann equation: scattering rates ``moments of collision integral" are treated as
independent variables, and as an alternative to moments of the distribution function, to describe the
rarefied gas near local equilibrium. Triangle version of the entropy maximum principle is used to
derive the Grad-like description in terms of a finite number of scattering rates. The equations are
compared to the Grad moment system in the heat nonconductive case. Estimations for hard spheres
demonstrate, in particular, some 10\% excess of the viscosity coefficient resulting from the
scattering rate description, as compared to the Grad moment estimation \cite{GKPRE96}.

In the remainder of this section we deal solely with one-moment near-equilibrium closures:
$M_{I}=M_0$, (i.\ e. $W_I^*=W_{eq}$), and the set $M_{II}$ contains a single moment $M=\int mWdx$,
$m(x)\neq 1$. We shall specify notations for the near--equilibrium FPE , writing the distribution
function as $W=W_{eq}(1+\Psi)$, where the function $\Psi$ satisfies an equation:
\begin{equation}
\label{FPLIN}
\partial_t \Psi=W_{eq}^{-1}\hat{J}\Psi,
\end{equation}
where $\hat{J}=\partial_x\cdot[W_{eq}D\cdot\partial_x]$. The triangle one-moment quasi-equilibrium
function reads:
\begin{equation}
\label{initial} W^{(0)}=W_{eq}\left[1+\Delta M m^{(0)}\right]
\end{equation}
where
\begin{equation}
\label{INDIR} m^{(0)}= [\langle m m\rangle-\langle m \rangle^2]^{-1} [m-\langle m \rangle].
\end{equation}
Here brackets $\langle\dots\rangle=\int W_{eq}\dots dx$ denote equilibrium averaging. The superscript
$(0)$ indicates that the triangle quasi-equilibrium function (\ref{initial}) will be considered as an
initial approximation to a procedure which we address below. Projector for the approximation
(\ref{initial}) has the form
\begin{equation}
\label{P0} \Pi^{(0)}\bullet=W_{eq}\frac{m^{(0)}} {\langle m^{(0)}m^{(0)}\rangle}\int
m^{(0)}(x)\bullet dx.
\end{equation}
Substituting the function (\ref{initial}) into the FPE (\ref{FPLIN}), and applying the projector
(\ref{P0}) on both the sides of the resulting formal expression, we derive an equation for $M$:
\begin{equation}
\label{MACRO0} \dot{M}=-\lambda_0\Delta M,
\end{equation}
where $1/\lambda_0$ is an effective time of relaxation of the moment $M$ to its equilibrium value, in
the quasi-equilibrium approximation (\ref{initial}):
\begin{equation}
\label{TIME0} \lambda_0=\langle m^{(0)}m^{(0)} \rangle^{-1} \langle \partial_x m^{(0)}\cdot D
\cdot\partial_x m^{(0)}\rangle.
\end{equation}

\addcontentsline{toc}{subsubsection}{The invariance equation for the Fokker-Planck equation}

\subsubsection*{\textbf{The invariance equation for the Fokker-Planck equation}}

Both the quasi-equilibrium and the triangle quasi-equilibrium closures are almost never invariants of
the FPE dynamics. That is, the moments $M$ of solutions to the FPE (\ref{FP}) vary in time
differently from the solutions to the closed moment equations like (\ref{GENERIC2}), and these
variations are generally significant even for the near--equilibrium dynamics. Therefore, we ask for
corrections to the quasi-equilibrium closures to finish with the invariant closures. This problem
falls precisely into the framework of the method of invariant manifold \cite{GKTTSP94}, and we shall
apply this method to the one--moment triangle quasi-equilibrium closing approximations.

First, the invariant one--moment closure is given by an unknown distribution function
$W^{(\infty)}=W_{eq}[1+\Delta M m^{(\infty)}(x)]$ which satisfyes an equation
\begin{equation}
[1-\Pi^{(\infty)}]\hat{J}m^{(\infty)}=0. \label{INV}
\end{equation}
Here $\Pi^{(\infty)}$ is a projector, associated with an unknown function $m^{(\infty)}$, and which
is also yet unknown. Eq.\ (\ref{INV}) is a formal expression of the invariance principle for a
one--moment near--equilibrium closure: considering $W^{(\infty)}$ as a manifold in the space of
distribution functions, parameterized with the values of the moment $M$, we require that the
microscopic vector field $\hat{J}m^{(\infty)}$ be equal to its projection,
$\Pi^{(\infty)}\hat{J}m^{(\infty)}$, onto the tangent space of the manifold $W^{(\infty)}$.

Now we turn our attention to solving the invariance equation (\ref{INV}) iteratively, beginning with
the triangle one--moment quasi-equilibrium approximation $W^{(0)}$ (\ref{initial}). We apply the
following iteration process to the Eq.\ (\ref{INV}):
\begin{equation}
[1-\Pi^{(k)}]\hat{J}m^{(k+1)}=0, \label{ITERATIONS}
\end{equation}
where $k=0, 1,\dots$, and where $m^{(k+1)}=m^{(k)}+\mu^{(k+1)}$, and the correction satisfies the
condition $\langle \mu^{(k+1)}m^{(k)}\rangle=0$. Projector is updated after each iteration, and it
has the form
\begin{equation}
\Pi^{(k+1)}\bullet=W_{eq}\frac{m^{(k+1)}} {\langle m^{(k+1)}m^{(k+1)}\rangle}\int m^{(k+1)}(x)\bullet
dx. \label{Pk}
\end{equation}
Applying $\Pi^{(k+1)}$ to the formal expression, $$W_{eq}m^{(k+1)}\dot{M}=\Delta
M[1-\Pi^{(k+1)}]m^{(k+1)},$$ we derive the $(k+1)$th update of the effective time (\ref{TIME0}):
\begin{equation}
\label{LAMBDAk} \lambda_{k+1}= \frac{\langle \partial_x m^{(k+1)}\cdot D \cdot\partial_x
m^{(k+1)}\rangle} {\langle m^{(k+1)}m^{(k+1)}\rangle}.
\end{equation}
Specializing to the one-moment near-equilibrium closures, and following a general argument
\cite{GKTTSP94}, solutions to the invariance equation (\ref{INV}) are eigenfunctions of the operator
$\hat{J}$, while the formal limit of the iteration process (\ref{ITERATIONS}) is the eigenfunction
which corresponds to the eigenvalue with the minimal nonzero absolute value.

\addcontentsline{toc}{subsubsection}{Diagonal approximation}

\subsubsection*{\textbf{Diagonal approximation}}

To obtain more explicit results, we shall now turn to an approximate solution to the problem
(\ref{ITERATIONS}) {\it at each iteration}. The correction $\mu^{(k+1)}$ satisfyes the condition
$\langle m^{(k)}\mu^{(k+1)}\rangle=0$, and can be decomposed as follows: $\mu^{(k+1)}=\alpha_k
e^{(k)}+e^{(k)}_{ort}$. Here $e^{(k)}$ is the variance of the $k$th approximation:
$e^{(k)}=W_{eq}^{-1}[1-\Pi^{(k)}]\hat{J}m^{(k)}=\lambda_k m^{(k)}+R^{(k)}$, where
\begin{equation}
R^{(k)}=W_{eq}^{-1}\hat{J}m^{(k)}. \label{Rk}
\end{equation}
The function $e^{(k)}_{ort}$ is orthogonal to both $e^{(k)}$ and $m^{(k)}$ ($\langle
e^{(k)}e^{(k)}_{ort}\rangle=0$, and $\langle m^{(k)}e^{(k)}_{ort}\rangle=0$).

Our {\it diagonal approximation} (DA) consists in disregarding the part $e^{(k)}_{ort}$. In other
words, we seek an improvement of the non--invariance of the $k$th approximation {\it along its
variance} $e^{(k)}$. Specifically, we consider the following ansatz at the $k$th iteration:
\begin{equation}
m^{(k+1)}=m^{(k)}+\alpha_k e^{(k)}. \label{ANSATZ}
\end{equation}
Substituting the ansatz (\ref{ANSATZ}) into the Eq.\ (\ref{ITERATIONS}), and integrating the latter
expression with the functon $e^{(k)}$ to evaluate the coefficient $\alpha_k$:

\begin{equation}
\alpha_k=\frac{A_k-\lambda_k^2}{\lambda_k^3-2\lambda_k A_k+B_k}, \label{ALPHA}
\end{equation}
where parameters $A_k$ and $B_k$ represent the following equilibrium averages:
\begin{eqnarray}
\label{CONSTANTS} A_k&=&\langle m^{(k)} m^{(k)}\rangle^{-1} \langle R^{(k)}R^{(k)}\rangle\\\nonumber
B_k&=&\langle m^{(k)} m^{(k)}\rangle^{-1} \langle \partial_x R^{(k)}\cdot D\cdot\partial_x
R^{(k)}\rangle.
\end{eqnarray}

Finally, putting together Eqs.\ (\ref{LAMBDAk}), (\ref{Rk}), (\ref{ANSATZ}), (\ref{ALPHA}), and
(\ref{CONSTANTS}), we arrive at the following DA recurrence solution, and which is our main result:

\begin{eqnarray}
m^{(k+1)}&=&m^{(k)}+\alpha_k[\lambda_k m^{(k)}+R^{(k)}],\\
\lambda_{k+1}&=&\frac{\lambda_k-(A_k-\lambda_k^2)\alpha_k} {1+(A_k-\lambda_k^2)\alpha_k^2}.
\label{RESULT}
\end{eqnarray}
Notice that the stationary points of the DA process (\ref{RESULT}) are the true solutions to the
invariance equation (\ref{INV}). What {\it may be} lost within the DA is the convergency to the true
limit of the procedure (\ref{ITERATIONS}), i.\ e.\ to the {\it minimal} eigenvalue.

To test the convergency of the DA process (\ref{RESULT}) we have considered two potentials $U$ in the
FPE (\ref{FP}) with a constant diffusion matrix $D$. The first test was with the square potential
$U=x^2/2$, in the three--dimensional configuration space, since for this potential the detail
structure of the spectrum is well known. We have considered two examples of initial one--moment
quasi-equilibrium closures with $m^{(0)}=x_1+100(x^2-3)$ (example 1), and $m^{(0)}=x_1+100x^6 x_2$
(example 2), in the Eq.\ (\ref{INDIR}). The result of performance of the DA for $\lambda_k$ is
presented in the Table \ref{TabFP}, together with the error $\delta_k$ which was estimated as the
norm of the variance at each iteration: $\delta_k= \langle e^{(k)}e^{(k)}\rangle/\langle
m^{(k)}m^{(k)}\rangle$. In both examples, we see a good monotonic convergency to the minimal
eigenvalue $\lambda_{\infty}=1$, corresponding to the eigenfunction $x_1$. This convergency is even
striking in the example 1, where the initial choice was very close to a different eigenfunction
$x^2-3$, and which can be seen in the non--monotonic behavior of the variance. Thus, we have an
example to trust the DA approximation as converging to the proper object.

For the second test, we have taken a one--dimensional potential $U=-50\ln(1-x^2)$, the configuration
space is the segment $|x|\leq 1$. Potentials of this type (so-called FENE potential) are used in
applications of the FPE to models of polymer solutions (\cite{Bird,Doi,HCO}). Results are given in
the Table \ref{TabFP1} for the two initial functions, $m^{(0)}=x^2+10x^4-\langle x^2+10x^4 \rangle$
(example 3), and $m^{(0)}=x^2+10x^8-\langle x^2+10x^8 \rangle$ (example 4). Both the examples
demonstrate a stabilization of the $\lambda_k$ at the same value after some ten iterations.

In conclusion, we have developed the principle of invariance to obtain moment closures for the
Fokker-Planck equation (\ref{FP}), and have derived explicit results for the one--moment
near-equilibrium closures, particularly important to get information about the spectrum of the FP
operator.

\begin{table}[t]
\caption{ Iterations $\lambda_k$ and the error $\delta_k$ for $U=x^2/2$.}

{\scriptsize

\begin{tabular}{|c|c|c|c|c|c|c|c|c|}
\hline & & $0$ & $1$ & $4$ & $8$ & $12$ & $16$  &  $20$ \\ \cline{2-9}  Ex.\ 1& $\lambda$ & 1.99998 &
1.99993 & 1.99575 & 1.47795 & 1.00356 & 1.00001 & 1.00000
\\\cline{2-9} & $\delta $
& $0.16 \cdot 10^{-4}$ & $ 0.66 \cdot 10^{-4}$ & $ 0.42 \cdot 10^{-2}$ & $0.24$ & $0.35 \cdot
10^{-2}$ & $ 0.13 \cdot 10^{-4}$ & $ 0.54 \cdot 10^{-7}$ \\ \hline & & $ 0$ & $1$ & $2$ & $3$ & $ 4$
& $ 5$  &  $6$
\\  \cline{2-9} Ex.\ 2 &  $\lambda$
& 3.399 & 2.437 & 1.586 & 1.088 & 1.010 & 1.001 & 1.0002 \\ \cline{2-9} & $\delta$ & $1.99$ & $1.42$
& $0.83$ &  $0.16$ & $0.29 \cdot 10^{-1}$ & $0.27 \cdot 10^{-2}$ & $0.57 \cdot 10^{-3}$ \\
 \hline
\end{tabular}

}

\label{TabFP}
\end{table}

\begin{table}[t]
\caption{Iterations $\lambda_k$ for $U=-50\ln(1-x^2)$. }

{\scriptsize

\begin{tabular}{|c|c|c|c|c|c|c|c|c|c|c|}
\hline
 & &
 $0$ & $1$ & $2$ & $3$ &
$ 4$ & $ 5$  &  $6$ &$7$& $8$ \\ \hline Ex.\ 3 &
 $\lambda$ &
213.17 &
 212.186 & 211.914 & 211.861 & 211.849 & 211.845 & 211.843 &
 211.842 & 211.841
\\ \hline
Ex.\ 4 &
 $\lambda$ &
 216.586 & 213.135 & 212.212 & 211.998 &
211.929 & 211.899 & 211.884 & 211.876 & 211.871
 \\ \hline
\end{tabular}

} \label{TabFP1}
\end{table}

\section{\textbf{Method of invariant grids}}

Elsewhere above in this paper, we considered immersions $F(y)$, and methods for their construction,
without addressing the question of how to implement $F$ in a constructive way. In most of the works
(of us and of other people on similar problems), analytic forms were needed to represent manifolds
(see, however, dual quasiequilibrium integrators \cite{IKOePhA02,IKOePhA03}). However, in order to
construct manifolds of a relatively low dimension, grid-based representations of manifolds become a
relevant option. The \textit{Method of invariant grids} (MIG) was suggested recently in
\cite{InChLANL}.

The principal idea of (MIG) is to find a mapping of finite-dimensional grids into the phase space of
a dynamic system. That is we construct not just a point approximation of the invariant manifold
$F^*(y)$, but an {\it invariant grid}. When refined, in the limit it is expected to converge, of
course, to $F^*(y)$, but it is a separate, independently  defined object.

Let's denote $L=R^n$, $G$ is a discrete subset of $R^n$. A natural choice would be a regular grid,
but, this is not crucial from the point of view of the general formalism. For every point $y \in G$,
a neighborhood of $y$ is defined: $V_y \subset G$, where $V_y$ is a finite set, and, in particular,
$y \in V_y$. On regular grids, $V_y$ includes, as a rule, the nearest neighbors of $y$. It may also
include next to nearest points.

For our purposes, one should define a grid differential operator. For every function, defined on the
grid, also all derivatives are defined:

\begin{equation} \label{diffgrid}
\left.\frac{\partial f}{\partial y_i}\right|_{y\in G} = \sum_{z \in V_y} q_i(z,y)f(z), i=1,\ldots n.
\end{equation}

\noindent where $q_i(z,y)$ are some coefficients.

Here we do not specify the choice of the functions $q_i(z,y)$. We just mention in passing that, as a
rule, equation (\ref{diffgrid}) is established using some interpolation of $f$ in the neighborhood of
$y$ in $R^n$ by some differentiable functions (for example, polynomial). This interpolation is based
on the values of $f$ at the points of $V_y$. For regular grids, $q_i(z,y)$ are functions of the
difference $z-y$. For some $y$s which are close to the edges of the grid, functions are defined only
on the part of $V_y$. In this case, the coefficients in (\ref{diffgrid}) should be modified
appropriately in order to provide an approximation using available values of $f$. Below we will
assume this modification is always done. We also assume that the number of points in the neighborhood
$V_y$ is always sufficient to make the approximation possible. This assumption restricts the choice
of the grids $G$. Let's call {\it admissible} all such subsets $G$, on which one can define
differentiation operator in every point.

Let $F$ be a given mapping of some admissible subset $G \subset R^n$ into $U$. For every $y \in V$ we
define tangent vectors:

\begin{equation}\label{tanspace}
T_y = Lin\{g_i\}_1^n,
\end{equation}

\noindent where vectors $g_i (i=1, \ldots n)$ are partial derivatives (\ref{diffgrid}) of the
vector-function $F$:

\begin{equation}\label{bastan}
g_i = \frac{\partial F}{\partial y_i} = \sum_{z \in V_y} q_i(z,y)F(z),
\end{equation}

or in the coordinate form:

\begin{equation}\label{bastanco}
(g_i)_j = \frac{\partial F_j}{\partial y_i} = \sum_{z \in V_y} q_i(z,y)F_j(z).
\end{equation}

\noindent Here $(g_i)_j$ is the $j$th coordinate of the vector $(g_i)$, and $F_j(z)$ is the $j$th
coordinate of the point $F(z)$.

The grid $G$ is {\it invariant}, if for every node $y \in G$ the vector field $J(F(y))$ belongs to
the tangent space $T_y$ (here $J$ is the right hand site of the kinetic equations (\ref{sys})).

So, the definition of the invariant grid includes:

\noindent 1) Finite admissible subset $G \subset R^n$;

\noindent 2) A mapping $F$ of this admissible subset $G$ into $U$ (where $U$ is the phase space for
kinetic equations (\ref{sys}));

\noindent 3) The differentiation formulae (\ref{diffgrid}) with given coefficients $q_i(z,y)$;

The grid invariance equation has a form of inclusion:

$$J(F(y)) \in T_y \: \mbox{for every} \: y \in G,$$

\noindent or a form of equation:

$$ (1-P_y)J(F(y))=0 \: \mbox{for every} \: y \in G,$$

\noindent where $P_y$ is the thermodynamic projector (\ref{projgen}).

The grid differentiation formulae (\ref{diffgrid}) are needed, in the first place, to establish the
tangent space $T_y$, and the null space of the thermodynamic projector $P_y$ in each node. It is
important to realise that locality of construction of thermodynamic projector enables this without a
need for a global parametrization.

Basically, in our approach, the grid specifics are in:(a) differentiation formulae, (b) grid
 construction strategy (the grid can be extended, contracted, refined, etc.) The invariance
equations (\ref{defect}), equations of the film dynamics extension (\ref{tmapsubtr}), the iteration
Newton method (\ref{Nm1}), and the formulae of the relaxation approximation (\ref{terminator2}) do
not change at all. For convenience, let us repeat all these formulae in the grid context.

Let $x=F(y)$ be position of a grid's node $y$ immersed into $U$. We have set of tangent vectors
$g_i(x)$, defined in $x$ (\ref{bastan}), (\ref{bastanco}). Thus, the tangent space $T_y$ is defined
by (\ref{tanspace}). Also, one has entropy function $S(x)$, the linear functional $D_xS|_x$, and the
subspace $T_{0y}=T_y\bigcap\ker D_xS|_x$ in $T_y$. Let $T_{0y}\neq T_y$. In this case we have a
vector ${\bf e}_y\in T_y$, orthogonal to $T_{0y}$, $D_xS|_x({\bf e}_y)=1$. Then, the thermodynamic
projector is defined as:

\begin{equation}\label{projgengrid}
  P_y\bullet=P_{0y}\bullet+{\bf e}_y D_xS|_x\bullet,
\end{equation}

\noindent where $P_{0y}$ is the orthogonal projector on $T_{0y}$ with respect to the entropic scalar
product $\langle |\rangle_x$.

If $T_{0y} = T_y$, then the thermodynamic projector is the orthogonal projector on $T_y$ with respect
to the entropic scalar product $\langle |\rangle_x$.

For the Newton method with incomplete linearization, the equations for calculating new node position
$x'=x+\delta x$ are:

\begin{equation}\label{Nm1grid}
 \left\{\begin{array}{l}
   P_y\delta x=0 \\
   (1-P_y)(J(x)+DJ(x)\delta x)=0.\
 \end{array}\right.
\end{equation}

\noindent Here $DJ(x)$ is a matrix of derivatives of $J$, calculated in $x$. The self-adjoint
linearization may be useful too (see section 8).

Equation (\ref{Nm1grid}) is a system of linear algebraic equations. In practice, it is convenient to
choose some orthonormal (with respect to the entropic scalar product) basis ${\bf b_i}$ in $ker P_y$.
Let $r=dim(ker P_y )$. Then $\delta x = \sum_{i=1}^r{\delta_i {\bf b_i}}$, and the system looks like

\begin{equation}\label{Nm1grid1}
   \sum_{k=1}^{r}{\delta_k \langle {\bf b_i} \mid DJ(x){\bf b_k} \rangle _x
   = -  \langle J(x)\mid {\bf b_i} \rangle_x }, i = 1...r.
\end{equation}

Here $\langle |\rangle_x$ is the entropic scalar product. This is the system of linear equations for
adjusting the node position accordingly to the Newton method with incomplete linearization.

For the relaxation method, one needs to calculate the defect $\Delta_x = (1-P_y)J(x)$, and the
relaxation step

\begin{equation}\label{terminator2grid}
 \tau(x)=-\frac{\langle\Delta_x|\Delta_x\rangle_x}
 {\langle\Delta_x|DJ(x)\Delta_x\rangle_x}.
\end{equation}

Then, new node position $x'$ is calculated as

\begin{equation}\label{relaxgrid}
x' = x+\tau(x)\Delta_x.
\end{equation}

This is the equation for adjusting the node position according to the relaxation method.

\subsection{\textbf{Grid construction strategy}}

From all reasonable strategies of the invariant grid construction we will consider here the following
two: {\it growing lump} and {\it invariant flag}.

\subsubsection{\textbf{Growing lump}}

In this strategy one chooses as initial the equilibrium point $y^*$. The first approximation is
constructed as $F(y^*)=x^*$, and for some initial $V_0$ ($V_{y^*} \subset V_0$) one has
$F(y)=x^*+A(y-y^*)$, where $A$ is an isometric embedding (in the standard Euclidean metrics) of $R^n$
in $E$.

For this initial grid one makes a fixed number of iterations of one of the methods chosen (Newton's
method with incomplete linearization or the relaxation method), and, after that, puts
$V_1=\bigcup_{y\in V_0}V_y$ and extends $F$ from $V_0$ onto $V_1$ using linear extrapolation and the
process continues. One of the possible variants of this procedure is to extend the grid from $V_i$ to
$V_{i+1}$ not after a fixed number of iterations, but when  invariance defect $\Delta_y$ becomes
smaller than a given $\epsilon$ (in a given norm, which is entropic, as a rule), for all nodes $y\in
V_i$. The lump stops growing when it reaches the boundary and is within a given accuracy
$\|\Delta\|<\epsilon$.

\subsubsection{\textbf{Invariant flag}}

For the invariant flag one uses sufficiently regular grids $G$, in which many points are situated on
the coordinate lines, planes, etc. One considers the standard flag $R^0 \subset R^1 \subset R^2
\subset ... \subset R^n$ (every next space is constructed by adding one more coordinate). It
corresponds to a succession of grids $\{y\} \subset G^1 \subset G^2 ... \subset G^n$ , where $\{y^*\}
= R^0$, and $G^i$ is a grid in $R^i$.

First, $y^*$ is mapped in $x^*$ and further $F(y^*)=x^*$. Then an invariant grid is constructed on
$V^1\subset G^1$ (up to the boundaries $U$ and within a given accuracy $\|\Delta\|<\epsilon$). After
the neighborhoods in $G^2$ are added to the points $V^1$, and, using such extensions, the grid
$V^2\subset G^2$ is constructed (up to the boundaries and within a given accuracy) and so on, until
$V^n\subset G^n$ will be constructed.

We must underline here that, constructing the k-th grid $V^k\subset G^k$, the important role of the
grids of smaller dimension $V^0\subset ... \subset V^{k-1} \subset V^k$ embedded in it, is preserved.
The point $F(y^*)=x^*$ is preserved. For every $y\in V^q$ ($q<k$) the tangent vectors $g_1,...,g_q$
are constructed, using the differentiation operators (\ref{diffgrid}) on the whole $V^k$. Using the
tangent space $T_y = Lin\{g_1,..,g_q\}$, the projector $P_y$ is constructed, the iterations are
applied and so on. All this is done to obtain a succession of embedded invariant grids, given by the
same map $F$.

\subsubsection{\textbf{Boundaries check and the entropy}}

We construct grid mapping of $F$ onto the finite set $V\in G$. The technique of checking if the grid
still belongs to  the phase space $U$ of kinetic system $U$ ($F(V)\subset U$) is quite
straightforward: all the points $y\in V$ are checked to belong to $U$. If at the next iteration a
point $F(y)$ leaves $U$, then it is returned inside by a homothety transform with the center in
$x^*$. Since the entropy is a concave function, the homothety contraction with the center in $x^*$
increases the entropy monotonously. Another variant is cutting off the points leaving $U$.

By the way it was constructed, (\ref{projgen}), the kernel of the entropic projector is annulled by
the entropy differential. Thus, in the first order, steps in the Newton method with incomplete
linearization (\ref{Nm1}) as well as in the relaxation methods
(\ref{terminator1}),(\ref{terminator2}) do not change the entropy. But, if the steps are quite large,
then the increasing of the entropy can become essential and the points are returned on their entropy
level by the homothety contraction with the center in the equilibrium point.

\subsection{\textbf{Instability of fine grids}}

When one reduces the grid step (spacing between the nodes) in order to get a finer grid, then,
starting from a definite step, it is possible to face the problem of the Courant instability
\cite{Cour1,Cour2,Cour3}. Instead of converging, at the every iteration the grid becomes entangled
(see Fig.~\ref{diver}).

The way to get rid off this instability is well-known. This is decreasing the time step. Instead of
the real time step, we have a shift in the Newtonian direction. Formally, we can assign for one
complete step in the Newtonian direction a value $h=1$. Let us consider now the Newton method with an
arbitrary $h$. For this, let us find $ \delta x=\delta F(y)$ from (\ref{Nm1grid}), but we will change
$\delta x$ proportionally to $h$: the new value of $x_{n+1}=F_{n+1}(y)$ will be equal to

\begin{equation}
F_{n+1}(y) = F_n(y)+h_n\delta F_n(y)
\end{equation}

\noindent where the lower index $n$ denotes the step number.

One way to choose the $h$ step value is to make it adaptive, controlling the average value of the
invariance defect $\|\Delta_y\|$ at every step. Another way is the convergence control: then $\sum
h_n$ plays a role of time.

Elimination of Courant instability for the relaxation method can be made quite analogously.
Everywhere the step $h$ is maintained as big as it is possible without convergence problems.

\subsection{\textbf{What space is the most appropriate for the grid construction?}}

For the kinetics systems there are two distinguished representations of the phase space:

\begin{itemize}
\item The densities space (concentrations, energy or probability densities, etc.)
\item The spaces of conjugate intensive quantities, potentials (temperature, chemical potentials, etc.)
\end{itemize}

The density space is convenient for the construction of quasi-chemical representations. Here the
balance relations are linear and the restrictions are in the form of linear inequalities (the
densities themselves or some linear combinations of them must be positive).

The conjugate variables space is convenient in the sense that the equilibrium conditions, given the
linear restrictions on the densities, are in the linear form (with respect to the conjugate
variables). In these spaces the quasiequilibrium manifolds exist in the form of linear subspaces and,
vise versa, linear balance equations turns out to be equations of the conditional entropy maximum.

The duality we've just described is very well-known and studied in details in many works on
thermodynamics and Legendre transformations \cite{Leg1,Leg2}. This viewpoint of nonequilibrium
thermodynamics unifies many well-established mesoscopic dynamical theories, as for example the
Boltzmann kinetic theory and the Navier-Stokes-Fourier hydrodynamics \cite{GRMEL}. In the previous
section, the grids were constructed in the density space. But the procedure of constructing them in
the space of the conjugate variables seems to be more consistent. The principal argument for this is
the specific role of quasiequilibrium, which exists as a linear manifold. Therefore, linear
extrapolation gives a thermodynamically justified quasiequilibrium approximation. Linear
approximation of the slow invariant manifold in the neighborhood of the equilibrium in the conjugate
variables space already gives the global quasiequilibrium manifold, which corresponds to the motion
separation (for slow and fast motions) in the neighborhood of the equilibrium point.

For the mass action law, transition to the conjugate variables is simply the logarithmic
transformation of the coordinates.

\subsection{\textbf{Carleman's formulae in the analytical invariant manifolds approximations.
First benefit of analyticity: superresolution}}

When constructing invariant grids, one must define the differential operators (\ref{diffgrid}) for
every grid node. For calculating the differential operators in some point $y$, an interpolation
procedure in the neighborhood of $y$ is used. As a rule, it is an interpolation by a low-order
polynomial, which is constructed using the function values in the nodes belonging to the
neighbourhood of $y$ in $G$. This approximation (using values in the closest nodes) is natural for
smooth functions. But, we are looking for the {\it analytical} invariant manifold (see discussion in
the section: ``Film extension: Analyticity instead of the boundary conditions"). Analytical functions
have much more ``rigid" structure than the smooth ones. One can change a smooth function in the
neighborhood of any point in such a way, that outside this neighborhood the function will not change.
In general, this is not possible for analytical functions: a kind of ``long-range" effect takes place
(as is well known) .

The idea is to use this effect and to reconstruct some analytical function $f_G$ using function given
on $G$. There is one important requirement: if these values on $G$ are values (given at the points of
$G$) of some function $f$ which is analytical in the given neighborhood $U$, then if the $G$ is
refined ``correctly", one must have $f_G \rightarrow f$. The sequence of reconstructed function $f_G$
should should converge to the ``proper" function $f$.

What is the ``correct refinement"? For smooth functions for the convergence $f_G \rightarrow f$  it
is necessary and sufficient that,  in the course of refinement, $G$ would approximate the whole $U$
with arbitrary accuracy. For analytical functions it is necessary only that, under the refinement,
$G$ would approximate some uniqueness set \footnote{Let's remind to the reader that $A \subset U$ is
called {\it uniqueness set} in $U$ if for analytical in $U$ functions $\psi$ and $\varphi$ from
$\psi|_A \equiv \varphi|_A$ it follows $\psi=\varphi$.} $A \subset U$. Suppose we have a sequence of
grids $G$, each next is finer than previous, which approximates a set $A$. For smooth functions,
using function values defined on the grids, one can reconstruct the function in $A$. For analytical
functions, if the analyticity area $U$ is known, and $A$ is a uniqueness set in $U$, then one can
reconstruct the function in $U$. The set $U$ can be essentially bigger than $A$; because of this such
extension was named as {\it superresolution effects} \cite{Aiz}. There exist constructive formulae
for construction of analytical functions $f_G$ for different areas $U$, uniqueness sets $A\subset U$
and for different ways of discrete approximation of $A$ by a sequence of fined grids $G$ \cite{Aiz}.
Here we provide only one Carleman's formula which is the most appropriate for our purposes.

Let area $U=Q^{n}_\sigma \subset C^n$ be a product of strips $Q_\sigma\subset C$,
$Q_\sigma=\{z|\mbox{Im} z<\sigma\}$. We will construct functions holomorphic in $Q^{n}_\sigma$. This
is effectively equivalent to the construction of real analytical functions $f$ in whole $R^n$ with a
condition on the convergence radius $r(x)$ of the Taylor series for $f$ as a function of each
coordinate: $r(x) \geq \sigma$ in every point $x \in R^n$.

The sequence of fined grids is constructed as follows: let for every $l=1,...,n$ a finite sequence of
distinct points $N_l\subset Q_\sigma$ be defined:

\begin{equation}
N_l = \{x_{lj}|j = 1,2,3 ... \}, x_{lj} \neq x_{li} \hspace{5pt} for \hspace{5pt} i\neq j
\end{equation}

The uniqueness set $A$, which is approximated by a sequence of fined finite grids, has the form:

\begin{equation}
A = N_1 \times N_2 \times ... \times N_n = \{(x_{1i_1},x_{2i_2},..,x_{ni_n})|i_{1,..,n}=1,2,3,...\}
\end{equation}

The grid $G_m$ is defined as the product of initial fragments $N_l$ of length $m$:

\begin{equation}
G_m = \{(x_{1i_1},x_{2i_2}...x_{ni_n})|1\leq i_{1,..,n}\leq m\}
\end{equation}

Let's denote $\lambda=2\sigma/\pi$ ($\sigma$ is a half-width of the strip $Q_\sigma$). The key role
in the construction of the Carleman's formula is played by the functional $\omega_m^\lambda(u,p,l)$
of 3 variables: $u \in U = Q^n_\sigma$, $p$  is an integer, $1\leq p \leq m$, $l$ is an integer,
$1\leq p \leq n$. Further $u$ will be the coordinate value at the point where the extrapolation is
calculated, $l$ will be the coordinate number, and $p$  will be an element of multi-index
$\{i_1,...,i_n\}$ for the point $(x_{1i_1},x_{2i_2},...,x_{ni_n})\in G$:

\begin{equation}\label{Carleman}
\omega_m^\lambda(u,p,l) = \frac{(e^{\lambda x_{lp}}+e^{\lambda \bar{x}_{lp}})(e^{\lambda
u}-e^{\lambda x_{lp}})} {\lambda(e^{\lambda u}+e^{\lambda \bar{x}_{lp}})(u-x_{lp})e^{\lambda
x_{lp}}}\times \prod_{j=1 j\neq p}^m{\frac{(e^{\lambda x_{lp}}+e^{\lambda \bar{x}_{lj}})(e^{\lambda
u}-e^{\lambda x_{lj}})}{(e^{\lambda x_{lp}}-e^{\lambda x_{lj}})(e^{\lambda u}+e^{\lambda
\bar{x}_{lj}})}}
\end{equation}

For real-valued $x_{pk}$ formula (\ref{Carleman}) becomes simpler:

\begin{equation}\label{RealCarleman}
\omega_m^\lambda(u,p,l) = 2 \frac{e^{\lambda u}-e^{\lambda x_{lp}}} {\lambda(e^{\lambda u}+e^{\lambda
x_{lp}})(u-x_{lp})}\times \prod_{j=1 j\neq p}^m{\frac{(e^{\lambda x_{lp}}+e^{\lambda
x_{lj}})(e^{\lambda u}-e^{\lambda x_{lj}})}{(e^{\lambda x_{lp}}-e^{\lambda x_{lj}})(e^{\lambda
u}+e^{\lambda x_{lj}})}}
\end{equation}

The Carleman's formula for extrapolation from $G_M$ on $U=Q^n_\sigma$ ($\sigma=\pi\lambda/2$) has the
form ($z=(z_1,...,z_n)$):

\begin{equation}\label{Extr}
f_m(z) = \sum_{k_1,...,k_n=1}^m {f(x_k)\prod_{j=1}^n\omega^\lambda_m(z_j,k_j,j)},
\end{equation}

where $k={k_1,..,k_n}$, $x_k=(x_{1k_1},x_{2k_2},...,x_{nk_n})$.

There exists a theorem \cite{Aiz}:

{\bf If $f\in H^2(Q_\sigma^n)$, then $f(z) = lim_{m\rightarrow \infty} f_m(z)$, where
$H^2(Q_\sigma^n)$ is the Hardy class of holomorphic in $Q_\sigma^n$ functions. }

It is useful to present the asymptotics of (\ref{Extr}) for big $|\mbox{Re}z_j|$. For this we will
consider the asymptotics of (\ref{Extr}) for big $|\mbox{Re}u|$:

\begin{equation}\label{Asy}
|\omega_m^\lambda(u,p,l)| = \left|\frac{2}{\lambda u} \prod_{j=1 j\neq p}^m \frac{e^{\lambda
x_{lp}}+e^{\lambda x_{lj}}}{e^{\lambda x_{lp}}-e^{\lambda x_{lj}}}\right|+o(|\mbox{Re}u|^{-1}).
\end{equation}

From the formula (\ref{Extr}) one can see that for the finite $m$ and $|\mbox{Re}z_j|\rightarrow
\infty$ function $|f_m(z)|$ behaves like $const\cdot \prod_j |z_j|^{-1}$.

This property (zero asymptotics) must be taken into account when using the formula (\ref{Extr}). When
constructing invariant manifolds $F(W)$, it is natural to use (\ref{Extr}) not for the immersion
$F(y)$, but for the deviation of $F(y)$ from some analytical ansatz  $F_0(y)$ \cite{GapsGRW,GR99}.

The analytical ansatz $F_0(y)$ can be obtained using Taylor series, just as in the Lyapunov auxiliary
theorem \cite{Lya} (also see above in the sections about the film extensions). Another variant is
using Taylor series for the construction of Pade-approximations.

It is natural to use approximations (\ref{Extr}) in dual variables as well, since there exists for
them (as the examples demonstrate) a simple and very effective linear ansatz for the invariant
manifold. This is the slow invariant subspace $E_{\mbox{\scriptsize slow}}$ of the operator of
linearized system (\ref{sys}) in dual variables in the equilibrium point. This invariant subspace
corresponds to the the set of ``slow" eigenvalues (with small $|\mbox{Re}\lambda|$, $\mbox{Re}
\lambda<0$). In the initial space (of concentrations or densities) this invariant subspace is the
quasiequilibrium manifold. It consist of the maximal entropy points on the affine manifolds of the
$x+E_{\mbox{\scriptsize fast}}$ form, where $E_{\mbox{\scriptsize fast}}$ is the ``fast" invariant
subspace of the operator of linearized system (\ref{sys}) in the initial variables in the equilibrium
point. It corresponds to the ``fast" eigenvalues (big $|\mbox{Re}\lambda|$, $\mbox{Re}\lambda<0$).

In the problem of invariant grids constructing we can use the Carleman's formulae in two moments:
first, for the definition grid differential operators (\ref{diffgrid}), second, for the analitical
continuation the manifold from the grid.

\clearpage

\addcontentsline{toc}{subsection}{\textbf{Example 8: Two-step catalytic reaction}}

\subsection*{\textbf{Example 8: Two-step catalytic reaction}}

Let us consider a two-step four-component reaction with one catalyst $A_2$:

\begin{equation}\label{tscatreact}
A_1+A_2 \leftrightarrow A_3 \leftrightarrow A_2+A_4
\end{equation}

We assume the Lyapunov function of the form $S=-G=-\sum_{i=1}^4c_i[ln(c_i/c_i^{eq})-1]$. The kinetic
equation for the four-component vector of concentrations, ${\bf c}=(c_1,c_2,c_3,c_4)$, has the form

\begin{equation}
\dot{{\bf c}} = \gamma_1W_1+\gamma_2W_2.
\end{equation}

Here $\gamma_{1,2}$ are stoichiometric vectors,

\begin{equation}
\gamma_1 = (-1,-1,1,0), \ \gamma_2=(0,1,-1,1),
\end{equation}

while functions $W_{1,2}$ are reaction rates:

\begin{equation}
W_1 = k_1^+c_1c_2-k_1^-c_3, \  W_2 = k_2^+c_3-k_2^-c_2c_4.
\end{equation}

Here $k_{1,2}^\pm$ are reaction rate constants. The system under consideration has two conservation
laws,

\begin{equation}
c_1+c_3+c_4 = B_1, \ c_2+c_3=B_2,
\end{equation}

or $\langle {\bf b_{1,2}},{\bf c}\rangle = B_{1,2}$, where ${\bf b_1} = (1,0,1,1)$ and ${\bf b_1} =
(0,1,1,0)$. The nonlinear system (\ref{tscatreact}) is effectively two-dimensional, and we consider a
one-dimensional reduced description. For our example, we chose the following set of parameters:

\begin{equation}
\begin{array}{lll}
k_1^+ = 0.3, \  k_1^- = 0.15, \ k_2^+ = 0.8, \ k_2^- = 2.0; \\ c_1^{eq} = 0.5, \ c_2^{eq} = 0.1,
c_3^{eq} = 0.1, \ c_4^{eq} = 0.4;
\\ B_1 = 1.0, \ B_2 = 0.2
\end{array}
\end{equation}

In Fig.~\ref{4d1dgrid} one-dimensional invariant grid is shown in the ($c_1$,$c_4$,$c_3$)
coordinates. The grid was constructed by growing the grid, as described above. We used Newtonian
iterations to adjust the nodes. The grid was grown up to the boundaries of the phase space.

The grid derivatives for calculating tangent vectors $g$ were taken as simple as $g(x_i) =
(x_{i+1}-x_{i-1})/|| x_{i+1}-x_{i-1} ||$ for the internal nodes and $g(x_1) = (x_{1}-x_{2})/||
x_{1}-x_{2} ||$, $g(x_n) = (x_{n}-x_{n-1})/|| x_{n}-x_{n-1} ||$ for the grid's boundaries. Here $x_i$
denotes the vector of the $i$th node position, $n$ is the number of nodes in the grid.

Close to the phase space boundaries we had to apply an adaptive algorithm for choosing the time step
$h$: if, after the next growing step and applying $N=20$ complete Newtonian steps, the grid did not
converged, then we choose a new $h_{n+1}=h_{n}/2$ and recalculate the grid. The final value for $h$
was $h \approx 0.001$.

The nodes positions are parametrized with entropic distance to the equilibrium point measured in the
quadratic metrics given by ${\bf H_c} = -||\partial^2S({\bf c})/\partial c_i\partial c_j||$ in the
equilibrium $c^{eq}$. It means that every node is on a sphere in this quadratic metrics with a given
radius, which increases linearly. On this figure the step of the increase is chosen to be 0.05. Thus,
the first node is on the distance 0.05 from the equilibrium, the second is on the distance 0.10 and
so on. Fig.~\ref{4d1dgraphs} shows several basic values which facilitate understanding of the object
(invariant grid) extracted. The sign on the x-axis of the graphs at Fig.~\ref{4d1dgraphs} is
meaningless, since the distance is always positive, but in this situation it denotes two possible
directions from the equilibrium point.

Fig.~\ref{4d1dgraphs}a,b effectively represents the slow one-dimensional component of the dynamics of
the system. Given any initial condition, the system quickly finds the corresponding point on the
manifold and starting from this point the dynamics is given by a part of the graph on the
Fig.~\ref{4d1dgraphs}a,b.

One of the useful values is shown on the Fig.~\ref{4d1dgraphs}c. It is the relation between the
relaxation times ``toward" and ``along" the grid ($\lambda_2/\lambda_1$, where
$\lambda_1$,$\lambda_2$ are the smallest and the second smallest by absolute value non-zero
eigenvalue of the system, symmetrically linearized at the point of the grid node). It shows that the
system is very stiff close to the equilibrium point, and less stiff (by one order of magnitude) on
the borders. This leads to the conclusion that the reduced model is more adequate in the neighborhood
of the equilibrium where fast and slow motions are separated by two orders of magnitude. On the very
end of the grid which corresponds to the positive absciss values, our one-dimensional consideration
faces with definite problems (slow manifold is not well-defined).

\addcontentsline{toc}{subsection}{\textbf{Example 9: Model hydrogen burning reaction}}

\subsection*{\textbf{Example 9: Model hydrogen burning reaction}}

In this section we consider a more interesting illustration, where the phase space is 6-dimensional,
and the system is 4-dimensional. We construct an invariant flag which consists of 1- and
2-dimensional invariant manifolds.

We consider chemical system with six species called (provisionally) $H_2$ (hydrogen), $O_2$ (oxygen),
$H_2O$ (water), $H$, $O$, $OH$ (radicals). We assume the Lyapunov function of the form
$S=-G=-\sum_{i=1}^6c_i[ln(c_i/c_i^{eq})-1]$. The subset of the hydrogen burning reaction and
corresponding (direct) rate constants have been taken as:

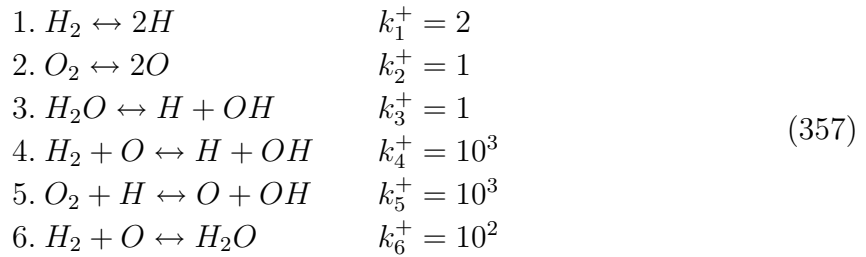
\begin{equation}
\begin{array}{llllll}
1.\hspace{0.1cm} H_2 \leftrightarrow 2H \hspace{0.5cm} & k_1^+ = 2
\\

2.\hspace{0.1cm} O_2 \leftrightarrow 2O \hspace{0.5cm} & k_2^+ = 1
\\

3.\hspace{0.1cm} H_2O \leftrightarrow H+OH \hspace{0.5cm} & k_3^+ = 1 \\

4.\hspace{0.1cm} H_2+O \leftrightarrow H+OH \hspace{0.5cm} & k_4^+ = 10^3 \\

5.\hspace{0.1cm} O_2+H \leftrightarrow O+OH \hspace{0.5cm} & k_5^+ = 10^3 \\

6.\hspace{0.1cm} H_2+O \leftrightarrow H_2O \hspace{0.5cm} & k_6^+ = 10^2
\end{array}
\end{equation}

The conservation laws are:

\begin{equation}
\begin{array}{ll}
2c_{H_2}+2c_{H_2O}+c_H+c_{OH} = b_H \\

2c_{O_2}+c_{H2O}+c_O+c_{OH} = b_O
\end{array}
\end{equation}

For parameter values we took  $b_H=2$, $b_O=1$, and the equilibrium point:

\begin{equation}
\begin{array}{llllll}
c_{H_2}^{eq}=0.27 & c_{O_2}^{eq} = 0.135 & c_{H_2O}^{eq}=0.7 & c_H^{eq}=0.05 & c_O^{eq}=0.02 &
c_{OH}^{eq}=0.01
\end{array}
\end{equation}

Other rate constants $k_i^{-},i=1..6$ were calculated from ${\bf c^{eq}}$ value and $k_i^{+}$. For
this system the stoichiometric vectors are:

\begin{equation}
\begin{array}{llllll}
\gamma_1 = (-1,0,0,2,0,0) & \gamma_2 = (0,-1,0,0,2,0) \\

\gamma_3 = (0,0,-1,1,0,1) & \gamma_4 = (-1,0,0,1,-1,1) \\

\gamma_5 = (0,-1,0,-1,1,1) & \gamma_6 = (-1,0,1,0,-1,0)
\end{array}
\end{equation}

We stress here once again that the system under consideration is fictional in that sense that the
subset of equations corresponds to the simplified picture of this physical-chemical process and the
constants do not correspond to any measured ones, but reflect only basic orders of magnitudes of the
real-world system. In this sense we consider here a qualitative model system, which allows us to
illustrate the invariant grids method without excessive complication. Nevertheless, modeling of real
systems differs only in the number of species and equations. This leads, of course, to
computationally harder problems, but not the crucial ones, and the efforts on the modeling of
real-world systems are on the way.

Fig.~\ref{6d1dgrid}a presents a one-dimensional invariant grid constructed for the system.
Fig.~\ref{6d1dgrid}b shows the picture of reduced dynamics along the manifold (for the explanation of
the meaning of the $x$-coordinate, see the previous subsection). On Fig.~\ref{6d1dgrid}c the three
smallest by absolute value non-zero eigen values of the symmetrically linearized system $A^{sym}$
have been shown. One can see that the two smallest values ``exchange" on one of the grid end. It
means that one-dimensional "slow" manifold has definite problems in this region, it is just not
defined there. In practice, it means that one has to use at least two-dimensional grids there.

Fig.~\ref{6d2dgrid}a gives a view onto the two-dimensional invariant grid, constructed for the
system, using the ``invariant flag" strategy. The grid was grown starting from the 1D-grid
constructed at the previous step. At the first iteration for every node of the initial grid, two
nodes (and two edges) were added. The direction of the step was chosen as the direction of the
eigenvector of the matrix $A^{sym}$ (at the point of the node), corresponding to the second
``slowest" direction. The value of the step was chosen to be $\epsilon=0.05$ in terms of entropic
distance. After several Newtonian iterations done until convergence, new nodes were added in the
direction ``ortogonal" to the 1D-grid. This time it is done by linear extrapolation of the grid on
the same step $\epsilon=0.05$. When some new nodes have one or several negative coordinates (the grid
reaches the boundaries) they were cut off. If a new node has only one edge, connecting it to the
grid, it was excluded (since it does not allow calculating 2D-tangent space for this node). The
process continues until the expansion is possible (after this, every new node has to be cut off).

Strategy of calculating tangent vectors for this regular rectangular 2D-grid was chosen to be quite
simple. The grid consists of {\it rows}, which are co-oriented by construction to the initial
1D-grid, and {\it columns} that consist of the adjacent nodes in the neighboring rows. The direction
of ``columns" corresponds to the second slowest direction along the grid. Then, every row and column
is considered as 1D-grid, and the corresponding tangent vectors are calculated as it was described
before: $$g_{row}(x_{k,i}) = (x_{k,i+1}-x_{k,i-1})/\| x_{k,i+1}-x_{k,i-1} \|$$ for the internal nodes
and $$g_{row}(x_{k,1}) = (x_{k,1}-x_{k,2})/\| x_{k,1}-x_{k,2}\|, g_{row}(x_{k,n_k}) =
(x_{k,n_k}-x_{k,n_k-1})/\| x_{k,n_k}-x_{k,n_k-1} \|$$ for the nodes which are close to the grid's
edges. Here $x_{k,i}$ denotes the vector of the node in the $k$th row, $i$th column; $n_k$ is the
number of nodes in the $k$th row. Second tangent vector $g_{col}(x_{k,i})$ is calculated completely
analogously. In practice, it is convenient to orthogonalize $g_{row}(x_{k,i})$ and
$g_{col}(x_{k,i})$.

Since the phase space is four-dimensional, it is impossible to visualize the grid in one of the
coordinate 3D-views, as it was done in the previous subsection. To facilitate visualization one can
utilize traditional methods of multi-dimensional data visualization. Here we make use of the
principal components analysis (see, for example, \cite{princ}), which constructs a three-dimensional
linear subspace with maximal dispersion of the othogonally projected data (grid nodes in our case).
In other words, method of principal components constructs in multi-dimensional space such a
three-dimensional box inside which the grid can be placed maximally tightly (in the mean square
distance meaning). After projection of the grid nodes into this space, we get more or less adequate
representation of the two-dimensional grid embedded into the six-dimensional concentrations space
(Fig.~\ref{6d2dgrid}b). The disadvantage of the approach is that the axes now do not have explicit
meaning, being some linear combinations of the concentrations.

One attractive feature of two-dimensional grids is the possibility to use them as a screen, on which
one can display different functions $f({\bf c})$ defined in the concentrations space. This technology
was exploited widely in the non-linear data analysis by the elastic maps method \cite{GorbZinElMaps}.
The idea is to ``unfold" the grid on a plane (to present it in the two-dimensional space, where the
nodes form a regular lattice). In other words, we are going to work in the internal coordinates of
the grid. In our case, the first internal coordinate (let's call it $s_1$) corresponds to the
direction, co-oriented with the one-dimensional invariant grid, the second one (let's call it $s_2$)
corresponds to the second slow direction. By how it was constructed, $s_2=0$ line corresponds to the
one-dimensional invariant grid. Units of $s_1$ and $s_2$ are entropic distances in our case.

Every grid node has two internal coordinates $(s_1,s_2)$ and, simultaneously, corresponds to a vector
in the concentration space. This allows us to map any function $f({\bf c})$ from the
multi-dimensional concentration space to the two-dimensional space of the grid. This mapping is
defined in a finite number of points (grid nodes), and can be interpolated (linearly, in the simplest
case) in between them. Using coloring and isolines one can visualize the values of the function in
the neighborhood of the invariant manifold. This is meaningful, since, by the definition, the system
spends most of the time in the vicinity of the invariant manifold, thus, one can visualize the
behaviour of the system. As a result of applying the technology, one obtains a set of color
illustrations (a stack of information layers), put onto the grid as a map. This allows applying all
the methods, working with stack of information layers, like geographical information systems (GIS)
methods, which are very well developed.

In short words, the technique is a useful tool for exploration of dynamical systems. It allows to see
simultaneously many different scenarios of the system behaviour, together with different system's
characteristics.

The simplest functions to visualize are the coordinates: $c_i({\bf c}) = c_i$. On
Fig.~\ref{6d2dgridcolor} we displayed four colorings, corresponding to the four arbitrarily chosen
concentrations functions (of $H_2$, $O$, $H$ and $OH$; Fig.~\ref{6d2dgridcolor}a-d). The qualitative
conclusions that can be made from the graphs are that, for example, the concentration of $H_2$
practically does not change during the first fast motion (towards the 1D-grid) and then, gradually
changes to the equilibrium value (the $H_2$ coordinate is ``slow"). The $O$ coordinate is the
opposite case, it is ``fast" coordinate which changes quickly (on the first stage of motion) to the
almost equilibrium value, and then it almost does not change. Basically, the slope angles of the
coordinate isolines give some presentation of how ``slow" a given concentration is.
Fig.~\ref{6d2dgridcolor}c shows interesting behaviour of the $OH$ concentration. Close to the 1D grid
it behaves like ``slow coordinate", but there is a region on the map where it has clear ``fast"
behaviour (middle bottom of the graph).

The next two functions which one can want to visualize are the entropy $S$ and the entropy production
$\sigma({\bf c})=-dG/dt({\bf c}) = \sum_i{\ln(c_i/c^{eq}_i){\dot c_i}}$. They are shown on
Fig.~\ref{6d2dgridcolor1}a,b.

Finally, we visualize the relation between the relaxation times of the fast motion towards the
2D-grid and along it. This is given on the Fig.~\ref{6d2dgridcolor1}c. This picture allows to make a
conclusion that two-dimensional consideration can be appropriate for the system (especially in the
``high $H_2$, high $O$" region), since the relaxation times ``towards" and ``along" the grid are
definitely separated. One can compare this to the Fig.~\ref{6d2dgridcolor1}d, where the relation
between relaxation times towards and along the 1D-grid is shown.

\section{\textbf{Method of natural projector}}\label{np}

Ehrenfest suggested in 1911 a model of dynamics with a coarse-graining of the original conservative
system in order to introduce irreversibility \cite{Ehrenfest}. The idea of Ehrenfest is the
following: One partitions the phase space of the Hamiltonian system into cells. The density
distribution of the ensemble over the phase space evolves it time according  to the Liouville
equation within the time segments $n\tau<t<(n+1)\tau$, where $\tau$ is the fixed coarse-graining time
step. Coarse-graining is executed at discrete times $n\tau$, densities are averaged over each cell.
This alternation of the regular flow with the averaging describes the irreversible behavior of the
system.

The formally most general construction extending the Ehrenfest idea is given below. Let us stay with
notation of section \ref{diff}, and let a submanifold $F(W)$ be defined in the phase space $U$.
Furthermore, we assume a map (a projection) is defined, $\Pi: U\rightarrow W$, with the properties:
\begin{equation}\label{projEhr}
  \Pi\circ F=1,\ \Pi(F(y))=y.
\end{equation}
In addition, one requires some mild properties of regularity, in particular, surjectivity of the
differential, $D_x\Pi: E\rightarrow L$, in each point $x\in U$.

Let us fix the coarse-graining time $\tau>0$, and consider the following problem: Find a vector field
$\Psi$ in $W$,
\begin{equation}\label{macro}
  \frac{dy}{dt}=\Psi(y),
\end{equation}
such that, for every $y\in W$,
\begin{equation}\label{match}
  \Pi(T_{\tau}F(y))=\Theta_{\tau}y,
\end{equation}
where $T_{\tau}$ is the shift operator for the system (\ref{sys}), and $\Theta_{\tau}$ is the (yet
unknown!) shift operator for the system in question (\ref{macro}).

Equation (\ref{match}) means that one projects not the vector fields but segments of trajectories.
Resulting vector field $\Psi(y)$ is called \textit{the natural projection} of the vector field
$J(x)$.

Let us assume that there is a very stiff hierarchy of relaxation times in the system (\ref{sys}): The
motions of the system tend very rapidly to a slow manifold, and next proceed slowly along it. Then
there is a smallness parameter, the ratio of these times. Let us take $F$ for the initial condition
to the film equation (\ref{tmapsubtr}). If the solution $F_t$ relaxes to the positively invariant
manifold $F_{\infty}$, then, in the limit of a very stiff decomposition of motions, the natural
projection of the vector field $J(x)$ tends to the usual infinitesimal projection of the restriction
of $J$ on $F_{\infty}$, as $\tau\rightarrow \infty$:

\begin{equation}\label{limit}
  \Psi_{\infty}(y)=D_x\Pi|_{x=F_{\infty}(y)}J(F_{\infty}(y)).
\end{equation}

For stiff dynamic systems, the limit (\ref{limit}) is qualitatively almost obvious: After some
relaxation time $\tau_0$ (for $t>t_0$), the motion $T_{\tau}(x)$ is located in an
$\epsilon$-neighborhood of $F_{\infty}(W)$. Thus, for $\tau\gg\tau_0$, the natural projection $\Psi$
(equations (\ref{macro}) and (\ref{match})) is defined by the vector field attached to $F_{\infty}$
with any predefined accuracy. Rigorous proofs requires existence and uniqueness theorems, as well as
homogeneous continuous dependence of solutions on initial conditions and right hand sides of
equations.

The method of natural projector is applied not only to dissipative systems but also (and even mostly)
to conservative systems. One of the methods to study the natural projector is based on series
expansion\footnote{In the well known work of Lewis \cite{Lew}, this expansion was executed
incorrectly (terms of different orders were matched on the left and on the right hand sides of
equation (\ref{match}). This created an obstacle in a development of the method. See more detailed
discussion in the section Example 10.} in powers of $\tau$. Various other approximation schemes like
Pade approximation are possible too.

The construction of natural projector was rediscovered in completely different context by Chorin,
Hald and Kupferman \cite{Raz}. They constructed the optimal prediction methods for estimation the
solution of nonlinear time-dependent problems when that solution is too complex to be fully resolved
or when data are missing. The initial conditions for the unresolved components of the solution are
drawn from a probability distribution, and their effect on a small set of variables that are actually
computed is evaluated via statistical projection. The formalism resembles the projection methods of
irreversible statistical mechanics, supplemented by the systematic use of conditional expectations
and methods of solution for the orthogonal dynamics equation, needed to evaluate a non-Markovian
memory term. The result of the computations is close to the best possible estimate that can be
obtained given the partial data.

Most of the methods of invariant manifold can be discussed as development of the Chapman--Enskog
method. The idea is to construct the manifold of distribution functions, where the slow dynamics
occurs. The change-over from solution of the Boltzmann equation to construction of an invariant
manifold was a crucial idea of Enskog and Chapman. On the other hand, the method of natural projector
gives development of ideas of the Hilbert method. This method was historically the first in the
solution of the Boltzmann equation. The Hilbert method is not very popular now, nevertheless, for
some purposes it may be more convenient than the Chapman--Enskog method, for example, for a studying
of stationary solutions \cite{Sone}. In the method of natural projector we are looking for a
solutions of kinetic equations with quasiequilibrium initial state (and in Hilbert method we start
from the local equilibrium too). The main new element in the method of natural projector with respect
to the Hilbert method is construction of the macroscopic equation (\ref{match}). In the next Example
the solution for the matching condition (\ref{match}) will be found in a form of Taylor expansion.

\clearpage

\addcontentsline{toc}{subsection}{\textbf{Example 10: From reversible dynamics to Navier-Stokes and
post-Navier-Stokes hydrodynamics by natural projector}}

\subsection*{\textbf{Example 10: From reversible dynamics to Navier-Stokes and
post-Navier-Stokes hydrodynamics by natural projector}}

The starting point of our construction are microscopic equations of motion. A traditional example  of
the microscopic description is the  Liouville equation for classical particles. However, we need to
stress that the distinction between ``micro'' and ``macro'' is always context dependent. For example,
Vlasov's equation describes the dynamics of the one-particle distribution function. In one statement
of the problem, this is a microscopic dynamics in comparison to the evolution of hydrodynamic moments
of the distribution function. In a different setting, this equation itself  is a result of reducing
the description from the microscopic Liouville equation.

The problem of reducing the description includes a definition of the microscopic dynamics, and of the
macroscopic variables of interest, for which equations of the reduced description must be found. The
next step is the construction of the initial approximation. This is the well known quasiequilibrium
approximation, which is the solution to the variational problem, $S\to {\rm max}$, where $S$ in the
entropy, under given constraints. This solution assumes that the microscopic distribution functions
depend on time only through their dependence on the macroscopic variables. Direct substitution of the
quasiequilibrium distribution function  into the microscopic equation of motion gives the initial
approximation to the macroscopic dynamics. All further corrections can be obtained from a more
precise approximation of the microscopic as well as of the macroscopic trajectories within a given
time interval $\tau$ which is the parameter of our method.

The method described here has several clear advantages:

(i)  It allows to  derive complicated macroscopic equations, instead of writing them {\it ad hoc}.
This fact is especially significant for the description of complex fluids. The method gives explicit
expressions for relevant variables with one unknown parameter ($\tau$). This parameter can be
obtained from the experimental data.

(ii)  Another advantage of the method is its simplicity. For example, in the case where the
microscopic  dynamics is given by the Boltzmann equation, the approach avoids evaluation of Boltzmann
collision integral.

(iii) The most significant advantage of this formalization is that it is applicable to nonlinear
systems. Usually, in the classical approaches to reduced description, the  microscopic equation of
motion is linear. In that case, one can formally write the evolution operator in the exponential
form. Obviously, this does not work for nonlinear systems, such as, for example, systems with mean
field interactions. The method which we are presenting here is based on mapping the expanded
microscopic trajectory into the consistently expanded macroscopic trajectory. This does not require
linearity. Moreover, the order-by-order recurrent construction can be, in principle, enhanced by
restoring to other types of approximations, like Pad\'e approximation, for example, but we do not
consider these options here.

In the present section we discuss in detail applications of the method of natural projector
\cite{GKIOeNONNEWT2001,GKOeTPRE2001,KTGOePhA2003} to derivations of macroscopic equations in various
cases, with and without mean field interaction potentials, for various choices of macroscopic
variables, and demonstrate how computations are performed in the higher orders of the expansion. The
structure of the Example is as follows: In the next subsection,  for the sake of completeness, we
describe briefly the formalization of Ehrenfest's approach \cite{GKIOeNONNEWT2001,GKOeTPRE2001}. We
stress the role of the quasiequilibrium approximation  as the starting point for the constructions to
follow. We derive explicit expressions for the correction to the quasiequilibrium dynamics, and
conclude this section with the entropy production formula and its discussion. In section 3, we begin
the discussion  of applications. We use the present formalism in order to derive hydrodynamic
equations. Zeroth approximation of the scheme is the  Euler equations of the compressible nonviscous
fluid. The first approximation leads to the system of Navier-Stokes equations. Moreover, the approach
allows to obtain the next correction, so-called post-Navier-Stokes equations. The latter example is
of particular interest. Indeed, it is well known that post-Navier-Stokes equations as derived from
the Boltzmann kinetic equation by the Chapman-Enskog method (Burnett and super-Burnett hydrodynamics)
suffer from unphysical instability already in the linear approximation \cite{Bob}. We demonstrate it
by the explicit computation that the linearized higher-order hydrodynamic equations derived within
our method are free from this drawback.

\addcontentsline{toc}{subsubsection}{General construction}

\subsubsection*{\textbf{General construction}}

Let us consider a microscopic dynamics given by an equation,

\begin{equation}
\label{DYN} \dot{f}=J(f),
\end{equation}

\noindent where $f(x,t)$ is a distribution function over the phase space $x$ at  time $t$, and where
operator $J(f)$ may be linear or nonlinear. We consider linear macroscopic variables $M_k=\mu_k(f)$,
where operator $\mu_k$ maps $f$ into $M_k$. The problem is to obtain closed macroscopic equations of
motion, $\dot {M_k}=\phi_k(M)$. This is achieved in two steps: First, we construct an initial
approximation to the macroscopic dynamics and, second, this approximation is further corrected on the
basis of the coarse-gaining.

The initial  approximation is the quasiequilibrium approximation, and it is based on the entropy
maximum principle under fixed constraints \cite{Janes1,G1}:

\begin{equation}
\label{SMAX} S(f)\to {\rm max},\ \mu(f)=M,
\end{equation}

\noindent where $S$ is the entropy functional, which is assumed to be strictly concave, and $M$ is
the set of the macroscopic variables $\{M\},$ and $\mu$ is the set of the corresponding operators. If
the solution to the problem (\ref{SMAX}) exists, it is unique thanks to the concavity of the entropy
functionals. Solution to equation (\ref{SMAX}) is called the quasiequilibrium state, and it will be
denoted as $f^*(M)$. The classical example is the local equilibrium of the ideal gas: $f$ is the
one-body distribution function, $S$ is the Boltzmann entropy, $\mu$ are five linear operators,
$\mu(f)=\int\{1,{\vv},v^2\}fd{\vv}$, with ${\vv}$ the particle's velocity; the corresponding $f^*(M)$
is called the local Maxwell distribution function.

If the microscopic dynamics is given by equation (\ref{DYN}), then the quasiequilibrium dynamics of
the variables $M$ reads:

\begin{equation}
\label{QEdyn} \dot{M}_k=  \mu_k(J(f^*(M))=\phi^*_k.
\end{equation}

The quasiequilibrium approximation has  important property, it conserves the type of the dynamics: If
the entropy monotonically increases (or not decreases) due to equation (\ref{DYN}), then the same is
true for the quasiequilibrium entropy, $S^*(M)=S(f^*(M))$, due to the quasiequilibrium dynamics
(\ref{QEdyn}). That is, if
\[\dot{S}={\partial S(f)\over \partial f}\dot{f}={\partial S(f)\over
\partial f}J(f)\ge0,\]
then

\begin{equation}
\label{proof} \dot{S}^*=\sum_k\frac{\partial S^*}{\partial M_k}\dot{M}_k =\sum_k\frac{\partial
S^*}{\partial M_k} \mu_k(J(f^*(M)))\ge0.
\end{equation}

Summation in $k$ always implies summation or integration over the set of labels of the macroscopic
variables.

Conservation of the type of dynamics by the quasiequilibrium approximation is a simple yet a general
and useful fact. If the entropy $S$ is an integral of motion of  equation (\ref{DYN}) then $S^*(M)$
is the integral of motion for the quasiequilibrium equation (\ref{QEdyn}). Consequently, if we start
with a system which conserves the entropy (for example, with the Liouville equation) then we end up
with the quasiequilibrium  system which conserves the quasiequilibrium entropy. For instance, if $M$
is the one-body distribution function, and (\ref{DYN}) is the (reversible)
 Liouville equation, then (\ref{QEdyn}) is the Vlasov
equation which  is reversible, too. On the other hand, if the entropy was monotonically increasing on
solutions to equation (\ref{DYN}), then the quasiequilibrium entropy also increases monotonically on
solutions to the quasiequilibrium dynamic equations (\ref{QEdyn}). For instance, if equation
(\ref{DYN}) is the Boltzmann equation for the one-body distribution function, and $M$ is a finite set
of moments (chosen in such a way that the solution to the problem (\ref{SMAX}) exists), then
(\ref{QEdyn}) are closed moment equations for $M$ which increase the quasiequilibrium entropy (this
is the essence of a well known generalization of Grad's moment method).

\addcontentsline{toc}{subsubsection}{Enhancement of quasiequilibrium approximations for
entropy-conserving dynamics}

\subsubsection*{\textbf{Enhancement of quasiequilibrium approximations for entropy-conserving dynamics}}

The goal of the present  subsection is to describe the simplest analytic implementation,  the
microscopic motion with periodic coarse-graining. The notion of coarse-graining was introduced by P.\
and T.\ Ehrenfest's in their seminal work \cite{Ehrenfest}: The phase space is partitioned into
cells, the coarse-grained variables are the amounts of the phase density inside the cells. Dynamics
is described by the two processes, by the Liouville equation for $f$, and by periodic
coarse-graining, replacement of $f(x)$ in each cell by its average value in this cell. The
coarse-graining operation means forgetting the microscopic details, or of the history.

From the perspective of general quasiequilibrium approximations, periodic coarse-graining amounts to
the return of the true microscopic trajectory on the quasiequilibrium manifold with the preservation
of the macroscopic variables. The motion starts at the quasiequilibrium state $f^*_i$. Then the true
solution $f_i(t)$ of the microscopic equation (\ref{DYN})
 with the initial condition $f_i(0)=f^*_i$
is coarse-grained  at a fixed time $t=\tau$, solution  $f_i(\tau)$ is replaced by the
quasiequilibrium function $f^*_{i+1}=f^*(\mu(f_i(\tau)))$. This process is sketched in Fig.
\ref{Fig1LAR}.

From the features of the quasiequilibrium approximation it follows that for the motion with periodic
coarse-graining, the inequality is valid,
\begin{equation}
\label{shaking} S(f_i^*)\le S(f^*_{i+1}),
\end{equation}
the equality occurs if and only if the quasiequilibrium is the invariant manifold of the dynamic
system (\ref{DYN}). Whenever the quasiequilibrium is {\it not} the solution to equation (\ref{DYN}),
the strict inequality in (\ref{shaking}) demonstrates the entropy increase.

In other words, let us assume that the trajectory begins at the quasiequilibrium manifold, then it
takes off from  this manifold according to the  microscopic evolution equations. Then, after some
time $\tau$, the trajectory is coarse-grained, that is the, state is brought back on the
quasiequilibrium manifold keeping the values of the macroscopic variables. The irreversibility is
born in the latter process, and this construction clearly rules out quasiequilibrium manifolds which
are invariant with respect to the microscopic dynamics, as candidates for a coarse-graining. The
coarse-graining indicates the way to derive equations for macroscopic variables from the condition
that the macroscopic trajectory, $M(t)$,  which governs the motion of the quasiequilibrium states,
$f^*(M(t))$, should match precisely the same points on the quasiequilibrium manifold,
$f^*(M(t+\tau))$, and this matching should be independent of both the initial time, $t$, and the
initial condition $M(t)$. The problem is then how to derive the continuous time macroscopic dynamics
which would be consistent with this picture. The simplest realization suggested in the Ref.\
\cite{GKIOeNONNEWT2001,GKOeTPRE2001} is based on using  an expansion of both the microscopic and the
macroscopic trajectories. Here we present this construction to the third order accuracy, in a general
form, whereas only the second-order accurate construction has been discussed in
\cite{GKIOeNONNEWT2001,GKOeTPRE2001}.

Let us write down the solution to the microscopic equation (\ref{DYN}), and approximate this solution
by the polynomial of third oder in $\tau$. Introducing notation, $J^*=J(f^*(M(t)))$, we write,
\begin{equation}
f(t+\tau)=f^*+\tau J^*+{\tau^2\over 2}{\partial J^*\over \partial f} J^*+{\tau^3\over
3!}\left({\partial J^*\over \partial f}{\partial J^*\over \partial f} J^*+{\partial^2 J^*\over
\partial f^2}J^* J^*\right)+ o(\tau^3).\label{mic}
\end{equation}

Evaluation of the macroscopic variables on the function (\ref{mic}) gives
\begin{eqnarray}
M_k(t+\tau )&=& M_k+\tau \phi_k^*+{\tau^2\over 2}\mu_k\left({\partial J^*\over
\partial f} J^*\right)\nonumber\\* &+& {\tau^3\over 3!}\left\{\mu_k\left({\partial
J^*\over \partial f} {\partial J^*\over \partial f} J^*\right)+ \mu_k\left({\partial^2 J^*\over
\partial f^2} J^* J^*\right)\right\}+ o(\tau^3),\label{mic1}
\end{eqnarray}
where $\phi^*_k=\mu_k(J^*)$ is the quasiequilibrium macroscopic vector field (the right hand side of
equation (\ref{QEdyn})), and all the functions and derivatives are taken in the quasiequilibrium
state at time $t$.

We shall now establish the macroscopic dynamic by matching the macroscopic and the microscopic
dynamics. Specifically, the macroscopic dynamic equations (\ref{QEdyn}) with the right-hand side not
yet defined, give the following third-order result:
\begin{eqnarray}
M_k(t+\tau)&=&M_k+\tau\phi_k+{\tau^2\over 2}\sum_j {\partial \phi_k\over
\partial M_j}\phi_j \nonumber\\*&+&{\tau^3\over 3!} \sum_{ij}\left(
{\partial^2\phi_k\over
\partial M_iM_j}\phi_i\phi_j +{\partial \phi_k\over \partial
M_i}{\partial \phi_i\over \partial M_j}\phi_j \right)+o(\tau^3).\label{mak}
\end{eqnarray}

Expanding functions $\phi_k$ into the series $\phi_k=R_k^{(0)}+\tau R_k^{(1)}+\tau^2R_k^{(2)}+...$,
($R_k^{(0)}=\phi^*$),  and requiring that the microscopic and the macroscopic dynamics coincide to
the order of $\tau^3$, we obtain the sequence of corrections for the right-hand side of the equation
for the macroscopic variables. Zeroth order is the quasiequilibrium approximation to the macroscopic
dynamics. The first-order correction gives:
\begin{eqnarray}
R_k^{(1)}={1\over 2}\left\{\mu_k\left({\partial J^*\over \partial f} J^*\right)-\sum_j {\partial
\phi_k^*\over
\partial M_j}\phi_j^*\right\}\label{Rk1}.
\end{eqnarray}
The next, second-order correction has the following explicit form:
\begin{eqnarray}
R_k^{(2)}&=&{1\over 3!} \left\{\mu_k \left({\partial J^*\over
\partial f}{\partial J^*\over \partial f} J^*\right) + \mu_k\left({\partial^2
J^*\over
\partial f^2}J^* J^*\right)\right\} - {1\over 3!} \sum_{ij} \left( {\partial
\phi_k^*\over \partial M_i} {\partial \phi_i^*\over \partial M_j}\phi_j^*\right)\nonumber\\*
&-&{1\over 3!}\sum_{ij}\left({\partial^2 \phi_k^*\over
\partial M_i\partial M_j} \phi_i^*\phi_j^* \right)-{1 \over 2 }
\sum_j \left( {\partial \phi_k^*\over \partial M_j} R_j^{(1)}+ {\partial R_j^{(1)}\over \partial M_j}
\phi_j^*\right),\label{26}
\end{eqnarray}
Further corrections are found by the same token. Equations (\ref{Rk1})--(\ref{26})  give explicit
closed expressions for corrections to the quasiequilibrium dynamics to the order of accuracy
specified above. They are used below in various specific examples.

\addcontentsline{toc}{subsubsection}{Entropy production}

\subsubsection*{\textbf{Entropy production}}

The most important consequence of the above construction is that the resulting continuous time
macroscopic equations retain the dissipation property of the discrete time coarse-graining
(\ref{shaking}) on each order of approximation $n\ge 1$. Let us first consider the entropy production
formula for the first-order approximation. In order to shorten notations, it is convenient to
introduce the quasiequilibrium projection operator,

\begin{equation}
P^*g= \sum_k{\partial f^*\over
\partial M_k}\mu_k(g).\end{equation}
It has been demonstrated in \cite{GKOeTPRE2001} that the entropy production,

\[\dot{S}^*_{(1)}=\sum_k\frac{\partial S^*}{\partial M_k}(R_k^{(0)}+\tau
R_k^{(1)}),\] equals
\begin{eqnarray}
\dot{S}^*_{(1)} =-\frac{\tau}{2}(1-P^*)J^*\left.{\partial ^2 S^*\over
\partial f
\partial f
}\right|_{f^*}(1-P^*)J^*.\label{RESULT2}
\end{eqnarray}

Equation (\ref{RESULT2}) is nonnegative definite due to concavity of the entropy. Entropy production
(\ref{RESULT2}) is equal to zero only if the quasiequilibrium approximation is the true solution to
the microscopic dynamics, that is, if $(1-P^*)J^*\equiv0$. While quasiequilibrium approximations
which solve the Liouville equation are uninteresting objects (except, of course, for the equilibrium
itself), vanishing of the entropy production in this case is a simple test of consistency of the
theory. Note that the entropy production (\ref{RESULT2}) is proportional to $\tau$. Note also that
projection operator does not appear in our consideration a priory, rather, it is the result of
exploring the coarse-graining condition in the previous section.

Though equation (\ref{RESULT2}) looks very natural, its existence is rather subtle. Indeed, equation
(\ref{RESULT2}) is a difference of the two terms, $\sum_k\mu_k( J^*\partial J^*/\partial f)$
(contribution of the second-order approximation to the microscopic trajectory), and
$\sum_{ik}R_i^{(0)}\partial R_k^{(0)}/\partial M_i$ (contribution of the derivative of the
quasiequilibrium vector field). Each of these expressions separately gives a positive contribution to
the entropy production, and equation (\ref{RESULT2}) is the difference of the two positive definite
expressions. In the higher order approximations, these subtractions are more involved, and explicit
demonstration of the entropy production formulae becomes a formidable task. Yet, it is possible to
demonstrate the increase-in-entropy without explicit computation, though at a price of smallness of
$\tau$. Indeed, let us denote $\dot{S}^*_{(n)}$ the time derivative of the entropy on the $n$th order
approximation. Then

\begin{eqnarray}
\int_t^{t+\tau}\dot{S}^*_{(n)}(s)ds=S^*(t+\tau)-S^*(t)+O(\tau^{n+1}),\nonumber
\end{eqnarray}

where $S^*(t+\tau)$ and $S^*(t)$ are true values of the entropy at the adjacent states of the
$H$-curve. The difference $\delta S=S^*(t+\tau)-S^*(t)$ is strictly positive for any fixed $\tau$,
and, by equation (\ref{RESULT2}), $\delta S\sim \tau^2$ for small $\tau$. Therefore, if $\tau$ is
small enough, the right hand side in the above expression is positive, and $$\tau
\dot{S}_{(n)}^*(\theta_{(n)})>0,$$ where $t\le\theta_{(n)}\le t+\tau$. Finally, since
$\dot{S}_{(n)}^*(t)=\dot{S}_{(n)}^*(s)+O(\tau^n)$ for any $s$ on the segment $[t,t+\tau]$, we can
replace $\dot{S}^*_{(n)}(\theta_{(n)})$ in the latter inequality by $\dot{S}^*_{(n)}(t)$. The sense
of this consideration is as follows: Since the entropy production formula (\ref{RESULT2}) is valid in
the leading order of the construction, the entropy production will not collapse in the higher orders
at least if the coarse-graining time is small enough. More refined estimations can be obtained only
from the explicit analysis of the higher-order corrections.

\addcontentsline{toc}{subsubsection}{Relation to the work of Lewis}

\subsubsection*{\textbf{Relation to the work of Lewis}}

Among various realizations of the coarse-graining procedures, the work of Lewis \cite{Lew} appears to
be most close to our approach. It is therefore pertinent to discuss the differences. Both methods are
based on the coarse-graining condition,
\begin{equation}\label{CG}
  M_k(t+\tau)=\mu_k\left(T_{\tau}f^*(M(t))\right),
\end{equation}
where $T_{\tau}$ is the formal solution operator of the microscopic dynamics. Above, we applied a
consistent expansion of both, the left hand side and the right hand side of the coarse-graining
condition (\ref{CG}), in terms of the coarse-graining time $\tau$. In the work of Lewis \cite{Lew},
it was suggested, as a general way to exploring the condition (\ref{CG}), to write the first-order
equation for $M$ in the form of the differential pursuit,
\begin{equation}\label{Lewis}
  M_k(t)+\tau\frac{dM_k(t)}{dt}\approx\mu_k\left(T_{\tau}f^*(M(t))\right).
\end{equation}
In other words, in the work of Lewis \cite{Lew}, the expansion to the first order was considered on
the left (macroscopic) side of equation (\ref{CG}), whereas the right hand side containing the
microscopic trajectory $T_{\tau}f^*(M(t))$ was not treated on the same footing. Clearly, expansion of
the right hand side to first order in $\tau$ is the only equation which is common in both approaches,
and this is the quasiequilibrium dynamics. However, the difference occurs already in the next,
second-order term (see Ref.\ \cite{GKIOeNONNEWT2001,GKOeTPRE2001} for details). Namely, the expansion
to the second order of the right hand side of Lewis' equation (\cite{Lew}) results in a dissipative
equation (in the case of the Liouville equation, for example) which remains dissipative even if the
quasiequilibrium approximation is the exact solution to the microscopic dynamics, that is, when
microscopic trajectories once started on the quasiequilibrium manifold belong to it in all the later
times, and thus no dissipation can be born by any coarse-graining.

On the other hand, our approach assumes a certain smoothness of trajectories so that application of
the low-order expansion bears physical significance. For example, while using lower-order truncations
it is not possible to derive the Boltzmann equation because in that case the relevant
quasiequilibrium manifold ($N$-body distribution function is proportional to the product of one-body
distributions, or uncorrelated states, see next section) is almost invariant during the long time (of
the order of the mean free flight of particles), while the trajectory steeply leaves this manifold
during the short-time pair collision. It is clear that in such a case lower-order expansions of the
microscopic trajectory do not lead to useful results. It has been clearly stated by Lewis \cite{Lew},
that the exploration of the condition (\ref{CG}) depends on the physical situation, and how one makes
approximations. In fact, derivation of the Boltzmann equation given by Lewis on the basis of the
condition (\ref{CG}) does not follow the differential pursuit approximation: As is well known, the
expansion in terms of particle's density of the solution to the BBGKY hierarchy is singular, and
begins with the \textit{linear} in time term. Assuming the quasiequilibrium approximation for the
$N$-body distribution function under fixed one-body distribution function, and that collisions are
well localized in space and time, one gets on the right hand side of equation (\ref{CG}),
\[f(t+\tau)=f(t)+n\tau J_B(f(t))+o(n),\]
where $n$ is particle's density, $f$ is the one-particle distribution function, and $J_B$ is the
Boltzmann's collision integral. Next, using the mean-value theorem on the left hand side of the
equation (\ref{CG}), the Boltzmann equation is derived (see also a recent elegant
renormalization-group argument for this derivation \cite{RG}).

We stress that our approach of matched expansion for exploring the coarse-graining condition
(\ref{CG}) is, in fact, the exact (formal) statement that the unknown macroscopic dynamics which
causes the shift of $M_k$ on the left hand side of equation (\ref{CG}) can be reconstructed
order-by-order to any degree of accuracy, whereas the low-order truncations may be useful for certain
physical situations. A thorough study of the cases beyond the lower-order truncations is of great
importance which is left for future work.

\addcontentsline{toc}{subsubsection}{Equations of hydrodynamics for simple fluid}

\subsubsection*{\textbf{Equations of hydrodynamics for simple fluid}}

The method discussed above enables one to establish in a simple way the form of equations of the
macroscopic dynamics to various degrees of approximation.  In this section, the microscopic dynamics
is given by the Liouville equation, similar to the previous case. However, we take another set of
macroscopic variables: density, average velocity, and average temperature of the fluid. Under this
condition the solution to the problem (\ref{SMAX}) is the local Maxwell distribution. For the
hydrodynamic equations, the zeroth (quasiequilibrium) approximation is given by Euler's equations of
compressible nonviscous fluid. The next order approximation are the Navier-Stokes equations which
have dissipative terms.

Higher-order approximations to the hydrodynamic equations, when they are derived from the  Boltzmann
kinetic equation (so-called Burnett approximation), are subject to various difficulties, in
particular, they exhibit an  instability of sound waves at sufficiently short wave length (see, e.\
g. \cite{KGAnPh2002} for a recent review). Here we demonstrate how model hydrodynamic equations,
including post-Navier-Stokes approximations, can be derived on the basis of coarse-graining idea, and
investigate the linear  stability of the obtained equations. We will find that the resulting
equations are stable.

Two points need a clarification before we proceed further \cite{GKOeTPRE2001}. First, below we
consider the simplest Liouville equation for the one-particle distribution, describing a free moving
particle without interactions. The procedure of coarse-graining we use is an implementation of
collisions leading to dissipation. If we had used the full interacting $N$-particle Liouville
equation, the result would be different, in the first place, in the expression for the local
equilibrium pressure. Whereas in the present case we have the ideal gas pressure, in the $N$-particle
case the non-ideal gas pressure would arise.

Second, and more essential is that, to the order of the Navier-Stokes equations, the {\it result} of
our method is identical to the lowest-order Chapman-Enskog method as applied to the Boltzmann
equation with a single relaxation time model collision integral (the Bhatnagar-Gross-Krook model
\cite{BGK}). However, this happens only at this particular order of approximation, because already
the next, post-Navier-Stokes approximation,  is different from the Burnett hydrodynamics as derived
from the BGK model (the latter is linearly unstable).

\addcontentsline{toc}{subsubsection}{Derivation of the Navier-Stokes equations}

\subsubsection*{\textbf{Derivation of the Navier-Stokes equations}}\label{ChH}

Let us assume that reversible microscopic dynamics is given by the one-particle Liouville equation,
\begin{eqnarray}
{\partial f\over \partial t}=-v_i{\partial f\over \partial r_i},\label{li}
\end{eqnarray}
where $f=f(\rr,\vv,t)$ is the one-particle distribution function, and index $i$ runs over spatial
components $\{x,\ y,\ z \}$. Subject to appropriate boundary conditions which we assume, this
equation conserves the Boltzmann entropy $S=-k_{\rm B}\int f\ln f d\vv d\rr$.

We introduce the following hydrodynamic moments as the macroscopic variables: $M_0=\int
fd\vv,\,M_i=\int v_ifd\vv,\,M_4=\int v^2fd\vv $. These variables are related to the more conventional
density, average velocity and temperature, $n,\,\uu,\,T$ as follows:
\begin{eqnarray}
&&M_0=n,\hspace{0.5cm}M_i=nu_i,\hspace{0.7cm}M_4={3nk_BT\over m}+nu^2,\nonumber\\
&&n=M_0,\hspace{0.5cm}u_i=M_0^{-1}M_i,\hspace{0.5cm}T={m\over 3k_BM_0}(M_4-M_0^{-1}M_iM_i).\label{M}
\end{eqnarray}
The quasiequilibrium distribution function  (local Maxwellian) reads:
\begin{eqnarray}
f_0=n\left({m\over 2\pi k_BT}\right)^{3/2}\exp\left({-m(v-u) ^2\over 2k_BT}\right).\label{QEM}
\end{eqnarray}
Here and below, $n,\ \uu,$ and $T$ depend on $\rr$ and $t.$

Based on the microscopic dynamics (\ref{li}), the set of macroscopic variables (\ref{M}), and the
quasiequilibrium (\ref{QEM}), we can derive the  equations of the macroscopic motion.

A specific feature of the present example is that the quasiequilibrium equation for the density (the
continuity equation),
\begin{eqnarray}
{\partial n\over
\partial t}&=&-{\partial nu_i\over \partial r_i},\label{n}
\end{eqnarray}
should be excluded out of the further corrections. This rule should be applied generally: If a part
of the chosen macroscopic variables (momentum flux $n\uu$ here) correspond to fluxes of other
macroscopic variables, then the quasiequilibrium equation for the latter is already exact, and has to
be exempted of corrections.

The quasiequilibrium approximation for the rest of the macroscopic variables is derived in the usual
way. In order to derive the  equation for the velocity, we substitute the local Maxwellian into the
one-particle Liouville equation, and act with the operator $\mu_k=\int v_k \cdot d\vv$ on both the
sides of the equation (\ref{li}). We have:
\begin{eqnarray}
{\partial n u_k\over \partial t}=-{\partial \over
\partial r_k}{nk_BT\over m}-{\partial nu_ku_j\over \partial r_j}.\nonumber
\end{eqnarray}

Similarly,  we derive the equation for the energy density, and the complete system of equations of
the quasiequilibrium approximation reads (Euler equations):
\begin{eqnarray}
{\partial n\over
\partial t}&=&-{\partial nu_i\over \partial r_i},\label{eu1}\\*
{\partial nu_k\over \partial t}&=& -{\partial \over \partial r_k}{nk_BT\over m}-{\partial nu_ku_j
\over \partial r_j},\nonumber\\ {\partial \varepsilon\over
\partial t}&=&-{\partial \over \partial r_i}\left({5k_BT\over
m}nu_i+u^2nu_i\right).\nonumber
\end{eqnarray}

Now we are going to derive the next order approximation to  the macroscopic dynamics (first order in
the coarse-graining time $\tau$). For the velocity equation we have:
\begin{eqnarray}
R_{nu_k}={1\over 2}\left(\int v_kv_iv_j{\partial^2 f_0\over
\partial r_i\partial r_j}d\vv -\sum_{j}{\partial \phi_{nu_k}\over
\partial M_j}\phi_j\right),\nonumber
\end{eqnarray}
where $\phi_j$ are the corresponding right hand sides of the Euler equations (\ref{eu1}). In order to
take derivatives with respect to macroscopic moments $\{M_0,M_i,M_4\}$, we need to rewrite equations
(\ref{eu1}) in terms of these variables instead of $\{n,u_i,T\}$. After some computation, we obtain:
\begin{eqnarray}
R_{nu_k} ={1\over 2}{\partial \over
\partial r_j}\left({nk_BT\over m}\left[{\partial u_k\over \partial
r_j}+{\partial u_j\over \partial r_k}-{2\over 3} {\partial u_n\over \partial
r_n}\delta_{kj}\right]\right).\label{NS1}
\end{eqnarray}

For the energy we obtain:
\begin{eqnarray}
R_{\varepsilon}={1\over 2}\left(\int v^2v_iv_j{\partial^2 f_0\over
\partial r_i\partial r_j}d\vv -\sum_{j}{\partial \phi_{\varepsilon }\over
\partial M_j}\phi_j\right)={5\over 2}{\partial \over \partial
r_i}\left({nk_B^2T\over m^2 }{\partial T\over \partial r_i}\right).\label{NS2}
\end{eqnarray}

Thus,  we get the system of the Navier-Stokes equation in the following form:
\begin{eqnarray}
{\partial n\over
\partial t}&=&-{\partial nu_i\over \partial r_i},\nonumber\\
{\partial nu_k\over \partial t}&=&-{\partial \over \partial r_k}{nk_BT\over m}-{\partial nu_ku_j
\over \partial r_j}+\nonumber\\* &&{\tau\over 2} {\partial \over \partial r_j}{nk_BT\over
m}\left({\partial u_k\over
\partial r_j}+{\partial u_j\over \partial r_k}-{2\over 3}
{\partial u_n\over \partial r_n}\delta_{kj}\right),\label{NS}\\* {\partial \varepsilon\over
\partial t}&=-&{\partial \over \partial r_i}\left({5k_BT\over
m}nu_i+u^2nu_i\right)+\tau{5\over 2}{\partial \over \partial r_i}\left({nk_B^2T\over m^2}{\partial
T\over
\partial r_i}\right).\nonumber
\end{eqnarray}
We see that kinetic coefficients (viscosity and heat conductivity) are proportional to the
coarse-graining time $\tau$. Note that they are identical with kinetic coefficients as derived from
the Bhatnagar-Gross-Krook model \cite{BGK}  in the first approximation of the Chapman-Enskog method
\cite{Chapman} (also, in particular, no bulk viscosity) .

\addcontentsline{toc}{subsubsection}{Post-Navier-Stokes equations}

\subsubsection*{\textbf{Post-Navier-Stokes equations}}

Now we are going to obtain the  second-order approximation to the hydrodynamic equations in the
framework of the present approach. We will compare qualitatively the result with the Burnett
approximation. The comparison concerns stability of the hydrodynamic modes near global equilibrium,
which is violated for the Burnett approximation. Though the derivation is straightforward also in the
general, nonlinear case, we shall consider only the linearized equations which is appropriate to our
purpose here.

Linearizing the local Maxwell  distribution function, we obtain:
\begin{eqnarray}
f&=&n_0\left({m\over 2\pi k_BT_0}\right)^{3/2}\left({n\over n_0}+{mv_n\over
k_BT_0}u_n+\left({mv^2\over 2k_BT_0}-{3\over 2 }\right){T\over T_0}\right)e^{-mv^2\over
2k_BT_0}=\nonumber\\* &=&\left\{(M_0+2M_ic_i+\left({2\over 3}M_4-M_0\right)\left(c^2-{3\over
2}\right)\right\}e^{-c^2},\label{mL}
\end{eqnarray}
where we have introduced dimensionless variables: $c_i=v_i/v_T$, $v_T=\sqrt{2k_BT_0/m}$ is the
thermal velocity, $M_0=\delta n/n_0,$ $M_i=\delta u_i/v_T$, $M_4=(3/2)(\delta n/n_0+\delta T/T_0)$.
Note that  $\delta n$, and $\delta T$ determine deviations of these variables from their equilibrium
values, $n_0,$ and $T_0$.

The linearized Navier-Stokes equations read:
\begin{eqnarray}
{\partial M_0\over \partial t}&=&-{\partial M_i\over
\partial r_i},\nonumber\\*
{\partial M_k\over \partial t}&=&-{1\over 3}{\partial M_4\over
\partial r_k}+{\tau\over 4}{\partial \over
\partial r_j }\left({\partial M_k\over
\partial r_j}+{\partial M_j\over \partial r_k}-{2\over 3}
{\partial M_n\over \partial r_n}\delta_{kj}\right),\label{linNS}\\* {\partial M_4\over
\partial t}&=&-{5\over 2}{\partial M_i\over
\partial r_i}+\tau {5\over 2}{\partial^2 M_4\over \partial r_i\partial r_i}.\nonumber
\end{eqnarray}

Let us first compute  the post-Navier-Stokes correction to the velocity equation. In accordance with
the equation (\ref{26}), the first part of this term under linear approximation is:
\begin{eqnarray}
&&{1\over 3!} \mu_k \left({\partial J^*\over \partial f}{\partial J^*\over \partial f} J^*\right) -
{1\over 3!} \sum_{ij} \left( {\partial \phi_k^*\over \partial M_i} {\partial \phi_i^*\over \partial
M_j}\phi_j^*\right)=\nonumber\\* &=&-{1\over 6}\int c_k{\partial^3\over \partial r_i\partial
r_j\partial r_n}c_ic_jc_n\left\{(M_0+2M_ic_i+\left({2\over 3}M_4-M_0\right)\left(c^2-{3\over
2}\right)\right\}e^{-c^2}d^3c\nonumber\\* &+&{5\over 108 }{\partial \over \partial r_i }{\partial^2
M_4\over
\partial r_s\partial r_s}={1\over 6}{\partial \over \partial r_k }\left({3\over 4}
{\partial^2 M_0\over
\partial r_s\partial r_s }-{\partial^2 M_4\over
\partial r_s\partial r_s }\right)
+{5\over 108 }{\partial \over \partial r_k }{\partial^2 M_4\over
\partial r_s\partial r_s}\nonumber\\*
&=&{1\over 8}{\partial \over \partial r_k }{\partial^2 M_0\over
\partial r_s\partial r_s }-{13\over 108}{\partial \over \partial r_k }{\partial^2 M_4\over
\partial r_s\partial r_s }.\label{3u1}
\end{eqnarray}

The part of equation (\ref{26}) proportional to the first-order correction is:
\begin{eqnarray} &-&{1 \over 2 } \sum_j \left(
{\partial \phi_k^*\over \partial M_j} R_j^{(1)}+ {\partial R_k^{(1)}\over
\partial M_j} \phi_j^*\right)={5\over
6}{\partial \over \partial r_k }{\partial^2 M_4\over
\partial r_s\partial r_s }+{1\over 9}{\partial \over \partial r_k }{\partial^2 M_4\over
\partial r_s\partial r_s }.\label{3u2}
\end{eqnarray}
Combining together terms (\ref{3u1}), and (\ref{3u2}), we obtain:
\begin{eqnarray} R_{M_k}^{(2)}={1\over 8}{\partial \over \partial r_k }{\partial^2 M_0\over
\partial r_s\partial r_s }+{89\over 108}{\partial \over \partial r_k }{\partial^2 M_4\over
\partial r_s\partial r_s }.\nonumber
\end{eqnarray}

Similar calculation for the energy equation leads to the following result:
\begin{eqnarray}
&&-\int c^2{\partial^3\over \partial r_i\partial r_j\partial
r_k}c_ic_jc_k\left\{(M_0+2M_ic_i+\left({2\over 3}M_4-M_0\right)\left(c^2-{3\over
2}\right)\right\}e^{-c^2}d^3c +\nonumber\\* &+&{25\over 72}{\partial\over
\partial r_i}{\partial^2 M_i\over
\partial r_s\partial r_s}=-{1\over 6}\left({21\over 4}{\partial\over
\partial r_i}{\partial^2 M_i\over
\partial r_s\partial r_s}+{25\over 12}{\partial\over
\partial r_i}{\partial^2 M_i\over
\partial r_s\partial r_s}\right)=-{19\over 36}{\partial\over
\partial r_i}{\partial^2 M_i\over
\partial r_s\partial r_s}.\nonumber
\end{eqnarray}

The term proportional to the first-order corrections gives:
\begin{eqnarray} {5\over 6}\left({\partial^2\over
\partial r_s\partial r_s}{\partial M_i\over
\partial r_i}\right)+{25\over 4}\left({\partial^2\over
\partial r_s\partial r_s}{\partial M_i\over
\partial r_i}\right).\nonumber
\end{eqnarray}

Thus,  we obtain:
\begin{eqnarray}
R^{(2)}_{M_4}={59\over 9}\left({\partial^2\over
\partial r_s\partial r_s}{\partial
M_i\over \partial r_i}\right).
\end{eqnarray}

Finally, combining  together all the terms, we obtain the following system of linearized hydrodynamic
equations:
\begin{eqnarray}
{\partial M_0\over \partial t}&=&-{\partial M_i\over
\partial r_i},\nonumber\\*
{\partial M_k\over \partial t}&=&-{1\over 3}{\partial M_4\over
\partial r_k}+{\tau\over 4}{\partial \over
\partial r_j }\left({\partial M_k\over
\partial r_j}+{\partial M_j\over \partial r_k}-{2\over 3}
{\partial M_n\over \partial r_n}\delta_{kj}\right)+\nonumber\\*&&\tau^2\left\{{1\over 8}{\partial
\over
\partial r_k }{\partial^2 M_0\over
\partial r_s\partial r_s }+{89\over 108}{\partial \over \partial r_k }{\partial^2 M_4\over
\partial r_s\partial r_s }\right\},\label{3order}\\*
{\partial M_4\over
\partial t}&=&-{5\over 2}{\partial M_i\over
\partial r_i}+\tau {5\over 2}{\partial^2 M_4\over \partial r_i\partial r_i}+{\tau^2}{59\over 9}\left({\partial^2\over
\partial r_s \partial r_s}{\partial
M_i\over \partial r_i}\right).\nonumber
\end{eqnarray}

Now we are in a position to investigate the dispersion relation of this system. Substituting
$M_i=\tilde{M}_i\exp(\omega t+i(\kk,\rr))$ $(i=0,\,k,\,4)$ into equation (\ref{3order}), we reduce
the problem to finding the spectrum of the matrix:
\begin{eqnarray}\left(
\begin{array}{ccccc}
  0 & -ik_x & -ik_y & -ik_z & 0 \\
  -ik_x{k^2\over 8} & -{1\over 4}k^2-{1\over 12}k_x^2 &
-{k_xk_y\over 12}& -{k_xk_z\over 12}& -ik_x\left({1\over 3 }+{89k^2\over 108}\right) \\
  -ik_y{k^2\over 8} & -{k_xk_y\over 12} & -{1\over
4}k^2-{1\over 12}k_y^2&  -{k_yk_z\over 12} & -ik_y\left({1\over 3}+ {89k^2\over 108}\right) \\
  -ik_z{k^2\over 8} &  -{k_xk_z\over 12} &  -{k_yk_z\over 12}
&-{1\over 4}k^2-{1\over 12}k_z^2 & -ik_z\left({1\over 3}+{89k^2\over 108}\right)
\\
  0 &-ik_x\left({5\over 2}+{59k^2\over 9}\right) & -ik_y\left({5\over
2}+{59k^2\over 9}\right) & -ik_z\left({5\over 2}+{59k^2\over 9}\right) & -{5\over 2}k^2
\end{array}\right)\nonumber
\end{eqnarray}

This matrix has five eigenvalues. The real parts of these eigenvalues responsible for the decay rate
of the corresponding modes  are shown in Fig.\ref{Fig2LAR} as functions of the wave vector $k$. We
see that {\it all real parts of all the eigenvalues are non-positive for any wave vector}. In other
words, this means that the present system is linearly stable. For the Burnett hydrodynamics as
derived from the Boltzmann or from the single relaxation time Bhatnagar-Gross-Krook model, it is well
known that the decay rate of the acoustic  becomes positive after some value of the wave vector
\cite{Bob,KGAnPh2002} which leads to the instability. While the method suggested here is clearly
semi-phenomenological (coarse-graining time $\tau$ remains unspecified), the consistency of the
expansion with the entropy requirements, and especially the latter result of the linearly stable
post-Navier-Stokes correction strongly indicates that it might be more suited  to establishing models
of highly nonequilibrium hydrodynamics.

\clearpage

\addcontentsline{toc}{subsection}{\textbf{Example 11: Natural projector for the Mc Kean model}}

\subsection*{\textbf{Example 11: Natural projector for the Mc Kean model}}

In this section the fluctuation-dissipation formula recently derived by the method of natural
projector \cite{GKMex2001} is illustrated by the explicit computation for McKean's kinetic model
\cite{McKean}. It is demonstrated that the result is identical, on the one hand, to the sum of the
Chapman-Enskog expansion, and, on the other hand, to the exact solution of the invariance equation.
The equality between all the three results holds up to the crossover from the hydrodynamic to the
kinetic domain.

\addcontentsline{toc}{subsubsection}{General scheme}

\subsubsection*{\textbf{General scheme}}

Let us consider a microscopic dynamics (\ref{sys}) given by an equation for the distribution function
$f(x,t)$ over a configuration space $x$:

\begin{equation}
\label{Li}
\partial_t f=J(f),
\end{equation}

\noindent where operator $J(f)$ may be linear or nonlinear. Let $\mm(f)$ be a set of linear
functionals whose values, $\MM=\mm(f)$, represent the macroscopic variables, and also let $f(\MM,x)$
be a set of distribution functions satisfying the consistency condition,

\begin{equation}
\label{cons} \mm(f(\MM))=\MM.
\end{equation}

\noindent The choice of the relevant distribution functions is the point of central importance which
we discuss later on but for the time being we need only specification (\ref{cons}).

The starting point has been the following observation \cite{GKIOeNONNEWT2001,GKOeTPRE2001}: Given a
finite time interval $\tau$, it is possible to reconstruct uniquely the macroscopic dynamics from a
single condition.  For the sake of completeness, we shall formulate this condition here. Let us
denote as $\MM(t)$ the initial condition at the time $t$ to the {\it yet unknown} equations of the
macroscopic motion, and let us take $f(\MM(t),x)$ for the initial condition of the microscopic
equation (\ref{Li}) at the time $t$.  Then the condition for the reconstruction of the macroscopic
dynamics reads as follows:  For every initial condition $\{\MM(t),t\}$, solutions to the macroscopic
dynamic equations at the time $t+\tau$ are equal to the values of the macroscopic variables on the
solution to equation (\ref{Li}) with the initial condition $\{f(\MM(t),x),t\}$:

\begin{equation}
\label{exact} \MM(t+\tau)=\mm\left(T_{\tau}f(\MM(t))\right),
\end{equation}

\noindent where $T_{\tau}$ is the formal solution operator of the microscopic equation (\ref{Li}).
The right hand side of equation (\ref{exact}) represents an operation on trajectories of the
microscopic equation (\ref{Li}), introduced in a particular form by Ehrenfest's \cite{Ehrenfest} (the
coarse-graining):  The solution at the time $t+\tau$ is replaced by the state on the manifold
$f(\MM,x)$.  Notice that the coarse-graining time $\tau$ in equation (\ref{exact}) is finite, and we
stress the importance of the required independence from the initial time $t$, and from the initial
condition at $t$.

The essence of the reconstruction of the macroscopic equations from the condition just formulated is
in the following \cite{GKIOeNONNEWT2001,GKOeTPRE2001}: Seeking the macroscopic equations in the form,

\begin{equation}
\label{exact1}
\partial_t \MM=\RR(\MM,\tau),
\end{equation}

\noindent we proceed with Taylor expansion of the unknown functions $\RR$ in terms of powers
$\tau^n$, where $n=0,1,\dots$, and require that each approximation, $\RR^{(n)}$, of the order $n$, is
such that resulting macroscopic solutions satisfy the condition (\ref{exact1}) to the order
$\tau^{n+1}$.  This process of successive approximation is solvable.  Thus, the unknown macroscopic
equation (\ref{exact1}) can be reconstructed to any given accuracy.

Coming back to the problem of choosing the distribution function $f(\MM,x)$, we recall that many
physically relevant cases of the microscopic dynamics (\ref{Li}) are characterized by existence of a
concave functional $S(f)$ (the entropy functional; discussions of $S$ can be found in
\cite{Wehrl78,Schloegl80,G1}). Traditionally, two cases are distinguished, the conservative
[$dS/dt\equiv 0$ due to equation (\ref{Li})], and the dissipative [$dS/dt\ge 0$ due to equation
(\ref{Li}), where equality sign corresponds to the stationary solution].  The approach (\ref{exact})
and (\ref{exact1}) is applicable to both these situations.  In both of these cases, among the
possible sets of distribution functions $f(\MM,x)$, the distinguished role is played by the well
known quasiequilibrium approximations, $f^*(\MM,x)$, which are maximizers of the functional $S(f)$
for fixed $\MM$.  We recall that, due to convexity of the functional $S$, if such maximizer exist
then it is unique.  The special role of the quasiequilibrium approximations is due to the well known
fact that they preserve the type of dynamics:  If $dS/dt\ge0$ due to equation (\ref{Li}), then
$dS^*/dt\ge0$ due to the quasiequilibrium dynamics, where $S^{*}(\MM)=S(f^*(\MM))$ is the
quasiequilibrium entropy, and where the quasiequilibrium dynamics coincides with the zeroth order in
the above construction, $\RR^{(0)}=\mm(J(f^*(\MM))$.  We notice it in passing that, since the well
known work of Jaynes \cite{Janes1}, the usefulness of quasiequilibrium approximations is well
understood in various versions of projection operator formalism for the conservative case
\cite{Robertson,Grabert,Zubarev,Eva}, as well as for the dissipative dynamics
\cite{G1,GKAMSE92,GKTTSP94,GK1}. Relatively less studied remains the case of open or externally
driven systems, where invariant quasiequilibrium manifolds may become unstable \cite{IK00}.  The use
of the quasiequilibrium approximations for the above construction has been stressed in
\cite{GKIOeNONNEWT2001,GKOeTPRE2001,GKGeoNeo}. In particular, the strict increase in the
quasiequilibrium entropy has been demonstrated for the first and higher order
approximations\cite{GKOeTPRE2001}. Examples have been provided \cite{GKOeTPRE2001}, focusing on the
conservative case, and demonstrating that several well known dissipative macroscopic equations, such
as the Navier-Stokes equation and the diffusion equation for the one-body distribution function, are
derived as the lowest order approximations of this construction.

The advantage of the approach \cite{GKIOeNONNEWT2001,GKOeTPRE2001} is the locality of construction,
because only Taylor series expansion of the microscopic solution is involved. This is also its
natural limitation.  From the physical standpoint, finite and fixed coarse-graining time $\tau$
remains a phenomenological device which makes it possible to infer the form of the macroscopic
equations by a non-complicated computation rather than to derive a full form thereof.  For instance,
the form of the Navier-Stokes equations can be derived from the simplest model of free motion of
particles, in which case the coarse-graining is a substitution for collisions.  Going away from the
limitations imposed by the finite coarse graining time \cite{GKIOeNONNEWT2001,GKOeTPRE2001} can be
recognized as the major problem of a consistent formulation of the nonequilibrium statistical
thermodynamics. Intuitively, this requires taking the limit $\tau\to\infty$, allowing for all the
relevant correlations to be developed by the microscopic dynamics, rather than to be cut off at the
finite $\tau$. Indeed, in the case of the dissipative dynamics, in particular, for the linearized
Boltzmann equation, one typically expects an initial layer \cite{Cercignani} which is completely cut
off in the short-memory approximation, whereas those effects can be made small by taking $\tau$ large
enough. A way of doing this in the general nonlinear setting for entropy-conserving systems still
requires further work at the time of this writing.

\addcontentsline{toc}{subsubsection}{Natural projector for linear systems}

\subsubsection*{\textbf{Natural projector for linear systems}}

However, there is one important exception when the `$\tau\to\infty$ problem' is readily solved
\cite{GKOeTPRE2001,GKMex2001}.  This is the case where equation (\ref{Li}) is linear,

\begin{equation}
\label{LiLi}
\partial_t f=Lf,
\end{equation}

\noindent and where the quasiequilibrium is a linear function of $\MM$.  This is, in particular, the
classical case of linear irreversible thermodynamics where one considers the linear macroscopic
dynamics near the equilibrium, $f^{\rm eq}$, $Lf^{\rm eq}=0$.  We assume, for simplicity of
presentation, that the macroscopic variables $\MM$ vanish at equilibrium, and are normalized in such
a way that $\mm(f^{\rm eq}\mm^{\dag})=\UNIT$, where $^{\dag}$ denotes transposition, and $\UNIT$ is
an appropriate identity operator.  In this case, the linear dynamics of the macroscopic variables
$\MM$ has the form,

\begin{equation}
\partial_t \MM=\RR\MM,
\label{macro2}
\end{equation}

\noindent where the linear operator $\RR$ is determined by the coarse-graining condition
(\ref{exact}) in the limit $\tau\to\infty$:

\begin{equation}
\label{GK} \RR=\lim_{\tau\to\infty}\frac{1}{\tau}\ln\left[ \mm\left( e^{\tau L}f^{\rm
eq}\mm^{\dag}\right) \right].
\end{equation}

\noindent Formula (\ref{GK}) has been already briefly mentioned in \cite{GKOeTPRE2001}, and its
relation to the Green-Kubo formula has been demonstrated in \cite{GKMex2001}.  In our case, the
Green-Kubo formula reads:

\begin{equation}
\label{Green-Kubo} \RR_{\rm GK}=\int^{\infty}_0\langle \dot{\mm}(0)\dot{\mm}(t)\rangle,
\end{equation}

\noindent where angular brackets denote equilibrium averaging, and where $\dot{\mm}=L^{\dag}\mm$. The
difference between the formulae (\ref{GK}) and (\ref{Green-Kubo}) stems from the fact that condition
(\ref{exact}) does not use an a priori hypothesis of the separation of the macroscopic and the
microscopic time scales.  For the classical $N$-particle dynamics, equation (\ref{GK}) is a
complicated expression, involving a logarithm of non-commuting operators.  It is therefore very
desirable to gain its understanding in simple model situations.

\addcontentsline{toc}{subsubsection}{Explicit example of the the fluctuation-dissipation formula}

\subsubsection*{\textbf{Explicit example of the the fluctuation-dissipation formula}}

In this section we want to give explicit example of the formula (\ref{GK}). In order to make our
point, we consider here dissipative rather than conservative dynamics in the framework of the well
known toy kinetic model introduced by McKean \cite{McKean} for the purpose of testing various ideas
in kinetic theory. In the dissipative case with a clear separation of time scales, existence of the
formula (\ref{GK}) is underpinned by the entropy growth in both the rapid and the slow parts of the
dynamics. This physical idea underlies generically the extraction of the slow (hydrodynamic)
component of motion through the concept of normal solutions to kinetic equations, as pioneered by
Hilbert \cite{Hilbert}, and has been discussed by many authors, e.\ g.\ .
\cite{Cercignani,Uhlenbeck,Grad}. Case studies for linear kinetic equation help clarifying the
concept of this extraction \cite{Hauge,Titulaer,McKean}.

Therefore, since for the dissipative case there exist well established approaches to the problem of
reducing the description, and which are exact in the present setting, it is very instructive to see
their relation to the formula (\ref{GK}). Specifically, we compare the result with the exact sum of
the Chapman-Enskog expansion \cite{Chapman}, and with the exact solution in the framework of the
method of invariant manifold \cite{GKAMSE92,GKTTSP94,GK1}.  We demonstrate that both the three
approaches, different in their nature, give the same result as long as the hydrodynamic and the
kinetic regimes are separated.

The McKean model is the kinetic equation for the two-component vector function
$\ff(r,t)=(f_{+}(r,t),f_{-}(r,t))^{\dag}$:

\begin{eqnarray}
\label{McKeq}
\partial_tf_{+}&=&-\partial_rf_{+}+\epsilon^{-1}\left(\frac{f_{+}+f_{-}}{2}-
f_{+}\right),\\\nonumber
\partial_tf_{-}&=&\partial_rf_{-}+\epsilon^{-1}\left(\frac{f_{+}+f_{-}}{2}-
f_{-}\right).
\end{eqnarray}

\noindent Equation (\ref{McKeq}) describes the one-dimensional kinetics of particles with velocities
$+1$ and $-1$ as a combination of the free flight and a relaxation with the rate $\epsilon^{-1}$ to
the local equilibrium.  Using the notation, $(\xx,\yy)$, for the standard scalar product of the
two-dimensional vectors, we introduce the fields, $n(r,t)=(\nn,\ff)$ [the local particle's density,
where $\nn=(1,1)$], and $j(r,t)=(\jj,\ff)$ [the local momentum density, where $\jj=(1,-1)$]. Equation
(\ref{McKeq}) can be equivalently written in terms of the moments,

\begin{eqnarray}
\partial_t n&=&-\partial_r j,\\\nonumber
\partial_t j&=&-\partial_r n-\epsilon^{-1}j.
\label{Grad}
\end{eqnarray}

\noindent The local equilibrium,

\begin{equation}
\label{loceq} \ff^*(n)=\frac{n}{2}\nn,
\end{equation}

\noindent is the conditional maximum of the entropy,

\[S=-\int (f_{+}\ln f_{+}+f_{-}\ln f_{-})dr,\]

\noindent under the constraint which fixes the density, $(\nn,\ff^*)=n$.  The quasiequilibrium
manifold (\ref{loceq}) is linear in our example, as well as is the kinetic equation.

The problem of reducing the description for the model (\ref{McKeq}) amounts to finding the closed
equation for the density field $n(r,t)$.  When the relaxation parameter $\epsilon^{-1}$ is small
enough (the relaxation dominance), then the first Chapman-Enskog approximation to the momentum
variable, $j(r,t)\approx -\epsilon\partial_r n(r,t)$, amounts to the standard diffusion
approximation. Let us consider now how the formula (\ref{GK}), and other methods, extend this result.

Because of the linearity of the equation (\ref{McKeq}), and of the local equilibrium, it is natural
to use the Fourier transform, $h_{k}=\int\exp(ikr)h(r)dr$. Equation (\ref{McKeq}) is then written,

\begin{equation}
\partial_t\ff_{k}=\LL_k\ff_{k},
\end{equation}

\noindent where

\begin{equation}
\LL_k=\left(
\begin{array}{cc}
-ik-\frac{1}{2\epsilon} & \frac{1}{2\epsilon}\\ \frac{1}{2\epsilon} & ik-\frac{1}{2\epsilon}
\end{array}\right).
\end{equation}

\noindent Derivation of the formula (\ref{GK}) in our example goes as follows: We seek the
macroscopic dynamics of the form,

\begin{equation}
\label{macroMK}
\partial_t n_k=R_k n_k,
\end{equation}

\noindent where the function $R_k$ is yet unknown. In the left-hand side of equation (\ref{exact}) we
have:

\begin{equation}
\label{left} n_k(t+\tau)=e^{\tau R_k}n_k(t).
\end{equation}

\noindent In the right-hand side of equation (\ref{exact}) we have:

\begin{equation}
\label{right} \left(\nn,e^{\tau\LL_k}\ff^*(n_k(t))\right)=\frac{1}{2}
\left(\nn,e^{\tau\LL_k}\nn\right)n_k(t).
\end{equation}

\noindent After equating the expressions (\ref{left}) and (\ref{right}), we require that the
resulting equality holds in the limit $\tau\to\infty$ independently of the initial data $n_k(t)$.
Thus, we arrive at the formula (\ref{GK}):

\begin{equation}
\label{GK1} R_k=\lim_{\tau\to\infty}\frac{1}{\tau}\ln\left[ \left(\nn,e^{\tau\LL_k}\nn\right)\right].
\end{equation}

\noindent Equation (\ref{GK1}) defines the macroscopic dynamics (\ref{macroMK}) within the present
approach. Explicit evaluation of the expression (\ref{GK1}) is straightforward in the present model.
Indeed, operator $\LL_k$ has two eigenvalues, $\Lambda^{\pm}_k$, where

\begin{equation}
\Lambda^{\pm}_k=-\frac{1}{2\epsilon}\pm\sqrt{\frac{1}{4\epsilon^2}- k^2}
\end{equation}

\noindent Let us denote as $\ee^{\pm}_k$ two (arbitrary) eigenvectors of the matrix $\LL_k$,
corresponding to the eigenvalues $\Lambda_k^{\pm}$.  Vector $\nn$ has a representation,
$\nn=\alpha_k^+\ee^{+}_k+\alpha^{-}_k\ee^{-}_{k}$, where $\alpha_k^{\pm}$ are complex-valued
coefficients.  With this, we obtain in equation (\ref{GK1}),

\begin{equation}
\label{GK2} R_k=\lim_{\tau\to\infty}\frac{1}{\tau}
\ln\left[\alpha_k^{+}(\nn,\ee^+_k)e^{\tau\Lambda^{+}_k}
+\alpha_{k}^{-}(\nn,\ee_{k}^-)e^{\tau\Lambda^{-}_k}\right].
\end{equation}

\noindent For $k\le k_{\rm c}$, where $k_{\rm c}^2=4\epsilon$, we have $\Lambda^{+}_k>\Lambda^{-}_k$.
Therefore,

\begin{equation}
R_k=\Lambda^{+}_k, \ {\rm for}\ k<k_{\rm c}. \label{exactGK}
\end{equation}

\noindent As was expected, formula (\ref{GK}) in our case results in the exact hydrodynamic branch of
the spectrum of the kinetic equation (\ref{McKeq}).  The standard diffusion approximation is
recovered from equation (\ref{exactGK}) as the first non-vanishing approximation in terms of the
$(k/k_{\rm c})^2$.

At $k=k_{\rm c}$, the crossover from the extended hydrodynamic to the kinetic regime takes place, and
${\rm Re}\Lambda^+_k={\rm Re}\Lambda^-_k$.  However, we may still extend the function $R_k$ for $k\ge
k_{\rm c}$ on the basis of the formula (\ref{GK1}):

\begin{equation}
R_k={\rm Re}\ \Lambda^{+}_k\ {\rm for}\ k\ge k_{\rm c} \label{extension}
\end{equation}

\noindent Notice that the function $R_k$ as given by equations (\ref{exactGK}) and (\ref{extension})
is continuous but non-analytic at the crossover.

\addcontentsline{toc}{subsubsection}{Comparison with the Chapman-Enskog method and solution of
invariance equation}

\subsubsection*{\textbf{Comparison with the Chapman-Enskog method and solution of
invariance equation}}

Let us now compare this result with the Chapman-Enskog method. Since the exact Chapman-Enskog
solution for the systems like equation (\ref{Grad}) has been recently discussed in detail elsewhere
\cite{GKPRL96,KGAnPh2002,K2,K3,Slem1,Slem2}, we shall be brief here. Following the Chapman-Enskog
method, we seek the momentum variable $j$ in terms of an expansion,

\begin{equation}
\label{CE1} j^{\rm CE}=\sum_{n=0}^{\infty}\epsilon^{n+1}j^{(n)}
\end{equation}

\noindent The Chapman-Enskog coefficients, $j^{(n)}$, are found from the recurrence equations,

\begin{equation}
\label{CE2} j^{(n)}=-\sum_{m=0}^{n-1}\partial_t^{(m)}j^{(n-1-m)},
\end{equation}

\noindent where the Chapman-Enskog operators $\partial_t^{(m)}$ are defined by their action on the
density $n$:

\begin{equation}
\label{CE3}
\partial_t^{(m)}n=-\partial_rj^{(m)}.
\end{equation}

\noindent The recurrence equations (\ref{CE1}), (\ref{CE2}), and (\ref{CE3}), become well defined as
soon as the aforementioned zero-order approximation $j^{(0)}$ is specified,

\begin{equation}
\label{F} j^{(0)}=-\partial_r n.
\end{equation}

\noindent From equations (\ref{CE2}), (\ref{CE3}), and (\ref{F}), it follows that the Chapman-Enskog
coefficients $j^{(n)}$ have the following structure:

\begin{equation}
\label{CE4} j^{(n)}=b_n\partial_r^{2n+1}n,
\end{equation}

\noindent where coefficients $b_n$ are found from the recurrence equation,

\begin{equation}
\label{CE5} b_n=\sum_{m=0}^{n-1}b_{n-1-m}b_m,\ b_0=-1.
\end{equation}

\noindent Notice that coefficients (\ref{CE5}) are real-valued, by the sense of the Chapman-Enskog
procedure.  The Fourier image of the Chapman-Enskog solution for the momentum variable has the form,

\begin{equation}
\label{CE6} j_k^{\rm CE}=ikB_k^{\rm CE}n_k,
\end{equation}

\noindent where

\begin{equation}
\label{CE7} B_k^{\rm CE}=\sum_{n=0}^{\infty}b_n(-\epsilon k^2)^n.
\end{equation}

\noindent Equation for the function $B$ (\ref{CE7}) is easily found upon multiplying equation
(\ref{CE5}) by $(-k^2)^n$, and summing in $n$ from zero to infinity:

\begin{equation}
\label{CE8} \epsilon k^2B_k^2+B_k+1=0.
\end{equation}

\noindent Solution to the latter equation which respects condition (\ref{F}), and which constitutes
the exact Chapman-Enskog solution (\ref{CE7}) is:

\begin{equation}
\label{CE9} B_k^{\rm CE}=\left\{
\begin{array}{ll}
k^{-2}\Lambda^{+}_k,& k<k_{\rm c}\\ {\rm none},&k\ge k_{\rm c}
\end{array}\right.
\end{equation}

\noindent Thus, the exact Chapman-Enskog solution derives the macroscopic equation for the density as
follows:

\begin{equation}
\label{CE10}
\partial_t n_k=-ikj_k^{\rm CE}
=R_k^{\rm CE}n_k,
\end{equation}

\noindent where

\begin{equation}
\label{CE_final} R_k^{\rm CE}=\left\{
\begin{array}{ll}
\Lambda^+_k,& k<k_{\rm c}\\ {\rm none},&k\ge k_{\rm c}
\end{array}\right.
\end{equation}

\noindent The Chapman-Enskog solution does not extends beyond the crossover at $k_{\rm c}$.  This
happens because the full Chapman-Enskog solution appears as a continuation the diffusion
approximation, whereas formula (\ref{GK1}) is not based on such an extension a priori.

Finally, let us discuss briefly the comparison with the solution within the method of invariant
manifold \cite{GKAMSE92,GKTTSP94,GK1}. Specifically, the momentum variable $j_k^{\rm inv}=ikB_k^{\rm
inv}n_k$ is required to be invariant of both the microscopic and the macroscopic dynamics, that is,
the time derivative of $j_k^{\rm inv}$ due to the macroscopic subsystem,

\begin{equation}
\label{macro_dot} \frac{\partial j_k^{\rm inv}}{\partial n_k}\partial_t n_k= ikB_k^{\rm
inv}(-ik)[ikB_k^{\rm inv}],
\end{equation}

\noindent should be equal to the derivative of $j_k^{\rm inv}$ due to the microscopic subsystem,

\begin{equation}
\label{micro_dot}
\partial_t j_k^{\rm inv} =-ikn_k-\epsilon^{-1}ikB_k^{\rm inv}n_k,
\end{equation}

\noindent and that the equality between the equations (\ref{macro_dot}) and (\ref{micro_dot}) should
hold independently of the specific value of the macroscopic variable $n_k$.  This amounts to a
condition for the unknown function $B_k^{\rm inv}$, which is essentially the same as equation
(\ref{CE8}), and it is straightforward to show that the same selection procedure of the hydrodynamic
root as above in the Chapman-Enskog case results in equation (\ref{CE_final}).

In conclusion, in this Example we have given the explicit illustration for the formula (\ref{GK}).
The example considered above demonstrates that the formula (\ref{GK}) gives the exact macroscopic
evolution equation, which is identical to the sum of the Chapman-Enskog expansion, as well as to the
invariance principle. This identity holds up to the point where the hydrodynamics and the kinetics
cease to be separated.  Whereas the Chapman-Enskog solution does not extend beyond the crossover
point, the formula (\ref{GK}) demonstrates a non-analytic extension.  The example considered adds to
the confidence of the correctness of the approach suggested in
\cite{GKIOeNONNEWT2001,GKOeTPRE2001,GKMex2001,GKPRE02}.

\section{\textbf{ Slow invariant manifold for a closed system has been found. What next?}}

Suppose that the slow invariant manifold is found for a dissipative system. {\bf What have we
constructed it for?}

{\it First of all, for solving the Cauchy problem, in order to separate motions.} This means that the
Cauchy problem is divided in the following two subproblems:

\begin{itemize}
\item{Reconstruct the ``fast" motion from the initial conditions
to the slow invariant manifold ({\it the initial layer problem}).}
\item{Solve the Cauchy problem for the ``slow" motions on the manifold.}
\end{itemize}

Thus, solving the Cauchy problem becomes easier (and in some complicated cases it just becomes
possible).

Let us stress here that for any sufficiently reliable solution of the Cauchy problem one must solve
not only the reduced Cauchy problem for the slow motion, but also the initial layer problem for fast
motions.

While solving the latter problem it was found to be surprisingly effective to use piece-wise linear
approximations with smoothing or even without it \cite{GKZNPhA96,GKZTTSP99}. This method was used for
the Boltzman equation, for chemical kinetics equations, and for the Pauli equation.

There exists a different way to model the initial layer in kinetics problems: it is the route of
model equations. For example, the Bhatnagar, Gross, Krook (BGK) equation \cite{BGK} is the simplest
model for the Boltzmann equation. It describes relaxation into a small neighborhood of the local
Maxwell distribution. There are many types and hierarchies of the model equations
\cite{BGK,Cercignani,GKMod,InChLANL,AKMod}. The principal idea of any model equation is to replace
the fast processes by a simple relaxation term. As a rule, it has a form
$dx/dt=\ldots-(x-x_{sl}(x))/\tau$, where $x_{sl}(x)$ is a point of the approximate slow manifold.
Such form is used in the BGK-equation, or in the quasi-equilibrium models \cite{GKMod}. It also can
take a gradient form, like in the gradient models \cite{InChLANL,AKMod}. These simplifications not
only allows to study the fast motions separately but it also allows to zoom in the details of the
interaction of fast and slow motions in the vicinity of the slow manifold.

What concerns solving the Cauchy problem for the ``slow" motions, this is the basic problem of the
hydrodynamics, of the gas dynamics (if the initial ``big" systems describes kinetics of a gas or a
fluid), etc. Here invariant manifold methods provide equations for a further study. However, even a
preliminary consideration of the practical aspects of these studies shows a definite shortcoming. In
practice, obtained equations are exploited not only for ``closed" systems. The initial equations
(\ref{sys}) describe a dissipative system that approaches the equilibrium. The equations of slow
motion describe dissipative system too. Then these equations are supplied with {\it various forces
and flows,} and after that they describe systems with more or less complex dynamics.

Because of this, there is a different answer to our question, {\it what  have we constructed the
invariant manifold for?}

{\it First of all, in order to construct models of open system dynamics in the neighborhood of the
slow manifold.}

Various approaches to this modeling are described in the following subsections.

\subsection{\textbf{Slow dynamics in open systems. Zero-order approximation
and the thermodynamic projector}}

Let the initial dissipative system (\ref{sys}) be ``spoiled" by an additional term (``external vector
field" $J_{ex}(x,t)$):

\begin{equation}\label{opsys}
\frac{dx}{dt} = J(x)+J_{ex}(x,t), x \subset U.
\end{equation}

\noindent For this new system the entropy does not increase everywhere. In the new system
(\ref{opsys}) different dynamic effects are possible, such as a non-uniqueness of stationary states,
auto-oscillations, etc. The ``inertial manifold" effect is well-known: solutions of (\ref{opsys})
approach some relatively low-dimensional manifold on which all the non-trivial dynamics takes place
\cite{IneManFSTe88,IneManTe88,IneManCFTe88}. This ``inertial manifold" can have a finite dimension
even for infinite-dimensional systems, for example, for the ``reaction+diffusion" systems
\cite{Vishik}.

In the theory of nonlinear control of partial differential equation systems a strategy based on
approximate inertial manifolds \cite{JonTiti} is proposed to facilitate the construction of
finite-dimensional systems of ordinary differential equations (ODE), whose solutions can be
arbitrarily close to the solutions of the infinite--dimensional system \cite{Christo}.

It is natural to expect that the inertial manifold of the system (\ref{opsys}) is located somewhere
close to the slow manifold of the initial dissipative system (\ref{sys}). This hypothesis has the
following basis. Suppose that the vector field $J_{ex}(x,t)$ is sufficiently small. Let's introduce,
for example, a small parameter $\varepsilon>0$, and consider $\varepsilon J_{ex}(x,t)$ instead of
$J_{ex}(x,t)$. Let's assume that for the system (\ref{sys}) a separation of motions into ``slow" and
``fast" takes place. In this case, there exists such interval of positive $\varepsilon$ that
$\varepsilon J_{ex}(x,t)$ is comparable to $J$ only in a small neighborhood of the given slow motion
manifold of the system (\ref{sys}). Outside this neighborhood, $\varepsilon J_{ex}(x,t)$ is
negligibly small in comparison with $J$ and only negligibly influences the motion (for this statement
to be true, it is important that the system (\ref{sys}) is dissipative and every solution comes in
finite time to a small neighborhood of the given slow manifold).

Precisely this perspective on the system (\ref{opsys}) allows to exploit slow invariant manifolds
constructed for the dissipative system (\ref{sys}) as the ansatz and the zero-order approximation in
a construction of the inertial manifold of the open system (\ref{opsys}). In the zero-order
approximation, the right part of the equation (\ref{opsys}) is simply projected onto the tangent
space of the slow manifold.

The choice of the projector is determined by the motion separation which was described above: fast
motion is taken from the dissipative system (\ref{sys}). A projector which is suitable for all
dissipative systems with given entropy function is unique. It is constructed in the following way
(detailed consideration of this is given above in the sections ``Entropic projector without a priori
parametrization" and in ref. \cite{UNIMOLD}). Let a point $x\in U$ be defined and some vector space
$T$, on which one needs to construct a projection ($T$ is the tangent space to the slow manifold at
the point $x$). We introduce the entropic scalar product $\langle \mid \rangle_x$:

\begin{equation}\label{es1}
\langle a \mid b\rangle_x = -(a,D_x^2S(b)).
\end{equation}

Let us consider $T_0$ that is a subspace of $T$ and which is annulled by the differential $S$ at the
point $x$.

\begin{equation}
T_0 = \{a\in T|D_xS(a)=0\}
\end{equation}

Suppose\footnote{If $T_0 = T$, then the thermodynamic projector is the orthogonal projector on $T$
with respect to the entropic scalar product $\langle |\rangle_x$.} that $T_0\neq T$. Let $e_g\in T$,
$e_g \perp T_0$ with respect to the entropic scalar product $\langle \mid \rangle_x$, and
$D_xS(e_g)=1$. These conditions define vector $e_g$ uniquely.

The projector onto $T$ is defined by the formula

\begin{equation}\label{ep}
P(J)=P_0(J)+e_gD_xS(J)
\end{equation}

\noindent where $P_0$ is the orthogonal projector onto $T_0$ with respect to the entropic scalar
product $\langle \mid \rangle_x$. For example, if $T$ a finite-dimensional space, then the projector
(\ref{ep}) is constructed in the following way. Let $e_1,..,e_n$ be a basis in $T$, and for
definiteness, $D_xS(e_1)\neq 0$.

\noindent 1) Let us construct a system of vectors

\begin{equation}
b_i=e_{i+1}-\lambda_i e_1, (i=1,..,n-1),
\end{equation}

\noindent where $\lambda_i=D_xS(e_{i+1})/D_xS(e_{1})$, and hence $D_xS(b_i)=0$. Thus,
$\{b_i\}_1^{n-1}$ is a basis in $T_0$.

\noindent2) Let us orthogonalize $\{b_i\}_1^{n-1}$ with respect to the entropic scalar product
$\langle \mid \rangle_x$ (\ref{sys}). We thus derived an orthonormal with respect to $\langle \mid
\rangle_x$ basis $\{g_i\}_1^{n-1}$ in $T_0$.

\noindent3) We find $e_g\in T$ from the conditions:

\begin{equation}
\langle e_g \mid g_i \rangle _x = 0, (i=1,..,n-1), D_xS(e_g)=1.
\end{equation}

\noindent and, finally we get

\begin{equation}\label{pfin}
P(J) = \sum_{i=1}^{n-1}g_i \langle g_i\mid J \rangle_x+e_g D_xS(J).
\end{equation}

If $D_xS(T)=0$, then the projector $P$ is simply the orthogonal projector with respect to the
$\langle \hspace{1pt}| \rangle_x$ scalar product. This is possible if $x$ is the global maximum of
entropy point (equilibrium). Then

\begin{equation}\label{peq}
P(J) = \sum_{i=1}^n{g_i\langle g_i|J \rangle_x}, \langle g_i|g_j \rangle = \delta_{ij}.
\end{equation}

{\it Remark}. In applications, the equation (\ref{sys}) often has additional linear balance
constraints such as numbers of particles, momentum, energy, etc. Solving the closed dissipative
system (\ref{sys}) we simply choose balance values and consider the dynamics of (\ref{sys}) on the
corresponding affine balance subspace.

For driven system (\ref{opsys}) the balances can be violated. Because of this, for the open system
(\ref{opsys}) the natural balance subspace includes the balance subspace of (\ref{sys}) with
different balance values. For every set of balance values there is a corresponding equilibrium. Slow
invariant manifold of the dissipative systems that is applied to the description of the driven
systems (\ref{opsys}) is usually the {\it union} of slow manifolds for all possible balance values.
The equilibrium of the dissipative closed system corresponds to the entropy maximum {\it given the
balance values are fixed}. In the phase space of the driven system (\ref{opsys}) the entropy gradient
in the equilibrium points of the system (\ref{sys}) is not necessarily equal to zero.

In particular, for the Boltzmann entropy in the local finite-dimensional case one gets the
thermodynamic projector in the following form.

\begin{eqnarray}
&&  S=-\int{f(v)(ln(f(v))-1)dv}, \nonumber \\ &&  D_fS(J)=-\int{J(v)\ln f(v) dv}, \nonumber \\ &&
\langle \psi \mid \varphi \rangle_f = -(\psi,D^2_fS(\varphi)) =
\int{\frac{\psi(v)\varphi(v)}{f(v)}dv} \nonumber \\&&
  P(J) = \sum_{i=1}^{n-1}g_i(v)\int{\frac{g_i(v)J(v)}{f(v)}dv} -
  e_g(v)\int{J(v)\ln f(v)dv},
\end{eqnarray}

\noindent where $g_i(v)$ and $e_g(v)$ are constructed accordingly to the scheme described above,
\begin{eqnarray}
&& \int{\frac{g_i(v)g_j(v)}{f(v)}dv}=\delta_{ij},\\ &&\int{g_i(v)\ln f(v)dv}=0,\\
&&\int{g_i(v)e_g(v)dv}=0,\\ && \int{e_g(v)\ln f(v)dv}=1.
\end{eqnarray}

If for all $g\in T$ we have $\int{g(v)\ln f(v)dv}=0$, then the projector $P$ is defined as the
orthogonal projector with respect to the $\langle \hspace{1pt}| \rangle_f$ scalar product.

\subsection{\textbf{Slow dynamics in open systems. First-order approximation}}

Thermodynamic projector (\ref{ep}) defines a "slow and fast motions" duality: if $T$ is the tangent
space of the slow motion manifold then $T = {\rm im} P$, and $\mbox{ker} P$ is the plane of fast
motions. Let us denote by $P_x$ the projector at a point $x$ of a given slow manifold.

The vector field $J_{ex}(x,t)$ can be decomposed in two components:

\begin{equation}
J_{ex}(x,t) = P_xJ_{ex}(x,t)+(1-P_x)J_{ex}(x,t).
\end{equation}

Let us denote $J_{ex \, s}=P_xJ_{ex}$, $J_{ex \, f}=(1-P_x)J_{ex}$. The slow component $J_{ex \, s}$
gives a correction to the motion along the slow manifold. This is a zero-order approximation. The
"fast" component shifts the slow manifold in the fast motions plane. This shift changes $P_xJ_{ex}$
accordingly. Consideration of this effect gives a first-order approximation. In order to find it, let
us rewrite the invariance equation taking $J_{ex}$ into account:

\begin{equation}
 \left\{\begin{array}{l}
   (1-P_x)(J(x+\delta x)+\varepsilon J_{ex}(x,t)) = 0 \\
 P_x \delta x = 0
 \end{array}\right.
\end{equation}

The first iteration of the Newton method subject to incomplete linearization gives:

\begin{equation}\label{Nm1a}
 \left\{\begin{array}{l}
   (1-P_x)(D_xJ(\delta x)+\varepsilon J_{ex}(x,t)) = 0 \\
   P_x \delta x = 0.\
 \end{array}\right.
\end{equation}

\begin{equation}\label{perpri}
(1-P_x)D_xJ(1-P_x)J(\delta x) = -\varepsilon J_{ex}(x,t).
\end{equation}

Thus, we have derived a linear equation in the space $\mbox{ker} P$. The operator $(1-P)D_xJ(1-P)$ is
defined in this space.

Utilization of the self-adjoint linearization instead of the traditional linearization $D_xJ$
operator (see "Decomposition of motions, non-uniqueness of selection..." section) considerably
simplifies solving and studying equation (\ref{perpri}). It is necessary to take into account here
that the projector $P$ is a sum of the orthogonal projector with respect to the $\langle
\hspace{1pt}|\rangle_x$ scalar product and a projector of rank one.

Assume that the first-order approximation equation (\ref{perpri}) has been solved and the following
function has been found:
\begin{equation}\label{persol}
\delta_1 x(x,\varepsilon J_{ex \, f}) = - [(1-P_x)D_xJ(1-P_x)]^{-1}\varepsilon J_{ex \, f},
\end{equation}

\noindent where $D_xJ$ is either the differential of $J$ or symmetrized differential of $J$
(\ref{Kx}).

Let $x$ be a point on the initial slow manifold. At the point $x+\delta x(x,\varepsilon J_{ex \, f})$
the right-hand side of equation (\ref{opsys}) in the first-order approximation is given by
\begin{equation}\label{righth}
J(x) + \varepsilon J_{ex}(x,t)+D_xJ(\delta x(x,\varepsilon J_{ex \, f})).
\end{equation}

Due to the first-order approximation (\ref{righth}), the motion of a point projection onto the
manifold is given by the following equation
\begin{equation}\label{sloweq}
\frac{dx}{dt} = P_x(J(x)+\varepsilon J_{ex}(x,t)+D_xJ(\delta x(x,\varepsilon J_{ex \, f}(x,t)))).
\end{equation}

Note that, in equation (\ref{sloweq}), the vector field $J(x)$ enters only in the form of projection,
$P_xJ(x)$. For the invariant slow manifold it holds $P_xJ(x)=J(x)$, but actually we always deal with
approximately invariant manifolds, hence, it is necessarily to use the projection $P_xJ$ instead of
$J$ in (\ref{sloweq}).

{\it Remark}. The notion "projection of a point onto the manifold" needs to be specified. For every
point $x$ of the slow invariant manifold $M$ there are defined both the thermodynamic projector $P_x$
(\ref{ep}) and the fast motions plane $\mbox{ker} P_x$. Let us define a projector $\Pi$ of some
neighborhood of $M$ onto $M$ in the following way:
\begin{equation}\label{proman}
\Pi(z)=x, \;\mbox{ if} \;  P_x(z-x)=0.
\end{equation}

Qualitatively, it means that $z$, after all fast motions took place, comes into a small neighborhood
of $x$. The operation (\ref{ep}) is defined uniquely in some small neighborhood of the manifold $M$.

A derivation of slow motions equations requires not only an assumption that $\varepsilon J_{ex}$ is
small but it must be slow as well: $\frac{d}{dt}(\varepsilon J_{ex})$ must be small too.

One can get the further approximations for slow motions of the system (\ref{opsys}), taking into
account the time derivatives of $J_{ex}$. This is an alternative to the usage of the projection
operators methods \cite{Grabert}. This is considered in a more detail in the Example 12 for a
particularly interesting driven system of dilute polymeric solutions. A short scheme description is
given in the next subsection.

\subsection{\textbf{Beyond the first-order approximation: higher-order dynamical
corrections, stability loss and invariant manifold explosion}}

Let us pose formally the invariance problem for the driven system (\ref{opsys}) in the neighborhood
of the slow manifold $M$ of the initial (dissipative) system.

Let for a given neighborhood of $M$ an operator $\Pi$ (\ref{proman}) be defined. One needs to define
the function $\delta x(x,...) = \delta x(x,J_{ex},\dot{J}_{ex},\ddot{J}_{ex},...)$, $x\in M$, with
the following properties:

\begin{eqnarray}\label{extinv}
&&P_x(\delta x(x,...))=0, \nonumber \\ &&J(x+\delta x(x,...))+J_{ex}(x+\delta x(x,...),t) = \nonumber
\\ &&\dot{x}_{sl}+D_x\delta x(x,...)\dot{x}_{sl}+\sum_{n=0}^{\infty}{D_{J_{ex}^{(n)}}\delta
x(x,...)J_{ex}^{(n+1)}},
\end{eqnarray}

\noindent where $\dot{x}_{sl} = P_x(J(x+\delta x(x,...))+J_{ex}(x+\delta x(x,...),t))$,
$J_{ex}^{(n)}=\frac{d^nJ_{ex}}{dt^n},$ $D_{J_{ex}^{(n)}}\delta x(x,...)$ is a partial differential of
the function $\delta x(x,J_{ex},\dot{J}_{ex},\ddot{J}_{ex},\ldots , J_{ex}^{(n)}, \ldots)$ with
respect to the variable $J_{ex}^{(n)}$. One can rewrite equations (\ref{extinv}) in the following
form:
\begin{eqnarray}\label{extinvk}
&&(1-P_x-D_x\delta x(x,...))(J(x+\delta x(x,...))+J_{ex}(x+\delta x(x,...),t)) = \nonumber \\ &&
\sum_{n=0}^{\infty}{D_{J_{ex}^{(n)}}\delta x(x,...)J_{ex}^{(n+1)}}.
\end{eqnarray}

For solving the equation (\ref{extinvk}) one can use an iteration method and take into account
smallness consideration. The series in the right hand side of equation (\ref{extinvk}) can be
rewritten as
\begin{equation}\label{ryad}
{\rm R.H.S.}=\sum_{n=0}^{k-1}{\varepsilon^{n+1}D_{J_{ex}^{(n)}}\delta x(x,...)J_{ex}^{(n+1)}}
\end{equation}
\noindent at the $k$th iteration, considering expansion terms only to order less than $k$. The first
iteration equation was solved in the previous subsection. On second iteration one gets the following
equation:
\begin{eqnarray}\label{second}
&&(1-P_x-D_x\delta_1x(x,J_{ex}))(J(x+\delta_1 x(x,J_{ex})) + \nonumber \\ &&D_zJ(z)|_{z=x+\delta_1
x(x,J_{ex})}\cdot(\delta_2x-\delta_1x(x,J_{ex}))+J_{ex}) =  \nonumber \\ &&
D_{J_{ex}}\delta_1x(x,J_{ex})\dot{J_{ex}}.
\end{eqnarray}

This is a linear equation with respect to $\delta_2x$. The solution
$\delta_2x(x,J_{ex},\dot{J}_{ex})$ depends linearly on $\dot{J}_{ex}$, but non-linearly on $J_{ex}$.
Let us remind that the first iteration equation solution depends linearly on $J_{ex}$.

In all these iteration equations the field $J_{ex}$ and it's derivatives are included in the formulas
as if they were functions of time $t$ only. Indeed, for any solution $x(t)$ of the equations
(\ref{opsys}) $J_{ex}(x,t)$ can be substituted for $J_{ex}(x(t),t)$. The function $x(t)$ will be a
solution of the system (\ref{opsys}) in which $J_{ex}(x,t)$ is substituted for $J_{ex}(t)$ in this
way.

However, in order to obtain the macroscopic equations (\ref{sloweq}) one must return to
$J_{ex}(x,t)$. For the first iteration such return is quite simple as one can see from
(\ref{righth}). There $J_{ex}(x,t)$ is calculated in points of the initial slow manifold. For general
case, suppose that $\delta x=\delta x(x,J_{ex},\dot{J}_{ex},..,J_{ex}^{(k)})$ has been found. The
motion equations for $x$ (\ref{sloweq}) have the following form:
\begin{equation}\label{slow1}
\frac{dx}{dt}=P_x(J(x+\delta x)+J_{ex}(x+\delta x,t)).
\end{equation}

In these equations the shift $\delta x$ must be a function of $x$ and $t$ (or a function of
$x,t,\alpha$, where $\alpha$ are external fields, see example 12, but from the point of view of this
consideration dependence on the external fields is not essential). One calculates the shift $\delta
x(x,t)$ using the following equation:
\begin{equation}\label{linimm}
J_{ex}=J_{ex}(x+\delta x(x,J_{ex},\dot{J}_{ex},...,J_{ex}^{(k)}),t).
\end{equation}

It can be solved, for example, by the iterative method, taking $J_{ex 0} = J_{ex}(x,t)$:
\begin{equation}\label{itim}
J_{ex (n+1)} = J_{ex}(x+\delta x(x,J_{ex (n)},\dot{J}_{ex (n)},...,J_{ex (n)}^{(k)}),t).
\end{equation}

We hope that using $J_{ex}$ in the equations (\ref{linimm}) and (\ref{itim}) both as a variable and
as a symbol of unknown function $J_{ex}(x,t)$ will not lead to a confusion.

In all the constructions introduced above it was assumed that $\delta x$ is sufficiently small and
the driven system (\ref{opsys}) will not deviate too far from the slow invariant manifold of the
initial system. However, a stability loss is possible: solutions of the equation (\ref{opsys}) can
deviate arbitrarily far given some perturbations level. The invariant manifold can loose it's
stability. Qualitatively, this effect of {\it invariant manifold explosion} can be represented as
follows.

Suppose that $J_{ex}$ includes the parameter $\varepsilon$: one has $\varepsilon J_{ex}$ in the
equation (\ref{opsys}). When $\varepsilon$ is small, system motions are located in a small
neighborhood of the initial manifold. This neighborhood grows monotonically and continuously with
increase of $\varepsilon$, but after some $\varepsilon_0$ a sudden change happens ("explosion") and
the neighborhood, in which the motion takes place, becomes significantly wider at
$\varepsilon>\varepsilon_0$ than at $\varepsilon<\varepsilon_0$. The stability loss is not
necessarily connected with the invariance loss. In Example 13 it is shown how the invariant manifold
(which is at the same time the quasi-equilibrium manifold in the example) can loose it's stability.
This "explosion" of the invariant manifold leads to essential physical consequences (see example 13).

\subsection{\textbf{Lyapunov norms, finite-dimensional asymptotic and volume contraction}}

In a general case, it is impossible to prove the existence of a global Lyapunov function on the basis
of local data. We can only verify or falsify the hypothesis about a given function, is it a global
Lyapunov function, or is it not. On the other hand, there exists a more strict stability property
which can be verified or falsified (in principle) on the base of local data analysis. This is a
Lyapunov norm existence.

A norm $\|\bullet\|$ is the {\it Lyapunov norm} for the system (\ref{opsys}), if for any two
solutions $x^{(1)}(t), \: x^{(2)}(t), \: t \geq 0,$ the function $\|x^{(1)}(t) - x^{(2)}(t)\|$ is
non-increasing in time.

Linear operator $A$ is {\it dissipative} with respect to a norm $\|\bullet\|$, if $\exp(At) \:(t\geq
0)$ is a semigroup of contractions: $\|\exp(At)x\|  \leq  \|x\|$ for any $x$ and $t\geq 0$. The
family of linear operators $\{A_{\alpha} \}_{\alpha \in K}$ is {\it simultaneously dissipative}, if
all operators $A_{\alpha}$  are dissipative with respect to some norm $\|\bullet\|$ (it should be
stressed that there exists {\bf one} norm for {\bf all} $A_{\alpha}, \: \alpha \in K$). The
mathematical theory of simultaneously dissipative operators for finite-dimensional spaces was
developed in refs. \cite{Ocherki,Ver1,Ver2,Ver3,Ver4}.

Let the system (\ref{opsys}) be defined in a convex set $U \subset E$, and $A_x$ be Jacobi operator
at the point $x$: $A_x=D_x(J(x)+J_{ex}(x))$ . This system has a Lyapunov norm, if the family of
operators $\{A_{x} \}_{x \in U}$ is simultaneously dissipative. If one can choose such $\varepsilon >
0$ that for all $A_{x}$, $t>0$, any vector $z$, and this Lyapunov norm $\| \exp(A_{x}t)z \|\leq
\exp(- \varepsilon t)\|z\|$, then for any two solutions $x^{(1)}(t), \: x^{(2)}(t), \: t \geq 0$ of
equations (\ref{opsys}) $\|x^{(1)}(t) - x^{(2)}(t)\|\leq  \exp(- \varepsilon t)\|x^{(1)}(0) -
x^{(2)}(0)\|$.

The simplest class of nonlinear kinetic (open) systems with Lyapunov norms was described in the paper
\cite{without}. These are reaction systems without interactions of  various substances. The
stoichiometric equation of each elementary reaction has a form
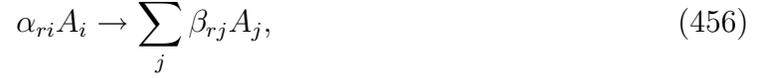
\begin{equation}\label{without}
\alpha_{ri}A_i\rightarrow \sum_j \beta_{rj}A_j,
\end{equation}
\noindent where $r$ enumerates reactions, $ \alpha_{ri}, \: \beta_{rj} $ are nonnegative
stoichiometric coefficients (usually they are integer), $A_i$ are symbols of substances.

In the right hand part of equation (\ref{without}) there is one initial reagent, though
$\alpha_{ri}>1$ is possible (there may be several copies of $A_i$, for example $3A \rightarrow
2B+C$).

Kinetic equations for reaction system (\ref{without}) have a Lyapunov norm \cite{without}. This is
$l^1$ norm with weights: $\|x\|=\sum_i w_i |x_i|$, $ w_i>0$. There exists no quadratic Lyapunov norm
for reaction systems without interaction of various substances.

Existence of the Lyapunov norm is a very strong restriction on nonlinear systems, and such systems
are not wide spread in applications. But if we go from distance contraction to contraction of
$k$-dimensional volumes ($k=2,3, \ldots $) \cite{Ilyash}, the situation changes. There exist many
kinetic systems with a monotonous contraction of $k$-dimensional volumes for sufficiently big $k$
(see, for example, \cite{IneManFSTe88,IneManTe88,IneManCFTe88,Vishik}). Let $x(t), \: t\geq 0$ be a
solution of equation (\ref{opsys}). Let us write a first approximation equation for small deviations
from $x(t)$:
\begin{equation}\label{firstapp}
\frac{d\Delta x}{dt}=A_{x(t)}\Delta x.
\end{equation}
\noindent This is linear system with coefficients depending on $t$. Let us study how the system
(\ref{firstapp}) changes $k$-dimensional volumes. A $k$-dimensional parallelepiped with edges
$x^{(1)},x^{(2)}, \ldots x^{(k)}$ is an element of the $k$th exterior power:

$$x^{(1)}\wedge x^{(2)}\wedge \ldots \wedge  x^{(k)}\in \underbrace{E \wedge E \wedge \ldots \wedge
E}_k $$

\noindent (this is an antisymmetric tensor). A norm in the $k$th exterior power of the space $E$ is a
measure of $k$-dimensional volumes (one of the possible measures). Dynamics of parallelepipeds
induced by the system (\ref{firstapp}) can be described by equations

\begin{eqnarray}\label{volcon}
&&\frac{d}{dt}(\Delta x^{(1)}\wedge \Delta x^{(2)}\wedge \ldots \wedge \Delta x^{(k)}) = \nonumber
\\ &&(A_{x(t)}\Delta x^{(1)})\wedge \Delta x^{(2)}\wedge \ldots \wedge \Delta x^{(k)} + \Delta
x^{(1)}\wedge (A_{x(t)} \Delta x^{(2)})\wedge \ldots \wedge \Delta x^{(k)} + \ldots  \nonumber \\ &&
+ \Delta x^{(1)}\wedge \Delta x^{(2)}\wedge \ldots \wedge (A_{x(t)} \Delta x^{(k)})= A_{x(t)}^{D
\wedge k}(\Delta x^{(1)}\wedge \Delta x^{(2)}\wedge \ldots \wedge \Delta x^{(k)}) .
\end{eqnarray}

Here $A_{x(t)}^{D \wedge k}$ are operators of induced action of $A_{x(t)}$ on the $k$th exterior
power of $E$. Again, a decreasing of $\| \Delta x^{(1)}\wedge \Delta x^{(2)}\wedge \ldots \wedge
\Delta x^{(k)} \| $ in time is equivalent to dissipativity of all operators $A_{x(t)}^{D \wedge k},
\: t \geq 0$ in the norm $\| \bullet \| $. Existence of such norm for all $A_{x}^{D \wedge k} \: (x
\in U)$ is equivalent to decreasing of volumes of all parallelepipeds due to first approximation
system (\ref{firstapp}) for any solution $x(t)$ of equations (\ref{opsys}). If one can choose such
$\varepsilon > 0$ that for all $A_{x} \: (x\in U)$, any vector $z \in E \wedge E \wedge \ldots \wedge
E$, and this norm $\| \exp(A_{x}^{D \wedge k}t)z \|\leq \exp(- \varepsilon t)\|z\|$, then the volumes
of parallelepipeds decrease exponentially as $\exp(- \varepsilon t)$.

{\it For such systems we can estimate the Hausdorff dimension of the attractor} (under some
additional technical conditions about solutions boundedness): {\it it can not exceed $k$}. It is
necessary to stress here that this estimation of the Hausdorff dimension does not solve the problem
of construction of the invariant manifold containing this attractor, and one needs special technique
and additional restriction on the system to obtain this manifold (see
\cite{IneManTe88,GalTem,JonTiti,ComIMKopp}).

The simplest way for construction slow invariant manifold is possible for systems with a dominance of
linear part in highest dimensions. Let an infinite-dimensional system have a form: $\dot{u}+Au=R(u)$,
where $A$ is self-adjoint, and has discrete spectrum $\lambda_i \rightarrow \infty$ with sufficiently
big gaps between $\lambda_i$, and $R(u)$ is continuous. One can build the slow manifold as the graph
over a root space of $A$. Let the basis consists of eigenvectors of $A$. In this basis
$\dot{u}_i=-\lambda_i u_i + R_i(u)$, and it seems very plausible that for some $k$ and sufficiently
big $i$ functions $u_i(t)$ exponentially fast tend to $u_i(u_1(t), \ldots u_k(t))$, if $R_i(u)$ are
bounded and continuous in a suitable sense.

Different variants of rigorous theorems about systems with such a dominance of linear part in highest
dimensions linear  may be found in literature (see, for example, the textbook \cite{Chueshov}). Even
if all the sufficient conditions hold, it remains the problem of efficient computing of these
manifold, and different ways for calculation are proposed: from Euler method for manifold correction
\cite{Kev} to different algorithms of discretisations \cite{Roberts,Grids,Robi2}.

The simplest conditions of simultaneous dissipativity for the family of operators $\{A_{x}\}$ can be
created in a following way: let us take a norm $\| \bullet \| $. If all operators $A_{x}$ are
dissipative with respect to this norm, then the family $A_{x}$ is (evidently) simultaneously
dissipative in this norm. So, we can verify or falsify a hypothesis about simultaneous dissipativity
for a given norm. Simplest examples give us quadratic and $l^1$ norms.

For quadratic norm associated with a scalar product $\langle|\rangle$ dissipativity of operator $A$
is equivalent to nonpositivity of all points of spectrum $A+A^+$, where $A^+$ is the adjoint to $A$
operator with respect to scalar product $\langle|\rangle$.

For $l^1$ norm with weights $\|x\|=\sum_i w_i |x_i|, \: w_i>0$. The condition of operator $A$
dissipativity for this norm is the weighted diagonal predominance for columns of the $A$ matrix
$A=(a_{ij})$: $$a_{ii}<0, \: w_i|a_{ii}|\geq\sum_{j,\, j\neq i} w_j|a_{ji}|.$$

For exponential contraction it is necessary and sufficient a gap existence in the dissipativity
inequalities:

\noindent for quadratic norm $\sigma (A+A^+) < \varepsilon < 0$, where $\sigma (A+A^+)$ is the
spectrum of $A+A^+$;

\noindent for $l^1$ norm with weights $a_{ii}<0, \: w_i|a_{ii}|\geq\sum_{j,\, j\neq
i}w_j|a_{ji}|+\varepsilon$, $\varepsilon > 0$.

The sufficient conditions of simultaneous dissipativity can have another form (not only the form of
dissipativity checking  with respect to a given norm) \cite{Ver1,Ver2,Ver3,Ver4}, but the problem of
necessary and sufficient conditions in general case is open.

The dissipativity conditions for operators  $A_{x}^{D \wedge k}$ of induced action of $A_{x}$ on the
$k$th exterior power of $E$ have the similar form, for example, if we know the spectrum of $A+A^+$,
then it is easy to find the spectrum of $A_{x(t)}^{D \wedge k}+(A_{x(t)}^{D \wedge k})^+$: each
eigenvalue of this operator is a sum of $k$ distinct eigenvalues of $A+A^+$; the $A_{x(t)}^{D \wedge
k}+(A_{x(t)}^{D \wedge k})^+$ spectrum is a closure of set of sums of $k$ distinct points $A+A^+$
spectrum.

A basis the $k$th exterior power of $E$  can be constructed from the basis $\{ e_i \}$ of $E$: it is
$$\{e_{i_1i_2 \ldots i_k} \} = \{e_{i_1} \wedge e_{i_2} \wedge \ldots \wedge e_{i_k} \}, \: i_1<i_2<
\ldots <i_k.$$ For $l^1$ norm with weights in the $k$th exterior power of $E$ the set of weights is
$\{w_{i_1i_2 \ldots i_k}>0, \: i_1<i_2< \ldots <i_k \}$. The norm of a vector $z$ is
$$\|z\|=\sum_{i_1<i_2< \ldots <i_k}w_{i_1i_2 \ldots i_k}|z_{i_1i_2 \ldots i_k}|.$$ The dissipativity
conditions for operators $A^{D \wedge k}$ of induced action of $A$ in $l^1$ norm with weights have
the form:
\begin{eqnarray}\label{L1con}
&&a_{i_1i_1}+a_{i_2i_2}+ \ldots + a_{i_ki_k}<0, \nonumber \\ &&w_{i_1i_2 \ldots
i_k}|a_{i_1i_1}+a_{i_2i_2}+ \ldots + a_{i_ki_k}|\geq \sum_{l=1}^k \sum_{j, \, j\neq i_1,i_2, \ldots
i_k} w_{i_1i_2 \ldots i_k}^{l,j}|a_{ji_l}| \nonumber \\&& \mbox{for any} \: i_1<i_2< \ldots <i_k,
\end{eqnarray}

\noindent where $w_{i_1i_2 \ldots i_k}^{l,j}=w_{I}$,  multiindex $I$ consists of indexes $i_p \, (p
\neq l)$, and $j$.

For infinite dimensional systems the problem of volume contraction and Lyapunov norms for exterior
powers of $E$ consists of three parts: geometrical part concerning the choice of norm for
simultaneous dissipativity of operator families, topological part concerning topological
nonequivalence of constructed norms, and estimation of the bounded set containing compact attractor.

The difficult problem may concern the appropriate apriori estimations of the bounded convex
positively invariant set $V \subset U$ where the compact attractor is situated. It may be crucial to
solve the problem of simultaneous dissipativity for the most narrow family of operators $\{A_{x}, \:
x\in V \}$ (and their induced action on the $k$th exterior power of $E$).

The estimation of attractor dimension based on Lyapunov norms in the exterior powers is rather rough.
This is a local estimation. More exact estimations are based on global Lyapunov exponents (Lyapunov
or Kaplan-Yorke dimension \cite{KY,LY}). There are many different measures of dimension
\cite{GrassPro,HerPro}, and many efforts are applied to create good estimations for different
dimensions \cite{ChIl}. Estimations of attractor dimension was given for different systems: from
Navier-Stokes hydrodynamic \cite{Re9/4} to climate dynamics \cite{ClimAtt}. The introduction and
review of many results is given in the book \cite{IneManTe88}. But local estimations remain the main
tools for estimation of attractors dimension, because global estimations for complex systems are much
more complicated and often unattainable because of computation complexity.

\clearpage

\addcontentsline{toc}{subsection}{\textbf{Example 12: The universal limit in dynamics of dilute
polymeric solutions}}

\subsection*{\textbf{Example 12: The universal limit in dynamics of dilute polymeric solutions}}

The method of invariant manifold is developed for a derivation of reduced description in kinetic
equations of dilute polymeric solutions. It is demonstrated that this reduced description becomes
universal in the limit of small  Deborah  and  Weissenberg numbers, and it is represented by the
(revised) Oldroyd 8 constants constitutive equation for the polymeric stress tensor. Coefficients of
this constitutive equation are expressed in terms of the microscopic parameters. A systematic
procedure of corrections to the revised Oldroyd 8 constants equations is developed. Results are
tested with simple flows.

Kinetic equations arising in the theory of polymer dynamics constitute a wide class of microscopic
models of complex fluids. Same as in any branch of kinetic theory, the problem of reduced description
becomes actual as soon as the kinetic equation is established. However, in spite of an enormous
amount of work in the field of polymer dynamics \cite{Bird,Doi,Birdreview,HCO,Martin}, this problem
remains less studied as compared to other classical kinetic equations.

It is the purpose of this section to suggest a systematic approach to the problem of reduced
description for kinetic models of polymeric fluids. First, we would like to specify our motivation by
comparing the problem of the reduced description for that case with a similar problem in the familiar
case of the rarefied gas obeying the classical Boltzmann kinetic equation \cite{Cercignani,Chapman}.

The problem of reduced description begins with establishing a set of slow variables. For the
Boltzmann equation, this set is represented by five hydrodynamic fields (density, momentum and
energy) which are low--order moments of the distribution function, and which are conserved quantities
of the dissipation process due to particle's collisions. The reduced description is a closed system
of equations for these fields. One starts with the manifold of local equilibrium distribution
functions (local Maxwellians), and finds a correction by the Chapman--Enskog method \cite{Chapman}.
The resulting reduced description (the Navier--Stokes hydrodynamic equations) is universal in the
sense that the form of equations does not depend on details of particle's interaction whereas the
latter shows up explicitly only in the transport coefficients (viscosity, temperature conductivity,
etc.).

Coming back to the complex fluids, we shall consider the simplest case of dilute polymer solutions
represented by dumbbell models studied below. Two obstacles preclude an application of the
traditional techniques. First, the question which variables should be regarded as slow is at least
less evident because the dissipative dynamics in the dumbbell models has no nontrivial conservation
laws compared to the Boltzmann case. Consequently, a priori, there are no distinguished manifolds of
distribution functions like the local equilibria which can be regarded as a starting point. Second,
while the Boltzmann kinetic equation provides a self--contained description, the dumbbell kinetic
equations are coupled to the hydrodynamic equations. This coupling manifests itself as an external
flux in the kinetic equation.

The well known distinguished macroscopic variable associated with the dumbbell kinetic equations is
the polymeric stress tensor \cite{Bird,Martin}. This variable is not the conserved quantity but
nevertheless it should be treated as a relevant slow variable because it actually contributes to the
macroscopic (hydrodynamic) equations. Equations for the stress tensor are known as the constitutive
equations, and the problem of reduced description for the dumbbell models consists in deriving such
equations from the kinetic equation.

Our approach is based on the method of invariant manifold \cite{GKTTSP94}, modified for systems
coupled with external fields. This method  suggests constructing invariant sets (or manifolds) of
distribution functions that represent the asymptotic states of slow evolution of the kinetic system.
In the case of dumbbell models, the reduced description is  produced by equations which constitute
stress--strain relations, and two physical requirements are met by our approach: The first is the
principle of {\it frame--indifference} with respect to any time--dependent reference frame. This
principle requires that the resulting equations for the stresses contain  only frame--indifferent
quantities. For example, the frame--dependent vorticity tensor should not show up in these equations
unless being  presented in  frame--indifferent combinations with  another tensors. The second
principle is the {\it thermodynamic stability}: In the absence of the flow,  the constitutive model
should be purely dissipative, in other words, it should describe the relaxation of stresses to their
equilibrium values.

The physical picture addressed below takes into account two assumptions: (i) In the absence of the
flow, deviations from the equilibrium are small. Then the  invariant manifold is represented  by
eigenvectors corresponding to the slowest relaxation modes. (ii). When the external flow is taken
into account, it is assumed to cause a  small deformation of the invariant manifolds of the purely
dissipative dynamics. Two characteristic parameters are necessary to describe this deformation. The
first is the characteristic time variation of the external field. The second is the characteristic
intensity of the external field. For dumbbell  models,  the first parameter is associated with the
conventional Deborah number while the  second one is usually called the  Weissenberg number. An
iteration approach which involves these parameters is developed.

Two main results of the analysis are as follows: First, the lowest--order constitutive equations with
respect to the characteristic parameters mentioned above has the form of the revised phenomenological
{\it Oldroyd 8 constants model}. This result is interpreted as the macroscopic limit of the
microscopic dumbbell dynamics whenever the rate of the strain is low, and the Deborah number is
small. This limit is valid generically, in the absence or in the presence of the hydrodynamic
interaction, and for the arbitrary nonlinear elastic force. The phenomenological constants of the
Oldroyd model are expressed in a closed form in terms of the microscopic parameters of the model. The
universality of this limit is similar to that of the Navier--Stokes equations which are the
macroscopic limit of the Boltzmann equation at small Knudsen numbers for arbitrary hard--core
molecular interactions. The test calculation for the nonlinear FENE force demonstrates a good
quantitative agreement of the constitutive equations with solutions to the microscopic kinetic
equation within the domain of their validity.

The second result  is a regular procedure of finding corrections to the zero--order model. These
corrections extend the model into the domain of higher rates of the strain, and to flows which
alternate faster in time. Same as in the zero--order approximation, the  higher--order corrections
are linear in the stresses,  while their dependence on the gradients of the flow velocity and its
time derivatives becomes highly nonlinear.

The section is organized as follows: For the sake of completeness, we present the nonlinear dumbbell
kinetic models in the next subsection, ``The problem of reduced description in polymer dynamics". In
the section, ``The method  of invariant manifold for weakly driven  systems", we describe in details
our approach to the derivation of macroscopic equations for an abstract kinetic equation coupled to
external fields. This derivation is applied to the dumbbell models in the section, ``Constitutive
equations". The zero--order constitutive equation is derived and discussed in detail in this section,
as well as the structure of the first correction. Tests of the zero--order constitutive equation for
simple flow problems are given in the section, ``Tests on the FENE dumbbell model".

\addcontentsline{toc}{subsubsection}{The problem of reduced description in polymer dynamics}

\subsubsection*{\textbf{The problem of reduced description in polymer dynamics}}

\paragraph{\textbf{Elastic dumbbell models.}}

The elastic dumbbell model is the simplest microscopic model of polymer solutions \cite{Bird}. The
dumbbell model reflects the two features of real--world macromolecules to be orientable and
stretchable by a flowing solvent. The polymeric solution is represented by a set of identical elastic
dumbbells placed in an isothermal incompressible liquid. In this example we adopt notations used in
kinetic theory of polymer dynamics \cite{Bird}.  Let $\QQ$ be the connector vector between the beads
of a dumbbell, and $\Psi(\xx,\QQ,t)$ be the configuration distribution function which depends on the
location in the space $\xx$ at  time $t$. We assume that dumbbells are distributed uniformly, and
consider the normalization, $\int \!\Psi(\xx,\QQ,t)\dd\QQ=1$. The Brownian motion of beads in the
physical space causes a diffusion in the phase space described by the Fokker--Planck equation (FPE)
\cite{Bird}:
\begin{equation}
\label{FPnonhom} {D  \Psi \over D t} = -  {\partial \over \partial \QQ} \cdot \kk \cdot \QQ \Psi
+\frac{2k_{\Bm} T}{\xi} {\partial \over \partial \QQ} \cdot \DD \cdot \left({\partial \over \partial
\QQ} \Psi + \frac{ \FF}{k_{\Bm} T} \Psi \right).
\end{equation}
Here $D / D t = \partial / \partial t + \vv \cdot \nabla$ is the material derivative, $\nabla$ is the
spatial gradient, $\kk(\xx,t)=(\nabla\vv)^{\dag}$
 is the gradient of  the velocity of the
solvent $\vv$, $\dag$ denotes transposition of tensors, $\DD$  is the dimensionless diffusion matrix,
$k_{\Bm}$ is the Boltzmann constant,  $T$ is the temperature, $\xi$  is the dimensional coefficient
characterizing a friction  exerted by beads moving through  solvent  media  (friction coefficient
\cite{Bird,Doi}), and $\FF = \partial \phi / \partial \QQ$ is the elastic spring force defined by the
potential $\phi$. We consider forces of the form $\FF=H f(Q^2)\QQ$, where  $f(Q^2)$  is a
dimensionless function of the variable $Q^2=\QQ\cdot\QQ$, and $H$ is the dimensional constant.
Incompressibility of solvent implies $\sum_i k_{ii}=0$.

Let  us  introduce a  time dimensional  constant $$\lambda_{\rrm}=\frac{\xi}{4H}, $$ which coincides
with a  characteristic relaxation  time of  dumbbell  configuration in the  case when the  force
$\FF$ is  linear: $f(Q^2)=1$. It proves convenient to rewrite the FPE (\ref{FPnonhom}) in the
dimensionless form:
\begin{equation}
\label{FPnondim} {D  \Psi \over D \tth} = -  {\partial \over \partial \QQh} \cdot \kkh \cdot \QQh
\Psi + {\partial \over \partial \QQh} \cdot \DD  \cdot \left({\partial \over \partial \QQh} \Psi +
\FFh \Psi \right).
\end{equation}
Various dimensionless quantities used are: $\QQh= (H/k_{\Bm} T)^{1/2} \QQ$, $D/ D \tth = \partial
/\partial \tth + \vv \cdot \nablah$, $\tth=t/\lambda_{\rrm}$
 is the dimensionless time,
$\nablah = \lambda_{\rrm} \nabla $
 is the reduced space gradient,  and
$\kkh=\kk\lambda_{\rrm}= (\nablah \vv)^{\dag}$ is the dimensionless tensor of the gradients of the
velocity. In the sequel, only dimensionless quantities  $\QQh$  and $\FFh$ are used, and we keep
notations    $\QQ$ and $\FF$ for them for the sake of simplicity.

The quantity of interest is the stress tensor introduced by Kramers \cite{Bird}:
\begin{equation}
\label{tau_def} {\mbox{\boldmath $\tau$}} = - \nu_{\sm} \dotgam + nk_{\Bm} T (\1 -   \left\langle \FF
\QQ \right\rangle),
\end{equation}
where $\nu_{\sm}$  is the viscosity of the solvent, $\dotgam = \kk+ \kk^{\dag}$ is the
rate--of--strain tensor,  $n$ is the concentration of polymer molecules,  and  the  angle brackets
stand for the averaging with the distribution function $\Psi$: $\left\langle \bullet \right\rangle
\equiv \int\! \bullet \Psi(\QQ) \dd \QQ $. The tensor
\begin{equation}
\label{tau_p} \ttau_{\ppm}=nk_{\Bm}T (\1 - \la \FF \QQ \ra)
\end{equation}
gives a contribution to stresses caused by the presence of polymer molecules.

The stress tensor is required in order to write down a closed system of hydrodynamic equations:
\begin{equation}
\label{v_dyn} {D  \vv \over D t} = - \rho^{-1} \nabla p - \nabla \cdot {\mbox{\boldmath
$\tau$}}[\Psi].
\end{equation}
Here $p$ is the pressure,  and $\rho=\rho_{\sm}+\rho_{\ppm}$ is the mass density of the solution
where $\rho_{\sm}$ is the solvent, and $\rho_{\ppm}$ is the polymeric contributions.

Several models of the elastic force are known in the literature. The Hookean law is relevant to small
perturbations of the equilibrium configuration of the macromolecule:
\begin{equation}
\label{Hookean_force} \FF =  \QQ.
\end{equation}
In that case, the differential equation for $\mbox{\boldmath $\tau$}$ is easily derived from the
kinetic equation, and is the well known {\it Oldroyd--B} constitutive  model \cite{Bird}.

The second model, the FENE force law \cite{FENE}, was derived as an approximation to the inverse
Langevin force law \cite{Bird} for a more realistic description of the elongation of a polymeric
molecule  in a solvent:
\begin{equation}
\label{FENE} \FF = \frac{\QQ}{1-\QQ^2/\QQ^2_0}.
\end{equation}
This force law takes into account the nonlinear stiffness and the finite extendibility of dumbbells,
where $\QQ_0$ is the maximal extendibility.

The features of the diffusion matrix are important for both the microscopic and the macroscopic
behavior. The isotropic diffusion is represented by the simplest  diffusion matrix
\begin{equation}
\label{iso_diff} \DD_{\Imm} = \frac{1}{2} \II.
\end{equation}
Here  $\II$ is the unit matrix. When the hydrodynamic interaction between the beads is taken into
account, this results in an anisotropic contribution to the diffusion matrix (\ref{iso_diff}). The
original form of this contribution is the Oseen--Burgers tensor $\DD_{\Hm}$ \cite{Oseen,Oseen1}:
\begin{equation}
\label{hydr_diff} \DD = \DD_{\Imm} - \kappa \DD_{\Hm}, \qquad \DD_{\Hm} = \frac{1}{Q} \left( \II +
\frac{\QQ \QQ}{Q^2} \right),
\end{equation}
where   $$\kappa= \left(\frac{H}{k_{\Bm} T}\right)^{1/2} \frac{\xi}{16 \pi \nu_{\sm}}. $$ Several
modifications of the  Oseen--Burgers tensor can be found in the literature (the {\it
Rotne-Prager-Yamakawa}  tensor \cite{modif_of_OB1,modif_of_OB2}), but here  we  consider only the
classical version.

\paragraph*{\textbf{Properties of the Fokker--Planck operator.}} Let us review some of the properties of  the
Fokker--Planck operator $J$ in the right hand side of the Eq.\  (\ref{FPnondim}) relevant to what
will follow. This operator can be written as $J=J_{\dm}+J_{\hm}$,  and it represents two processes.

The first term,  $J_{\dm}$, is the dissipative  part,
\begin{equation}
\label{J_d} J_{\dm} = {\partial \over \partial \QQ} \cdot \DD \cdot \left({\partial \over \partial
\QQ} + \FF  \right).
\end{equation}
This part is responsible for the diffusion and  friction which affect internal  configurations of
dumbbells, and it drives the system to the unique equilibrium state, $$ \Psi_{\eqm}=c^{-1} \exp
(-\phi(Q^2)),$$ where $c=\int\! \exp (- \phi ) \dd \QQ$ is the normalization constant.

The second part, $J_{\hm}$, describes the hydrodynamic drag of the beads in the flowing solvent:
\begin{equation}
\label{J_h} J_{\hm} =- {\partial \over \partial \QQ} \cdot \kkh \cdot \QQ.
\end{equation}

The dissipative nature of the operator $J_{\dm}$ is reflected by its spectrum. We assume that this
spectrum consists of real--valued nonpositive eigenvalues, and that the zero eigenvalue is not
degenerated. In the sequel, the following scalar product will be useful: $$\la g, h \ra_{\sm} =
\int\! \Psi_{\eqm}^{-1} gh\dd\QQ. $$
 The operator $J_{\dm}$ is symmetric and
nonpositive definite in this  scalar product:
\begin{equation}
\label{J_dsymm} \la J_{\dm}g, h \ra_{\sm}=\la g, J_{\dm}h \ra_{\sm}, \ \mbox{  and   }\ \la
J_{\dm}g,g \ra_{\sm} \leq 0.
\end{equation}
Since $$\la J_{\dm}g,g \ra_{\sm} = -\int\! \Psi_{\eqm}^{-1} (\partial {\textstyle g} /
\partial \QQ) \cdot \Psi_{\eqm} \DD \cdot (\partial {\textstyle  g} / \partial \QQ) \dd
\QQ,$$ the above inequality is valid if the diffusion matrix $\DD$ is positive semidefinite. This
happens if $\DD=\DD_{\Imm}$ (\ref{iso_diff}) but is not generally valid in the presence of the
hydrodynamic interaction  (\ref{hydr_diff}). Let us split the operator $J_{\dm}$  according to the
splitting of the diffusion matrix $\DD$: $$J_{\dm}=J^{\Imm}_{\dm}-\kappa J^{\Hm}_{\dm}, \:
\mbox{where} \: J^{\Imm,\Hm}_{\dm} = \partial /
\partial \QQ \cdot \DD_{\Imm,\Hm}  \cdot \left(\partial / \partial \QQ + \FF  \right).$$
Both the operators $J^{\Imm}_{\dm}$ and $J^{\Hm}_{\dm}$ have nondegenerated eigenvalue $0$ which
corresponds to their common eigenfunction $\Psi_{\eqm}$: \mbox{$J^{\Imm,\Hm}_{\dm}\Psi_{\eqm}=0$},
while the rest of the spectrum of both  operators belongs to the nonpositive real semi--axis. Then
the spectrum of the operator \mbox{$J_{\dm}=J^{\Imm}_{\dm}-\kappa J^{\Hm}_{\dm}$} remains nonpositive
for sufficiently small values of the parameter $\kappa$. The  spectral  properties of both  operators
$J^{\Imm,\Hm}_{\dm}$ depend
 only on the  choice of the spring  force $\FF$.
Thus,  in  the sequel we    assume  that the   hydrodynamic interaction parameter $\kappa$  is
sufficiently   small so that the  {\it thermodynamic  stability} property (\ref{J_dsymm})  holds.

We  note that  the  scalar product $\la \bullet, \bullet \ra_{\sm}$ coincides with the  second
differential $D^2 S\bigr|_{\Psi_{\eqm}}$
 of an   entropy  functional  $S[\Psi]$:
$ \la \bullet,\bullet \ra_{\sm}= -  D^2 S\bigr|_{\Psi_{\eqm}}[\bullet,\bullet]$, where the entropy
has the  form:
\begin{equation}
\label{entropy1} S[\Psi] = - \int\! \Psi \ln \left( \frac{\Psi}{\Psi_{\eqm}} \right) \dd \QQ =
-\left< \ln \left( \frac{\Psi}{\Psi_{\eqm}}  \right) \right>.
\end{equation}
The  entropy $S$  grows   in the course of dissipation: $$ D S [J_{\dm}\Psi] \geq 0. $$ This
inequality similar to inequality (\ref{J_dsymm}) is satisfied  for  sufficiently  small $\kappa$.
Symmetry and  nonpositiveness of operator $J_{\dm}$  in the   scalar  product defined  by  the second
differential of the entropy  is  a common  property  of linear  dissipative systems.

\paragraph{\textbf{Statement of the problem.}}

Given the kinetic equation (\ref{FPnonhom}), we aim at deriving differential equations for the stress
tensor $\ttau$ (\ref{tau_def}). The latter includes the moments $\left< \FF \QQ\right>= \int\! \FF
\QQ \Psi \dd \QQ $.

In general, when the diffusion matrix is non--isotropic and/or the spring force is nonlinear, closed
equations for these moments are not available, and approximations are required. With this, any
derivation  should  be consistent  with  the  three requirements:

(i). {\it Dissipativity or thermodynamic stability:} The  macroscopic dynamics should  be dissipative
in the absence of the flow.

(ii). {\it Slowness}: The macroscopic equations should represent  slow degrees of freedom of the
kinetic equation.

(iii). {\it Material frame indifference}: The form of equations for the stresses should be invariant
with respect to the Eucluidian, time dependent transformations  of the  reference  frame
\cite{Bird,Noll}.

While  these  three  requirements should  be  met  by  any  approximate derivation, the  validity  of
our approach  will  be  restricted  by  two additional assumptions:

(a). Let us denote $\theta_1$ the  inertial time of the  flow,  which we define via characteristic
value of the gradient of the flow velocity: $\theta_1=|\nabla \vv|^{-1}$,  and $\theta_2$   the
characteristic time of the variation of the flow velocity. We assume that the characteristic
relaxation time of the molecular configuration  $\theta_{\rrm}$ is small as compared to both the
characteristic times $\theta_1$ and $\theta_2$:
\begin{equation}
\label{validity}
 \theta_{\rrm}\ll\theta_1\  {\rm and}\ \theta_{\rrm} \ll\theta_2.
\end{equation}

(b). In the absence of the flow, the initial deviation of the distribution function from the
equilibrium is small so that the linear approximation is valid.

While the assumption (b) is merely of a technical nature, and it is intended to simplify the
treatment of the dissipative part of the Fokker--Planck operator  (\ref{J_d}) for elastic forces of a
complicated form, the assumption (a) is crucial for taking into account the flow in an adequate way.
We have assumed that the two parameters characterizing the composed system `relaxing polymer
configuration + flowing solvent' should be small: These  two parameters  are:
\begin{equation}
\label{two_epsilon} \varepsilon_1=\theta_{\rrm}/\theta_1,\quad \varepsilon_2=\theta_{\rrm}/\theta_2.
\end{equation}

The characteristic relaxation time of the polymeric configuration is defined via the coefficient
$\lambda_{\rrm}$: $\theta_{\rrm}=c\lambda_{\rrm}$, where $c$  is some positive dimensionless constant
which is  estimated by the  absolute value of  the lowest nonzero eigenvalue of the operator
$J_{\dm}$. The first parameter $\varepsilon_1$  is usually  termed the {\it Weissenberg  number}
while the second  one  $\varepsilon_2$  is  the   {\it Deborah number}  (cf.\ Ref.\ \cite{Marrucci},
sec. 7--2).

\addcontentsline{toc}{subsubsection}{The method  of invariant manifold for weakly driven systems}

\subsubsection*{\textbf{The method  of invariant manifold for weakly driven  systems}}

\paragraph{\textbf{The Newton iteration scheme.}}

In this section we introduce an extension of the method of  invariant manifold \cite{GKTTSP94}  onto
systems coupled with external fields. We consider a  class of dynamic systems of the form
\begin{equation}
\label{dynsys} \frac{\dd \Psi}{\dd t}= J_{\dm}\Psi+ J_{\exm}(\alpha)\Psi,
\end{equation}
where $J_{\dm}$ is a linear operator representing the dissipative  part  of the  dynamic vector
field, while $J_{\exm}(\alpha)$ is a linear operator which represents an  external flux  and depends
on a set of external fields $\alpha=\{\alpha_1,\ldots,\alpha_k \}$. Parameters $\alpha$ are either
known functions of the time, $\alpha=\alpha(t)$, or they obey a set of equations,
\begin{equation}
\label{alphadyn} \frac{\dd \alpha }{\dd t}=\Phi(\Psi,\alpha).
\end{equation}
Without any restriction, parameters $\alpha$ are adjusted  in such a way that
$J_{\exm}(\alpha=0)\equiv 0$. Kinetic equation (\ref{FPnondim}) has the  form (\ref{dynsys}), and
general results of this section will be applied to  the dumbbell models below in a straightforward
way.

We assume that the  vector field   $J_{\dm}\Psi$ has the  same  dissipative properties as  the
Fokker--Planck  operator  (\ref{J_d}). Namely there exists the  globally  convex entropy  function
$S$  which  obeys: $D S [J_{\dm}\Psi]\geq 0$,  and  the  operator $J_{\dm}$  is  symmetric and
nonpositive in the  scalar product $\la \bullet, \bullet \ra_{\sm}$  defined by  the  second
differential of the  entropy: $\la g,h \ra_{\sm}=-D^2 S[g,h]$. Thus, the vector  field $J_{\dm}\Psi$
drives
 the system irreversibly   to the  unique equilibrium  state $\Psi_{\eqm}$.

We consider a set of $n$ real--valued functionals, $M^*_i[\Psi]$ (macroscopic variables),  in the
phase space $\cal F$ of the system (\ref{dynsys}). A  macroscopic description is obtained once we
have derived a closed  set of equations for the variables $M^*_i$.

Our approach is based on constructing a relevant invariant manifold  in phase space ${\cal F}$. This
manifold is thought as a  finite--parametric set of  solutions  $\Psi(M)$ to Eqs. (\ref{dynsys})
which depends   on time implicitly via the $n$ variables $M_i[\Psi]$. The latter may differ from the
macroscopic variables $M_i^*$. For  systems with external fluxes (\ref{dynsys}),  we assume that the
invariant manifold depends also on the parameters $\alpha$, and  on their  time  derivatives taken to
arbitrary order: $\Psi(M,{\cal A })$,  where ${\cal A}=\{\alpha,\alpha^{(1)},\ldots \}$ is the set of
time derivatives\ \mbox{$\alpha^{(k)}=\dd^k \alpha/\dd t^k$}. It is  convenient to consider time
derivatives  of $\alpha$  as independent parameters. This assumption is important because then we do
not need an explicit form of the Eqs.\ (\ref{alphadyn}) in the course of construction of the
invariant manifold.

By a definition, the dynamic invariance postulates the equality  of the ``macroscopic'' and the
``microscopic'' time derivatives:
\begin{equation}
\label{invariance} J\Psi(M,{\cal A})=\sum_{i=1}^{n}\frac{\partial \Psi(M,{\cal A})}{\partial M_i}
\frac{\dd M_i}{\dd t} + \sum_{n=0}^{\infty} \sum_{j=1}^k\frac{\partial \Psi(M,{\cal A})}{\partial
\alpha^{(n)}_j} \alpha_j^{(n+1)},
\end{equation}
where $J=J_{\dm}+J_{\exm}(\alpha)$. The time derivatives of the macroscopic variables, $\dd M_i / \dd
t$, are  calculated as follows:
\begin{equation}
\frac{\dd M_i}{\dd t} = D M_i [J\Psi(M,{\cal A})],
\end{equation}
 where  $D M_i$  stands for  differentials of the functionals $M_i$.

Let us introduce the projector operator associated  with  the parameterization  of the  manifold
$\Psi(M,{\cal A})$  by the  values  of the  functionals $M_i[\Psi]$.:
\begin{equation}
\label{proj_M} P_{M}=\sum_{i=1}^n \frac{\partial \Psi(M,{\cal A})}{\partial M_i} D M_i[\bullet]
\end{equation}
It projects  vector  fields from the  phase  space ${\cal F}$  onto  tangent bundle  $T\Psi(M,{\cal
A})$   of  the  manifold $\Psi(M,{\cal A})$. Then Eq.\  (\ref{invariance}) is rewritten as  the {\it
invariance  equation}:
\begin{equation}
\label{invequation} (1-P_{M} )J\Psi(M,{\cal A})= \sum_{n=0}^{\infty} \sum_{j=1}^k \frac{\partial
\Psi}{\partial \alpha^{(n)}_j}
 \alpha^{(n+1)}_j,
\end{equation}
which has the invariant manifolds  as its solutions.

Furthermore, we assume the following: (i). The external flux $J_{\exm} (\alpha)\Psi$  is  small  in
comparison  to  the  dissipative part $J_{\dm}\Psi$,  i.e. with  respect to some  norm  we  require:
\mbox{$|J_{\exm} (\alpha)\Psi| \ll  |J_{\dm}\Psi|$}. This allows us to introduce a small parameter
$\varepsilon_1$, and to replace the operator $J_{\exm}$  with $\varepsilon_1  J_{\exm}$ in the Eq.\
(\ref{dynsys}). Parameter $\varepsilon_1$ is proportional  to the characteristic value of the
external variables $\alpha$. (ii). The characteristic time $\theta_{\alpha}$ of the variation of the
external fields  $\alpha$ is large in comparison to the characteristic relaxation time
$\theta_{\rrm}$, and the second small parameter is $\varepsilon_2=\theta_{\rrm}/\theta_{\alpha} \ll
1$. The parameter $\varepsilon_2$  does not  enter  the  vector  field $J$ explicitly  but it shows
up in the invariance equation. Indeed, with a substitution, $\alpha^{(i)} \to  \varepsilon_2^i
\alpha^{(i)}$, the invariance equation (\ref{invariance}) is rewritten in a  form which incorporates
both the parameters $\varepsilon_1$  and  $\varepsilon_2$:
\begin{equation}
\label{inveq_2par} (1-P_M )\{J_{\dm}+\varepsilon_1 J_{\exm}\} \Psi = \varepsilon_2
\sum_{i}\sum_{j=1}^k \frac{\partial \Psi}{\partial \alpha^{(i)}_j} \alpha^{(i+1)}_j
\end{equation}

We  develop  a modified  Newton  scheme  for  solution  of this  equation. Let us  assume that we
have  some  initial  approximation  to  desired manifold $\Psi_{(0)}$. We  seek  the  correction of
the  form $ \Psi_{(1)}=\Psi_{(0)}+\Psi_{1}$. Substituting  this  expression  into Eq. \
(\ref{inveq_2par}),  we  derive:
\begin{eqnarray}
\label{Newtonfull} (1-P^{(0)}_M )\{J_{\dm}+\varepsilon_1 J_{\exm}\} \Psi_1 - \varepsilon_2
\sum_{i}\sum_{j=1}^k \frac{\partial \Psi_1}{\partial \alpha^{(i)}_j} \alpha^{(i+1)}_j =& &
\phantom{aaaaaaaaaaa}\nonumber\\ -(1-P^{(0)}_M )J \Psi_{(0)}+\varepsilon_2 \sum_{i} \sum_{j=1}^k
\frac{\partial \Psi_{(0)}}{\partial \alpha^{(i)}_j} \alpha^{(i+1)}_j.&&
\end{eqnarray}
Here $P^{(0)}_M$  is a  projector  onto  tangent  bundle  of  the  manifold $\Psi_{(0)}$. Further, we
neglect  two terms in the left hand side  of  this equation, which  are  multiplied by   parameters
$\varepsilon_1$  and $\varepsilon_2$,  regarding them   small in  comparison to  the first term.  In
the result  we arrive at the  equation,
\begin{equation}
\label{Newton0} (1-P^{(0)}_M ) J_{\dm} \Psi_1= - (1-P^{(0)}_M )J \Psi_{(0)}+ \varepsilon_2 \sum_{i}
\sum_{j=1}^k \frac{\partial \Psi_{(0)}}{\partial \alpha^{(i)}_j} \alpha^{(i+1)}_j.
\end{equation}
For $(n+1)$-th  iteration  we  obtain:
\begin{equation}
\label{Newton} (1-P^{(n)}_M ) J_{\dm} \Psi_{n+1}= - (1-P^{(0)}_M ) J \Psi_{(n)}+
 \varepsilon_2 \sum_{i} \sum_{j=1}^k \frac{\partial
\Psi_{(n)}}{\partial \alpha^{(i)}_j} \alpha^{(i+1)}_j,
\end{equation}
where $\Psi_{(n)}=\sum_{i=0}^{n} \Psi_i$  is  the   approximation  of $n$-th order and $P^{(n)}_M$ is
the  projector onto its   tangent bundle.

It  should  be  noted  that  deriving   equations  (\ref{Newton0})  and (\ref{Newton}) we have  not
varied the  projector $P_M$ with respect to yet unknown term $\Psi_{n+1}$, i.e. we have kept
$P_M=P_M^{(n)}$  and have neglected  the contribution from the  term  $\Psi_{n+1}$. The motivation
for this  action   comes from  the original paper  \cite{GKTTSP94}, where it was shown that such
modification generates   iteration   schemes  properly converging to   slow invariant manifold.

In  order to gain the solvability of Eq. (\ref{Newton}) an  additional condition is  required:
\begin{equation}
\label{proj_cond} P^{(n)}_M\Psi_{n+1}=0.
\end{equation}
This  condition is  sufficient  to provide  the existence of the  solution to linear system
(\ref{Newton}), while the   additional   restriction onto the  choice of the  projector is required
in  order to  guarantee the uniqueness of the  solution. This condition  is
\begin{equation}
\label{uniq_cond} \ker [ (1-P^{(n)}_M)J_{\dm}] \cap \ker P^{(n)}_M = \0.
\end{equation}
Here $\ker $  denotes  a null space of the  corresponding  operator. How this  condition can be  met
is  discussed in the  next  subsection.

It is  natural to begin  the  iteration  procedure (\ref{Newton}) starting from the invariant
manifold of the  non--driven system. In  other  words, we  choose   the  initial  approximation
$\Psi_{(0)}$ as the  solution of the invariance equation (\ref{inveq_2par}) corresponding to
$\varepsilon_1=0$ and  $\varepsilon_2=0$:
\begin{equation}
\label{man0ord} (1-P^{(0)}_M) J_{\dm} \Psi_{(0)}=0.
\end{equation}
We shall  return to the  question how to construct   solutions to this equation in the   subsection
``Linear zero-order equations".

The  above recurrent equations (\ref{Newton}),  (\ref{proj_cond}) are simplified  Newton method  for
the  solution of invariance equation (\ref{inveq_2par}), which involves the   small parameters. A
similar procedure for Grad equations of the Boltzmann kinetic theory was used recently in the Ref.
\cite{KGDNPRE98}. When these parameters  are  not  small, one should proceed   directly  with
equations (\ref{Newtonfull}).

Above, we have focused our attention on how to organize the iterations to construct invariant
manifolds of weakly driven systems. The only question we have not yet  answered   is  how  to  choose
projectors in    iterative equations  in a consistent way. In the next subsection we discuss the
problem of derivation of  the reduced  dynamics and its  relation to the problem  of the choice of
projector.

\paragraph*{\textbf{Projector and reduced dynamics.}}
Below   we suggest the   projector   which is   equally  applicable for constructing    invariant
manifolds by the iteration  method (\ref{Newton}), (\ref{proj_cond})  and  for generating macroscopic
equations  based on given manifold.

Let  us  discuss  the  problem  of constructing   closed equations  for macroparameters. Having some
approximation to the invariant manifold,  we nevertheless deal with a non-invariant manifold and we
face the  problem how to construct the dynamics on it. If  the $n-$dimensional   manifold
$\widetilde{\Psi}$  is  found  the  macroscopic  dynamics is induced by any projector $P$ onto the
tangent bundle of $\widetilde{\Psi}$ as follows \cite{GKTTSP94}:
\begin{equation}
\label{ind_dynamics} \frac{\dd M^*_i}{\dd t}= D M_i^* \bigr|_{\tilde{\Psi}} \left[ P J
\widetilde{\Psi} \right].
\end{equation}
To specify the projector we involve the  two above mentioned principles: dissipativity  and slowness.
The  dissipativity  is required to have the unique and stable equilibrium solution for macroscopic
equations, when the external  fields  are  absent ($\alpha=0$).  The  slowness condition requires
the {\it induced}  vector  field $PJ\Psi$ to  match  the  slow modes  of the original vector  field
$J\Psi$.

Let us consider the   parameterization of  the  manifold $\widetilde{\Psi}(M)$  by the   parameters
$M_i[\Psi]$. This parameterization  generates associated  projector $P=P_{M}$ by the Eq.
(\ref{proj_M}). This  leads  us  to  look  for the admissible parameterization of this  manifold,
where by  admissibility we  understand the  concordance with the dissipativity and the  slowness
requirements. We solve the  problem of the  admissible parameterization in the  following way. Let us
define the functionals  $M_i$ $i=1,\ldots,n$  by  the set  of the  lowest eigenvectors $\varphi_i$ of
the operator $J_{\dm}$: $$ M_i[\widetilde{\Psi}]=\la \varphi_i,\widetilde{\Psi} \ra_{\sm}, $$ where
$J_{\dm}\varphi_i=\lambda_i \varphi_i$. The  lowest eigenvectors $\varphi_1,\ldots,\varphi_n$ are
taken as a join of basises   in  the eigenspaces of the  eigenvalues with  smallest absolute values:
$0<|\lambda_1| \leq |\lambda_2|\leq \ldots \leq |\lambda_n|$. For simplicity we  shall work with the
orthonormal set  of  eigenvectors: $\la \varphi_i, \varphi_j  \ra_{\sm}= \delta_{ij}$ with
$\delta_{ij}$ the Kronecker symbol. Since the function $\Psi_{\eqm}$  is  the eigenvector of the zero
eigenvalue we  have: $M_i[\Psi_{\eqm}]=\la \varphi_i, \Psi_{\eqm} \ra_{\sm}=0$.

Then  the associated  projector $P_M$, written as:
\begin{equation}
\label{proj_eigen}
 P_M=\sum^{n}_{i=1}\frac{\partial
\widetilde{\Psi}}{\partial M_i} \la \varphi_i,\bullet  \ra_{\sm},
\end{equation}
 will  generate the  equations in  terms   of the parameters  $M_i$ as  follows:
$$\dd M_i/\dd t= \la \varphi_i P_M J\widetilde{\Psi} \ra_{\sm}= \la \varphi_i J\widetilde{\Psi}
\ra_{\sm}.$$ Their  explicit form is
 \begin{equation}
\label{lin_dyn} \frac{\dd M_i}{\dd t }= \lambda_i M_i+ \la J^+_{\exm}(\alpha)g_i,\widetilde{\Psi}(M)
\ra_{\sm},
\end{equation}
where the  $J^+_{\exm}$  is  the adjoint to operator $J_{\exm}$  with respect   to the  scalar
product $\la \bullet,\bullet \ra_{\sm}$.

Apparently,  in the absence  of  forcing ($\alpha \equiv 0$)  the macroscopic  equations $\dd M_i /
\dd t = \lambda_i M_i$   are thermodynamically  stable. They  represent   the  dynamics  of  slowest
eigenmodes  of equations $\dd \Psi/\dd t = J_{\dm}\Psi$. Thus, the projector (\ref{proj_eigen})
complies   with  the above  stated requirements of dissipativity and slowness   in the absence
external flux.

To  rewrite the macroscopic equations (\ref{lin_dyn})  in  terms  of  the required  set of
macroparameters, $M_i^*[\Psi]=\la m^*_i, \Psi \ra_{\sm}$, we  use the   formula (\ref{ind_dynamics})
which is equivalent to the change of variables $\{M \} \to \{M^*(M)\}$,\ $M^*_i =\la m^*_i,
\widetilde{\Psi} (M)\ra_{\sm}$ in  the equations (\ref{lin_dyn}). Indeed,  this  is  seen from   the
relation: $$ D M^*_i \bigr|_{\tilde{\Psi}} \left[ P_M J \widetilde{\Psi} \right]=\sum_j\frac{\partial
M^*_i}{\partial M_j}D M_j \bigr|_{\tilde{\Psi}}[J\widetilde{\Psi}]. $$

We have  constructed the  dynamics with the  help  of the  projector $P_M$ associated with the lowest
eigenvectors of the  operator $J_{\dm}$. It is directly  verified  that such  projector
(\ref{proj_eigen})  fulfills   the condition (\ref{proj_cond}) for arbitrary manifold
 $\Psi_{(n)}=\widetilde{\Psi}$.
For this  reason   it  is  natural to  use the  projector (\ref{proj_eigen}) for  both  procedures,
constructing  the invariant  manifold,  and deriving the  macroscopic  equations.

We have to  note  that the above described  approach to defining   the dynamics via the  projector is
different from the  concept of ``thermodynamic parameterization'' proposed in the Refs.
\cite{GKTTSP94,GK1}. The latter  was applicable for arbitrary dissipative systems including nonlinear
ones, whereas the present  derivations  are applied  solely  for linear systems.

\paragraph*{\textbf{Linear zero-order equations.}}

In this section we focus our attention on  the  solution of  the zero--order invariance equation
(\ref{man0ord}). We seek the {\it linear} invariant manifold of the form
\begin{equation}
\label{lin_manifold} \Psi_{(0)}(a)= \Psi_{\eqm} + \sum_{i=1}^{n} a_i m_i,
\end{equation}
where $a_i$  are  coordinates on this manifold. This manifold can be considered as an expansion of
the relevant slow manifold near the equilibrium  state. This limits the domain of validity of the
manifolds (\ref{lin_manifold}) because they are not generally  positively definite. This remark
indicates  that nonlinear invariant manifolds should be considered for large deviations from the
equilibrium but this goes beyond the scope of this Example.

The linear $n-$dimensional manifold representing the slow motion for the linear dissipative system
(\ref{dynsys}) is   associated with $n$ slowest eigenmodes. This manifold  should be built up as the
linear hull of the eigenvectors $ \varphi_i$ of the operator  $J_{\dm}$, corresponding to the lower
part of its spectrum. Thus we  choose $m_i= \varphi_i$.

Dynamic equations for the macroscopic variables $M^*$ are derived in two steps. First, following the
subsection, ``Projector and reduced dynamics", we parameterize the linear manifold $\Psi_{(0)}$ with
the values of the moments $M_i[\Psi]=\la \varphi_i,\Psi \ra_{\sm}$. We obtain that  the
parameterization of the manifold (\ref{lin_manifold}) is  given by
 $a_i=M_i$,  or:
$$ \Psi_{(0)}(M)= \Psi_{\eqm} + \sum_{i=1}^{n} M_i \varphi_i, $$ Then the reduced dynamics in terms
of variables $M_i$  reads:
\begin{equation}
\label{lin_dyn0} \frac{\dd M_i}{\dd t }= \lambda_i M_i+ \sum_j \la J_{\exm}^+\varphi_i, \varphi_j
\ra_{\sm} M_j + \la  J_{\exm}^+\varphi_i, \Psi_{\eqm} \ra_{\sm},
\end{equation}
where $\lambda_i=\la \varphi_i,J_{\dm} \varphi_i \ra_{\sm}$  are eigenvalues which correspond  to
eigenfunctions  $\varphi_i$.

Second, we switch from the variables $M_i$ to the variables $M^*_i(M)=\la m^*_i,\Psi_{(0)}(M)
\ra_{\sm}$   in the Eq. (\ref{lin_dyn0}). Resulting equations for the variables $M^*$ are also
linear:
\begin{eqnarray}
\label{lin_transreddyn} \frac{\dd  M^*_i}{\dd t }= \sum_{jkl} (B^{-1})_{ij}\Lambda_{jk} B_{kl} \Delta
M^*_l+ \sum_{jk} (B^{-1})_{ij}\la J_{\exm}^+\varphi_j, \varphi_k \ra_{\sm} \Delta M^*_k  && \nonumber
\\
 + \sum_j (B^{-1})_{ij}\la J_{\exm}^+ \varphi_j, \Psi_{\eqm} \ra_{\sm}. &&
\end{eqnarray}
Here $\Delta M^*_i=M^*_i-M^*_{\eqm|i}$ is the deviation of the  variable $M^*_i$  from its
equilibrium value $M^*_{\eqm|i}$, and $B_{ij}=\la m_i^*,\varphi_j \ra$  and  $\Lambda_{ij}=\lambda_i
\delta_{ij}$.

\subsubsection*{\textbf{Auxiliary formulas. 1. Approximations  to  eigenfunctions
of the Fokker-Planck operator}}

In this  subsection  we  discuss the question  how  to  find the lowest eigenvectors $\Psi_{\eqm}
m_0(Q^2)$  and  $\Psi_{\eqm}m_1(Q^2)\oQQ$  of the operator $J_{\dm}$  (\ref{J_d}) in the classes of
functions having  a form:  $w_0(Q)$  and  $w_1(Q)\oQQ$. The  results presented in this subsection
 were used in the subsections: ``Constitutive equations" and ``Tests on the FENE
dumbbell model". It is directly verified that:
\begin{eqnarray*}
&& J_{\dm}w_0 =G_0^{\hm} w_0,  \\ && J_{\dm }w_1\oQQ=(G_1^{\hm}w_1)\oQQ,
\end{eqnarray*}
where  the operators
 $G_0^{\hm}$  and  $G_1^{\hm}$  are  given by:
\begin{equation}
\label{perturb_operator} G_0^{\hm}=G_0 -\kappa H_0,  \qquad G_1^{\hm} = G_1 - \kappa H_1.
\end{equation}
The operators  $G_{0,1}$  and $H_{0,1}$  act in the space  of isotropic functions (i.e.  dependent
only  on  $Q=(\QQ\cdot\QQ)^{1/2}$) as  follows:
\begin{eqnarray}
\label{G_1} &&G_0=\frac{1}{2}\left(\frac{\partial^2}{ \partial Q^2} - f Q \frac{\partial}{\partial Q}
+ \frac{2}{Q} \frac{\partial }{\partial Q} \right), \\ \label{G_2} &&G_1 = \frac{1}{2}\left(
\frac{\partial^2}{\partial Q^2}  - f Q \frac{\partial}{\partial Q} + \frac{6}{Q}
\frac{\partial}{\partial Q } -2 f \right), \\ &&H_0 =\frac{2}{ Q }\left( \frac{\partial^2 }{\partial
Q^2} - f Q\frac{\partial }{\partial Q} + \frac{2}{Q} \frac{\partial }{\partial Q}   \right),\\ &&H_1
=\frac{2}{ Q}\left( \frac{\partial^2 }{\partial Q^2} - f  Q\frac{\partial }{\partial Q} + \frac{5}{Q}
\frac{\partial }{\partial Q} - 2 f  + \frac{1}{Q^2} \right).
\end{eqnarray}

The following two  properties  of the operators $G_{0,1}^{\hm}$ are important  for our  analysis: Let
us  define two  scalar products $\la \bullet,\bullet  \ra_{0}$ and  $\la \bullet,\bullet  \ra_{1}$:
$$\la y,x \ra_{0} = \la xy \ra_{\emm},$$ $$ \la y,x  \ra_{1} = \la xy Q^4 \ra_{\emm}. $$ Here $\la
\bullet \ra_{\emm}$ is the equilibrium average as defined in (\ref{average}). Then  we state that for
sufficiently  small $\kappa$   the operators $G_{0}^{\hm}$ and  $G_{1}^{\hm}$  are  symmetric and
nonpositive in the scalar products $\la \bullet,\bullet \ra_0$ and   $\la \bullet,\bullet \ra_1$
respectively. Thus for  obtaining  the  desired eigenvectors of the operator $J_{\dm}$  we  need  to
find the eigenfunctions $m_0$  and $m_1$  related to the   lowest   nonzero eigenvalues   of  the
operators $G_{0,1}^{\hm}$.

Since we  regard the parameter  $\kappa$   small  it is  convenient, first, to find  lowest
eigenfunctions $g_{0,1}$  of the  operators  $G_{0,1}$ and,  then,  to  use standard perturbation
technique  in  order to obtain $m_{0,1}$. For the  perturbation of  the  first order one  finds
\cite{Kato}:
\begin{eqnarray}
\label{g1g2_hydr} m_0=g_0 + \kappa h_0, &\ \ & h_0=- g_0 \frac{\la g_0 H_0 G_0 g_0\ra_0}{\la
g_0,g_0\ra_0}- G_0 H_0  g_0;  \nonumber \\ m_1=g_1 + \kappa h_1, &\ \ &  h_1= - g_1\frac{\la g_1 H_1
G_1 g_1\ra_1}{\la g_1,g_1 \ra_1}- G_1 H_1  g_1.
\end{eqnarray}

For  the  rest of this subsection we  describe  one  recurrent   procedure for obtaining  the
functions $m_{0}$ and $m_1$ in a  constructive way. Let us solve    this  problem  by  minimizing the
functionals $\LLambda_{0,1}$:
\begin{equation}
\label{minim_eigen} \Lambda_{0,1}[m_{0,1}]= -\frac{\la m_{0,1}, G^{\hm}_{0,1} m_{0,1}\ra_{0,1}} {\la
m_{0,1}, m_{0,1} \ra_{0,1}} \longrightarrow \min,
\end{equation}
by  means  of the {\it gradient descent  method}.

Let us  denote $e_{0,1}$  the eigenfunctions of the zero  eigenvalues  of the  operators
$G_{0,1}^{\hm}$. Their explicite values   are $e_0=1$  and $e_1=0$. Let the initial approximations
$m_{0,1}^{(0)}$  to the lowest eigenfunctions  $m_{0,1}$  be  chosen so that $\la
m_{0,1}^{(0)},e_{0,1}\ra_{0,1}=0$. We  define the  variation  derivative $\delta\Lambda_{0,1}/\delta
m_{0,1} $  and  look  for the correction in  the form:
\begin{equation}
\label{appA:corr} m_{0,1}^{(1)}=m_{0,1}^{(0)}+\delta m_{0,1}^{(0)}, \qquad \delta m_{0,1}^{(0)}=
\alpha \frac{\delta\Lambda_{0,1}}{\delta m_{0,1}},
\end{equation}
where  scalar  parameter $\alpha <0$  is  found from the  condition: $$ \frac{\partial
\Lambda_{0,1}[m_{0,1}^{(1)}(\alpha)] }{\partial \alpha}=0. $$

In the  explicit form the  result reads: $$ \delta
m_{0,1}^{(0)}=\alpha^{(0)}_{0,1}\Delta^{(0)}_{0,1}, $$ where
\begin{eqnarray}
\Delta^{(0)}_{0,1}&=& \frac{2}{\la m^{(0)}_{0,1}, m^{(0)}_{0,1}\ra_{0,1} } \left( m^{(0)}_{0,1}
\lambda_{0,1}^{(0)}-G_{0,1}^{\hm} m^{(0)}_{0,1} \right), \nonumber \\
 \lambda_{0,1}^{(0)} &=&\frac{\la m^{(0)}_{0,1},
G^{\hm}_{0,1} m^{(0)}_{0,1} \ra_{0,1}}{\la m^{(0)}_{0,1}, m^{(0)}_{0,1} \ra_{0,1}}, \nonumber \\
\alpha^{(0)}_{0,1} &=& q_{0,1}-\sqrt{q_{0,1}^2+ \frac{\la  m^{(0)}_{0,1}, m^{(0)}_{0,1}
\ra_{0,1}}{\la  \Delta^{(0)}_{0,1}, \Delta^{(0)}_{0,1}  \ra_{0,1}}}, \nonumber  \\
 q_{0,1}&=&\frac{1}{\la \Delta^{(0)}_{0,1}, \Delta^{(0)}_{0,1}  \ra_{0,1}}
 \left( \frac{\la
m^{(0)}_{0,1} , G^{\hm}_{0,1} m^{(0)}_{0,1} \rangle_{0,1}} {\la m^{(0)}_{0,1} ,m^{(0)}_{0,1}
\rangle_{0,1}} - \frac{\la \Delta^{(0)}_{0,1} , G^{\hm}_{0,1} \Delta^{(0)}_{0,1} \rangle_{0,1}} {\la
\Delta^{(0)}_{0,1} , \Delta^{(0)}_{0,1}  \rangle_{0,1}}  \right).
\end{eqnarray}

Having the  new  correction $m^{(1)}_{0,1}$  we can repeat the procedure and eventually generate  the
recurrence   scheme.  Since by  the  construction all  iterative approximations $m_{0,1}^{(n)}$
remain  orthogonal  to zero eigenfunctions  $e_{0,1}$: $\la m_{0,1}^{(n)},e_{0,1}\ra_{0,1}=0$ we
avoid the convergence of this recurrence procedure to the  eigenfunctions $e_{0,1}$.

The  quantities $\delta_{0,1}^{(n)}$: $$ \delta_{0,1}^{(n)}=\frac{\la
\Delta^{(n)}_{0,1},\Delta^{(n)}_{0,1} \ra_{0,1}}{\la m^{(n)}_{0,1},m^{(n)}_{0,1} \ra_{0,1}} $$
 can  serve  as  relative  error
parameters   for  controlling the   convergence of the iteration procedure (\ref{appA:corr}).

\addcontentsline{toc}{subsubsection}{Integration  formulas}

\subsubsection*{\textbf{Auxiliary formulas. 2. Integral relations}}

Let $\Omega $  be a sphere in $\mbox{\bf R}^3$ with the center at the origin of the coordinate system
or be the entire  space $\mbox{\bf R}^3$. For any function $s(x^2)$, where $x^2=\xx\cdot\xx$, $\xx
\in \mbox{\bf R}^3$,  and any square $3 \times 3 $ matrices  $\Ab$, $\BB$, $\CC$ independent of $\xx$
the following integral relations are valid:
\begin{eqnarray}
&& \int_{\Omega} s(x^2) \oxx (\oxx  \ :\Ab) \dd \xx = \frac{2}{15} \oA \int_{\Omega} s x^4 \dd \xx;
\nonumber \\ && \int_{\Omega} s(x^2) \oxx (\oxx \ :\Ab)  (\oxx  \ :\BB) \dd \xx =  \frac{4}{105}
\stackrel{\circ}{(\Ab \cdot \BB + \BB\cdot \Ab)} \int_{\Omega} s x^6 \dd \xx; \nonumber \\
&&\int_{\Omega} s(x^2) \oxx (\oxx \ :\Ab)(\oxx \ :\BB) (\oxx \ :\CC)  \dd \xx = \nonumber \\
&&\phantom{aaaaaaaaaaaaa}\frac{4}{315} \left\{\oA(\BB:\CC)+ \oB(\Ab:\CC)+  \oC (\Ab:\BB)\right\}
\int_{\Omega} s x^8 \dd \xx. \nonumber
\end{eqnarray}

\addcontentsline{toc}{subsubsection}{Microscopic derivation of constitutive equations}

\subsubsection*{\textbf{Microscopic derivation of constitutive equations}}

\paragraph*{\textbf{Iteration scheme.}}

In  this  section  we  apply the above developed formalism to the elastic dumbbell model
(\ref{FPnondim}). External field  variables $\alpha$ are the components of the tensor $\kkh$.

Since we  aim at constructing a   closed description  for the stress  tensor $\ttau$ (\ref{tau_def})
with the six independent components,  the relevant manifold in our problem should be
six--dimensional. Moreover,  we allow  a dependence of the manifold  on  the  material derivatives of
the tensor $\kkh$: $\kkh^{(i)} = D^i \kk /D t^i$. Let $\Psi^*(M,{\cal K})$ \  ${\cal
K}=\{\kkh,\kkh^{(1)},\ldots  \}$  be the desired  manifold parameterized by the six variables $M_i$ \
$i=1,\ldots,6$ and the independent components (maximum eight  for each $\kkh^{(l)}$ ) of the tensors
$\kkh^{(l)}$. Small parameters $\varepsilon_1$  and $\varepsilon_2$,  introduced in  the section:
``The problem of reduced description in polymer dynamics", are established by Eq.
(\ref{two_epsilon}). Then we define  the invariance equation:
\begin{equation}
\label{inv_condition_ext} (1-P_M)(J_{\dm}+\varepsilon_1 J_{\hm})\Psi = \varepsilon_2
\sum_{i=0}^{\infty} \sum_{lm}\frac{\partial \Psi}{\partial \kh^{(i)}_{lm}}  \kh_{lm}^{(i+1)},
\end{equation}
where $P_M=(\partial \Psi / \partial M_i) D M_i[\bullet]$  is the projector associated with chosen
parameterization and summation indexes $l,m$  run only  eight independent  components of tensor
$\kkh$.

Following  the  further procedure we straightforwardly obtain the recurrent equations:
\begin{eqnarray}
\label{norder_invariance} (1-P^{(n)}_M )J_{\dm} \Psi_{n+1} &=& - (1- P^{(n)}_M
)[J_{\dm}+\varepsilon_1 J_{\hm}] \Psi_{(n)} + \varepsilon_2 \sum_{i}\sum_{lm} \frac{\partial
\Psi_{(n)}}{\partial \kh_{lm}^{(i)}} \kh_{lm}^{(i+1)},  \\  P^{(n)}_M\Psi_{n+1}&=&0,
\end{eqnarray}
where $\Psi_{n+1}$ is the   correction  to  the  manifold $\Psi_{(n)}=\sum_{i=0}^n \Psi_i$.

The   zero-order manifold is  found as the   relevant  solution  to equation:
\begin{equation}
\label{man0ord1} (1-P^{(0)}_M ) J_{\dm} \Psi_{(0)}=0
\end{equation}
We  construct  zero-order manifold $\Psi_{(0)}$  in the subsection, ``Zero-order  constitutive
equation".

\paragraph*{\textbf{The dynamics in  general form.}}
Let us  assume that some approximation  to invariant  manifold $\widetilde{\Psi}(a,{\cal K})$ is
found  (here $a=\{a_1,\ldots,a_6 \}$ are some coordinates on this  manifold). The next  step is
constructing the macroscopic   dynamic equations.

In  order to comply with    dissipativity  and slowness  by means  of the recipe from the  previous
section  we need to find six lowest eigenvectors of the operator $J_{\dm }$. We shall  always  assume
in a sequel that the hydrodynamic interaction parameter $\kappa$  is  small enough that the
dissipativity  of  $J_{\dm}$ (\ref{J_dsymm}) is  not  violated.

Let  us  consider two  classes of  functions: ${\cal C}_1=\{w_0(Q^2) \}$ and ${\cal C}_2=\{w_1(Q^2)
\oQQ \}$,  where $w_{0,1}$  are   functions  of $Q^2$ and the  notation  $\circ$  indicates traceless
parts of tensor or matrix, e.g. for the  {\it dyad} $\QQ\QQ$:  $$(\oQQ)_{ij} = Q_i Q_j - \frac{1}{3}
\delta_{ij} Q^2 .$$ Since the  sets ${\cal C}_1$  and  ${\cal C}_2$  are invariant with respect to
operator $J_{\dm}$, i.e. $J_{\dm}{\cal C}_1 \subset {\cal C}_1$  and $J_{\dm}{\cal C}_2 \subset {\cal
C}_2$,  and densities  $\FF \QQ=f\oQQ + (1/3) \1 fQ^2$ of the moments  comprising the stress  tensor
$\ttau_p$  (\ref{tau_p})  belong   to the  space ${\cal C}_1+{\cal C}_2$,  we  shall seek  the
desired eigenvectors  in the classes ${\cal C}_1$  and  ${\cal C}_2$. Namely,  we   intend to find
one lowest  isotropic eigenvector $\Psi_{\eqm}m_0(Q^2)$  of  eigenvalue $-\lambda_0$ ($\lambda_0 >0$)
and  five nonisotropic eigenvectors $m_{ij}=\Psi_{\eqm} m_1(Q^2)(\oQQ)_{ij}$ of another  eigenvalue
$-\lambda_1$ ($\lambda_1>0$). The method of  derivation and  analytic evaluation of these eigenvalues
are   discussed in the subsection  ``Auxiliary formulas, 1". For a  while   we assume that these
eigenvectors are known.

In the  next step we  parameterize given manifold  $\widetilde{\Psi}$  by the  values  of the
functionals:
\begin{eqnarray}
&&M_0=\la \Psi_{\eqm} m_0,\widetilde{\Psi} \ra_{\sm} = \int\! m_0 \widetilde{\Psi} \dd \QQ, \nonumber
\\ &&\oMM=\la \Psi_{\eqm} m_1 \oQQ, \widetilde{\Psi} \ra_{\sm} = \int\! m_1 \oQQ \widetilde{\Psi} \dd
\QQ.
\end{eqnarray}
Once a  desired parameterization  $\widetilde{\Psi}(M_0,\oMM,{\cal K})$ is obtained,  the dynamic
equations are found  as:
\begin{eqnarray}
\label{dyn_anyorder} \frac{D M_0}{D \tth} + \lambda_0 M_0 &  = & \left\langle (\dotgamh  : \,\oQQ)
m'_0 \right\rangle  \nonumber \\ \oMM_{[1]} + \lambda_1 \oMM  &=& - \frac{1}{3} \II \dotgamh: \,\oMM
-   \frac{1}{3} \dotgamh \left\langle   m_1 Q^2 \right\rangle +   \left\langle  \oQQ (\dotgamh  : \,
\oQQ)  m'_1 \right\rangle,
\end{eqnarray}
where all  averages  are  calculated with the  d.f. $\widetilde{\Psi}$, i.e. $\la \bullet \ra=\int\!
\bullet \widetilde{\Psi} \dd \QQ$, \mbox{$m'_{0,1}=\dd m_{0,1}(Q^2)/ \dd (Q^2)$}  and subscript $[1]$
represents  the  upper convective derivative  of tensor: $$\LLambda_{[1]}=\frac{D \LLambda}{D \tth} -
\left\{ \kkh \cdot \LLambda + \LLambda \cdot \kkh^{\dag} \right\}. $$ The parameters $\lambda_{0,1}$,
which are absolute  values of  eigenvalues of operator $J_{\dm}$, are calculated by formulas (for
definition of  operators $G_1$  and  $G_2$ see subsection ``Auxiliary formulas, 1"):
\begin{eqnarray}
&& \lambda_{0}= - \frac{\left\langle m_0 G_0 m_0 \right\rangle_{\emm}}{\la m_0 m_0 \ra_{\emm}} >0,
\\ &&\lambda_{1} = - \frac{\la Q^4 m_1  G_1 m_1 \ra_{\emm}}{\la m_1  m_1 Q^4\ra_{\emm}} >0,
\label{lambda_01}
\end{eqnarray}
where we  have  introduced the  notation  of the   equilibrium average:
\begin{equation}
\label{average} \la y  \ra_{\emm}=\int\! \Psi_{\eqm} y \dd \QQ.
\end{equation}

Equations on   components of the  polymeric  stress  tensor $\ttau_{\ppm}$ (\ref{tau_p}) are
constructed  as a  change of  variables $\{ M_0,\oMM \} \to \ttau_{\ppm}$. The  use of the projector
$\widetilde{P}$  makes this operation straightforward:
\begin{equation}
\label{tau_dyn} \frac{D {\ttau_{\ppm}}}{D \tth} =- nk_{\Bm} T \int\! \FF\QQ \widetilde{P} J
\widetilde{\Psi}(M_0(\ttau_{\ppm},{\cal K}),\oMM(\ttau_{\ppm},{\cal K}),{\cal K}) \dd \QQ.
\end{equation}
Here,   the  projector $\widetilde{P}$  is  associated  with  the parameterization by the  variables
$M_0$  and $\oMM$:
\begin{equation}
\label{proj_tilde} \widetilde{P}=\frac{\partial \widetilde{\Psi}}{\partial M_0} \la \Psi_{\eqm}
m_0,\bullet \ra_{\sm}+ \sum_{kl}\frac{\partial \widetilde{\Psi}}{\partial \oMM_{kl}} \la \Psi_{\eqm}
m_1 (\oQQ)_{kl},\bullet \ra_{\sm}.
\end{equation}

We  note  that sometimes it is   easier  to make  transition to  the variables $\ttau_{\ppm}$  after
solving  the equations (\ref{dyn_anyorder}) rather than to construct explicitly  and  solve equations
in  terms of $\ttau_{\ppm}$. It allows to  avoid  inverting the functions $\ttau_{\ppm}( M_0,\oMM )$
and  to  deal with   simpler  equations.

\addcontentsline{toc}{subsubsection}{Approximations  to  eigenfunctions  of the Fokker-Planck
operator}

\paragraph*{\textbf{Zero-order  constitutive equation.}}

In this subsection we derive the  closure based on the   zero--order manifold   $\Psi_{(0)}$ found as
appropriate  solution to Eq. (\ref{man0ord1}). Following the  approach described in  subsection,
``Linear zero-order equations", we construct such a  solution as the linear expansion  near the
equilibrium state $\Psi_{\eqm}$  (\ref{lin_manifold}). After parameterization by the values of the
variables $M_0$ and $\oMM$ associated with the eigenvectors $\Psi_{\eqm}m_0$  and $\Psi_{\eqm} m_1
\oQQ$ we  find:
\begin{equation}
\label{Psi0_par} \Psi_{(0)} = \Psi_{\eqm} \left( 1+ M_0 \frac{m_0}{\left\langle  m_0
m_0\right\rangle_{\emm} }  + \frac{15}{2} \oMM :\, \oQQ \frac{m_1}{\left\langle  m_1  m_1 Q^4
\right\rangle_{\emm} } \right).
\end{equation}

Then with the  help of  the  projector (\ref{proj_tilde}):
\begin{equation}
\label{proj_0} P^{(0)}_M= \Psi_{\eqm} \left\{ \frac{m_0}{\la   m_0 m_0\ra_{\emm} }  \la  m_0, \bullet
\ra_{\emm}  + \frac{15}{2} \frac{m_1}{\la  m_1  m_1 Q^4 \ra_{\emm}} \oQQ \ : \la m_1 \oQQ, \bullet
\ra_{\emm} \right\}
\end{equation}
by the  formula  (\ref{tau_dyn})  we  obtain:
\begin{eqnarray}
\label{0order_dyn} \frac{D \tr {\mbox{\boldmath $\tau$}}_{\ppm}}{D \tth} + \lambda_0 \tr
{\mbox{\boldmath $\tau$}}_{\ppm} & =& a_0 \left( \otau_{\ppm}\,: \dotgamh \right),  \\ \otau_{p[1]} +
\lambda_0 \otau_{\ppm} & = &b_0 \left[\otau_{\ppm} \cdot \; \dotgamh + \dotgamh \; \cdot
\otau_{\ppm}\right] - \frac{1}{3} \II  (\otau_{\ppm}\, : \dotgamh ) + ( b_1  \tr {\mbox{\boldmath
$\tau$}}_{\ppm} - b_2 nk_{\Bm}T ) \dotgamh, \nonumber
\end{eqnarray}
where the constants $b_i$,\  $a_0$  are
\begin{eqnarray}
&&a_0 =  \frac{\la f m_0 Q^2\ra_{\emm} \la m_0 m_1 Q^4  m'_1 \ra_{\emm}}{\left\langle fm_0
Q^4\right\rangle_{\emm} \left\langle m_0^2\right\rangle_{\emm}}, \nonumber\\ &&b_0= \frac{2}{7}
\frac{\la m_1 m'_2  Q^6 \ra_{\emm}}{\la  m_1^2 Q^4\ra_{\emm}}, \nonumber \\ && b_1= \frac{1}{15}
\frac{\la fm_1 Q^4 \ra_{\emm}}{\left\langle fm_0 Q^2\right\rangle_{\emm}} \left\{ 2
\frac{\left\langle m_0 m'_2 Q^4 \right\rangle_{\emm}}{\left\langle m_1^2 Q^4 \right\rangle_{\emm}} +
5 \frac{\left\langle m_0 m_1 Q^2 \right\rangle_{\emm}}{\left\langle m_1 m_1 Q^4 \right\rangle_{\emm}}
\right\}, \nonumber \\ && b_2= \frac{1}{15} \frac{\left\langle fm_1 Q^4
\right\rangle_{\emm}}{\left\langle m_1 m_1 Q^4\right\rangle_{\emm}} \left\{ 2 \left\langle m'_2  Q^4
\right\rangle_{\emm} + 5 \left\langle m_1  Q^2 \right\rangle_{\emm} \right\}.
\label{constants_0order}
\end{eqnarray}
We remind that   $m'_{0,1}= \partial m_{0,1}/ \partial (Q^2)$. These formulas were  obtained using
the auxiliary results from  subsection  ``Auxiliary formulas, 2".

\paragraph*{\textbf{Revised Oldroyd 8 constant constitutive equation for the stress.}}
It is remarkable that being rewritten in terms of the  full stresses ${\mbox{\boldmath
$\tau$}}=-\nu_{\sm} \dotgam + {\mbox{\boldmath $\tau$}}_{\ppm}$ the dynamic system (\ref{0order_dyn})
takes a form:
\begin{eqnarray}
\label{Oldroyd9} &&{\mbox{\boldmath $\tau$}} + c_1 {\mbox{\boldmath $\tau$}}_{[1]} + c_3 \left\{
\dotgam \cdot {\mbox{\boldmath $\tau$}} + {\mbox{\boldmath $\tau$}} \cdot \dotgam \right\} + c_5 (\tr
{\mbox{\boldmath $\tau$}}) \dotgam +  \II \left( c_6 {\mbox{\boldmath $\tau$}}: \dotgam + c_8 \tr
{\mbox{\boldmath $\tau$}} \right) = \nonumber \\ && \phantom{ssssssssssssss} - \nu \left\{ \dotgam +
c_2 \dotgam_{[1]} +  c_4 \dotgam \cdot \dotgam + c_7(\dotgam:\dotgam ) \II \right\},
\end{eqnarray}
where the parameters $\nu$, $c_i$  are given by  the  following  relationships:
\begin{eqnarray}
&&\nu=\lambda_{\rrm} \nu_{\sm} \mu ,  \qquad    \mu=1 +  nk_{\Bm}T \lambda_1 b_2/\nu_{\sm}, \nonumber
\\ && c_1=\lambda_{\rrm}/\lambda_1, \qquad c_2 = \lambda_{\rrm}/(\mu \lambda_1), \nonumber
\\ && c_3 = - b_0 \lambda_{\rrm} /\lambda_0, \qquad c_4= - 2  b_0\lambda_{\rrm}/(\mu\lambda_1),
\nonumber  \\ && c_5= \frac{\lambda_{\rrm}}{3 \lambda_1} ( 2 b_0 - 3b_1 - 1 ), \qquad c_6 =
\frac{\lambda_{\rrm}}{\lambda_1}  ( 2 b_0 + 1 - a_0 ),  \nonumber \\ && c_7=
\frac{\lambda_{\rrm}}{\lambda_1 \mu} ( 2 b_0 +1 - a_0), \qquad c_8 = \frac{1}{3} (\lambda_0
/\lambda_1 -1 ). \label{Oldroyd9_const}
\end{eqnarray}
In the  last  two  formulas we  returned  to the  original dimensional quantities:  time $t$ and
gradient of velocity tensor $\kk=\nabla \vv$, and  at  the  same time  we  kept the old  notations
for the  dimensional convective derivative $\LLambda_{[1]}=D \LLambda/D t-\kk\cdot \LLambda-\LLambda
\cdot \kk^{\dag}$.

Note that all the parameters (\ref{Oldroyd9_const}) are related to the entropic spring law $f$ due to
equation (\ref{constants_0order}). Thus, the constitutive relation for the stress $\ttau$
(\ref{Oldroyd9}) is fully derived from the microscopic kinetic model.

If the constant $c_8$ were equal to zero, then the form  of  Eq. (\ref{Oldroyd9}) would be recognized
as  {\it  the  Oldroyd 8 constant} model \cite{Oldroyd}, proposed by Oldroyd about 40 years ago on a
phenomenological basis. Nonzero $c_8$ indicates  a presence of difference between
$\lambda_{\rrm}/\lambda_0$  and  $\lambda_{\rrm}/\lambda_1$  which are relaxation  times of trace
$\tr \ttau$ and traceless  components $\otau$ of the  stress  tensor $\ttau$.

\paragraph*{\textbf{Higher-order constitutive equations.}}

In this subsection  we  discuss  the  properties  of   corrections   to  the zero-order model
(\ref{Oldroyd9}). Let  $P^{(0)}_M$ (\ref{proj_0})  be  the projector onto the zero-order manifold
$\Psi_{(0)}$  (\ref{Psi0_par}). The invariance equation (\ref{norder_invariance})  for the
first-order correction  $\Psi_{(1)}=\Psi_{(0)}+\Psi_1$  takes a  form:
\begin{eqnarray}
\label{1firstorder} &&L\Psi_{1} =-(1-P^{(0)}_M)(J_{\dm} + J_{\hm}) \Psi_{(0)} \\ && P^{(0)}_M\Psi_{1}
=0  \nonumber
\end{eqnarray}
where  $L=(1-P^{(0)}_M)J_{\dm}(1-P^{(0)}_M)$  is the   symmetric operator. If the  manifold
$\Psi_{(0)}$  is   parameterized  by the  functionals  $M_0=\int\! g_0\Psi_{(0)}\dd \QQ $  and $\oMM
= \int\! m_1\oQQ\Psi_{(0)}\dd \QQ$,   where  $\Psi_{\eqm}m_0$   and $\Psi_{\eqm}\oQQ m_1$   are
lowest eigenvectors  of  $J_{\dm}$, then  the  general  form  of the  solution is  given by:
 \begin{eqnarray}
\label{solution_1order} \Psi_1 & = &  \Psi_{\eqm} \left\{ z_0 M_0 ( \dotgam : \,\oQQ ) + z_1 (\oMM :
\oQQ )( \dotgam :\, \oQQ )  + \right. \nonumber \\ && \left. z_2 \{  \dotgam \cdot \oMM + \oMM\cdot
\dotgam \} :\, \oQQ + z_3 \dotgam  : \,\oMM  + \frac{1}{2} \dotgam :\, \oQQ \right\}.
\end{eqnarray}
The  terms   $z_0$  through  $z_3$   are  the  functions  of $Q^2$ found as the  solutions to some
linear differential  equations.

We   observe  two  features of the  new   manifold: first,  it   remains {\it linear}   in  variables
$M_0$   and  $\oMM$  and   second it contains the  dependence on the   rate  of  strain tensor
$\dotgam$. As the consequence,  the  transition  to variables $\ttau$  is  given by  the linear
relations:
  \begin{eqnarray}
\label{tau_MM_1order} &&- \frac{\otau_{\ppm}}{nk_{\Bm} T}=   r_0 \oMM + r_1 M_0 \dotgam + r_2
\{\stackrel{\circ}{\dotgam \cdot \oMM +\oMM \cdot\dotgam }\} + r_3 \stackrel{\circ}{\dotgam \cdot
\dotgam },
 \\
&&- \frac{\tr {\mbox{\boldmath $\tau$}}_{\ppm}}{nk_{\Bm} T}  =  p_0 M_0  + p_1 \dotgam : \oMM,
 \nonumber
\end{eqnarray}
where  $r_i$  and  $p_i$   are  some  constants. Finally the equations   in terms of $\ttau$  should
be  also  linear. Analysis  shows that  the first-order correction to the modified Oldroyd 8
constants  model (\ref{Oldroyd9}) will be  transformed   into the  equations of the following general
structure:
\begin{eqnarray}
\label{genmodel} {\mbox{\boldmath $\tau$}} + c_1 {\mbox{\boldmath $\tau$}}_{[1]} + \left\{ \GGamma_1
\cdot {\mbox{\boldmath $\tau$}}\cdot \GGamma_2 + \GGamma_2^{\dag}\cdot {\mbox{\boldmath $\tau$}}
\cdot \GGamma_1^{\dag} \right\} + \GGamma_3 (\tr {\mbox{\boldmath $\tau$}}) + \GGamma_4
(\GGamma_5:{\mbox{\boldmath $\tau$}}) = - \nu_0 \GGamma_6,
\end{eqnarray}
where  $\GGamma_1$  through $\GGamma_6$   are  tensors  dependent on the rate-of-strain  tensor
$\dotgam$  and  its  first convective derivative $\dotgam_{[1]}$, constant $c_1$ is  the same as  in
Eq. (\ref{Oldroyd9_const})   and   $\nu_0$  is a  positive constant.

Because the  explicit form  of  the tensors  $\GGamma_i$  is quite extensive we do not present  them
in this section. Instead we give several general remarks about the   structure of the first- and
higher-order corrections:

1) Since the  manifold  (\ref{solution_1order}) does  not  depend  on the vorticity tensor
$\omega=\kk-\kk^{\dag} $ the latter enters  the equations (\ref{genmodel}) only  via  convective
derivatives  of $\ttau$ and $\dotgam$. This is sufficient to acquire  the  frame indifference
feature, since all  the  tensorial quantities  in dynamic equations are
 indifferent in  any  time dependent reference frame \cite{Marrucci}.

2) When $\kk=0$ the  first order  equations (\ref{genmodel}) as  well as equations for  any order
reduce  to  linear relaxation   dynamics of slow modes:
\begin{eqnarray}
\frac{D \otau}{D t} + \frac{\lambda_1}{\lambda_{\rrm}} \otau =0, \nonumber
\\ \frac{D \tr \ttau}{D t} + \frac{\lambda_0}{\lambda_{\rrm}}  \tr \ttau=0,
\nonumber
\end{eqnarray}
which   is  obviously  concordant  with   the dissipativity  and  the slowness  requirements.

3) In all higher-order corrections one  will be  always left with linear manifolds if the projector
associated with  functionals $M_0[\Psi]$  and $\oMM[\Psi]$ is  used in every step. It follows  that
the  resulting constitutive equations will always take  a linear form   (\ref{genmodel}), where  all
tensors $\GGamma_i$   depend on higher order convective derivatives of  $\dotgam$  (the highest
possible order is limited  by the order of the  correction). Similarly to   the  first  and  zero
orders   the frame indifference is guaranteed  if the  manifold  does  not  depend  on the vorticity
tensor unless the latter is incorporated in any  frame invariant time  derivatives. It is  reasonable
to  eliminate the  dependence on vorticity (if  any)  at  the stage    of constructing the  solution
to iteration   equations (\ref{norder_invariance}).

4) When  the  force  $\FF$  is  linear $\FF = \QQ$   our approach is  proven to be also correct since
it leads the  Oldroyd--B model (Eq. (\ref{Oldroyd9}) with $c_i=0$  for  $i=3,\ldots,8$). This follows
from  the fact that the spectrum of the  corresponding  operator $J_{\dm}$ is  more degenerated,  in
particular  $\lambda_0=\lambda_1=1$  and  the corresponding lowest  eigenvectors comprise a  simple
dyad $\Psi_{\eqm}\QQ\QQ$.

\addcontentsline{toc}{subsubsection}{Tests on the FENE dumbbell model}

\subsubsection*{\textbf{Tests on the FENE dumbbell model}}

In this section we specify the  choice of the  force law as the FENE springs (\ref{FENE}) and present
results of test calculations for the revised Oldroyd 8  constants (\ref{0order_dyn}) equations on the
examples of two simple viscometric flows.

We  introduce the  extensibility parameter of FENE dumbbell model $b$:
\begin{equation}
\label{b_parameter} b=\QQh^2_0=\frac{H \QQ_0^2}{k_{\Bm} T}.
\end{equation}
It was estimated \cite{Bird}  that $b$  is proportional to the  length of polymeric molecule and has
a meaningful variation interval 50--1000. The limit $b \to \infty$  corresponds to the Hookean case
and therefore to the Oldroyd-B constitutive relation.

In our  test calculations  we  will  compare our  results  with the Brownian dynamic (BD)  simulation
data  made on FENE  dumbbell equations \cite{Herrchen},  and also with  one popular  approximation to
the FENE model known  as {\it FENE--P} (FENE--Peterelin)   model \cite{Martin,Bird,FENEP}. The latter
is obtained  by selfconsistent approximation to  FENE  force:
\begin{equation}
\label{FENE-P} \FF = \frac{1}{1-\left\langle\QQ^2 \right\rangle/b} \QQ.
\end{equation}
This force law  like  Hookean  case  allows  for the exact moment closure leading to  nonlinear
constitutive equations  \cite{Bird,FENEP}. Specifically we  will use the  modified variant  of FENE-P
model,  which matches   the  dynamics of original FENE in  near equilibrium  region  better than the
classical variant. This modification  is   achieved by a slight modification of Kramers definition of
the  stress tensor:
\begin{equation}
\label{stress_for_fenep} {\mbox{\boldmath $\tau$}}_{\ppm} =  nk_{\Bm} T (1- \theta b)\1  -
\left\langle \FF \QQ \right\rangle.
\end{equation}
The case $\theta =0$  gives the classical definition of FENE-P, while more thorough estimation
\cite{Birdreview,FENEP} is $\theta =(b(b+2))^{-1}$.

{\it Constants}

The specific feature  of the FENE  model is that the length of  dumbbells $\QQ$  can vary   only in a
bounded domain of $\mbox{\bf  R}^3$, namely inside a sphere $S_b=\left\{ Q^2 \leq b \right\}$. The
sphere  $S_b$ defines the domain of integration for averages $\left\langle \bullet
\right\rangle_{\emm} = \int_{S_b}\! \Psi_{\eqm} \bullet \dd \QQ$, where the equilibrium distribution
reads  $\Psi_{\eqm}=c^{-1} \left(1-Q^2/b \right)^{b/2}$, $c=\int_{S_b}  \left(1-Q^2/b \right)^{b/2}
\dd \QQ $.

In order to find   constants for the  zero-order  model (\ref{0order_dyn}) we do  the  following:
First  we  analytically compute  the lowest eigenfunctions of operator $J_{\dm}$: $g_1(Q^2)\oQQ$  and
$g_0(Q^2)$ without  account  of the  hydrodynamic  interaction ($\kappa=0$). The functions $g_0$ and
$g_1$  are   computed by  a   procedure presented in subsection ``Auxiliary formulas, 1" with the
help of the symbolic manipulation software {\it Maple V.3} \cite{Maple}. Then  we  calculate the
perturbations  terms $h_{0,1}$ by   formulas (\ref{g1g2_hydr}) introducing the  account  of
hydrodynamic interaction. The Table \ref{Tab:const}  presents the  constants $\lambda_{0,1}$,\ $a_i$,
\   $b_i$  (\ref{lambda_01}) (\ref{constants_0order}) of the  zero-order  model (\ref{0order_dyn})
without inclusion of hydrodynamic  interaction  $\kappa=0$ for several values of extensibility
parameter $b$. The relative error $\delta_{0,1}$ (see subsection  ``Auxiliary formulas, 1") of
approximation for these calculations did  not exceed the value $0.02$. The  Table
\ref{Tab:corrections}  shows the linear correction  terms  for constants from Tab. \ref{Tab:const}
which  account a  hydrodynamic interaction effect: \mbox{$\lambda^{\hm}_{0,1}=\lambda_{0,1}(1+\kappa
( \delta \lambda_{0,1}))$}, \mbox{$a^{\hm}_i=a_i(1+\kappa (\delta a_i))$},
\mbox{$b^{\hm}_i=b_i(1+\kappa (\delta b_i))$}. The latter   are calculated   by  substituting the
perturbed functions \mbox{$m_{0,1}=g_{0,1}+\kappa h_{0,1}$}   into (\ref{lambda_01})  and
(\ref{constants_0order}), and  expanding  them up to first-order  in $\kappa $. One can observe,
since $\kappa>0$, the effect of hydrodynamic interaction results in the  reduction   of the
relaxation times.

{\it Dynamic problems }

The rest  of  this section concerns the computations for two particular flows. The shear flow is
defined by
\begin{equation}
\kk(t) = \dotga (t) \left[
\begin{array}{ccc} 0& 1& 0 \\ 0& 0& 0 \\ 0 & 0 & 0
\end{array}
\right],
\end{equation}
where $\dotga (t)$ is the  shear rate, and the elongation flow corresponds to the  choice:
\begin{equation}
\kk (t)  = \doteps (t) \left[
\begin{array}{ccc} 1& 0& 0 \\ 0& -1/2& 0 \\ 0 & 0 & -1/2
\end{array}
\right],
\end{equation}
where $\doteps (t)$ is the elongation rate.

In test computations we will look  at  so  called  viscometric  material functions  defined through
the components of the polymeric part of the stress tensor $\ttau_{\ppm}$. Namely, for shear flow they
are the  shear viscosity $\nu$, the first and  the  second normal stress coefficients $\psi_1$,
$\psi_2$, and for elongation flow the only function is the elongation viscosity $\bar{\nu}$. In
dimensionless form they are written as:
\begin{eqnarray}
\label{nu} && \nuhat = \frac{\nu-\nu_{\sm}}{nk_{\Bm}T \lambda_{\rrm} } = - \frac{{\mbox{\boldmath
$\tau$}}_{\ppm,12}}{\dotgamhat  nk_{\Bm}T},  \\ \label{psi_1} && \widehat{\psi}_1 =
\frac{\psi_1}{nk_{\Bm}T \lambda^2_{\rrm}} = \frac{{\mbox{\boldmath
$\tau$}}_{\ppm,22}-{\mbox{\boldmath $\tau$}}_{\ppm,11}}{\dotgamhat^2 nk_{\Bm}T}, \\ \label{psi_2} &&
\widehat{\psi}_2= \frac{\psi_2}{nk_{\Bm}T \lambda^2_{\rrm}} = \frac{{\mbox{\boldmath
$\tau$}}_{p,33}-{\mbox{\boldmath $\tau$}}_{p,22}}{\dotgamhat^2 nk_{\Bm}T}, \\ \label{nuelon} &&
\nuelo = \frac{\bar{\nu} - 3\nu_{\sm}}{nk_{\Bm}T \lambda_{\rrm}} = \frac{{\mbox{\boldmath
$\tau$}}_{\ppm,22}-{\mbox{\boldmath $\tau$}}_{\ppm,11}}{\dotepshat nk_{\Bm}T},
\end{eqnarray}
where  $\dotgamhat = \dotga \lambda_{\rrm} $  and $\dotepshat= \doteps \lambda_{\rrm}$  are
dimensionless shear and elongation rates. Characteristic values of latter   parameters  $\dotgamhat$
and $\dotepshat$ allow to  estimate  the parameter  $\varepsilon_1$ (\ref{two_epsilon}). For all
flows considered  below  the second flow  parameter (Deborah number) $\varepsilon_2$  is equal to
zero.

\begin{table}[t]
\begin{center}
\caption{\label{Tab:const} Values of constants to the revised Oldroyd 8 constants model computed on
the base of the FENE dumbbells model} \vspace{5mm}
\begin{tabular}{|c|cccccc|}
\hline $ b$ &  $ \lambda_0$ & $\lambda_1$ &   $ b_0$ & $b_1$ &  $b_2$  & $a_0$
\\ \hline
20  &    1.498 &  1.329  &     -0.0742  &    0.221   &     1.019  &  0.927 \\ 50  &    1.198 &  1.135
&   -0.0326  &  0.279   &      1.024&  0.982  \\
100 &  1.099 &  1.068 &   -0.0179 &   0.303 &   1.015 &  0.990  \\
200 & 1.050 &  1.035 &   0.000053 &   0.328 &  1.0097  &  1.014 \\ $\infty$ & 1     & 1         & 0 &
1/3     & 1      & 1 \\ \hline
\end{tabular}
\end{center}
\end{table}

\begin{table}[t]
\begin{center}
\caption{\label{Tab:corrections} Corrections due to hydrodynamic interaction to the constants of the
revised  Oldroyd 8 constants model based on FENE force} \vspace{5mm}
\begin{tabular}{|c|cccccc|}
\hline $ b$ & $\delta  \lambda_0$ &  $\delta \lambda_1$ &  $\delta b_0$ & $\delta b_1$ &  $\delta
b_2$ & $\delta a_0$ \\ \hline
  20 &   -0.076 &      -0.101&  0.257  &   -0.080 &  -0.0487 &   -0.0664  \\
  50 &  -0.0618 &  -0.109  &   -0.365  &  0.0885  &  -0.0205  &   -0.0691  \\
  100 & -0.0574 &  -0.111  &    -1.020  &  0.109  & -0.020  &  -0.0603  \\ \hline
\end{tabular}
\end{center}
\end{table}

\begin{table}[t]
\begin{center}
\caption{\label{Tab:lim_values} Singular  values of   elongation rate} \vspace{5mm}
\begin{tabular}{|c|cccccc|}
\hline $b$ &  20 &  50 &   100  & 120 & 200& $\infty$ \\ \hline $\dotepshat^*$ &  0.864  &  0.632  &
0.566  &  0.555 &  0.520 & 0.5  \\  \hline
\end{tabular}
\end{center}
\end{table}

Let us  consider the steady state values of viscometric  functions  in steady shear and elongation
flows: $\dotga=const$,  $\doteps=const$. For the shear flow the steady values of  these  functions
are  found from  Eqs. (\ref{0order_dyn})  as  follows: $$\nuhat = b_2/(\lambda_1 - c \dotgamhat^2),
\qquad \widehat{\psi}_1=2 \nuhat/\lambda_1,   \qquad \widehat{\psi}_2=2b_0\nuhat/\lambda_1,$$ where
$c=2/3(2 b^2_0 + 2 b_0 -1)/\lambda_1 + 2 b_1 a_0/\lambda_0$. Estimations for the constants  (see
Table I)  shows   that $c \leq 0$ for all values of $b$ (case $c=0$ corresponds to $b=\infty$), thus
all three functions are monotonically decreasing  in absolute value with increase of quantity
$\dotgamhat$, besides  the case when  $b=\infty$. Although  they qualitatively correctly predict the
shear thinning for large shear rates due  to  power law, but the exponent $-2$ of power dependence in
the limit of large $\dotgamhat$ from the  values $-0.66$ for parameter $\nuhat$  and $-1.33$  for
$\widehat{\psi}_1$ observed in Brownian dynamic  simulations \cite{Herrchen}. It is  explained  by
the fact that slopes of shear  thining lie  out  of the  applicability  domain  of  our  model. A
comparison with  BD simulations  and modified FENE-P  model is depicted in Fig. \ref{fig1Deb}.

The  predictions  for the  second normal stress coefficient indicate one more  difference between
revised Oldroyd 8 constant equation   and  FENE--P model. FENE--P  model   shows  identically  zero
values  for $\widehat{\psi}_2$  in any  shear  flow, either  steady  or  time dependent, while the
model (\ref{0order_dyn}),  as well as BD  simulations (see Fig. 9 in Ref. \cite{Herrchen})  predict
small,  but  nonvanishing   values for this  quantity. Namely, due to  the  model  (\ref{0order_dyn})
in  shear flows the following relation   $\widehat{\psi}_2=b_0 \widehat{\psi}_1$  is always  valid,
with proportionality  coefficient  $b_0$ small and mostly negative,   what leads to small and  mostly
negative values of $\widehat{\psi}_2$.

In the elongation flow the steady state value to $\nuelo$  is found as:
\begin{equation}
\nuelo = \frac{3b_2}{\lambda_1 - \frac{5}{6}(2b_0 + 1)\dotepshat - 7 b_1 a_0 \dotepshat^2/\lambda_0
}.
\end{equation}
 The denominator has  one
root on positive semi--axis
\begin{equation}
\label{limit_eps} \dotepshat_* = - \frac{5\lambda_0(2b_0+1)}{84 b_1 a_0} +
\left(\left(\frac{5\lambda_0(2b_0+1)}{84 b_1 a_0}\right)^2 + \frac{\lambda_1\lambda_0}{7 b_1
a_0}\right)^{1/2},
\end{equation}
which defines a  singularity point for the dependence $\nuelo (\dotepshat)$. The BD  simulation
experiments \cite{Herrchen} on the  FENE dumbbell models shows  that there is no divergence of
elongation viscosity for all  values of elongation rate   (see Fig. \ref{fig2Deb}). For Hookean
springs $\dotepshat_*=1/2$ while in our model (\ref{0order_dyn}) the singularity point shifts  to
higher values  with respect to  decreasing  values  of  $b$ as it is demonstrated in  Table
\ref{Tab:lim_values}.

The Figure  \ref{fig3Deb}  gives an  example of dynamic behavior for elongation viscosity   in the
instant start-up  of the elongational  flow. Namely  it  shows the evolution of initially  vanishing
polymeric stresses after  instant  jump  of elongation rate at the time  moment $t=0$  from the value
$\dotepshat=0$  to the value $\dotepshat=0.3$.

It is possible to conclude that the revised Oldroyd 8 constants model (\ref{0order_dyn}) with
estimations given by (\ref{constants_0order}) for small and  moderate rates of strain up to
$\varepsilon_1=\lambda_{\rrm}|\dotgam|/(2\lambda_1) \sim 0.5$   yields a good approximation to
original FENE dynamics. The quality of the approximation in this interval is the  same or   better
than the one of the  nonlinear FENE-P model.

\addcontentsline{toc}{subsubsection}{The main results of this Example}

\subsubsection*{\textbf{The main results of this Example are as follows:}}

\begin{itemize}

\item[(i)] We have developed a systematic method of constructing  constitutive
equations from  the  kinetic dumbbell models for the polymeric solutions. The method is free from
a'priori assumptions about the form of the  spring force  and  is consistent with basic physical
requirements: frame invariance and  dissipativity of  internal motions of fluid. The  method extends
so-called  the method of invariant manifold  onto equations   coupled  with external  fields. Two
characteristic  parameters  of   fluid   flow   were distinguished   in  order to account for the
effect of the  presence of external fields. The iterative Newton scheme    for obtaining  a  slow
invariant manifold of the system  driven by the  flow with   relatively  low values of both
characteristic parameters was  developed.

\item[(ii)]
We demonstrated that the revised phenomenological Oldroyd 8 constants constitutive equations
represent  the slow dynamics of microscopic  elastic dumbbell model with any nonlinear spring force
in the limit  when the rate of strain and frequency of time variation of the flow are sufficiently
small and  microscopic   states at initial time of evolution  are  taken   not  far  from the
equilibrium.

\item[(iii)]
The corrections to the  zero-order manifold  lead generally  to  linear  in stresses  equations  but
with  highly nonlinear dependence on the  rate of strain  tensor  and its convective  derivatives.

\item[(iv)]
The zero-order  constitutive equation  is  compared to the direct Brownian dynamics simulation for
FENE dumbbell model  as well as  to predictions of FENE-P model. This comparison shows  that the
zero-order  constitutive equation  gives the correct predictions  in the domain of its  validity, but
does not  exclude  qualitative discrepancy occurring out of this domain, particularly in  elongation
flows.
\end{itemize}

This discrepancy calls for a further development, in particular, the use of nonlinear manifolds for
derivation of zero-order model.  The reason is in the necessity to provide  concordance  with the
requirement of  the positivity of  distribution function. It may  lead  to nonlinear constitutive
equation on  any order of  correction. These issues are currently under consideration and results
will be reported separately.

\clearpage \addcontentsline{toc}{subsection}{\textbf{Example 13:  Explosion of invariant manifold,
limits of macroscopic description for polymer molecules, molecular individualism, and multimodal
distributions}}

\subsection*{\textbf{Example 13: Explosion of invariant manifold, limits of macroscopic description
for polymer molecules,  molecular individualism, and multimodal distributions}}

Derivation of macroscopic equations from the simplest dumbbell models is revisited \cite{IK00}. It is
demonstrated that the onset of the macroscopic description is sensitive to the flows. For the FENE-P
model, small deviations from the Gaussian solution undergo a slow relaxation before the macroscopic
description sets on. Some consequences of these observations are discussed. A new class of closures
is discussed, the kinetic multipeak polyhedra. Distributions of this type are expected in  kinetic
models with multidimensional instability as universally, as the Gaussian distribution appears for
stable systems. The number of possible relatively stable states of a nonequilibrium system grows as
$2^m$, and the number of macroscopic parameters is in order $mn$, where $n$ is the dimension of
configuration space, and $m$ is the number of independent unstable directions in this space. The
elaborated class of closures and equations pretends to describe the effects of ``molecular
individualism".

\addcontentsline{toc}{subsubsection}{Dumbbell models and the problem of the classical Gaussian
solution stability}

\subsubsection*{\textbf{Dumbbell models and the problem of the classical Gaussian solution stability}}

We shall consider the simplest case of dilute polymer solutions represented by dumbbell models. The
dumbbell model reflects the two features of real--world macromolecules to be orientable and
stretchable by a flowing solvent \cite{Bird}.

Let us consider the simplest one-dimensional kinetic equation for the configuration distribution
function $\Psi(q,t)$, where $q$ is the reduced vector connecting the beads of the dumbbell. This
equation is slightly different from the usual Fokker-Planck equation. It is nonlinear, because of the
dependence of potential energy $U$ on the moment $M_{2}[\Psi]=\int q^{2}\Psi(q) dq$. This dependence
allows us to get the exact quasiequilibrium equations on $M_{2}$, but this equations are not solving
the problem: this quasiequilibrium manifold may become unstable when the flow is present \cite{IK00}.
Here is this model:
\begin{equation}\label{530}
\partial_{t}\Psi=-\partial_{q}\{\alpha(t)q\Psi\}+\frac{1}{2}\partial^{2}_{q}\Psi.
\end{equation}
Here
\begin{equation}\label{531}
\alpha(t)=\kappa(t)-\frac{1}{2}f(M_{2}(t)),
\end{equation}
$\kappa(t)$ is the given time-independent velocity gradient, $t$ is the reduced time, and the
function $-fq$ is the reduced spring force. Function $f$ may depend on the second moment of the
distribution function $M_{2}=\int q^{2}\Psi(q,t)dq$. In particular, the case $f\equiv1$ corresponds
to the linear Hookean spring, while $f=[1-M_{2}(t)/b]^{-1}$ corresponds to the self-consistent finite
extension nonlinear elastic spring (the FENE-P model, first introduced in \cite{FENEP}). The second
moment $M_{2}$ occurs in the FENE-P force $f$ as the result of the pre-averaging approximation to the
original FENE model (with nonlinear spring force $f=[1-q^{2}/b]^{-1}$). The parameter $b$ changes the
characteristics of the force law from Hookean at small extensions to a confining force for
$q^{2}\rightarrow b$. Parameter $b$ is roughly equal to the number of monomer units represented by
the dumbell and should therefore be a large number. In the limit $b\rightarrow\infty$, the Hookean
spring is recovered. Recently, it has been demonstrated that FENE-P model appears as first
approximation within a systematic self-confident expansion of nonlinear forces
\cite{GKIOeNONNEWT2001}.

Equation (\ref{530}) describes an ensemble of non-interacting dumbells subject to a
pseudo-elongational flow with fixed kinematics. As is well known, the Gaussian distribution function,
\begin{equation}\label{532}
\Psi^{G}(M_{2})=\frac{1}{\sqrt{2\pi M_{2}}}\exp\left[-\frac{q^{2}}{2M_{2}}\right],
\end{equation}
solves equation (\ref{530}) provided the second moment $M_{2}$ satisfies
\begin{equation}\label{533}
\frac{dM_{2}}{dt}=1+2\alpha(t)M_{2}.
\end{equation}
Solution (\ref{532}) and (\ref{533}) is the valid macroscopic description if all other solutions of
the equation (\ref{530}) are rapidly attracted to the family of Gaussian distributions (\ref{532}).
In other words \cite{GKTTSP94}, the special solution (\ref{532}) and (\ref{533}) is the macroscopic
description if equation (\ref{532}) is the stable invariant manifold of the kinetic equation
(\ref{530}). If not, then the Gaussian solution is just a member of the family of solutions, and
equation (\ref{533}) has no meaning of the macroscopic equation. Thus, the complete answer to the
question of validity of the equation (\ref{533}) as the macroscopic equation requires a study of
dynamics in the neighborhood of the manifold (\ref{532}). Because of the simplicity of the model
(\ref{530}), this is possible to a satisfactory level even for $M_{2}$-dependent spring forces.

\addcontentsline{toc}{subsubsection}{Dynamics of the moments and explosion of the Gaussian manifold}

\subsubsection*{\textbf{Dynamics of the moments and explosion of the Gaussian manifold}}

In the paper \cite{IK00} it was shown, that there is a possibility of ``explosion" of the Gaussian
manifold: with the small initial deviation from it, the solutions of the equation (\ref{530}) are
very fast going far from, and then slowly come back to the stationary point which is located on the
Gaussian manifold. The distribution function $\Psi$ is stretched fast, but looses the Gaussian form,
and after that the Gaussian form recovers slowly with the new value of $M_{2}$. Let us describe
briefly the results of \cite{IK00}.

Let $M_{2n}=\int q^{2n}\Psi dq$ denote the even moments (odd moments vanish by symmetry). We consider
deviations $\mu_{2n}=M_{2n}-M_{2n}^{\rm G}$, where $M_{2n}^{\rm G}=\int q^{2n} \Psi^{\rm G}dq$ are
moments of the Gaussian distribution function (\ref{532}). Let $\Psi(q,t_0)$ be the initial condition
to the Eq.\ (\ref{530}) at time $t=t_0$. Introducing functions,
\begin{equation}
\label{result0} p_{2n}(t,t_0)=\exp\left[4n\int_{t_0}^{t}\alpha(t')dt'\right],
\end{equation}
where $t\ge t_0$, and $2n \ge 4$, the {\it exact} time evolution of the deviations $\mu_{2n}$ for
$2n\ge 4$ reads
\begin{equation} \label{result1}
    \mu_4(t)=p_4(t,t_0)\mu_4(t_0),
\end{equation}
and
\begin{equation} \label{result2}
    \mu_{2n}(t)=\left[ \mu_{2n}(t_0) + 2n(4n-1)\int_{t_0}^t
    \mu_{2n-2}(t')p_{2n}^{-1}(t',t_0)dt' \right] p_{2n}(t,t_0),
\end{equation}
for $2n\ge 6$. Equations (\ref{result0}), (\ref{result1}) and (\ref{result2}) describe evolution near
the Gaussian solution for arbitrary initial condition $\Psi(q,t_0)$. Notice that explicit evaluation
of the integral in the Eq.\ (\ref{result0}) requires solution to the moment equation (\ref{533})
which is not available in the analytical form for the FENE-P model.

It is straightforward to conclude that any solution with a non-Gaussian initial condition converges
to the Gaussian solution asymptotically as $t\to\infty$ if

\begin{equation}
\label{result3} \lim_{t\to\infty}\int_{t_0}^t\alpha(t')dt'<0.
\end{equation}
However, even if this asymptotic condition is met, deviations from the Gaussian solution may survive
for considerable {\it finite} times. For example, if for some finite time $T$, the integral in the
Eq.\ (\ref{result0}) is estimated as $\int_{t_0}^t\alpha(t')dt'>\alpha (t-t_0)$, $\alpha>0$, $t\le
T$, then the Gaussian solution becomes exponentially unstable during this time interval. If this is
the case, the moment equation (\ref{533}) cannot be regarded as the macroscopic equation. Let us
consider specific examples.

For the Hookean spring ($f\equiv 1$) under a constant elongation ($\kappa={\rm const}$), the Gaussian
solution is exponentially stable for $\kappa<0.5$, and it becomes exponentially unstable for
$\kappa>0.5$. The exponential instability in this case is accompanied by the well known breakdown of
the solution to the Eq.\ (\ref{533}) due to infinite stretching of the dumbbell. The situation is
much more interesting for the FENE-P model because this nonlinear spring force does not allow the
infinite stretching of the dumbbell.

Eqs.\ (\ref{533}) and (\ref{result1}) were integrated  by the 5-th order Runge-Kutta method with
adaptive time step. The FENE-P parameter $b$ was set equal to 50. The initial condition was
$\Psi(q,0)=C(1-q^2/b)^{b/2}$, where $C$ is the normalization (the equilibrium of the FENE model,
notoriously close to the FENE-P equilibrium \cite{Herrchen}). For this initial condition, in
particular, $\mu_4(0)=-6b^2/[(b+3)^2(b+5)]$ which is about 4$\%$ of the value of $M_4$ in the
Gaussian equilibrium for $b=50$. In Fig.~\ref{EPJ713_fig} we demonstrate deviation $\mu_4(t)$ as a
function of time for several values of the flow. Function $M_2(t)$ is also given for comparison. For
small enough $\kappa$ we find an adiabatic regime, that is $\mu_4$ relaxes exponentially to zero. For
stronger flows, we observe an initial {\it fast runaway} from the invariant manifold with $|\mu_4|$
growing over three orders of magnitude compared to its initial value. After the maximum deviation has
been reached, $\mu_4$ relaxes to zero. This relaxation is exponential as soon as the solution to Eq.\
(\ref{533}) approaches the steady state. However, the time constant for this exponential relaxation
$|\alpha_{\infty}|$ is very small. Specifically, for large $\kappa$,
\begin{equation}
\label{alpha_lim} \alpha_{\infty}=\lim_{t\to\infty}\alpha(t)=-\frac{1}{2b}+O(\kappa^{-1}).
\end{equation}
Thus, the steady state solution is unique and Gaussian but the stronger is the flow, the larger is
the initial runaway from the Gaussian solution, while the return to it thereafter becomes
flow-independent. Our observation demonstrates that, though the stability condition (\ref{result3})
is met, {\it significant deviations from the Gaussian solution persist over the times when the
solution of Eq.}\ (\ref{533}) {\it is already reasonably close to the stationary state.} If we accept
the usually quoted physically reasonable minimal value of parameter $b$ of the order $20$ then the
minimal relaxation time is of order $40$ in the reduced time units of Fig.~\ref{EPJ713_fig}. We
should also stress that the two limits, $\kappa\to\infty$ and $b\to\infty$, are not commutative, thus
it is not surprising that the estimation (\ref{alpha_lim}) does not reduce to the above mentioned
Hookean result as $b\to\infty$. Finally, peculiarities of convergence to the Gaussian solution are
even furthered if we consider more complicated (in particular, oscillating) flows $\kappa(t)$.
Further numerical experiments are presented in \cite{drugs}. The statistics of FENE-P solutions with
random strains was studied recently by J.-L. Thiffeault \cite{JLT}

\addcontentsline{toc}{subsubsection}{Two-peak approximation for polymer stretching in flow and
explosion of the Gaussian manifold}

\subsubsection*{\textbf{Two-peak approximation for polymer stretching in flow and explosion of the Gaussian
manifold}}

In accordance with \cite{Legendre} the ansatz for $\Psi$ can be suggested in the following form:
\begin{equation}\label{Anz}
\Psi^{An}(\{\sigma,\varsigma\},q)=
\frac{1}{2\sigma\sqrt{2\pi}}\left(e^{-\frac{(q+\varsigma)^{2}}{2\sigma^{2}}}+
e^{-\frac{(q-\varsigma)^{2}}{2\sigma^{2}}}\right).
\end{equation}
Natural inner coordinates on this manifold are $\sigma$ and $\varsigma$. Note, that now
$\sigma^{2}\neq M_{2}$. The value $\sigma^{2}$ is a dispersion of one of the Gaussian summands in
(\ref{Anz}),
\begin{eqnarray*}
M_{2}(\Psi^{An}(\{\sigma,\varsigma\},q))=\sigma^{2}+\varsigma^{2}.
\end{eqnarray*}
To build the thermodynamic projector on the manifold (\ref{Anz}), the thermodynamic Lyapunov function
is necessary. It is necessary to emphasize, that equations (\ref{530}) are nonlinear. For such
equations, the arbitrarity in the choice of the thermodynamic Lyapunov function is much smaller than
for the linear Fokker Planck equation. Nevertheless, such a function exists. It is the free energy
\begin{equation}\label{Free}
F=U(M_{2}[\Psi])-TS[\Psi],
\end{equation}
where
\begin{eqnarray*}
S[\Psi]=-\int\Psi(\ln\Psi-1)dq,
\end{eqnarray*}
$U(M_{2}[\Psi])$ is the potential energy in the mean field approximation, $T$ is the temperature
(further we assume that $T=1$).

 Note, that Kullback-form entropy \cite{Kull}
$S_{k}=-\int\Psi\ln\left(\frac{\Psi}{\Psi^{*}}\right)$ also has the form $S_{k}=-F/T$:
\begin{eqnarray*}
\Psi^{*}=\exp(-U),\\ S_{k}[\Psi]=-\langle U\rangle-\int\Psi\ln\Psi dq.
\end{eqnarray*}
If $U(M_{2}[\Psi])$ in the mean field approximation is the convex function of $M_{2}$, then the free
energy (\ref{Free}) is the convex functional too.

For the FENE-P model $U=-\ln[1-M_{2}/b]$.

In accordance to the thermodynamics the vector $I$ of flow of $\Psi$ must be proportional to the
gradient of the corresponding chemical potential $\mu$:
\begin{equation}\label{Flux}
I=-B(\Psi)\nabla_{q}\mu,
\end{equation}
where $\mu=\frac{\delta F}{\delta\Psi}$, $B\geq0$. From the equation (\ref{Free}) it follows, that
\begin{eqnarray}\label{muflux}
\mu=\frac{d U(M_{2})}{d M_{2}}\cdot q^{2}+\ln\Psi\nonumber\\ I=-B(\Psi)\left[2\frac{dU}{dM_{2}}\cdot
q+\Psi^{-1}\nabla_{q}\Psi\right].
\end{eqnarray}
If we suppose here $B=\frac{D}{2}\Psi$, then we get
\begin{eqnarray}\label{TDeq}
I=-D\left[\frac{dU}{dM_{2}}\cdot q\Psi+\frac{1}{2}\nabla_{q}\Psi\right]\nonumber\\
\frac{\partial\Psi}{\partial t}={\rm div}_{q}I=D\frac{d U(M_{2})}{d
M_{2}}\partial_{q}(q\Psi)+\frac{D}{2}\partial^{2}q\Psi,
\end{eqnarray}
When $D=1$ this equations coincide with (\ref{530}) in the absence of the flow: due to equation
(\ref{TDeq}) $dF/dt\leq0$.

Let us construct the thermodynamic projector with the help of the thermodynamic Lyapunov function $F$
(\ref{Free}). Corresponding entropic scalar product at the point $\Psi$ has the form
\begin{equation}\label{Scal}
\left.\langle f|g\rangle_{\Psi}=\frac{d^{2}U}{dM_{2}^{2}}\right|_{M_{2}=M_{2}[\Psi]}\cdot\int
q^{2}f(q)dq\cdot\int q^{2}g(q)dq+\int\frac{f(q)g(q)}{\Psi(q)}dq
\end{equation}
During the investigation of the ansatz (\ref{Anz}) the scalar product (\ref{Scal}), constructed for
the corresponding point of the Gaussian manifold with $M_{2}=\sigma^{2}$, will be used. It will let
us to investigate the neighborhood of the Gaussian manifold (and to get all the results in the
analytical form):
\begin{equation}\label{ScalG}
\left.\langle f|g\rangle_{\sigma^{2}}=\frac{d^{2}U}{dM_{2}^{2}}\right|_{M_{2}=\sigma^{2}}\cdot\int
q^{2}f(q)dq\cdot\int q^{2}g(q)dq+\sigma\sqrt{2\pi}\int e^{\frac{q^{2}}{2\sigma^{2}}}f(q)g(q)dq
\end{equation}
Also we will need to know the functional $DF$ at the point of Gaussian manifold:
\begin{equation}\label{Prod}
\left.DF_{\sigma^{2}}(f)=\left(\frac{d U(M_{2})}{dM_{2}}\right|_{M_{2}=\sigma^{2}}
-\frac{1}{2\sigma^{2}}\right)\int q^{2}f(q)dq,
\end{equation}
\noindent (with the condition $\int f(q)dq=0$). The point
\begin{eqnarray*}
\left.\frac{d U(M_{2})}{dM_{2}}\right|_{M_{2}=\sigma^{2}}=\frac{1}{2\sigma^{2}},
\end{eqnarray*}
corresponds to the equilibrium.

The tangent space to the manifold (\ref{Anz}) is spanned by the vectors
\begin{eqnarray}\label{basis}
&&f_{\sigma}=\frac{\partial\Psi^{An}}{\partial(\sigma^{2})}; \:
f_{\varsigma}=\frac{\partial\Psi^{An}}{\partial(\varsigma^{2})};\nonumber\\
f_{\sigma}&=&\frac{1}{4\sigma^{3}\sqrt{2\pi}}\left[e^{-\frac{(q+\varsigma)^{2}}{2\sigma^{2}}}
\frac{(q+\varsigma)^{2}-\sigma^{2}}{\sigma^{2}}+e^{-\frac{(q-\varsigma)^{2}}{2\sigma^{2}}}
\frac{(q-\varsigma)^{2}-\sigma^{2}}{\sigma^{2}} \right];\\
f_{\varsigma}&=&\frac{1}{4\sigma^{2}\varsigma\sqrt{2\pi}}\left[-e^{-\frac{(q+\varsigma)^{2}}{2\sigma^{2}}}
\frac{q+\varsigma}{\sigma}+e^{-\frac{(q-\varsigma)^{2}}{2\sigma^{2}}} \frac{(q-\varsigma)}{\sigma}
\right];\nonumber
\end{eqnarray}
The Gaussian entropy (free energy) production in the directions $f_{\sigma}$ and $f_{\varsigma}$
(\ref{Prod}) has a very simple form:
\begin{eqnarray}\label{Fpro}
\left.DF_{\sigma^{2}}(f_{\varsigma})=DF_{\sigma^{2}}(f_{\sigma})=\frac{d
U(M_{2})}{dM_{2}}\right|_{M_{2}=\sigma^{2}}-\frac{1}{2\sigma^{2}}.
\end{eqnarray}
The linear subspace $\ker DF_{\sigma^{2}}$ in $lin\{f_{\sigma},f_{\varsigma}\}$ is spanned by the
vector $f_{\varsigma}-f_{\sigma}$.

Let us have the given vector field $d\Psi/dt=J(\Psi)$ at the point $\Psi(\{\sigma,\varsigma\})$. We
need to build the projection of $J$ onto the tangent space $T_{\sigma,\varsigma}$ at the point
$\Psi(\{\sigma,\varsigma\})$:
\begin{equation}\label{Prosigma}
P^{th}_{\sigma,\varsigma}(J)=\varphi_{\sigma}f_{\sigma}+\varphi_{\varsigma}f_{\varsigma}.
\end{equation}
This equation means, that the equations for $\sigma^{2}$ and $\varsigma^{2}$ will have the form
\begin{equation}\label{eqsigma}
\frac{d\sigma^{2}}{dt}=\varphi_{\sigma};\:\: \frac{d\varsigma^{2}}{dt}=\varphi_{\varsigma}
\end{equation}
Projection $(\varphi_{\sigma},\varphi_{\varsigma})$ can be found from the following two equations:
\begin{eqnarray}\label{psieq}
\varphi_{\sigma}+\varphi_{\varsigma}=\int q^{2}J(\Psi)(q)dq\nonumber;\\ \langle
\varphi_{\sigma}f_{\sigma}+\varphi_{\varsigma}f_{\varsigma}|f_{\sigma}-f_{\varsigma}\rangle_{\sigma^{2}}
=\langle J(\Psi)|f_{\sigma}-f_{\varsigma}\rangle_{\sigma^{2}},
\end{eqnarray}
where $\langle f|g\rangle_{\sigma^{2}}=\langle J(\Psi)|f_{\sigma}-f_{\varsigma}\rangle_{\sigma^{2}}$,
(\ref{Scal}). First equation of (\ref{psieq}) means, that the time derivative $dM_{2}/dt$ is the same
for the initial and the reduced equations. Due to the formula for the dissipation of the free energy
(\ref{Prod}), this equality is equivalent to the persistence of the dissipation in the neighborhood
of the Gaussian manifold. Indeed, in according to (\ref{Prod}) $dF/dt=A(\sigma^{2})\int q^2
J(\Psi)(q)dq= A(\sigma^{2}) dM_2/dt$, where $A(\sigma^{2})$ does not depend of $J$. On the other
hand, time derivative of $M_2$ due to projected equation (\ref{eqsigma}) is
$\varphi_{\sigma}+\varphi_{\varsigma}$, because $M_2=\sigma^2+\varsigma^2$.

The second equation in (\ref{psieq}) means, that $J$ is projected orthogonally on $\ker DS\bigcap
T_{\sigma,\varsigma}$. Let us use the orthogonality with respect to the entropic scalar product
(\ref{ScalG}). The solution of equations (\ref{psieq}) has the form
\begin{eqnarray}\label{projphi}
\frac{d\sigma^{2}}{dt}=\varphi_{\sigma}=\frac{\langle
J|f_{\sigma}-f_{\varsigma}\rangle_{\sigma^{2}}+M_{2}(J)(\langle
f_{\varsigma}|f_{\varsigma}\rangle_{\sigma^{2}}-\langle
f_{\sigma}|f_{\varsigma}\rangle_{\sigma^{2}})}{\langle
f_{\sigma}-f_{\varsigma}|f_{\sigma}-f_{\varsigma}\rangle_{\sigma^{2}}}\nonumber,\\\\
\frac{d\varsigma^{2}}{dt}=\varphi_{\varsigma}=\frac{-\langle
J|f_{\sigma}-f_{\varsigma}\rangle_{\sigma^{2}}+M_{2}(J)(\langle
f_{\sigma}|f_{\sigma}\rangle_{\sigma^{2}}-\langle
f_{\sigma}|f_{\varsigma}\rangle_{\sigma^{2}})}{\langle
f_{\sigma}-f_{\varsigma}|f_{\sigma}-f_{\varsigma}\rangle_{\sigma^{2}}}\nonumber,
\end{eqnarray}
where $J=J(\Psi)$, $M_{2}(J)=\int q^{2}J(\Psi)dq$.

It is easy to check, that the formulas (\ref{projphi}) are indeed defining the projector: if
$f_{\sigma}$ (or $f_{\varsigma}$) is substituted there instead of the function $J$, then we will get
$\varphi_{\sigma}=1, \varphi_{\varsigma}=0$ (or $\varphi_{\sigma}=0, \varphi_{\varsigma}=1$,
respectively). Let us substitute the right part of the initial kinetic equations (\ref{530}),
calculated at the point $\Psi(q)=\Psi(\{\sigma,\varsigma\},q)$ (see the equation (\ref{Anz})) in the
equation (\ref{projphi}) instead of $J$. We will get the closed system of equations on $\sigma^{2},
\varsigma^{2}$ in the neighborhood of the Gaussian manifold.

This system describes the dynamics of the distribution function $\Psi$. The distribution function is
represented as the half-sum of two Gaussian distributions with the averages of distribution
$\pm\varsigma$ and mean-square deviations $\sigma$. All integrals in the right-hand part of
(\ref{projphi}) are possible to calculate analytically.

Basis $(f_{\sigma},f_{\varsigma})$ is convenient to use everywhere, except the points in the Gaussian
manifold, $\varsigma=0$, because if $\varsigma\rightarrow0$, then
\begin{eqnarray*}
f_{\sigma}-f_{\varsigma}=O\left(\frac{\varsigma^{2}}{\sigma^{2}}\right)\rightarrow0.
\end{eqnarray*}

Let us analyze the stability of the Gaussian manifold to the ``dissociation" of the Gaussian peak in
two peaks (\ref{Anz}). To do this, it is necessary to find first nonzero term in the Taylor expansion
in $\varsigma^{2}$ of the right-hand side of the second equation in the system (\ref{projphi}). The
denominator has the order of $\varsigma^{4}$, the numerator has, as it is easy to see, the order not
less, than $\varsigma^{6}$ (because the Gaussian manifold is invariant with respect to the initial
system).

With the accuracy up to $\varsigma^{4}$:
\begin{equation}\label{itog}
\frac{1}{\sigma^{2}}\frac{d\varsigma^{2}}{dt}=2\alpha\frac{\varsigma^{2}}{\sigma^{2}}+
o\left(\frac{\varsigma^{4}}{\sigma^{4}}\right),
\end{equation}
where $$\alpha=\kappa - \left. \frac{d U(M_{2})}{dM_{2}}\right|_{M_{2}=\sigma^2}.$$

So, if $\alpha>0$, then $\varsigma^{2}$ grows exponentially ($\varsigma\sim e^{\alpha t}$) and the
Gaussian manifold is unstable; if $\alpha<0$, then $\varsigma^{2}$ decreases exponentially and the
Gaussian manifold is stable.

Near the vertical axis $d\sigma^2/dt = 1+2\alpha \sigma^2$\footnote{Pavel Gorban calculated the
projector (\ref{projphi}) analytically without Taylor expansion and with the same, but exact result:
$d\varsigma^{2}/dt=2\alpha \varsigma^{2}$ $d\sigma^2/dt = 1+2\alpha \sigma^2$.} The form of the phase
trajectories is shown qualitative on Fig. \ref{figFENEP}. Note that this result completely agrees
with equation (\ref{result1}).

For the real equation FPE (for example, with the FENE potential) the motion in presence of the flow
can be represented as the motion in the effective potential well $\tilde{U}(q)=U(q)-\frac{1}{2}\kappa
q^{2}$. Different variants of the phase portrait for the FENE potential are present on Fig.
\ref{figFENE}. Instability and dissociation of the unimodal distribution functions (``peaks") for the
FPE is the general effect when the flow is present.

The instability occurs when the matrix $\partial^{2}\tilde{U}/\partial q_{i}\partial q_{j}$ starts to
have negative eigenvalues ($\tilde{U}$ is the effective potential energy, $\tilde{U}(q)=U(q)- {1
\over 2} \sum_{i,j}\kappa_{i,j}q_{i}q_{j}$).

\addcontentsline{toc}{subsubsection}{Polymodal polyhedron and molecular individualism}

\subsubsection*{\textbf{Polymodal polyhedron and molecular individualism}}

The discovery of the molecular individualism for  dilute polymers in the flow \cite{Chu} was the
challenge to theory from the very beginning. ``Our data should serve as a guide in developing
improved microscopic theories for polymer dynamics"... was the concluding sentence of the paper
\cite{Chu}. P. de Gennes invented the term ``molecular individualism" \cite{DeGenne}. He stressed
that in this case the usual averaging procedures are not applicable. At the highest strain rates
distinct conformation shapes with different dynamics were observed \cite{Chu}. Further works for
shear flow demonstrated not only shape differences, but different large temporal fluctuations
\cite{Chu2}.

Equation for the molecules in a flow are known. These are the Fokker-Planck equations with external
force. The theory of the molecular individualism is hidden inside these equations. Following the
logic of model reduction we should solve two problems: to construct the slow manifold, and to project
the equation on this manifold. The second problem is solved: the thermodynamic projector is necessary
for this projection.

How to solve the first problem? We can find a hint in previous subsections. The Gaussian
distributions form the invariant manifold for the FENE-P model of polymer dynamics, but, as it was
discovered in \cite{IK00}, this manifold can become unstable in the presence of a flow. We propose to
model this instability as dissociation of the Gaussian peak into two peaks. This dissociation
describes appearance of an unstable direction in the configuration space.

In the classical FENE-P model of polymer dynamics a polymer molecule is represented by one
coordinate: the stretching of molecule (the connector vector between the beads). There exists a
simple mean field generalized models for multidimensional configuration spaces of molecules. In these
models dynamics of distribution functions is described by the Fokker-Planck equation in a quadratic
potential well. The matrix of coefficients of this quadratic potential depends on the matrix of the
second order moments of the distribution function. The Gaussian distributions form the invariant
manifold for these models, and the first dissociation of the Gaussian peak after appearance of the
unstable direction in the configuration space has the same nature and description, as for the
one-dimensional models of molecules considered below.

At the highest strain there can appear new unstable directions, and corresponding dissociations of
Gaussian peaks form a {\it cascade} of dissociation. For $m$ unstable directions we get the Gaussian
parallelepiped: The distribution function is represented as a sum of $2^m$ Gaussian peaks located in
the vertixes of parallelepiped:

\begin{eqnarray}\label{parall}
&&\Psi(q)=\nonumber \\ &&{1 \over 2^m(2\pi)^{n/2}\sqrt{\det \Sigma}} \sum_{\varepsilon_i=\pm 1, \,
(i=1, \ldots, m)} \exp\left(-\frac{1}{2}\left(\Sigma^{-1}\left(q+\sum_{i=1}^m \varepsilon_i
\varsigma_i \right), \: q+\sum_{i=1}^m \varepsilon_i \varsigma_i\right)\right),
\end{eqnarray}
where $n$ is dimension of configuration space, $2\varsigma_i$ is the vector of the $i$th edge of the
parallelepiped, $\Sigma$ is the one peak covariance matrix (in this model $\Sigma$ is the same for
all peaks). The macroscopic variables for this model are:
\begin{enumerate}
\item The covariance matrix $\Sigma$ for one peak;
\item The set of vectors $\varsigma_i$ (or the parallelepiped edges).
\end{enumerate}

The stationary polymodal distribution for the Fokker-Planck equation corresponds to the persistence
of several local minima of the function $\tilde{U}(q)$. The multidimensional case is different from
one-dimensional because it has the huge amount of possible configurations. An attempt to describe
this picture quantitative meet the following obstacle: we do not know the potential $U$, on the other
hand, the effect of molecular individualism \cite{Chu,DeGenne,Chu2} seems to be universal in its
essence, without dependence of the qualitative picture on details of interactions. We should find a
mechanism that is as general, as the effect. The simplest dumbbell model which we have discussed in
previous subsection does not explain the effect, but it gives us a hint: the flow can violate the
stability of unimodal distribution. If we assume that the whole picture is hidden insight a
multidimensional Fokker-Planck equation for a large molecule in a flow, then we can use this hint in
such a way: when the flow strain grows there appears a sequence of bifurcations, and for each of them
a new unstable direction arises. For qualitative description of such a picture we can apply a
language of normal forms \cite{ArVarGZ1995-1998}, but with some modification.

The bifurcation  in dimension one with appearance of two point of minima from one point has the
simplest polynomial representation: $U(q, \alpha)= q^4+ \alpha q^2$. If $\alpha \geq 0$, then this
potential has one minimum, if $\alpha < 0$, then there are two points of minima. The normal form of
degenerated singularity is $U(q)=q^4$. Such polynomial forms as $q^4+ \alpha q^2$ are very simple,
but they have inconvenient asymptotic at $q \rightarrow \infty$. For our goals it is more appropriate
to use logarithms of convex combinations of Gaussian distributions instead of polynomials. It is the
same class of jets near the bifurcation, but with given quadratic asymptotic $q \rightarrow \infty$.
If one needs another class of asymptotic, it is possible just to change the choice of the basic peak.
All normal forms of the critical form of functions, and families of versal deformations  are well
investigated and known \cite{ArVarGZ1995-1998}.

Let us represent the deformation of the probability distribution under the strain in multidimensional
case as a cascade of peak dissociation. The number of peaks will duplicate on the each step. The
possible cascade of peaks dissociation is presented qualitatively on Fig. \ref{Cartoon}. The
important property of this qualitative picture is the linear complexity of dynamical description with
exponential complexity of geometrical picture. Let $m$ be the number of bifurcation steps in the
cascade. Then
\begin{itemize}
\item{For description of parallelepiped  it is sufficient to describe
$m$ edges;}
\item{There are $2^{m-1}$ geometrically different conformations associated with $2^{m}$
vertex of parallelepiped (central symmetry halved this number).}
\end{itemize}
Another important property is the {\it threshold} nature of each dissociation: It appears in points
of stability loss for new directions, in these points the dimension of unstable direction increases.

Positions of peaks correspond to parallelepiped vertices. Different vertices in configuration space
present different geometric forms. So, it seems {\it plausible}\footnote{We can not {\it prove} it
now, and it is necessary to determine the status of proposed qualitative picture: it is much more
general than a specific model, it is the mechanism which acts in a wide class of models. The cascade
of instabilities can appear and, no doubt, it appears for the Fokker-Planck equation for a large
molecule in a flow. But it is not proven yet that the effects observed in well-known experiments have
exactly this mechanism. This proof requires quantitative verification of a specific model. And now we
talk not about a proven, but about the plausible mechanism which typically appears for systems with
instabilities.} that observed different forms (``dumbbels", ``half-dumbbels", ``kinked", ``folded"
and other, not classified forms) correspond to these vertices of parallelepiped. Each vertex is a
metastable state of a molecule and has its own basin of attraction. A molecule goes to the vertex
which depends strongly on details of initial conditions.

The simplest multidimensional dynamic model is the Fokker-Planck equation with quadratic mean field
potential. This is direct generalization of the FENE-P model: the quadratic potential $U(q)$ depends
on the tensor of second moments $\MM_{2}= \langle q_iq_j \rangle$ (here the angle brackets denote the
averaging). This dependence should provide the finite extensibility. This may be, for example, a
simple matrix generalization of the FENE-P energy: $$U(q)=\sum_{ij}K_{ij}q_iq_j, \: \KK=\KK^0 +
\phi(\MM_{2}/b), \: \langle U(q)\rangle ={\rm tr}(\KK \MM_{2}/b) $$ where $b$ is a constant (the
limit of extensibility), $\KK^0$ is a constant matrix, $\MM_{2}$ is the matrix of second moments, and
$\phi$ is a positive analytical monotone increasing function of one variable on the interval $(0,1)$,
$\phi(x)\rightarrow \infty$ for $x \rightarrow  1$ (for example, $\phi(x)=-\ln(1-x)/x$, or
$\phi(x)=(1-x)^{-1}$).

For quadratic multidimensional mean field models persists the qualitative picture of Fig.
\ref{figFENEP}: there is {\it non-stationary moleqular individualism for stationary ``molecular
collectivism"}. The stationary distribution  is the Gaussian distribution, and on the way to this
stationary point there exists an unstable region, where the distribution dissociates onto $2^m$ peaks
($m$ is the number of unstable degrees of freedom).

Dispersion of individual peak in unstable region increases too. This effect can deform the observed
situation: If some of the peaks have  significant intersection, then these peaks join into new
extended classes of observed molecules. The stochastic walk of molecules between connected peaks can
be observed as ``large non-periodical fluctuations". This walk can be unexpected fast, because it can
be effectively a {\it motion in a low-dimensional space}, for example, in one-dimensional space (in a
neighborhood of a part of one-dimensional skeleton of the polyhedron).

We discussed the important example of ansatz: the multipeak models. Two examples of these type of
models demonstrated high efficiency during decades: the Tamm--Mott-Smith bimodal ansatz for shock
waves, and the the Langer--Bar-on--Miller \cite{Spinodal,Grant,Kum} approximation for spinodal
decomposition.

The multimodal polyhedron appears every time as an appropriate approximation for distribution
functions for systems with instabilities. We create such an approximation for the Fokker--Planck
equation for polymer molecules in a flow. Distributions of this type are expected to appear in each
kinetic model with multidimensional instability as universally, as Gaussian distribution appears for
stable systems. This statement needs a clarification: everybody knows that the Gaussian distribution
is stable with respect to convolutions, and the appearance of this distribution is supported by
central limit theorem. Gaussian polyhedra form a stable class: convolution of two Gaussian polyhedra
is a Gaussian polyhedron, convolution of a Gaussian polyhedron with a Gaussian distribution is a
Gaussian polyhedron with the same number of vertices. On the other hand, a Gaussian distribution in a
potential well appears as an exponent of a quadratic form which represents the simplest stable
potential (a normal form of a nondegenerated critical point). Families of Gaussian parallelepipeds
appear as versal deformations with given asymptotic for systems with cascade of simplest
bifurcations.

The usual point of view is: The shape of the polymers in a flow is either a coiled ball, or a
stretched ellipsoid, and the Fokker--Planck equation describes the stretching from the ball to the
ellipsoid. It is not the whole truth, even for the FENE-P equation, as it was shown in ref.
\cite{IK00,Legendre}. The Fokker-Planck equation describes the shape of a probability cloud in the
space of conformations. In the flow with increasing strain this shape changes from the ball to the
ellipsoid, but, after some thresholds, this ellipsoid transforms into a multimodal distribution which
can be modeled  as the peak parallelepiped. The peaks describe the finite number of possible molecule
conformations. The number of this distinct conformations grows for a parallelepiped as $2^{m}$ with
the number $m$ of independent unstable direction. Each vertex has its own basin of attraction. A
molecule goes to the vertex which depends strongly on details of initial conditions.

These models pretend to be the kinetic basis for the theory of molecular individualism. The detailed
computations will be presented in following works, but some of the qualitative features of the models
are in agreement with some of  qualitative features of the picture observed in experiment
\cite{Chu,DeGenne,Chu2}: effect has the threshold character, different observed conformations depend
significantly on the initial conformation and orientation.

Some general questions remain open:

\begin{itemize}
\item{Of course, appearance of $2^m$ peaks in the Gaussian
parallelepiped is possible, but some of these peaks can join in following dynamics, hence the first
question is: what is the typical number of significantly different peaks for a $m-$dimensional
instability?}
\item{How can we decide what scenario is more realistic from the experimental
point of view: the proposed universal kinetic mechanism, or the scenario with long living metastable
states (for example, the relaxation of knoted molecules in the flow can give an other picture than
the relaxation of unknoted molecules)?}
\item{The analysis of random walk of molecules from peak to peak should be
done, and results of this analysis should be compared with observed large fluctuations.}
\end{itemize}

The systematic discussion of a difference between the Gaussian elipsoid (and its generalizations) and
the Gaussian multipeak polyhedron (and its generalizations) seems to be necessary. This polyhedron
appears generically as the effective ansatz for kinetic systems with instabilities.

\section{\textbf{Accuracy estimation and postprocessing
in invariant manifolds construction}}

Assume that for the dynamical system (\ref{sys}) the approximate invariant manifold has been
constructed and the slow motion equations have been derived:

\begin{equation}\label{slag}
\frac{dx_{sl}}{dt} = P_{x_{sl}}(J(x_{sl})), x_{sl}\in M,
\end{equation}

\noindent where $P_{x_{sl}}$ is the corresponding projector onto the tangent space $T_{x_{sl}}$ of
$M$. Suppose that we have solved the system (\ref{slag}) and have obtained $x_{sl}(t)$. Let's
consider the following two questions:

\begin{itemize}
\item{How well this solution approximates the true solution $x(t)$
given the same initial conditions?}
\item{Is it possible to use the solution $x_{sl}(t)$ for it's
refinement?}
\end{itemize}

These two questions are interconnected. The first question states the problem of the {\it accuracy
estimation}. The second one states the problem of {\it postprocessing}.

The simplest (``naive") estimation is given by the ``invariance defect":

\begin{equation}\label{defag}
\Delta_{x_{sl}} = (1-P_{x_{sl}})J(x_{sl}),
\end{equation}

\noindent which can be compared with $J(x_{sl})$. For example, this estimation is given by $\epsilon
= \|\Delta_{x_{sl}}\|/\|J(x_{sl})\|$ using some appropriate norm.

Probably, the most comprehensive answer to this question can be given by solving the following
equation:

\begin{equation}\label{varia}
\frac{d(\delta x)}{dt}=\Delta_{x_{sl}(t)}+D_xJ(x)|_{x_{sl}(t)}\delta x.
\end{equation}

This linear equation describes the dynamics of the deviation $\delta x(t) = x(t) - x_{sl}(t)$ using
the linear approximation. The solution with zero initial condition $\delta x(0) = 0$ allows to
estimate the robustness of $x_{sl}$,  as well as the error value. Substituting $x_{sl}(t)$ for
$x_{sl}(t)+\delta x(t)$ gives the required solution refinement. This {\it dynamical postprocessing}
\cite{PostPro} allows to refine the solution substantially and to estimate its accuracy and
robustness. However, the price for this is solving the equation (\ref{varia}) with variable
coefficients. Thus, this dynamical postprocessing can be addressed by a whole hierarchy of
simplifications, both dynamical and static. Let us mention some of them, starting from the dynamical
ones.

1) {\bf Freezing the coefficients}. In the equation (\ref{varia}) the linear operator
$D_xJ(x)|_{x_{sl}(t)}$ is replaced by it's value in some distinguished point $x^*$ (for example, in
the equilibrium) or it is frozen somehow else. As a result, one gets the equation with constant
coefficients and the explicit integration formula:

\begin{equation}\label{duam}
\delta x(t) = \int_0^t{exp(D^*(t-\tau))\Delta_{x_{sl}(\tau)}d\tau},
\end{equation}

\noindent where $D^*$ is the ``frozen" operator and $\delta x(0)=0$.

Another important way of freezing is substituting (\ref{varia}) for some {\it model equation}, i.e.
substituting $D_xJ(x)$ for $-\frac{1}{\tau^*}$, where $\tau^*$ is the relaxation time. In this case
the formula for $\delta x(t)$ has very simple form:

\begin{equation}\label{duam1}
\delta x(t) = \int_0^t{e^{\frac{\tau-t}{\tau^*}}\Delta_{x_{sl}(\tau)}d\tau}.
\end{equation}

2) {\bf One-dimensional Galerkin-type approximation.} Another ``scalar" approximation is given by
projecting (\ref{varia}) on $\Delta(t)= \Delta_{x_{sl}(t)}$. Using the ansatz
\begin{equation}\label{1var1}
\delta x(t) = \delta(t)\cdot \Delta(t),
\end{equation}
substituting it into eq. (\ref{varia}) after orthogonal projection on $\Delta(t)$ we obtain
\begin{equation}\label{1var2}
\frac{d\delta(t)}{dt} = 1+\delta\frac{\langle\Delta|D\Delta\rangle -
\langle\Delta|\dot{\Delta}\rangle}{\langle\Delta|\Delta\rangle},
\end{equation}
\noindent where $\langle\hspace{1pt}|\rangle$ is an appropriate scalar product, which can depend on
the point $x_{sl}$ (for example, the entropic scalar product), $D=D_xJ(x)|_{x_{sl}(t)}$ or the
self-adjoint linearizarion of this operator, or some approximation of it,
$\dot{\Delta}=d\Delta(t)/dt$.

A ``hybrid" between equations (\ref{1var2}) and (\ref{varia}) has the simplest form (but it is more
difficult for computation than eq. (\ref{1var2})):

\begin{equation}\label{hybrid}
\frac{d(\delta
x)}{dt}=\Delta(t)+\frac{\langle\Delta|D\Delta\rangle}{\langle\Delta|\Delta\rangle}\delta x.
\end{equation}

\noindent Here one uses the normalized matrix element
$\frac{\langle\Delta|D\Delta\rangle}{\langle\Delta|\Delta\rangle}$ instead of the linear operator
$D=D_xJ(x)|_{x_{sl}(t)}$.

Both equations (\ref{1var2}) and (\ref{hybrid}) can be solved explicitly:

\begin{eqnarray}
\delta(t)&=&\int_0^t d \tau \exp\left(\int_{\tau}^t k(\theta)d\theta \right), \\ \delta x(t)&=&
\int_0^t \Delta(\tau)d \tau \exp\left(\int_{\tau}^t k_1(\theta)d\theta \right),
\end{eqnarray}
\noindent where $k(t)=\frac{\langle\Delta|D\Delta\rangle -
\langle\Delta|\dot{\Delta}\rangle}{\langle\Delta|\Delta\rangle},$
$k_1(t)=\frac{\langle\Delta|D\Delta\rangle}{\langle\Delta|\Delta\rangle}.$

The projection of $\Delta_{x_{sl}}(t)$ on the slow motion is equal to zero, hence, for the
post-processing analysis of the slow motion, the one-dimensional model (\ref{1var2}) should be
supplemented by one more iteration in order to find the first non-vanishing term in $\delta
x_{sl}(t)$:
\begin{eqnarray}
{d(\delta x_{sl}(t)) \over dt} = \delta(t) P_{x_{sl}(t)}(D_xJ(x)|_{x_{sl}(\tau)})(\Delta(t));
\nonumber \\ \delta x_{sl}(t)= \int_0^t \delta(\tau)
P_{x_{sl}(\tau)}(D_xJ(x)|_{x_{sl}(\tau)})(\Delta(\tau))d\tau.
\end{eqnarray}
where $\delta(t)$ is the solution of (\ref{1var2}).

 3) For a {\bf static post-processing} one uses stationary points of dynamical
equations (\ref{varia}) or their simplified versions (\ref{duam}),(\ref{1var2}). Instead of
(\ref{varia}) one gets:

\begin{equation}\label{stvar}
D_xJ(x)|_{x_{sl}(t)}\delta x = -\Delta_{x_{sl}(t)}
\end{equation}

\noindent with one additional condition $P_{x_{sl}}\delta x=0$. This is exactly the iteration
equation of the Newton's method in solving the invariance equation. A clarification is in order here.
Static post-processing (\ref{stvar}) as well as other post-processing formulas should not be confused
with the Newton method and other for correcting the approximately invariant manifold. Here, only a
single trajectory $x_{sl}(t)$ on the manifold is corrected, not the whole manifold.

The corresponding stationary problems for the model equations and for the projections of
(\ref{varia}) on $\Delta$ are evident. We only mention that in the projection on $\Delta$ one gets a
step of the relaxation method for the invariant manifold construction.

In the following Example 14 it will be demonstrated how one can use function $\Delta(x_{sl}(t))$ in
the accuracy estimation of macroscopic equations  on example of polymer solution dynamics.

\clearpage

\addcontentsline{toc}{subsection}{\textbf{Example 14: Defect of invariance estimation and switching
from the microscopic simulations to macroscopic equations}}

\subsection*{\textbf{Example 14: Defect of invariance estimation and switching from
the microscopic simulations to macroscopic equations}}

A method which recognizes the onset and breakdown of the macroscopic description in microscopic
simulations was developed in \cite{GKIOeNONNEWT2001,KIOeDev,IKar2}. The method is based on the
invariance of the macroscopic dynamics relative to the microscopic dynamics, and it is demonstrated
for a model of dilute polymeric solutions where it decides switching between Direct Brownian Dynamics
simulations and integration of constitutive equations.

\addcontentsline{toc}{subsubsection}{Invariance principle and micro-macro computations}

\subsubsection*{\textbf{Invariance principle and micro-macro computations}}

Derivation of reduced (macroscopic) dynamics from the microscopic dynamics is the dominant theme of
non--equilibrium statistical mechanics. At the present time, this very old theme demonstrates new
facets in view of a massive use of simulation techniques on various levels of description. A
two--side benefit of this use is expected: On the one hand, simulations provide data on molecular
systems which can be used to test various theoretical constructions about the transition from micro
to macro description. On the other hand, while the microscopic simulations in many cases are based on
limit theorems [such as, for example, the central limit theorem underlying the Direct Brownian
Dynamics simulations (BD)] they are extremely time--consuming in any real situation, and a timely
recognition of the onset of a macroscopic description may considerably reduce computational efforts.

In this section, we aim at developing a `device' which is able to recognize the onset and the
breakdown of a macroscopic description in the course of microscopic computations.

Let us first present the main ideas of the construction in an abstract setting. We assume that the
microscopic description is set up in terms of microscopic variables $\xi$. In the examples considered
below, microscopic variables are distribution functions over the configuration space of polymers. The
microscopic dynamics of variables $\xi$ is given by the microscopic time derivative
$d\xi/dt=\dot{\xi}(\xi)$. We also assume that the set of macroscopic variables $\setM$ is chosen.
Typically, the macroscopic variables are some lower-order moments if the microscopic variables are
distribution functions. The reduced (macroscopic) description assumes (a) The dependence
$\xi(\setM)$, and (b) The macroscopic dynamics $d\setM/dt=\dot{\setM}(\setM)$. We do not discuss here
in any detail the way one gets the dependence $\xi(\setM)$, however, we should remark that,
typically, it is based on some (explicit or implicit) idea about decomposition of motions into slow
and fast, with $M$ as slow variables. With this, such tools as maximum entropy principle,
quasi-stationarity, cumulant expansion etc. become available for constructing the dependence
$\xi(\setM)$.

Let us compare the microscopic time derivative of the function $\xi(\setM)$ with its macroscopic time
derivative due to the macroscopic dynamics:
\begin{equation}
\label{VAR} \Delta(\setM)=\frac{\partial\xi(\setM)}{\partial \setM}\cdot \dot{\setM}(\setM)-
\dot{\xi}(\xi(\setM)).
\end{equation}
If the {\it defect of invariance} $\Delta(\setM)$ (\ref{VAR}) is equal to zero on the set of
admissible values of the macroscopic variables $M$, it is said that the reduced description
$\xi(\setM)$ is invariant. Then the function $\xi(\setM)$ represents the invariant manifold in the
space of microscopic variables. The invariant manifold is relevant if it is stable. Exact invariant
manifolds are known in a very few cases (for example, the exact hydrodynamic description in the
kinetic Lorentz gas model \cite{Hauge}, in Grad's systems \cite{GKPRL96,KGAnPh2002}, and one more
example will be mentioned below). Corrections to the approximate reduced description through
minimization of the defect of invariance is a part of the so--called method of invariant manifolds
\cite{GKTTSP94}. We here consider a different application of the invariance principle for the purpose
mentioned above.

The time dependence of the macroscopic variables can be obtained in two different ways: First, if the
solution of the microscopic dynamics at time $t$ with initial data at $t_0$ is $\xi_{t,t_0}$, then
evaluation of the macroscopic variables on this solution gives $\setM_{t,t_0}^{\rm micro}$. On the
other hand, solving dynamic equations of the reduced description with initial data at $t_0$ gives
$\setM_{t,t_0}^{\rm macro}$. Let $\|\Delta\|$ be a value of defect of invariance with respect to some
norm, and $\epsilon>0$ is a fixed tolerance level. Then, if at the time $t$ the following inequality
is valid,
\label{INDICATOR}
\begin{equation}
\label{out-of-BD}
 \|\Delta(\setM_{t,t_0}^{\rm micro})\|<\epsilon,
\end{equation}
this indicates that  the accuracy provided by the reduced description is not worse than the true
microscopic dynamics (the macroscopic description {\it sets on}).
 On the other hand, if
\begin{equation}
\label{out-of-macro}
 \|\Delta(\setM_{t,t_0}^{\rm macro})\|>\epsilon,
\end{equation}
then the accuracy of the reduced description is insufficient (the reduced description {\it breaks
down}), and we must use the microscopic dynamics.

Thus, evaluating the defect of invariance  (\ref{VAR}) on the current solution to macroscopic
equations, and checking the inequality (\ref{out-of-macro}), we are able to answer the question
whether we can trust the solution without looking at the microscopic solution. If the tolerance level
is not exceeded then we can safely integrate the macroscopic equation. We now proceed to a specific
example of this approach.
We consider a well--known class of microscopic models of dilute polymeric solutions

\addcontentsline{toc}{subsubsection}{Application to dynamics of dilute polymer solution}

\subsubsection*{Application to dynamics of dilute polymer solution}

A well--known problem of the non--Newtonian fluids is the problem of establishing constitutive
equations on the basis of microscopic kinetic equations. We here consider a model introduced by
Lielens {\it et al.} \cite{Keu98}:
\begin{equation}
\label{FENE1} \dot{f}(q,t)=-\partial_q \left\{ \kappa(t)qf - \frac{1}{2}f\partial_qU(q^2) \right\} +
\frac{1}{2}\partial_q^2f .
\end{equation}
With the potential $U(x)=-(b/2)\ln (1-x/b)$ Eq.~(\ref{FENE1}) becomes the one--dimensional version of
the FENE dumbbell model which is used to describe the elongational behavior of dilute polymer
solutions.

The reduced description seeks a closed time evolution equation for the stress $\tau=\langle
q\partial_qU(q^2)\rangle-1$. Due to its non--polynomial character, the stress $\tau$ for the FENE
potential depends on all moments of $f$. We have shown in \cite{IKOe99} how such potentials can be
approximated systematically by a set of polynomial potentials $U_n(x)=\sum_{j=1}^n\frac{1}{2j}c_jx^j$
of degree $n$ with coefficients $c_j$ depending on the even moments $M_j=\langle q^{2j}\rangle$ of
$f$ up to order $n$, with $n=1,2,\dots$, formally converging to the original potential as $n$ tends
to infinity. In this approximation, the stress $\tau$ becomes a function of the first $n$ even
moments of $f$, $\tau(\setM)=\sum_{j=1}^nc_jM_j-1$, where the set of macroscopic variables is denoted
by $\setM=\{M_1,\ldots,M_n\}$.

The first two potentials approximating the FENE potential are:
\begin{eqnarray}
U_1(q^2)&=&U'(M_1)q^2\label{FENEP}\\ U_2(q^2)&=&\frac{1}{2}(q^4-2M_1q^2)U''(M_1)
+\frac{1}{2}(M_2-M_1^2)q^2U'''(M_1)\label{FENEP+1},
\end{eqnarray}
where $U'$, $U''$ and $U'''$ denote the first, second and third derivative of the potential U,
respectively. The potential $U_1$ corresponds to the well--known FENE--P model. The kinetic equation
(\ref{FENE1}) with the potential $U_2$ (\ref{FENEP+1}) will be termed the FENE--P+1 model below.
Direct Brownian Dynamics simulation (BD) of the kinetic equation (\ref{FENE1}) with the potential
$U_2$ for the flow situations studied in \cite{Keu98} demonstrates that it is a reasonable
approximation to the true FENE dynamics whereas the corresponding moment chain is of a simpler
structure. In \cite{GKIOeNONNEWT2001} this was shown for a periodic flow, while Fig.~\ref{FIG1} shows
results for the flow
\begin{equation} \label{kappa_t}
\kappa(t)=\left\{ \begin{array}{cc}
    100t(1-t)e^{-4t} & 0\leq t\leq 1  \\ 0 & {\rm else} \end{array} \right.
\end{equation}
The quality of the approximation indeed increases with the order of the polynomial.

For any potential $U_n$, the invariance equation can be studied directly in terms of the full set of
the moments, which is equivalent to studying the distribution functions. The kinetic equation
(\ref{FENE1}) can be rewritten equivalently in terms of moment equations,
\begin{eqnarray}
\label{CHAIN_FENEPn} \dot{M}_k & = & F_k(M_1,\ldots,M_{k+n-1})\\\nonumber F_k & = & 2k\kappa(t)M_k +
k(2k-1)M_{k-1} -k\sum_{j=1}^{n}c_jM_{k+j-1}.
\end{eqnarray}
We seek functions $M_k^{\rm macro}(\setM)$, $k=n+1,\dots$ which are form--invariant under the
dynamics:
\begin{equation}
\label{invFENEP} \sum_{j=1}^n\frac{\partial M_k^{\rm macro}(\setM)}{\partial M_j} F_j(\setM) =
F_k(M_1,\ldots,M_n,M_{n+1}(\setM),\ldots,M_{n+k}(\setM)).
\end{equation}
This set of invariance equations states the following: The time derivative of the form $M_k^{\rm
macro}(\setM)$ when computed due to the closed equation for $\setM$ (the first contribution on the
left hand side of Eq.~(\ref{invFENEP}), or the `macroscopic' time derivative) equals the time
derivative of $M_k$ as computed by true moment equation with the same form $M_k(\setM)$ (the second
contribution, or the `microscopic' time derivative), and this equality should hold whatsoever values
of the moments $\setM$ are.

Equations (\ref{invFENEP}) in case $n=1$ (FENE--P) are solvable exactly with the result
\[ M_k^{\rm macro} = a_k M_1^k,
\qquad \mbox{with}\quad a_k = (2k-1)a_{k-1},\ a_0=1.\] This dependence corresponds to the Gaussian
solution in terms of the distribution functions. As expected, the invariance principle give just the
same result as the usual method of solving the FENE--P model.

Let us briefly discuss  the potential $U_2$, considering a simple closure approximation
\begin{equation}
\label{ANSATZ0} M_k^{\rm macro}(M_1,M_2)=a_kM_1^k+b_kM_2M_1^{k-2},
\end{equation}
where $a_k=1-k(k-1)/2$ and $b_k=k(k-1)/2$. The function $M_3^{\rm macro}$ closes the moment equations
for the two independent moments $M_1$ and $M_2$. Note, that $M_3^{\rm macro}$ differs from the
corresponding moment $M_3$ of the actual distribution function by the neglect of the 6--th cumulant.
The defect of invariance of this approximation is a set of functions $\Delta_k$ where
\begin{equation}
\label{Delta3} \Delta_3(M_1,M_2)= \frac{\partial M_3^{\rm macro}}{\partial M_1}F_1+ \frac{\partial
M_3^{\rm macro}}{\partial M_2}F_2- F_3,
\end{equation}
and analogously for $k\ge 3$. In the sequel, we make all conclusions based on the defect of
invariance $\Delta_3$ (\ref{Delta3}).

It is instructive to plot the defect of invariance $\Delta_3$ versus time, assuming the functions
$M_1$ and $M_2$ are extracted from the BD simulation (see Fig.~\ref{FIG2}). We observe that the
defect of invariance is a nonmonotonic function of the time, and that there are three pronounced
domains: From $t_0=0$ to $t_1$ the defect of invariance is almost zero which means that the ansatz is
reasonable. In the intermediate domain, the defect of invariance jumps to high values (so the quality
of approximation is poor). However, after some time $t=t^\ast$, the defect of invariance again
becomes negligible, and remains so for later times. Such behavior is typical of so--called ``kinetic
layer".

Instead of attempting to improve the closure, the invariance principle can be used directly to switch
from the BD simulation to the solution of the macroscopic equation without loosing the accuracy to a
given tolerance. Indeed,  the defect of invariance is a function of $M_1$ and $M_2$, and it can be
easily evaluated both on the data from the solution to the macroscopic equation, and the BD data. If
the defect of invariance {\it exceeds} some given tolerance on the macroscopic solution this signals
to switch to the BD integration. On the other hand, if the defect of invariance becomes {\it less}
than the tolerance level on the BD data signals that the BD simulation is not necessary anymore, and
one can continue with the integration of the macroscopic equations. This reduces the necessity of
using BD simulations only to get through the kinetic layers. A realization of this hybrid approach is
demonstrated in Fig.~\ref{FIG3}: For the same flow we have used the BD dynamics only for the first
period of the flow while integrated the macroscopic equations in all the later times. The quality of
the result is comparable to the BD simulation whereas the total integration time is much shorter. The
{\it transient dynamics} at the point of switching from the BD scheme to the integration of the
macroscopic equations (shown in the inset in Fig.~\ref{FIG3}) deserves a special comment: The initial
conditions at $t^*$ are taken from the BD data. Therefore, we cannot expect that at the time $t^*$
the solution is already on the invariant manifold, rather, at best, close to it. Transient dynamics
therefore signals the {\it stability} of the invariant manifold we expect: Even though the
macroscopic solution starts not on this manifold, it nevertheless attracts to it. The transient
dynamics becomes progressively less pronounced if the switching is done at later times. The stability
of the invariant manifold in case of the FENE--P model is studied in detail in \cite{IK00}.

The present approach of combined microscopic and macroscopic simulations can be realized on the level
of moment closures (which then needs reconstruction of the distribution function from the moments at
the switching from macroscopic integration to BD procedures), or for parametric sets of distribution
functions if they are available \cite{Keu98}. It can be used for a rigorous construction of domain
decomposition methods in various kinetic problems.

\clearpage

\section{\textbf{Conclusion}}

{\bf To construct slow invariant manifolds is useful.} Effective model reduction becomes impossible
without them for complex kinetic systems.

Why to reduce description in the times of supercomputers?

First, in order to gain understanding. In the process of reducing the description one is often able
to extract the essential, and the mechanisms of the processes under study become more transparent.

Second, if one is given the detailed description of the system, then one should be able also to solve
the initial--value problem for this system. But what should one do in the case where the system is
representing just a small part of the huge number of interacting systems? For example, a complex
chemical reaction system may represent only a point in a three--dimensional flow.

Third, without reducing the kinetic model, it is impossible to construct this model. This statement
seems paradoxal only at the first glance: How come, the model is first simplified, and is constructed
only after the simplification is done? However, in practice, the typical for a mathematician
statement of the problem, (Let the system of differential equations be {\it given}, then ...) is
rather rarely applicable for detailed kinetics. Quite on the contrary, the thermodynamic data
(energies, enthalpies, entropies, chemical potentials etc) for sufficiently rarefied systems are
quite reliable. Final identification of the model is always done on the basis of comparison with the
experiment and with a help of fitting. For this purpose, it is extremely important to reduce the
dimension of the system, and to reduce the number of tunable parameters.

And, finally, for every supercomputer there exist too complicated problems. Model reduction makes
these problems less complicated and sometimes gives us the possibility to solve them.

{\bf It is useful to apply thermodynamics and the quasiequilibrium concept  while seeking slow
invariant manifolds.} Though the open systems are important for many applications, however, it is
useful to begin their study and model reduction with the analysis of closed (sub)sustems. Then the
thermodynamics equips these systems with Lyapunov functions (entropy, free energy, free enthalpy,
depending on the context). These Lyapunov functions are usually known much better than the right hand
sides of kinetic equations (in particular, this is the case in reaction kinetics). Using this
Lyapunov function, one constructs the initial approximation to the slow manifold, that is, the
quasiequilibrium manifold, and also one constructs the thermodynamic projector.

{\bf The thermodynamic projector} is the unique operator which transforms the arbitrary vector field
equipped with the given Lyapunov function into a vector field with the same Lyapunov function (and
also this happens on any manifold which is not tangent to the level of the Lyapunov function).

{\bf The quasi--chemical approximation} is an extremely rich toolbox for assembling equations. It
enables to construct and study wide classes of evolution equations equipped with prescribed Lyapunov
functions, with Onsager reciprocity relations and like.

Slow invariant manifolds of thermodynamically closed systems are useful for constructing slow
invariant manifolds of corresponding open systems. The necessary technic is developed.

{\bf Postprocessing} of the invariant manifold construction is important both for estimation of the
accuracy and for the accuracy improvement.

The main result of this work can be formulated as follows: {\bf It is possible indeed to construct
invariant manifolds.} The problem of constructing invariant manifolds can be formulated as the
invariance equation, subject to additional conditions of slowness (stability). The Newton method with
incomplete linearization, relaxation methods, the method of natural projector, and the method of
invariant grids enables educated approximations to the slow invariant manifolds.

Studies on invariant manifolds were initiated by A.\ Lyapunov \cite{Lya} and H.\ Poincare \cite{Lya2} (see \cite{Lya1}). Essential
stages of the development of these ideas in the XX century are reflected in the books
\cite{Lya1,HiPuSh,IneManCFTe88,IneManTe88}. It becomes more and more evident at the present time that the
constructive methods of invariant manifold are useful on a host of subjects, from applied
hydrodynamics \cite{ComIMJon} to physical and chemical kinetics.

\clearpage

\addcontentsline{toc}{section}{

\section*{Figures}

\begin{figure}
\centering{
\caption{Logical connections between sections. All the sections depend on Section 3. There are many
possible routes for reading, for example the following route gives the invariance equation and one of
the main methods for its solution with applications to the Boltzmann equation: Section 3 -- Section 5
(with Example 1) -- Section 6 (with Example 3). Another possibility gives the shortest way to
rheology applications: Section 3 -- Section 5 -- Section 6  -- Section 11 (with Example 12). The
formalization of the classical Ehrenfests idea of coarse--graining and its application for derivation
of the correct high-order hydrodynamic equations can be reached on such a way: Section 3 -- Section 5
-- Section 10 (with Examples 10, 11). The shortest road to numerical representation of invariant
manifolds and to the method of invariant grids is as follows: Section 3 -- (Section 5 -- Section 6)
--  (Section 4 -- Section 8) -- Section 9 (with Examples 8, 9) (For Introduction).
\label{flowchart}}}
\end{figure}

\begin{figure}
\centering{
\caption{Logical connections between sections and examples. Only one connection between examples is
significant: Example 3 depends on Example 1. All the examples depend on corresponding subsections of
Section 2 (For Introduction). \label{flowchart2}}}
\end{figure}

\begin{figure}
\begin{centering}
\caption {Approximations for hard spheres:
    bold line - function $S^{HS}$,
    solid line - approximation $S_a^{HS}$,
    dotted line - Grad moment approximation. (For Section5)}
    \label{Hard}
\end{centering}
\end{figure}

\begin{figure}
\centering{
\caption {Acoustic dispersion curves for approximation \ref{4.70} (solid line), for second (the
Burnett) approximation of the  Chapman-Enskog expansion  \cite{Bob} (dashed  line) and for  the
regularization  of   the   Burnett approximation via partial summing of the Chapman-Enskog  expansion
\cite{GKJETP91,GKTTSP92} (punctuated dashed line). Arrows indicate an increase of $k^{2}$. (For
Section6)}
    \label{Disp}}
\end{figure}

\begin{figure}
\centering{
\caption {Dependency of viscosity on compression for approximation (\ref{4.73a}) (solid line),  for
partial  summing  (\ref{4.76}) (punctuated dashed line), and for the  Burnett  approximation
\cite{KTTSP92,MBCh} (dashed line). The latter changes the sign at a regular point and, hence, nothing
prevents the flow to transfer into the  nonphysical region. (For Section6)}
    \label{Visc.eps}}
\end{figure}

\begin{figure}[t]
\centering{
\caption{Attenuation rate of sound waves. Dotts: Burnett approximation. Bobylev's instability occurs
when the curve intersects the horizontal axis. Solid: First iteration of the Newton method on the
invariance equation. (For Section6)} \label{lindisp}}
\end{figure}

\begin{figure}[t]
\centering{
\caption{ Attenuation ${\rm Re}\omega_{1,2}(k)$ (lower pair of curves), frequency ${\rm
Im}\omega_{1,2}(k)$ (upper pair of curves). Dashed lines - Grad case ($a=0$), drawn lines - dynamic
correction ($a=0.5$). (For Section6)} \label{Figmom}}
\end{figure}

\begin{figure}
\centering{
\caption{Grid instability. For small grid steps approximations in the calculation of grid derivatives
lead to the grid instability effect. On the figure several successive iterations of the algorithm
without adaptation of the time step are shown that lead to undesirable ``oscillations", which
eventually destruct the grid starting from one of it's ends. (For Section9)} \label{diver}}
\end{figure}

\begin{figure}
\centering{
\caption{One-dimensional invariant grid (circles) for two-dimensional chemical system. Projection
into the 3d-space of $c_1$, $c_4$, $c_3$ concentrations. The trajectories of the system in the phase
space are shown by lines. The equilibrium point is marked by square. The system quickly reaches the
grid and further moves along it. (For Section9)} \label{4d1dgrid}}
\end{figure}

\begin{figure}
\centering{



\caption{One-dimensional invariant grid for two-dimensional chemical system. a) Values of the
concentrations along the grid. b) Values of the entropy and the entropy production (-$dG/dt$) along
the grid. c) Relation of the relaxation times ``toward" and ``along" the manifold. The nodes
positions are parametrized with entropic distance measured in the quadratic metrics given by ${\bf
H_c} = -||\partial^2S({\bf c})/\partial c_i\partial c_j||$ in the equilibrium $c^{eq}$. Zero
corresponds to the equilibrium. (For Section9)}\label{4d1dgraphs}}
\end{figure}

\begin{figure}
\centering{




\caption{One-dimensional invariant grid for model hydrogen burning system. a) Projection into the
3d-space of $c_H$, $c_O$, $c_{OH}$ concentrations. b) Concentration values along the grid. c) three
smallest by absolute value non-zero eigen values of the symmetrically linearized system. (For
Section9) }\label{6d1dgrid}}
\end{figure}

\begin{figure}
\centering{



\caption{Two-dimensional invariant grid for the model hydrogen burning system. a) Projection into the
3d-space of $c_H$, $c_O$, $c_{OH}$ concentrations. b) Projection into the principal 3D-subspace.
Trajectories of the system are shown coming out from the every grid node. Bold line denotes the
one-dimensional invariant grid, starting from which the 2D-grid was constructed. (For Section9)}
\label{6d2dgrid}}
\end{figure}

\begin{figure}
\centering{

a) Concentration $H_2$ \hspace{2.5cm} b) Concentration $O$ \\
c) Concentration $OH$ \hspace{2.5cm} d) Concentration $H$

\caption{Two-dimensional invariant grid as a screen for visualizing different functions defined in
the concentrations space. The coordinate axes are entropic distances (see the text for the
explanations) along the first and the second slowest directions on the grid. The corresponding 1D
invariant grid is denoted by bold line, the equilibrium is denoted by square. (For
Section9)}\label{6d2dgridcolor}}
\end{figure}

\begin{figure}
\centering{

a) Entropy \hspace{3.5cm} b) Entropy Production \\
c) $\lambda_3/\lambda_2$ relation \hspace{3.5cm} d) $\lambda_2/\lambda_1$ relation

\caption{Two-dimensional invariant grid as a screen for visualizing different functions defined in
the concentrations space. The coordinate axes are entropic distances (see the text for the
explanations) along the first and the second slowest directions on the grid. The corresponding 1D
invariant grid is denoted by bold line, the equilibrium is denoted by square. (For Section9)
}\label{6d2dgridcolor1}}
\end{figure}

\begin{figure}
\begin{centering}

\caption{Coarse-graining scheme. $f$ is the space of microscopic variables, $M$ is the space of the
macroscopic variables, $f^*$ is the quasiequilibrium manifold, $\mu$ is the mapping from the
microscopic to the macroscopic space. (For Section10)} \label{Fig1LAR}
\end{centering}
\end{figure}

\begin{figure}
\begin{centering}
\caption{Attenuation rates  of various modes of the post-Navier-Stokes equations as functions of the
wave vector. Attenuation rate of the twice degenerated shear mode is curve 1. Attenuation rate of the
two sound modes is curve 2. Attenuation rate of the diffusion mode is curve 3. (For Section10)}
\label{Fig2LAR}
\end{centering}
\end{figure}

\begin{figure}
\centering{
\caption{Dimensionless shear viscosity $\protect\nuhat$  and first normal stress coefficient
$\protect\widehat{\psi}_1$   vs. shear rate: (\protect\rule[1mm]{1cm}{0.1mm})
 revised Oldroyd 8 constants model;
($\cdots\cdots$) FENE-P model; ($\circ\circ\circ$) BD  simulations on the FENE model;
($-\cdot-\cdot-$) Hookean dumbbell model. (For Section11)}\label{fig1Deb}}
\end{figure}

\begin{figure}
\centering{
\caption{ Dimensionless elongation viscosity vs. elongation rate: (\protect\rule[1mm]{1cm}{0.1mm})
revised Oldroyd 8 constants model, ($\cdots\cdots$) FENE-P model, ($\circ\circ\circ$) BD simulations
on the FENE model; ($-\cdot-\cdot-$) Hookean dumbbell model. (For Section11)}\label{fig2Deb}}
\end{figure}

\clearpage

\begin{figure}
\centering{
\caption{ Time evolution of  elongation viscosity  after inception of the elongation flow with
elongation rate $\dotepshat=0.3$: (\protect\rule[1mm]{1cm}{0.1mm}) revised Oldroyd 8 constants model,
($\cdots\cdots$) FENE-P model, ($-\ -\ -$) BD simulations on  FENE model; ($-\cdot-\cdot-$) Hookean
dumbbell model. (For Section11)} \label{fig3Deb}}
\end{figure}

\begin{figure}
\centering{
\caption{Deviations of reduced moments from the Gaussian solution as a function of reduced time $t$
in pseudo-elongation flow for the FENE-P model. Upper part: Reduced second moment $X=M_2/b$. Lower
part: Reduced deviation of fourth moment from Gaussian solution $Y=-\mu_4^{1/2}/b$. Solid:
$\kappa=2$, dash-dot: $\kappa=1$, dash: $\kappa=0.75$, long dash: $\kappa=0.5$. (The figure from the
paper \cite{IK00}, computed by P. Ilg.) (For Section11)} \label{EPJ713_fig}}
\end{figure}

\begin{figure}
\begin{centering}
\caption {Phase trajectories for the two-peak approximation, FENE-P model. The vertical axis
($\varsigma=0$) corresponds to the Gaussian manifold. The triangle with $\alpha(M_2)>0$ is the domain
of exponential instability. (For Section11)}
    \label{figFENEP}
\end{centering}
\end{figure}

\begin{figure}
\begin{centering}
\caption {Phase trajectories for the two-peak approximation, FENE model: {\bf a)} A stable
equilibrium on the vertical axis, one stable peak; {\bf b)} A stable equilibrium with $\varsigma>0$,
stable two-peak configuration. (For Section11)}
    \label{figFENE}
\end{centering}
\end{figure}

\begin{figure}
\begin{centering}
\caption {Cartoon representing the steps of molecular individualism. Black dots are vertices of
Gaussian parallelepiped. Zero, one, and four-dimensional polyhedrons are drawn. Presented is also the
three-dimensional polyhedron used to draw the four-dimensional object. Each new dimension of the
polyhedron adds as soon as the corresponding bifurcation occurs. Quasi-stable polymeric conformations
are associated with each vertex. First bifurcation pertinent to the instability of a dumbbell model
in elongational flow is described in the text. (For Section11)}
    \label{Cartoon}
\end{centering}
\end{figure}

\begin{figure}
\centering{
\caption{Stress $\tau$ versus time from direct Brownian dynamics simulation: symbols -- FENE, dashed
line -- FENE--P, solid line -- FENE-P+1. (For Section12)} \label{FIG1}}
\end{figure}

\begin{figure}
\centering{
\caption{Defect of invariance $\Delta_3/b^3$, Eq.~(\ref{Delta3}), versus time extracted from BD
simulation (the FENE--P+1 model) for the flow situation of Eq.~(\ref{kappa_t}). (For Section12)}
\label{FIG2}}
\end{figure}

\begin{figure}
\centering{
\caption{Switching from the BD simulations to macroscopic equations after the defect of invariance
has reached the given tolerance level (the FENE--P+1 model): symbols -- the BD simulation, solid line
-- the BD simulation from time $t=0$ up to time $t=t^\ast$, dashed line -- integration of the
macroscopic dynamics with initial data from BD simulation at time $t=t^\ast$. For comparison, the
dot--dashed line gives the result for the integration of the macroscopic dynamics with e quilibrium
conditions from $t=0$. Inset: Transient dynamics at the switching from BD to macroscopic dynamics on
a finer time scale. (For Section12)} \label{FIG3}}
\end{figure}

\end{document}